# THEORY OF ELECTRON COOLING

Yaroslav Derbenev*

*Thomas Jefferson National Accelerator Facility, Newport News, VA 23606, USA*

*Translated from Russian by V.S. Morozov, Jefferson Lab, VA 23606, USA*
*Translation supported by the U.S. Department of Energy, Office of Science, Office of Nuclear Physics under contract DE-AC05-06OR23177.*

# TABLE OF CONTENTS

# INTRODUCTION

Efficiency of charged particle storage rings, as physics research tools, to a large, if not decisive, degree is determined by how intense the accumulated beams can be, how much their density and monochromaticity can be enhanced, and, finally, what luminosity of the apparatus can be achieved. To solve these problems, it is highly desirable to possess some mechanism of fast damping of the phase space volume of a circulating beam, which would reduce the particle angular and energy spreads and, most importantly, would allow for multiple injections of new particle bunches from the source into the freed up regions of the storage ring's phase space.

We know that colliding $e^-e^-$ and $e^-e^+$ beams currently providing the most part of the fundamental information in the elementary particle physics became possible due to these particles having radiation friction in the ultra-relativistic energy range, which, when the average energy loss is compensated by an RF system, shrinks (cools) the beams to quite small sizes in a fraction of a second. In case of heavy particles, such a natural mechanism is not present, and this, for a long time, was limiting the effectiveness of colliding pp-beams and made it impractical to develop projects in the most promising combination of protons and antiprotons.

A hope to overcome this difficulty appeared in 1966 when G.I. Budker proposed an electron cooling technique [1]. The idea behind this method is very simple. An accompanying electron beam of a lower temperature is introduced into one of the straight sections of a heavy particle (proton, antiproton, ion) orbit. Due to Coulomb scattering of the particles, the ion gas is cooled in the electron one; after multiple passes of the interaction region, the size and energy spread of the ion beam decrease to some equilibrium values. In general terms, this process can be considered as relaxation of a two-component plasma with the difference that the electron component is a continuously replenished (or replaced, as the electrons get heated) flow playing the role of a thermostat. Estimates showed that this method allows one to obtain quite high luminosities with a simultaneous strong improvement of the beams' monochromaticities.

Also in 1966, a detailed study of electron cooling, its capabilities and practical feasibility was started. The theoretical investigations presented in this dissertation were done in two stages. In the first stage, completed by 1968, prior to the start of experiments, the goal was to find out the features introduced by the cyclic nature of heavy particle motion in a storage ring: focusing, coupling of the transverse and longitudinal degrees of freedom. It was shown that finiteness of proton motion leads to a number of conceptual constraints on the possible deviations of the electron beam state from the thermodynamic equilibrium (in the system moving with the cooled beam). The resulting requirements can be satisfied in practice without significant loss of electron cooling efficiency [2].

In 1974, a new experimental complex consisting of a model storage ring NAP-M and an EPOKHA system generating an electron beam provided first results on proton beam cooling confirming method's high efficiency [3, 4, 5]. This success stimulated further more in-depth theoretical and experimental research.

It was discovered that, under certain conditions, electron cooling possesses unique properties additionally increasing its efficiency. In case of cooling by a continuous electron flux passing the beam interaction region once, the longitudinal electron temperature is small compared to the transverse one (approximately equal to the cathode temperature) due to the electro-static nature of acceleration preserving the energy spread in the laboratory frame [3, 4]. Transverse electron motion is "frozen" by the magnetic field guiding the electron beam. Under these conditions, due to the long-range effect of Coulomb forces, the cooling rate in the ion temperature range $T < (M/m)\, T_e$ is no longer limited by the thermal spread of electron velocities, cooling becomes much faster, and the beam temperature may go down to values many times lower than the cathode temperature [6].

Discovery of the magnetization effect (an "anomalously" fast beam damping [7] was observed almost simultaneously and independently in experiments with an improved system), development of a kinetic theory for these special conditions, analysis of factors impeding or distorting the positive role of magnetization and low longitudinal electron temperature are the main contents of the second stage of work done in parallel with the experimental studies.

Due to the unique conditions that the interacting ion and electron beams may encounter, it was necessary to reformulate the collision integral not using known existing methods. The modified integral, besides for the incoherent electron Larmor precession, directly accounts for a finite time of flight through the region of interaction with the electron flux, non-stationary screening processes due to polarization of electron "plasma", spatial velocity gradients. The main practically important difference of the utilized collision integral from the one known in the plasma theory is that the radius of the non-equilibrium screening of the interaction is, in general case, determined by the relative velocity of ion and electron Larmor circles and can significantly exceed the Debye radius. The latter, under the considered conditions, can reach ultimately small values of the order of the average distance between electrons.

In the final stage of cooling, the effective relative velocities can become so small that applicability of the perturbation theory breaks down. Interaction then reaches saturation, friction decrement and diffusion rate stop growing with a further slow down of ions. Quantitative consideration in this region is difficult but one can obtain sufficiently reliable estimates of maximum decrements and minimum equilibrium temperatures achievable in electron cooling. They are determined by the electron density, their Larmor radii and the charge of the cooled beam particles. These minimum temperatures set the precision level (of the order of $10^{-4} - 10^{-5}$), which it makes sense to pursue when building and tuning a cooling system.

In case of magnetization and small longitudinal temperature of the electron flux, the cooling decrements, while generally increased (in the $T < (M/m)\, T_e$ region), strongly depend on factors causing deviation of the Larmor circle velocities from the ion velocities on equilibrium orbits. They include deviation of magnetic field lines, gradients of electrostatic potential, drift of circles, a transverse gradient of coherent Larmor velocity (leading to a gradient of longitudinal velocity). The dependence on the Larmor velocities themselves is relatively weak. An experimental study [8, 9, 32] of the effect of some of these factors on the cooling process at NAP-M with EPOKHA apparatus qualitatively confirms the main theoretical predictions. From the complete collection of data, one can draw a conclusion that the qualities of the cooling system with the achievable today precision of construction and control of its parameters are close to optimal. I would also like to note a high degree of reproducibility of the cooling effect in each experimental cycle.

Another category of effects related to interaction of heavy particles arises when switching to intense (dense) beams. The role of this interaction increases with reduction of temperature and can become especially significant at the final cooling stage in a magnetized electron flux. Besides space charge and coherent effects, mutual scattering of particles in the beam becomes substantial leading to either fast thermalization or self-heating (increase of equilibrium velocity spread at the expense of negligible overall beam slow down) depending on the particle energy in the storage ring and focusing strength. A study of intra-beam interaction effects is necessary for a self-consistent description of the cooling process of an intense beam and for proper optimization of the heavy particle accumulation regime.

With all the simplicity of the basic idea of electron cooling, its physics is quite rich in various phenomena. This has to do with the fact that electron cooling is a synergy of both the fields of accelerator beam physics and plasma physics. This work does not claim to be a comprehensive investigation of all aspects, which may be related to the cooling process. Our main goal was to understand the dynamics of heavy particle interaction with an electron beam and the cooling kinetics not using existing recipes as much as possible and with sufficiently weak assumptions about the parameters that the relaxation process is sensitive to. Therefore, we hope that our work will be useful for current studies of the method and its development in the near future.

In general, theoretical and experimental studies not only confirmed initial expectations but also discovered new positive aspects of electron cooling substantially enhancing its capabilities.

Since 1968, CERN and other laboratories around the world have developed and aplied another method for damping the incoherent particle motion, the so-called *stochastic cooling* invented by Van der Meer in 1968 based on the use of wideband incoherent feedback loops (see Addition II).

Success in realization of electron and stochastic cooling stimulated born of ideas of the *coherent electron cooling* [10, 11] as an organic alliance of EC and SC principles based on use of an electron beam, and also optical stochastic cooling (OSC) [12, 13, 14] based on use of a optics frequence range feedbacks.

Creation and developments of the beam cooling techniques resulted in promotion of heavy particle storage rings and colliders to a high level of performance in doing new class of critical experiments in medium and high energy physics. More developments and application to come in near future.

# Nomenclature

$e$     electron charge

$m$     electron mass

$\vec{v}_e = (\vec{v}_{e\perp}, v_{e\parallel})$     velocity in the co-moving frame

$w_e = (\gamma - 1)mc^2$     electron energy in the laboratory frame

$n'_e$     electron beam density in the co-moving frame

$n_e$     electron beam density in the laboratory frame

$j_e$     electron current density in the laboratory frame

$H$     accompanying longitudinal field

$\Omega = eH/(mc)$     Larmor frequency

$\omega_e = \sqrt{4\pi n'_e e^2/m}$     Langmuir frequency

$v_{et} = \sqrt{T_{e\perp}/m}$     thermal velocity

$T_k$     cathode temperature

$T_{e\perp}$     transverse electron temperature

$T_{e\parallel}$     longitudinal electron temperature

$\Delta_{e\perp}$     transverse spread of electron velocities

$\Delta_{e\parallel}$     longitudinal spread of electron velocities

$r_D = \Delta_{e\parallel}/\omega_e$     Debye radius of magnetized electrons

$r_{scr} = v/\omega_e$     screening radius

$r_L = v_{e\perp}/\Omega$     electron Larmor radius

$R_L$     radius of coherent Larmor motion

$\theta_e$     angular spread of electron velocities

$f(\vec{v}_e, r)$     electron phase-space distribution

$\varepsilon_{\vec{k}}(\omega)$     electric permittivity of electron beam

$\omega(k)$     plasma frequency accounting for dispersion

$r_e = e^2/mc^2$     classical electron radius

$F$     friction force acting on a heavy particle

$d_{\alpha\beta}$     electron scattering tensor

$\vec{v}$     velocity of a heavy particle

$\beta c$     beam velocity

$c$     speed of light

$\vec{u} = \vec{v} - \vec{v}_e$     relative particle velocity

$\vec{u}_A = \vec{v} - \vec{v}_{e\parallel}$     particle velocity with respect to a Larmor circle

$\theta$     angular deviation of the ion velocity from the closed orbit

$\vec{F}^L$     contribution to the friction force from the magnetization region

$\vec{F}^A$ friction due to adiabatic collisions
$L^0$ Coulomb log of fast collisions
$L^A$ Coulomb log of adiabatic collisions
$-Ze$ heavy particle charge ($Z = -1$ for antiprotons)
$M$ heavy particle mass
$\rho_{max}$ and $\rho_{min}$ maximum and minimum impact parameters
$\tau$ cooling time
$\lambda = \tau^{-1}$ cooling decrement
$\tau_{eff}$ interaction time in a collision
$t_{scr}, \tau_{scr}$ screening times of particle interaction
$T$ heavy particle temperature
$T_{st}$ equilibrium temperature
$v_0$ velocity oscillation amplitude
$I, \psi$ action-phase variables for heavy particles
$p$ momentum in the laboratory frame
$\omega_0$ circulation frequency of the storage ring
$2\pi R$ orbit circumference
$l$ length of the cooling section
$t_0 = l/\gamma\beta c$ time of flight in the co-moving frame
$\eta = l/2\pi R$ fraction of the cooling section in the total orbit length
$r_0$ electron beam radius
$\sigma_x, \sigma_z$ sizes of the heavy particle beam
$x, z$ transverse coordinates of heavy particles
$\psi$ psi function of the storage ring
$\nu_x, \nu_z, \nu$ transverse oscillation tunes
$\beta_x, \beta_z, \beta$ beta functions of the storage ring
$\vec{\alpha} = (\alpha_x, \alpha_z)$ angular deviation of the magnetic field accompanying electrons, from the ion closed orbit
$\vec{k} = (k_\parallel, \vec{k}_\perp)$ wave vector of a Fourier transformation
$J_0, J_l$ Bessel functions
$\delta$ delta function
$\mathcal{K}$ characteristic of friction
$\vec{V}(\vec{r})$ electron hydrodynamic velocity
$S$ electron path from the cathode
$N$ total number of heavy particles
$n_i$ density of the heavy particle beam

# I. GENERAL PROPERTIES OF ELECTRON COOLING

In this chapter we will consider some general properties of the method, which are generally not related to the cooling beam formation technique. The main attention will be concentrated on the aspects related to specific properties of the particle motion in storage rings. For simplicity, all consideration is done in the approximation of free electron motion in the region of beam interaction. However, all investigated phenomena are equally relevant to cooling by magnetized electrons.

## 1.1 Main processes and equations
## 1. Friction and diffusion in an electron beam

When a heavy particle (ion) moves in an electron beam, it experiences a force, which is a superposition of Coulomb interaction forces with individual electrons. As a function of time, this force is determined by initial positions and velocities of all particles before the start of the interaction. Due to the large difference in masses of ions and electrons, when calculating forces, one can consider ions moving uniformly, then plug these forces into the equations of motion and thus account for systematic reaction of the electron beam. Besides that, assuming the ion beam to have sufficiently low density, we will for now ignore the mutual influence of ions; effects of this kind are discussed in Chapters IV, V. The resulting electric field of electrons (we are working in terms of the reference frame co-moving with the beams) can be represented as:

$$\vec{E}(\vec{r}, \vec{v}, t) = \langle \vec{E}^0 \rangle(\vec{r}, t) + \langle \Delta \vec{E} \rangle(\vec{r}, \vec{v}, t) + \vec{E}^{fl}(\vec{r}, \vec{v}, t) \tag{1.1}$$

where $\langle \vec{E}^0 \rangle$ is the so-called space charge field obtained by averaging the sum of Coulomb fields over the electron probability distribution in the phase space unperturbed by the moving ion; $\langle \Delta \vec{E} \rangle$ is the deviation of the average field from the space charge field owing specifically to the perturbation of electron motion by the ion, $\vec{E}^{fl}$ is the statistical fluctuation of the total field. The space charge field plays the role of an external field in addition to the field of the storage ring. The field $\langle \Delta \vec{E} \rangle(\vec{r}, \vec{v})$ correlated with the ion motion and depending on its velocity determines the friction force $\vec{F}$ whose action reduces the phase space volume of the beam:

$$\vec{F} = -ze\langle \Delta \vec{E} \rangle(\vec{r}, \vec{v}, t)\big|_{\vec{r}=\vec{r}(t), \dot{\vec{r}}(t)=\vec{v}} \tag{1.2}$$

The fluctuation part of the field is responsible for the dispersion of ion momentum change appearing as a result of interaction with electrons:

$$\langle \Delta p_\alpha \Delta p_\beta \rangle = (ze)^2 \int_0^t dt_1 \int_0^t dt_2 \langle E_\alpha^{fl}(t_1) E_\beta^{fl}(t_2) \rangle \tag{1.3}$$

($t \leq 1/(\gamma \beta c)$ where $l$ is the length of the cooling section). The average change in the ion momentum of

$$\langle \Delta \vec{p} \rangle = \int_0^t \vec{F} dt_1 \tag{1.4}$$

and the dispersion in Eq. (1.3) completely determine the irreversible interaction effects at each ion passage through the cooling section.

A systematic method of calculating the friction force and scattering tensor would have to consist of solving the equations of motion of electrons interacting with the ion and between each other, substituting the trajectories into the total Coulomb field $-(\partial/\partial \vec{r})\sum e/|\vec{r} - \vec{r}_a(t)|$ and averaging $\Delta \vec{p}$ and $\Delta p_\alpha \Delta p_\beta$ over the initial conditions with a given, at $t=0$ distribution $D(\Gamma_1, \ldots \Gamma_a, \ldots)$. We will approach such a program in the next chapter when studying the properties of cooling in a magnetized flow with a low longitudinal temperature. For now, we limit ourselves to situations when effect of an external field on the electron motion in the interaction region is small while the average kinetic energy of relative particle motion is large compared to fluctuations of the interaction's potential energy. Interaction effects can then be described using the language of collisions characterized by the impact parameter and transferred momentum.

Let us provide a short derivation of the known expressions [15, 16] for the friction force and scattering tensor in the considered situation. First suppose that all electrons are moving with the same velocity $\vec{u}$ with respect to a proton. Let us denote by $\vec{q}$ the momentum gained by the ion as a result of a collision with an electron; then

$$\vec{F} = n'_e u \int \vec{q} d\sigma$$

$$d_{\alpha\beta} = \frac{d}{dt_1} \langle \Delta p_\alpha \Delta p_\beta \rangle = n'_e u \int q_\alpha q_\beta d\sigma \tag{1.5}$$

where $d\sigma$ is the differential scattering cross section, $n'_e$ is the electron flow density. Using the cylindrical symmetry of scattering with respect to the $\vec{u}$ direction, let us write the tensor structure of $\vec{F}$ and $d_{\alpha\beta}$ as

$$\langle \vec{q} \rangle = A \frac{\vec{u}}{u}, \quad \langle q_\alpha q_\beta \rangle = \frac{1}{2} q^2 \left( \delta_{\alpha\beta} - \frac{u_\alpha u_\beta}{u^2} \right) + B \left( \delta_{\alpha\beta} - 3 \frac{u_\alpha u_\beta}{u^2} \right)$$

To determine the scalars $A$ and $B$, we use the energy conservation law:

$$\frac{q^2}{2m} - \vec{u}\vec{q} = 0$$

Then

$$A = \frac{q^2}{2mu}, \quad B = -\frac{q^4}{8m^2 u^2}$$

For Coulomb interaction

$$d\sigma = \frac{8\pi z^2 e^4}{u^2}\frac{dq}{q^3}, \quad 0 \leq q \leq 2mu,$$

so that

$$\int q^2 d\sigma = \frac{8\pi z^2 e^4}{u^2}\int_{q_{min}}^{2mu}\frac{dq}{q} = \frac{8\pi z^2 e^4}{u^2}\ln\frac{2mu}{q_{min}} \equiv \frac{8\pi z^2 e^4}{u^2}L(u),$$

$$\int q^4 d\sigma = 16\pi z^2 e^4 m^2.$$

The final expressions for $\vec{F}$ and $d_{\alpha\beta}$ are obtained by integrating Eq. (1.5) over the electron velocities $\vec{v}_e$ with a distribution function $f(\vec{v}_e)$:

$$\vec{F} = -\frac{4\pi z^2 e^4 n'_e}{m}\int L(u)\frac{\vec{u}}{u^3}f(\vec{v}_e)d^3 v_e, \tag{1.6}$$

$$d_{\alpha\beta} = 4\pi z^2 e^4 n'_e \int\left[L(u)\frac{u^2\delta_{\alpha\beta} - u_\alpha u_\beta}{u^3} - \frac{1}{2}\frac{u^2\delta_{\alpha\beta} - 3u_\alpha u_\beta}{u^3}\right]f(\vec{v}_e)d^3 v_e. \tag{1.7}$$

The minimum transferred momentum $q_{min}$ in the logarithm $L(u)$ is determined by the impact parameter $\rho_{max}$ above which the Coulomb interaction is effectively cut off:

$$q_{min} \simeq \frac{2ze^2}{u\rho_{max}}$$

The quantities competing as $\rho_{max}$ are the transverse electron beam size $d$, impact parameter $\rho_l = u\tau_l = ul/(\gamma\beta c)$ determined by the time of particle flight to the cooling section length, and effective screening radius $r_{scr} = u/\omega_e$ where $\omega_e = \sqrt{4\pi n'_e e^2/m}$ is the Langmuir frequency:

$$\rho_{max} = \min\left\{d, \frac{ul}{\gamma\beta c}, \frac{u}{\omega_e}\right\}. \tag{1.8}$$

A characteristic size $d$ of an electron beam is on the order of 1 cm. The flight parameter $\rho_l$ with the relative particle velocity spread in the beams of $|u/(\gamma\beta c)| \simeq 2\cdot 10^{-3}$ and the cooling section length of $l = 1$ m, is $\simeq 2$ mm, the radius of plasma screening at the electron beam density of $n'_e \simeq 10^8$ cm$^{-3}$, is 0.3 mm. The specified parameter values are typical for the experimental conditions at the NAP-M storage ring and EPOKHA forming system.

One sometimes encounters an opinion that, in systems with space charge, the parameter $\rho_{max}$ is determined only by the size of the system (in our case, it is $d$ or $\rho_l$). Such a point of view is apparently based on the consideration that, in the absence of a compensating background, screening of the interaction is not possible. In practice, one has to keep in mind that one must talk about screening not of the average particle field but of only the fluctuations related to relative particle motion, by which means the collision interaction takes place. Screening of this interaction is a dynamic process and is of the same nature as that in neutral plasma. Note also that the screening radius in a general case is determined by the relative particle velocity and not simply by the electron thermal velocity spread as it comes out of the Lenard-Balescu theory [17, 18] or when using collision integral calculation techniques customary in plasma kinetics (see, for example, [16, 19]). The noted aspects are considered in more detail in Sections 2.3 and 3.1 for situations, in which they become of practical importance. In case of cooling in a flow of freely moving electrons, the Coulomb logarithm

$$\rho_{max} = \min\left\{d, \frac{ul}{\gamma\beta c}, \frac{u}{\omega_e}\right\} \qquad (1.9a)$$

under all conditions, changes in the range of 5-20. In practical cases, one usually has $d \gg u/\omega_e$, $ul/(\gamma\beta c)$ and then

$$L(u) = \ln\frac{mu^3\tau_{eff}}{ze^2}, \quad \tau_{eff} = \min\left\{\frac{1}{\omega_e}, \frac{l}{\gamma\beta c}\right\}. \qquad (1.9b)$$

The large value of the logarithm allows one to neglect the numerical uncertainty in the parameter $\rho_{max}$. In principle, one can improve the accuracy of Eqs. (1.6, 7) and eliminate the uncertainty if, when considering interactions at large distances, for example, $\rho > (n'_e)^{-1/3}$ where the perturbation theory is applicable, one accounts for dynamic polarization of the medium by a moving ion leading to screening of Coulomb potential at distances $\rho \gtrsim r_{scr}$ and match this region to the region of "small" distances ($\rho < (n'_e)^{-1/3}$) where interaction of electrons is not significant and one can use the exact differential cross section of pair-wise collisions. However, such an improvement only makes sense for a large value of the Coulomb logarithm and therefore presents only a theoretical interest if there are no special reasons to improve the accuracy of the collision integral.

With an accuracy of the order of $1/L$, one can neglect the non-logarithmic term in the scattering tensor (1.7) and, besides that, take the logarithms out of the integrals. Although, in some cases, the dependence $L(u)$ is significant for the details of the friction force behavior as a function of the ion velocity and it must then be taken into account.

Finally, the expressions for the friction force and scattering tensor can be generalized to a spatially non-uniform case using the fact of smallness of the region of effective particle interaction with electrons:

$$\vec{F} = -\frac{4\pi z^2 e^4}{m}\int L(u)\frac{\vec{u}}{u^3}f(\vec{v}_e, \vec{r})d^3v_e, \qquad (1.10)$$

$$\frac{d}{dt_1}\langle \Delta p_\alpha \Delta p_\beta \rangle = 4\pi z^2 e^4 \int L(u) \frac{u^2 \delta_{\alpha\beta} - u_\alpha u_\beta}{u^3} f(\vec{v}_e, \vec{r}) d^3 v_e, \qquad (1.11)$$

where we introduced a distribution function over velocities and coordinates in the co-moving frame normalized as $\int f(\vec{v}_e, \vec{r}) d^3 v_e = n'_e(\vec{r})$. The structure of the friction force as a function of the velocity $\vec{v}$ and electron distribution over velocities $f(\vec{v}_e)$, is remarkable in that it is analogous (with an accuracy of up to the dependence $L(u)$) to the field of attracting Coulomb centers with the "charge" distribution $\rho(\vec{v}_e)$ [16]. Such an analogy is useful to keep in mind when estimating the behavior of the friction force $\vec{F}(\vec{v})$ for a given distribution $f(\vec{v}_e)$. For example, it is immediately clear that, for a Maxwell distribution with a temperature $T_e = m v_{eT}^2$, the friction force is linear in velocity $\vec{v}$ in the region $v \ll v_{eT}$ and drops off as $1/v^2$ in the region $v \gg v_{eT}$ (Fig. 1):

$$\vec{F}(\vec{v}) = -\frac{4\sqrt{2\pi}}{3} \frac{z^2 n'_e e^4 L}{m v_{eT}^3} \vec{v} \quad \text{for} \quad v \ll v_{eT} \qquad (1.12)$$

$$\vec{F}(\vec{v}) = -\frac{4\pi z^2 n'_e e^4 L}{m v^3} \vec{v} \quad \text{for} \quad v \gg v_{eT}$$

The pointed out Coulomb analogy of the friction force manifests itself in a non-trivial way in a number of situations.

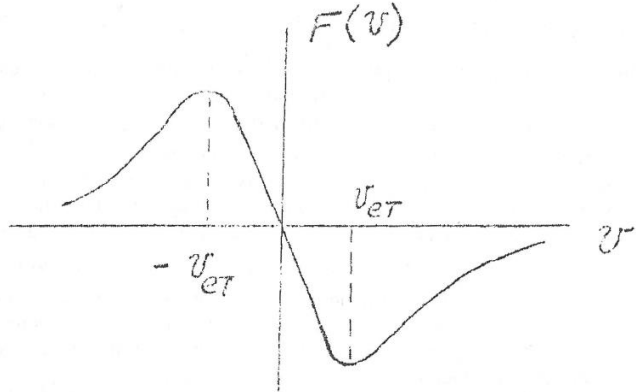

Fig. 1.

## 2. Electron cooling as plasma relaxation

To illustrate the main tendencies of the method, let us consider the cooling process neglecting the cyclic nature of the heavy particle motion in a storage ring and assuming that particle distributions over velocities in the co-moving frame are Maxwellian with temperatures $T$ and $T_e$.

Let us express the average rate of change of the heavy particle energy $d\langle \Delta W_c \rangle/dt'$ through the friction force and scattering tensor:

$$\frac{d}{dt'}\langle \Delta W_c \rangle = \vec{v}\vec{F} + \frac{1}{M}\frac{d}{dt'}\langle (\Delta \vec{p})^2 \rangle.$$

Integrating this equation using Maxwell distributions over velocities, we arrive at an equation for the temperature [15, 20]:

$$\frac{dT}{dt'} = -\frac{8\sqrt{2\pi}}{3}\eta \frac{n_e' z^2 e^4 L}{mM} \frac{T - T_e}{\left(\frac{T}{m} + \frac{T_e}{m}\right)^{3/2}} \qquad (1.13)$$

Here $\eta$ is the fraction of the ion orbit occupied by the electron beam.

In terms of the laboratory frame ($dt = \gamma dt'$, $n_e = \gamma n_e'$)

$$\frac{dT}{dt} = -\frac{8\sqrt{2\pi}}{3}\eta \frac{n_e z^2 e^4 L}{\gamma^2 mM} \frac{T - T_e}{\left(\frac{T}{m} + \frac{T_e}{m}\right)^{3/2}} \qquad (1.13a)$$

As one can see, there are two characteristic regions in the dependence of the cooling time $\tau$ on the ion velocity spread $v_T = \sqrt{T/M}$:

$$\tau_1 \approx \frac{3}{8\sqrt{2\pi}} \frac{\gamma^2 mM v_{eT}^3}{\eta n_e z^2 e^4 L} \qquad \text{when} \quad v_T < v_{eT} \qquad (1.14)$$

$$\tau_2 \approx \frac{1}{4\sqrt{2\pi}} \frac{\gamma^2 mM v_T^3}{\eta n_e z^2 e^4 L} \qquad \text{when} \quad v_T > v_{eT} \qquad (1.15)$$

In the first case, the time $\tau_1$ is defined as the inverse decrement of the exponential decay of the temperature $T$ to the equilibrium value $T_e$ while, in the second case, it has the meaning of the absolute time of the temperature change from its initial value to the value $T \simeq (M/m)T_e$.

For estimates, it can be convenient to express the temperatures through the angular spreads of the particle velocities $\theta$ and $\theta_e$:

$$T = \frac{1}{2}M(\gamma\beta c\theta)^2, \quad T_e = \frac{1}{2}M(\gamma\beta c\theta_e)^2.$$

Then ($r_e = e^2/(mc^2)$):

$$\tau_1 \approx \frac{3}{32\sqrt{\pi}} \frac{\gamma^5 (\beta\theta_e)^3}{\eta n_e z^2 r_e^2 cL} \frac{M}{m} \qquad \text{when} \quad \theta < \theta_e, \qquad (1.16)$$

$$\tau_2 \approx \frac{1}{16\sqrt{\pi}} \frac{\gamma^5 (\beta\theta)^3}{\eta n_e z^2 r_e^2 cL} \qquad \text{when} \quad \theta > \theta_e. \qquad (1.17)$$

The cooling time is proportional to the cube of the maximum of the velocity spreads of the two beams. For a given density of the electron beam in the laboratory frame and fixed angular spreads, the cooling time is proportional the ion kinetic energy in the storage ring to the power 3/2 in the non-relativistic region and to the fifth power in energy in the relativistic case. In the region $\theta < \theta_e$, if the electron temperature is fixed (temperature $T_k$ of the electron gun's cathode) the cooling time is proportional only to the square of the particle total energy $E = \gamma M c^2$.

Let us provide some numerical examples. Suppose the proton kinetic energy is $W = 65$ MeV, density $n_e = 10^8$ cm$^{-3}$, coefficient $\eta = 0.02$ ($l = 1$ m and $2\pi R = 50$ m), electron temperature $T_e = T_k = 2000° \approx 0.2$ eV. Then the time of cooling protons ($z = 1$) with an initial angular spread $\theta < 2 \cdot 10^{-3}$ is 1 s. At a proton energy of $W = 1$ GeV and therefore $\theta < 4 \cdot 10^{-4}$, the cooling time increases by a factor of 4 and becomes 4 s, however, already with an angular spread of $\theta = 2 \cdot 10^{-3}$, the cooling time is 100 s. The equilibrium angular spreads $\theta_{st}$ in these cases are $5 \cdot 10^{-5}$ and $10^{-5}$ rad while the transverse beam sizes (with a focal distance of $f \simeq R \simeq 10$ m) are 1 and 0.2 mm.

## 3. Scattering and energy loss in the residual gas

If there is some process expanding (heating) the ion beam then one should, first of all, make sure that the rate of this heating is small compared to the maximum rate of the thermal energy transfer to the electron beam. In angular variables

$$\left(\frac{d\theta^2}{dt}\right)_{ext} \ll \left|\frac{d\theta^2}{dt}\right|_{max} = \frac{64}{9}\sqrt{\frac{\pi}{3}}\eta \frac{n_e z^2 e^4 L}{\gamma^5 (\beta c)^3 m M \theta_e}. \tag{1.18}$$

The rate of external scattering can be considered independent of the beam angular spread. Then, if the indicated condition is met, there are two equilibrium temperatures: one is stable $T_1$ on the left slope of the $dT/dt(T)$ curve and the other is unstable $T_2$ on the right slope (Fig. 2). All particles with energies greater than $T_2$ will leave the beam while the others will assemble in the region of $T_1$.

A process always present but depending on the vacuum level in the storage ring's beam pipe is scattering on the residual gas. Then

$$\frac{d\theta^2}{dt} = \frac{8\pi(z_0 z e^2)^2 L_0 n_0}{\gamma^2 M^2 (\beta c)^3}, \tag{1.19}$$

where $z_0$ and $n_0$ are the nuclear charge and concentration of gas, $L_0$ is the corresponding logarithm (typically $L_0 \simeq 5 - 6$). Comparing this expression to Eq. (1.18), we get the ratio of the electron beam density to the residual gas density, with which cooling of the proton beam becomes possible:

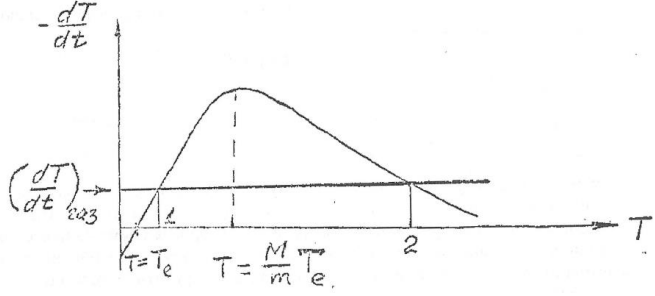

Fig. 2.

$$\frac{n_{cr}}{n_0} = 3\gamma^3 z_0^2 \frac{m}{M} \frac{L_0}{L} \frac{\theta_e}{\eta}. \tag{1.20}$$

When $n \gg n_{cr}$, the angular spread is established at

$$\theta_{st} = \theta\sqrt{0.4 n_{cr}/n}. \tag{1.21}$$

It should be noted that $\theta_{st}$ is not smaller than the value of $\theta_e\sqrt{m/M}$ corresponding to the temperature equality. The maximum angular spread captured into the damping mode is

$$\theta_{max} = \frac{4}{\sqrt{\pi}} \frac{\theta_e n}{n_{cr}}. \tag{1.22}$$

In a cooling mode without an RF field, there may be a question whether deceleration by the electrons of the residual gas will dominate over the pull of protons by the electron beam. The question is solved by comparing the ionization losses per unit length and maximum value of the average friction force in the electron flow:

$$\frac{dW}{dx} = -\frac{4\pi z^2 e^4 z_0 n_0 L_0}{\gamma m \beta^2 c^2}, \tag{1.23}$$

$$|F|_{max} \simeq \eta \frac{4\pi z^2 e^4 n L}{\gamma^3 m (\beta c \theta_e)^2}. \tag{1.24}$$

Then

$$\left(\frac{n_{cr}}{n_0}\right)_{decel} \simeq \frac{z_0 L_0}{\eta L} \gamma^2 \theta_e^2. \tag{1.25}$$

In practice, the criteria of Eqs. (1.20) and (1.25) do not differ significantly.

There are also energy loss fluctuations related to scattering on the electrons of the residual gas whose ratio to transverse scattering equals:

$$\frac{\langle(\Delta p_\parallel)^2\rangle}{\langle(\Delta \vec{p}_\perp)^2\rangle} = \frac{\gamma^2 + 1}{2z_0 L_0}. \qquad (1.26)$$

The loss fluctuations become dominant in a not too far ultrarelativistic regime: $\gamma > \sqrt{2z_0 L_0}$.

## 4. Relaxation equation in action variables. Averaging over phases

A realistic process of beam relaxation in a storage ring has to be described by equations including "external" field, force of interaction with the electron flow and possible "outside" sources of stochastic fields. According to the above estimates, the cooling time is long compared to the typical periods of finite particle motion in a storage ring. Under these conditions, it is natural to consider the beam evolution averaged over multiple passes of particles trough the electron beam. The most convenient dynamical variables are then the action-phase type variables, since one may always, with a certain accuracy, replace averaging over time with averaging over phases and thus transition to a description of kinetics in an external field equivalent to that in terms of momenta for free particle motion.

For arbitrary initial conditions, particles move with small oscillations about closed orbits determined by energies. The generalized azimuthal angle $\theta = 2\pi\sigma/\Pi_s$ is chosen as one of the particle coordinates where $\sigma$ is the path length along one of the closed orbit with energy $\varepsilon_s$ and circumference $\Pi_s = 2\pi R$ while the particle position vector $\vec{r}$ is represented as

$$\vec{r} = \vec{r}_s(\theta) + \vec{r}_\perp,$$

where $\vec{r}_s(\theta)$ is the "equilibrium" closed orbit ($\varepsilon = \varepsilon_s$), $\vec{r}_\perp$ is the transverse deviation from the orbit $\vec{r}_s(\theta)$, in a general case, having the following structure [21] (the prime denotes a derivative with respect to the azimuthal angle $\theta$ playing the role of time):

$$\begin{pmatrix} \vec{r}_\perp \\ \vec{r}'_\perp \end{pmatrix} = \begin{pmatrix} \vec{\psi}(\theta) \\ \vec{\psi}'(\theta) \end{pmatrix} \frac{\Delta p}{p} + \frac{1}{2}\left\{ A_1 \begin{pmatrix} \vec{f}_1(\theta) \\ \vec{f}'_1(\theta) \end{pmatrix} + A_2 \begin{pmatrix} \vec{f}_2(\theta) \\ \vec{f}'_2(\theta) \end{pmatrix} \right\},$$

where the function $\vec{\psi}(\theta)$ describes the dependence of the closed orbit on the energy (or momentum $p$) while the functions $\vec{f}_1(\theta)$, $\vec{f}_2(\theta)$ are the normal solutions of the equations of particle small free oscillations about the closed orbit (the Floquet solutions) possessing the property:

$$\vec{f}_{1,2}(\theta) = \vec{f}_{1,2}(\theta - 2\pi)e^{2\pi i \nu_{1,2}}$$

and normalized as:

$$\mathrm{Im}\, \vec{f}'_\alpha \vec{f}^*_\beta = \delta_{\alpha\beta},$$

besides, the betatron oscillations frequencies $\nu_1\omega_0$, $\nu_2\omega_0$ are generally not commensurate with the revolution frequency. The oscillation amplitudes $A_1$, $A_1^*$ and $A_2$, $A_2^*$ are determined by specifying the initial conditions for $x$, $x'$ and particle energy. As in the case of a harmonic oscillator, these amplitudes can be combined into canonically conjugate pairs $A_1$, $\bar{A}_1 = iA_1^*$ and $A_2$, $\bar{A}_2 = iA_2^*$ and action-phase variables $I$, $\varphi$ (see Appendix I):

$$A_{1,2} = \sqrt{2\frac{R}{p}I_{1,2}}e^{i\varphi_{1,2}}; \quad I_{1,2} = \frac{1}{2}\frac{p}{R}|A_{1,2}|^2. \tag{1.27}$$

Further on, we will consider the usual case of a nearly "perfect" magnetic system when the vertical ($z$) and radial ($x$) oscillations are uncoupled, at least, on the time scales of the order of the revolution period for flat closed orbits:

$$\vec{r} = \vec{r}_s(\theta) + \vec{i}_z z + \vec{i}_x[\psi(\theta)\frac{\Delta p}{p} + x_b];$$

$$\begin{pmatrix}z\\z'\end{pmatrix} = \frac{1}{2}A_z\begin{pmatrix}f_z\\f_z'\end{pmatrix} + c.c.; \quad \begin{pmatrix}x_b\\x_b'\end{pmatrix} = \frac{1}{2}A_x\begin{pmatrix}f_x\\f_x'\end{pmatrix} + c.c.;$$

$$\operatorname{Im} f_z^* f_z' = \operatorname{Im} f_x^* f_x' = 1;$$

$$\theta(t) = \omega_s t + \varphi + (\psi x_b' - \psi' x_b)/R^2; \quad \bar{\dot{\varphi}} = \frac{d\omega_0}{dp}\Delta p. \tag{1.28}$$

Effect of small distortions of the focusing field structure can be accounted for, when needed, using the perturbation theory. Far from resonances coupling the $x$ and $z$ motions, one can assume $A_x = const$ and $A_z = const$.

Having the structure of a magnetic system close to ideal does not exclude a strong dynamical coupling of the vertical and radial oscillations near the $\nu_z \simeq \nu_x + k$ resonances. If we introduce parameters specifying detuning $\varepsilon = \nu_z - \nu_x - k$ and amplitude beat tune $\Omega_M = \sqrt{\kappa^2 + \varepsilon^2}$ ($\kappa$ is the coupling parameter) then the amplitudes $A_x$ and $A_z$ in Eq. (1.28) modulated by coupling change according to:

$$A_x = A_1\sqrt{\frac{\Omega_M + \varepsilon}{2\Omega_M}}e^{-i\frac{\Omega_M-\varepsilon}{2}\omega_0 t} - A_2\sqrt{\frac{\Omega_M - \varepsilon}{2\Omega_M}}e^{i\frac{\Omega_M+\varepsilon}{2}\omega_0 t},$$

$$A_z = A_1\sqrt{\frac{\Omega_M - \varepsilon}{2\Omega_M}}e^{-i\frac{\Omega_M+\varepsilon}{2}\omega_0 t} + A_2\sqrt{\frac{\Omega_M + \varepsilon}{2\Omega_M}}e^{i\frac{\Omega_M-\varepsilon}{2}\omega_0 t}, \tag{1.29}$$

where $A_1$ and $A_2$ are the constant canonical amplitudes of the normal-mode oscillations. Coupling of the oscillations can be significant for the cooling kinetics since it leads to redistribution of the decrements.

Under the approximation of uniform motion of heavy particles in the interaction section

$$f_{x,z}(\theta) = f_{0x,z} + f'_{0x,z} \cdot \theta,$$

where $f_{0x,z}$ and $f'_{0x,z}$ are the values at the origin of the azimuthal angle coordinate. When $|f_{x,z}(\theta)|$ is symmetric with respect to the center of the cooling section, we have

$$\begin{aligned} f(\theta) &= \left(|f_0| + \frac{i\theta}{|f_0|}\right) e^{i\psi_k}, \quad (|f|'_0 = 0), \\ f' &= \frac{i}{|f_0|} e^{i\psi_k} = const, \quad |\theta| \leq \frac{l}{4\pi R}, \\ \psi_k &= 2\pi k \nu + \psi_0. \end{aligned} \qquad (1.30)$$

In a sufficiently long cooling section, the magnetic field guiding the electrons will cause twisting of the particle trajectories and may also lead to a strong coupling of the oscillations. The former effect by itself is not dangerous and can be easily included in the consideration. Coupling of the oscillations can be compensated by additional lenses in the adjacent sections. We will neglect the twisting and assume the coupling is small accounting for it phenomenologically according to Eq. (1.29). For example, for the NAP-M conditions ($H_\| = 1$ kG, $\beta = 0.3$, $l = 1$ m), the twist angle is $10^{-1}$ rad. This angle also determines the coupling parameter $\kappa$.

Effect of the space-charge field is relatively weak and can present a danger only due to its introduction of nonlinear resonances leading to a stochastic instability [22, 23, 24, 25]. If necessary this field can be compensated by positive ions.

For the longitudinal motion, the action-phase variables are the length $R\theta$ and momentum $p$. When working with RF field, the momentum and the phase offset from the equilibrium $\varphi = \theta - \omega_s t$ undergo slow synchrotron oscillations with a tune $\nu_s$ while the action variable is the area enclosed by the trajectory in the $(p,\varphi)$ phase space; in the approximation of linear oscillations

$$I_s = \frac{(p - p_s)^2}{2\nu_s \mu_s} + \frac{1}{2} \nu_s \mu_s \varphi^2.$$

## Averaging over phases

When momentum is changed by $\Delta\vec{p}$ in a short time, the action gets increased by

$$\Delta I = \frac{\partial I}{\partial \vec{p}} \Delta\vec{p} + \frac{1}{2} \frac{\partial^2 I}{\partial p_\alpha \partial p_\beta} \Delta p_\alpha \Delta p_\beta; \qquad (1.31)$$

the change of $I$ per unit time averaged over the collision parameters is

$$\dot{I}_i = \frac{\partial I}{\partial \vec{p}} \vec{F} + \frac{1}{2} \frac{\partial^2 I}{\partial p_\alpha \partial p_\beta} d_{\alpha\beta}. \qquad (1.32)$$

The scattering tensor in terms of $I$ is

$$D_{ik} = \frac{d}{dt}\langle \Delta I_i \Delta I_k \rangle = \frac{\partial I_i}{\partial p_\alpha} \frac{\partial I_k}{\partial p_\beta} d_{\alpha\beta} . \qquad (1.33)$$

Together with the particle coordinates and momenta, the coefficients of Eqs. (1.32), (1.33) are periodic functions of the phases $\Phi_i = \omega_i t + \varphi_i$. Considered as functions of $I$ and $\varphi$, they oscillate in time, moreover, oscillations with respect to the average level are not small. One can, however, replace $I_i$ and $D_{ik}$ with their time-averaged values if the relaxation time $\tau$ is long compared to the characteristic periods of motion in the external field $\tau_{ext}$.

In practice, averaging over time can almost always be replaced by an explicit averaging over the phases $\Phi_i$. Let us express, for example, $\dot{I}_i$ as a function of the phases in a Fourier series:

$$\dot{I} = \sum_{\{m\}} \dot{I}_m(I) \exp\{i \sum_k m_k \Phi_k\} .$$

The frequencies of the harmonic change in time are

$$\omega_{\{m\}} = \sum_k m_k \omega_k .$$

Besides the "zero-integer" harmonic $\{m_k\} = 0$ representing simply a phase-averaged value of $\dot{I}$, when averaging over time, a non-zero contribution to $\bar{\dot{I}}$ can come, generally speaking, from the harmonics with frequencies

$$\omega_{\{m\}} \lesssim \tau^{-1} . \qquad (1.34)$$

Formally, one can always select such a combination of $\{m\}$ that $\omega_{\{m\}}$ can be made arbitrarily small. However, condition (1.34) can, in general, be satisfied only for quite large $m$ considering that $\omega_k \tau \gg 1$ while the frequencies $\omega_k$ do not, typically, form rational relationships between themselves of low order. Therefore, the magnitudes of such harmonics will be negligibly small. In practice, one also needs to account for nonlinearity of particle oscillations, which leads to dependence of the frequencies on the amplitudes. Despite the relative weakness of the frequency change with amplitude:

$$\Delta \omega_i = \Delta I_k \frac{\partial \omega_i}{\partial I_k} \ll \omega_i ,$$

this dependence leads to violation of the "resonant" condition when, under the effect of collisions, $I_i$ gets an increase such that

$$|\Delta \omega_{\{m\}}| > \tau^{-1} .$$

This circumstance is especially significant when evaluating the role of resonating harmonics of low order, whose magnitudes can be comparable to the phase-averaged one. The exceptions are those cases when the resonant condition is maintained due to auto-phasing arising under the effect of some "external" field such as, for example, the field of an electron beam. For nonlinear resonances, this phenomenon can be neglected if the phase-space volume inside the resonance separatrix is relatively small [22].

Thus, under the condition $\tau \gg \tau_{ext}$, one can almost always assume that the coefficients $\bar{\dot{I}}_i$ and $\bar{D}_{ik}$ are independent of the phases $\varphi_i$. This allows one, after averaging coefficients (1.32) and (1.33) over the phases, to describe the relaxation process by the Fokker-Planck equation for a distribution function of the three variables $I_i$:

$$\frac{\partial f}{\partial t} + \frac{\partial}{\partial I_i}\left(\bar{\dot{I}}_i - \frac{1}{2}\frac{\partial}{\partial I_k}\bar{D}_{ik}\right)f = 0. \tag{1.35}$$

Using the property of canonicity of the transformation $I_i(\vec{p},\vec{r})$, $\Phi_i(\vec{p},\vec{r})$, one can show that there exists a relation (see Eq. (A.4.2) of Appendix 4):

$$\langle \bar{\dot{I}}_i \rangle = Q_i + \frac{1}{2}\frac{\partial D_{ik}}{\partial I_k} \equiv \bar{\dot{I}}_{fr} + \bar{\dot{I}}_{fl}, \tag{1.36}$$

where

$$Q_i = \overline{\frac{\partial I_i}{\partial \vec{p}}\vec{F}(\vec{p},\vec{r})}. \tag{1.37}$$

Using Eq. (1.36), Eq. (1.35) transforms to

$$\frac{\partial f}{\partial t} + \frac{\partial}{\partial I_i}\left(Q_i - \frac{1}{2}D_{ik}\frac{\partial}{\partial I_k}\right)f = 0. \tag{1.38}$$

In terms of physical meaning, $\bar{\dot{I}}_{ifr} = Q_i$ determines the rate of change of $I_i$ due to dissipative processes ($Q_i$ is the friction power) while $\bar{\dot{I}}_{ifl}$ describes the average growth rate of $I_i$ due to the absorbtion of the fluctuation part of the "medium's" energy. In the oscillation mode, when $I_i$ determines the oscillator energy, the quantity $\bar{\dot{I}}_i$ characterizes the direction of the kinetic process as a whole. For infinite motion, when $I_i \sim p_i$, moments (1.33) are also important in this sense. Let us provide explicit expressions for the moments of $\bar{\dot{I}}_i$ in terms of the friction force and scattering tensor $d_{\alpha\beta}$. In practice, it is somewhat more convenient to deal not with the action variables directly but with the amplitudes $v_{x0}$ and $v_{z0}$ of velocity oscillations $(v_x, v_\parallel, v_z)$ in the co-moving frame in the cooling section. For simplicity, we will assume the $\psi$ function to be constant in the cooling section and also neglect the relatively small change of the transverse coordinates with the azimuthal angle in the cooling section (small values of the $\beta$ functions $\beta_x = R|f_x|^2$ and $\beta_z = R|f_z|^2$ are not favorable). As a result, formulae are significantly simplified without loss of dependence on the main parameters of the heavy particle dynamics in the beam interaction region:

$$v_x = v_{x0} \sin \Phi_x, \quad x = \psi \frac{v_\parallel}{\beta c} - \frac{v_{x0}}{\gamma \beta c} \beta_x \cos \Phi_x \equiv \psi \frac{v_\parallel}{\beta c} + x_b,$$

$$v_z = v_{z0} \sin \Phi_z, \quad z = -\frac{v_{z0}}{\gamma \beta c} \beta_z \cos \Phi_z,$$

$$\gamma \frac{d}{dt} v_{x0}^2 = \frac{2}{M} \overline{v_x F_x} + \frac{d_{xx}}{M^2} - 2 \left(\frac{\gamma \beta c}{\beta_x}\right)^2 \frac{\psi}{M \beta c} \overline{x_b F_\parallel} + \left(\frac{\gamma \psi}{\beta_x}\right)^2 \frac{d_\parallel}{M^2}, \quad (1.39)$$

$$\gamma \frac{d}{dt} v_{z0}^2 = \frac{2}{M} \overline{v_z F_z} + \frac{\bar{d}_{zz}}{M^2},$$

$$\gamma \frac{d}{dt} v_\parallel^2 = \frac{2}{M} \overline{v_\parallel F_\parallel} + \frac{d_\parallel}{M^2}.$$

The additional terms in the first equation appear due to perturbation of the equilibrium position of the radial oscillations with an energy change under the effect of longitudinal forces.

With RF field, instead of the last equation, one needs to use a corresponding equation for the amplitude $v_{\parallel 0}$ ($v_\parallel = v_{\parallel 0} \sin \Phi_s$) with additional averaging over the synchrotron oscillation phase $\Phi_s$.

The equations become more complicated in the presence of a coupling resonance since one then needs to determine the average rate of change of the amplitudes $A_1$ and $A_2$ (1.29) depending on the three velocity components.

However, the role of coupling in specific cases is not hard to estimate using quantities such as the coupling parameter $\kappa$, detuning $\varepsilon$ and known friction parameters.

To illustrate Eqs. (1.39), let us estimate the character of relaxation for the Maxwellian distribution of electron velocities in a spatially uniform beam. Using the expressions for the friction force (1.12) and scattering tensor (1.11) in the $v < v_{eT}$ velocity region, we get

$$\frac{d}{dt} v_{x0}^2 = -\frac{1}{\tau_1} \left[\frac{1}{2} v_{x0}^2 - \frac{m}{M}\left(1 + \gamma^2 \frac{\psi^2}{\beta_x^2}\right) v_{eT}^2\right]; \quad T_{xs} = \left(1 + \gamma^2 \frac{\psi^2}{\beta_x^2}\right) T_e,$$

$$\frac{d}{dt} v_{z0}^2 = -\frac{1}{\tau_1} \left(\frac{1}{2} v_{z0}^2 - \frac{m}{M} v_{eT}^2\right); \quad T_{zs} = T_e, \quad (1.40)$$

$$\frac{d}{dt} v_\parallel^2 = -\frac{1}{\tau_1} \left(\frac{1}{2} v_\parallel^2 - \frac{m}{M} v_{eT}^2\right); \quad T_{\parallel s} = T_e,$$

where $\tau_1$ is the damping time (1.14). As one can see, for the transverse degrees of freedom, the rate of damping of the temperature to its equilibrium value reduces by a factor of two compared to (1.13), which has a simple explanation: thermal capacity of oscillators exceeds that of free particles by a factor of two. Next, the equilibrium temperature of the radial degree of freedom is relatively higher due to excitation of oscillations by jumps in energy (or, which is the same, in longitudinal velocity) when scattering on electrons. This effect disappears if the closed orbit position in the cooling region does not depend on the energy ($\psi = 0$).

For large initial amplitudes $v_0 \gg v_{eT}$, the friction characteristics during velocity oscillations go through all values in a large interval $(1/v_{eT}^3 - 1/v_0^3)$. Behavior of the resulting rate depends on the characteristics of the initial distribution in amplitudes. If one ignores the electron velocity spread then the partial rates have a universal form:

$$\frac{M}{2}\frac{dv_{x,z,0}^2}{dt} = -\frac{4\pi n_e z^2 e^4 L}{\gamma^2 m}\eta \langle\frac{v_{x,z}^2}{v^3}\rangle,$$
$$\frac{M}{2}\frac{dv_\parallel^2}{dt} = -\frac{4\pi n_e z^2 e^4 L}{\gamma^2 m}\eta \langle\frac{v_\parallel^2}{v^3}\rangle,$$
(1.41)

where the brackets $\langle ... \rangle$ denote averaging over the phases of betatron oscillations while diffusion can be neglected. For two-, three-dimensional excitations, or for excitation of only the longitudinal degree of freedom, the averaged expressions do not have non-integrable features and the damping time in these cases is equal to, for example:

$$\tau_2 = \frac{\gamma^2 mM v_i^3}{24\pi n_e z^2 e^4 L \eta} \quad \text{for} \quad [v_\parallel \gg v_{x,z,0}]_{t=t_i},$$
(1.42)

$$\tau_2 = \frac{\gamma^2 mM v_i^3}{24 n_e z^2 e^4 L \eta} \cdot \frac{\zeta}{\ln(1+\sqrt{2})} \; ; \quad \frac{2}{\pi} < \zeta < 1 \quad \text{for} \quad [v_{x0} = v_{z0} = v_\parallel]_{t=t_i}.$$
(1.43)

Since there is no unique way to compare the damping times of non-interacting particles and of an ensemble with a given distribution shape (Maxwellian one in (1.15)) then one can in general talk only about agreement of the times (1.42, 43) and (1.15) within an order of magnitude.

A certain special feature appears in case of one dimensional excitation of the oscillatory degree of freedom:

$$\frac{d}{dt}\frac{1}{2}Mv_{\alpha 0}^2 = -\frac{4\pi n_e z^2 e^4 L\eta}{\gamma^2 m v_{\alpha 0}}\int_{\psi_{min}}^{\frac{\pi}{2}}\frac{2}{\pi}\frac{d\psi}{|\sin\psi|} \; ;$$

the logarithmic divergence should be cut off at the value $\psi \simeq \sqrt{v_{tr}^2 + v_{eT}^2}/v_{\alpha 0}$ where $v_{tr}$ are the velocities of the weakly excited degrees of freedom. Thus, in this case

$$\tau_2 = \frac{\gamma^2 mM v_i^3}{48 n_e z^2 e^4 L\eta \ln(v_i/\sqrt{v_{tr}^2 + v_{eT}^2})}.$$
(1.44)

It should be noted that, when the average beam velocities are the same, damping of the large spread in practice very weakly depends on the shape of the electron distribution in velocities due to universality of the friction force behavior ($\sim 1/v^2$).

## 5. Diffusion due to non-thermal noise

Coulomb scattering may be not the only mechanism of diffusion (or heating) of the ion beam in the electron flow. Besides the static part, the space charge field may contain an irregular, stochastic part related to collective fluctuations of density and velocity excited by "external" sources (control voltage oscillations, cathode flickering). Since the times of collective fluctuations may not be small compared to the periods of motion in a storage ring, their effect should right away be considered accounting for the finiteness of motion, i.e. in terms of the action-phase variables. In the co-moving frame, the diffusion tensor $D_{ik}$ can be expressed through the fluctuations of electric field:

$$\Delta I_i = ze \int_{-\infty}^{t} \Delta \tilde{\vec{E}}(\vec{r}(t'), t') \frac{\partial I_i}{\partial \vec{p}(t')} dt', \qquad (1.45)$$

$$D_{ik} = \frac{d}{dt} \langle \Delta I_i \Delta I_k \rangle = z^2 e^2 \int_{-\infty}^{t} \langle \Delta \tilde{E}_{\alpha t} \Delta \tilde{E}_{\beta t'} \rangle \left( \frac{\partial I_i}{\partial p_{\alpha t}} \frac{\partial I_k}{\partial p_{\beta t'}} + \frac{\partial I_i}{\partial p_{\beta t'}} \frac{\partial I_k}{\partial p_{\alpha t}} \right) dt'. \qquad (1.46)$$

The average $\langle ... \rangle$ is, by definition, a field correlation function $K_{\alpha\beta}(\vec{r}/\vec{r}', t - t')$, which can be expressed through the spectral density of fluctuations:

$$K_{\alpha\beta}(\vec{r}/\vec{r}', t - t') = \int d\omega K_{\alpha\beta}^{\omega} e^{-i\omega(t-t')}.$$

Tensor (1.46) can, of course, represent a general expression of the diffusion coefficient, including the thermal part (1.11) as well. Non-equilibrium fluctuations are distinguished by the fact that they can be represented as functions of the fields set by "external" sources and the equilibrium state of the flow corresponding to the static component of the electron distribution. With specific assumptions about the nature of the fields, one must, generally speaking, solve the problem of forced collective motion in the electron beam, which gives the field $\Delta \tilde{\vec{E}}(\vec{r}, t)$ and one can then compute the diffusion coefficient (1.46) averaging over the source distribution parameters. Since it is assumed that perturbation of electron motion under the effect of "external" noise field is small, the latter can be significant compared to the thermal ones only in the case when they involve large groups of electrons, i.e. have long wave lengths.

As an illustration, let us consider the effect of noise oscillations of the electron accelerating potential $U$. In the co-moving frame, change of the transverse space-charge field follows $U(t)$:

$$\Delta \tilde{\vec{E}}_\perp = \vec{E}_\perp \frac{\tilde{U}}{U} \left( \kappa - \frac{\gamma}{\gamma + 1} \right),$$

where the parameter $\kappa = U dJ/J dU$ is determined by the dependence of the electron current on voltage $U$. Let us assume for simplicity that $\vec{E}_\perp$ near the closed orbit is independent of the transverse coordinates. According to (1.46) and (1.36), the diffusion rate of the square of amplitude of transverse momentum oscillations $p_0^2$ (in one of the two dimensions) is

$$\frac{d}{dt}p_0^2 = 8\frac{z^2e^2}{\gamma^2}\left(\frac{E_\perp}{U}\right)^2\left(\kappa + \frac{\gamma}{\gamma+1}\right)^2 \langle \tilde{U}^2 \rangle$$
$$\times \int_{-\infty}^{t} dt' K(t-t')\cos\omega_b(t-t') \sum_{n=0}^{\infty} \frac{\sin^2 n\pi\eta}{\pi^2 n^2}\cos n\omega_0(t-t'),$$

where we introduced a correlator $K(\tau)$:

$$\langle \tilde{U}(t)\tilde{U}(t') \rangle = \langle \tilde{U}^2 \rangle K(t-t'); \quad K(0) = \int K^\omega d\omega = 1$$

and averaged over the phases. Under stationary conditions, $K(\tau) = K(-\tau)$ and the integral over time is a Fourier harmonic of the correlator, i.e. the spectral density at $\omega = n\omega_0 \pm \omega_b$:

$$\frac{d}{dt}p_0^2 = 4\pi z^2 e^2 \left(\frac{\kappa}{\gamma} - \frac{1}{\gamma+1}\right)^2 E_\perp^2 \overline{\left(\frac{\tilde{U}}{U}\right)^2} \sum_{n=-\infty}^{\infty} K^{n\omega_0+\omega_b}\frac{\sin^2 \pi n\eta}{\pi^2 n^2}. \tag{1.47}$$

The final result depends on the relationship between the motion periods ($\omega_0^{-1}$, $\omega_b^{-1}$, $l/(\gamma\beta c)$) and characteristic time $\tau_c$, during which the correlator $K(\tau)$ is different from zero. For example, in the case

$$l/(\gamma\beta c) \ll \tau_c \ll 2\pi/\omega_0,$$

the sum in (1.47) is $\eta^2/\omega_0$ that has an obvious explanation. Let us estimate a permissible fluctuation level for this case using the condition that the diffusive amplitude growth does not exceed the friction strength for the initial ion momentum spread. The latter has an order of magnitude of

$$\left|\frac{dp_0^2}{dt}\right|_{fr} \simeq M\frac{4\pi z^2 e^4 L n_e \eta}{m\gamma^3 \beta c\theta}.$$

Comparing this expression with (1.42), we get:

$$\left|\frac{\tilde{U}}{U}\right|_{cr} \simeq \left(\frac{ML}{mn_e V\gamma\theta}\right)^{1/2}/\left|\kappa - \frac{\gamma}{\gamma+1}\right|,$$

where $V$ is the volume of the electron beam. Let us take, for example, the values of $L = 15$, $\gamma \approx 1$, $\theta = 3 \cdot 10^{-3}$, $n_e = 10^8$ cm$^{-3}$, $V = 3 \cdot 10^3$ cm$^3$, $\kappa = 3/2$. Then $|\tilde{U}/U|_{cr} \simeq 5 \cdot 10^{-3}$ while the stability level achieved in practice is about $10^{-5}$ [3, 4]. The considered type of fluctuations seems to give the strongest effect from the point of view of magnitude of the "scattering cross section". Slow $U$ oscillations ($\tau_c \gg 2\pi/\omega_0$, $2\pi/\omega_b$) are adiabatic with respect to the transverse oscillations and give no effect.

## On the criterion of stochasticity of coherent fluctuations

Above we used a conventional way of introducing the diffusion coefficient by means of a correlation function or by means of the spectral density of a "random field". Without a formal (explicit) definition of this function, a significant constructive aspect is that of averaging over the distribution of physical parameters determining the field as an exact function of time. Strictly speaking, in reality, parameters always form a discrete set. This discreteness is not significant if the fields themselves of separate elementary sources have continuous spectra as functions of time (for example, in collisions when motion is infinite). If the field spectrum is discrete then the applicability of the diffusion equation is limited to the times (from the start of the interaction)

$$t < \delta\omega_{ext}^{-1} \quad \text{if} \quad \delta\omega_{ext} > \sqrt{V_m \frac{\partial \omega_m}{\partial I}},$$

where $\delta\omega_{ext}$ is the average distance in frequency between the neighboring harmonics of perturbation $V$ ($\dot{I} = -\partial V/\partial \phi$), $\partial \omega_m/\partial I$ is the nonlinearity parameter determining dependence on the action of the $m$-th phase harmonic of interaction with the "random" field. For long times, if one does not limit oneself to the lowest order of the perturbation theory (as in (1.45)), one can see that the diffusion stops and statistical predictions lose definitiveness (the dispersion of physical averages becomes large) [26].

Fulfillment of the Chirikov criterion $\delta\omega_{ext} \ll \sqrt{V_m \partial \omega_m/\partial I}$ ensures applicability of the diffusion equation without limits on time [23, 24, 26].

## 6. Recombination

When cooling protons or positive ions by electrons (or antiprotons by positrons), the heavy particles can attach light ones through radiation of quanta or three-body collisions and thus leave the beam. The recombination process limits the duration of existence of a heavy particle beam in the stationary state after damping. On the other hand, observation of this process (count of exiting neutrals) can serve as a way to control the beam interaction parameters. And, finally, recombination when cooling antiprotons by positrons is presently the only conceivable way of producing a sufficiently intense beam of anti-atoms for physics experiments [13].

The cross section of radiative recombination has a substantially different behavior as a function of electron energy $W = mv^2/2$ depending on the relation between $W$ and the potential $I = z^2 e^4 m/(2\hbar^2)$. For the region of interest $W \ll I$, one can give a semi-classical estimate of the recombination cross section used in Ref. [2]. Let us consider recombination to the $n$-th level of an atom, $n \gg I$. Then it is natural to assume that electron motion is almost classical and one needs to account only for the intermittent nature of radiation of an electron accelerated by the nucleus (quantum fluctuations of radiation). The picture of the process looks as follows. Moving slowly from afar, being attracted by a nucleus, an electron comes close to it (the lower the velocity $v$ and impact parameter $\rho$ the closer) and, going around the nucleus with a turn of its (increased) velocity by almost 180°, the electron can radiate a quantum $\hbar\omega$. The limitation of the classical picture is the condition $M = mv\rho \gg \hbar$; the electron then does not approach the nucleus closer than the Bohr radius. This condition also ensures that the radiation is quasi-classical: $\hbar\omega \ll mv^2/2$ where $\omega \simeq v_{max}/r_{min}$, $v_{max} = ze^2/M$, $r_{min} = M^2/(ze^2 m)$. Let us write a recombination cross section summed over all quasi-classical levels:

$$\sigma_{rec} = \int_{\omega=\frac{mv^2}{2\hbar}}^{\omega<\frac{I}{\hbar}} \frac{d\omega}{\hbar\omega} \int_{\rho=\frac{\hbar}{mu}}^{\infty} J_\omega(\rho)\rho d\rho,$$

where $J_\omega(\rho)d\omega$ is the classical intensity of radiation in the interval $d\omega$ when moving on a trajectory with an impact parameter $\rho$. Integration of the known expression $J_\omega(\rho)$ for the radiation intensity when scattering on an attractive Coulomb center in the considered conditions leads to the result:

$$\sigma_{rec} = \frac{16\pi z^2 e^6}{3\sqrt{3} m^2 v^2 c^3 \hbar} \ln \frac{2I}{mv^2} \quad \text{for} \quad mv^2 \ll 2I.$$

For Maxwellian electron distribution in velocity, the recombination life time of a damped beam equals

$$\tau_{rec} \simeq 137 \frac{3}{32} \sqrt{\frac{6}{\pi}} \frac{v_{eT}(\gamma mc)^2}{z^2 e^4 n_e \eta \ln\left(\frac{z\alpha c}{v_{eT}}\right)^2} \;; \quad z\alpha c \gg v_{eT}.$$

Ratio of the life time to the cooling time, in practical situations, is large:

$$\frac{\tau_{rec}}{\tau} \simeq 137 \frac{\sqrt{3}}{4} \frac{m}{M} \frac{mc^2}{T_e} \frac{L}{\ln(z\alpha c/v_{eT})}.$$

If one writes this ratio in terms of the angular spread, it takes the form:

$$\frac{\tau_{rec}}{\tau} \simeq 137 \frac{\sqrt{3}}{2} \frac{mc^2}{(\gamma+1)W\theta^2} \frac{L}{\ln\left(\frac{z\alpha}{\gamma\beta\theta}\right)},$$

where $W = (\gamma - 1)Mc^2$ is the kinetic energy of the heavy particles, $\theta^2 = \theta_e^2 + \theta_i^2$. It is important to note that, in the region $\gamma\beta\theta < \alpha z$, the recombination time has essentially the same as the cooling time dependence on the heavy particle charge $\sim 1/z^2$. However, for given kinematic parameters $(\beta, \theta)$, the ratio $\tau_{rec}/\tau$ reduces when going to heavy nuclei due to the relative increase of the cooling time: $\tau \sim M$. Reduction of the life time with increase of $z$ may, of course, itself be undesirable for applications where it may be required to maintain the beam in a cooled state for an extended period of time using electron cooling. This difficulty can be overcome by switching to a "proton" cooling of heavy ions [13] (after initial cooling by electrons), i.e. by cooling an ion beam by a proton beam circulating in an adjacent storage ring and already cooled (or being cooled) by electrons. Despite the large proton mass in comparison to the electron one, such a technique even has an advantage in terms of the cooling decrement due to the mass parameter, since, at a given temperature (equal to the electron temperature), the cooling decrement increases by a factor of $\sqrt{M_p/m}$ when switching to cooling particles with a large mass:

$$\tau^{-1} = \frac{8\sqrt{2\pi}}{3} \frac{\eta n_p z^2 e^4 L \sqrt{M_p}}{\gamma^2 T_p^{3/2} M_i}, \quad \frac{T_p}{M_p} > \frac{T_i}{M_i}.$$

A realistic time ratio depends on the densities (or currents) off all three beams.

The recombination cross-section for large relative velocities $u \gg z\alpha c$ (or $\gamma\beta\theta > z\alpha$) is given in [27]:

$$\sigma_{rec} = \zeta(3) \frac{2^7 \pi}{3} z^5 \alpha^3 \left(\frac{e^2}{\hbar u}\right)^5 \left(\frac{\hbar}{me^2}\right)^2.$$

One can propose an interpolation formula [2] for the time ratio $\tau_{rec}/\tau$ connecting the two regions:

$$\frac{\tau_{rec}}{\tau} \simeq \frac{m}{M} L \left\{ \alpha(\gamma\beta\theta)^2 \ln\left[1 + \left(\frac{z\alpha}{\gamma\beta\theta}\right)^3\right] \right\}^{-1}.$$

It is curious to note that, despite the widespread of the recombination effect and its importance for many applications, until recently, there were no exact expressions for the probability of recombination to an arbitrary level. Just recently such formulae were obtained in [28].

The following estimate [28] of the recombination probability due to triple collisions can easily be obtained from elementary considerations:

$$\tau_3^{-1} \simeq \eta \left(n_e \sqrt{\frac{T_e}{m}}\right) \cdot \left(\frac{e^2}{T_e}\right)^2 \left[n_e \left(\frac{ze^2}{T_e}\right)^3\right] \frac{1}{\gamma^3} = \frac{\eta}{\gamma^3} z^3 n_e^2 \left(\frac{e^2}{T}\right)^5 \sqrt{\frac{T}{m}}.$$

The ratio of the radiative and collisional recombination times is of the order of magnitude of:

$$\frac{\tau_{rec}}{\tau_3} \simeq 20 z n_e r_e^3 \frac{(mc^2/T_e)^4}{\gamma \ln(I/T_e)}.$$

If one takes the factor of 20 in the numerator seriously then, for $T_e = 0.2$ eV, $n_e = 10^8$ cm$^{-3}$ and $\gamma = z = 1$, we get $\tau_{rec}/\tau_3 \simeq 10^{-3}$. However, with reduction in the temperature and increase in the density, the collisional recombination quite quickly becomes dominant. This particular process can possibly be used in the future for faster production of anti-hydrogen than by using radiative recombination.

## 1.2 Dependence of the relaxation process on the electron velocity distribution

In an experimental setting, the distribution $f(\vec{v}_e)$ will usually significantly deviate from a Maxwellian (isotropic) one either as a result of deformation at the beam "preparation" stage or in the autonomous mode due to processes determining the equilibrium state. Another independent parameter of non-equilibrium is the difference of the average beam velocities (generally speaking, local at each common point of the beam volumes).

### 1. Effect of anisotropy of velocity spread

Let us first estimate the effect of anisotropy assuming that there are no "macroscopic" relative flows. A characteristic situation may be when electron longitudinal velocity spread in the co-moving frame is small compared to the transverse ones (a disc-like distribution):

$$f(\vec{v}_e) = \left\{(2\pi)^{\frac{3}{2}}\Delta_{e\perp}^2 \Delta_{e\|} \exp\left(\frac{v_{e\perp}^2}{2\Delta_\perp^2} + \frac{v_{e\|}^2}{2\Delta_\|^2}\right)\right\}^{-1} ; \quad \Delta_\perp \gg \Delta_\| . \tag{1.48}$$

We are, of course, interested in the region $v < \Delta_{e\perp}$ since otherwise $\vec{F}(\vec{v})$ is independent of the distribution details. Calculations give the following answer (the disc field):

$$\vec{F}_\perp = -\pi\sqrt{2\pi}\frac{n_e' z^2 e^4 L(\Delta_{e\perp})}{m}\frac{\vec{v}_\perp}{\Delta_\perp^3}, \tag{1.49}$$

$$\vec{F}_\| = -\frac{4\pi n_e' z^2 e^4}{m\Delta_{e\perp}^2}\begin{cases}\frac{v_\|}{|v_\||}L(v_\|) - \frac{v_\|}{\Delta_\perp}\sqrt{\frac{\pi}{2}}L(\Delta_\perp); & v_\| > \Delta_\|, \\ \frac{v_\|}{\Delta_\|}\sqrt{\frac{2}{\pi}}L(\Delta_\|); & v_\| < \Delta_\|,\end{cases} \tag{1.50}$$

$$d_\perp = (2\pi)^{3/2} n_e' z^2 e^4 L(\Delta_\perp)/\Delta_\perp .$$

From here it follows (see Eq. (1.39)) that, without the effect of coupling of the decrements (due to the radial gradient of $F_\|$), the equilibrium temperatures are

$$T_{zst} = \frac{1}{2}T_{e\perp}, \quad T_{xst} = \left(1 + \frac{\gamma^2\psi^2}{\beta_x^2}\right)\frac{1}{2}T_{e\perp},$$

$$T_{\|st} = \begin{cases} \dfrac{\pi}{4}\sqrt{T_{e\|}T_{e\perp}}\dfrac{L(\Delta_\perp)}{L(\Delta_\|)}; & \dfrac{\Delta_\|}{\Delta_\perp} > \dfrac{m}{M}\sqrt{\dfrac{\pi}{8}}\dfrac{L(\Delta_\perp)}{L(\Delta_\|)}, \\ \dfrac{m}{M}T_{e\perp}\dfrac{\pi}{8}\left(\dfrac{L(\Delta_\perp)}{L(\Delta_\|)}\right)^2; & \dfrac{\Delta_\|}{\Delta_\perp} < \dfrac{m}{M}\sqrt{\dfrac{\pi}{8}}\dfrac{L(\Delta_\perp)}{L(\Delta_\|)}. \end{cases} \quad (1.51)$$

Due to the large magnitude of longitudinal friction when $v < \Delta_\perp$, the longitudinal temperature of heavy particles turns out to be small compared to $T_{e\perp}$.

In a more general case, the partial temperatures of an electron beam can be significantly different in all three degrees of freedom. Let us consider a "three-axis ellipsoid" model of the distribution with $\Delta_x \gg \Delta_z \gg \Delta_\|$. For the ion velocities lying inside the ellipsoid, the friction force is:

$$\vec{F} = -\frac{12\pi z^2 e^4 n'_e}{m}\left(\frac{v_x}{\Delta_x^3}\ln\frac{\Delta_x}{\Delta_z}, \frac{v_z}{\Delta_x\Delta_z^2}, \frac{v_\|}{\Delta_x\Delta_z\Delta_\|}\right); \quad (1.52)$$

while the diffusion is

$$d_{xx} \simeq \frac{6\pi z^2 e^4 n'_e}{\Delta_x}, \quad d_{zz} = d_\| = \Delta_{xx}\ln\frac{\Delta_x}{\Delta_z}.$$

The respective equilibrium temperatures are:

$$T_{xs} = \frac{1}{2}T_{xe}\left(1/\ln\frac{\Delta_x}{\Delta_z} + \frac{\gamma^2\psi^2}{\beta_x^2}\right);$$

$$T_{zs} = \frac{1}{2}T_{ze}\ln\frac{\Delta_x}{\Delta_z}; \quad T_{\|s} = \frac{1}{2}\sqrt{T_{ze}T_{\|e}}\ln\frac{\Delta_x}{\Delta_z}. \quad (1.53)$$

The considered examples are characteristic for an ultra-relativistic circulating electron beam: in the laboratory frame, the longitudinal and radial momentum spreads are usually of the same order of magnitude and then, in the co-moving frame, $v_{xe} \simeq \gamma v_{e\|}$. The vertical spread is determined by $x - z$ coupling and is also small compared to the radial one. However, Eqs. (1.51) and (1.53) do change if the electron distribution has a radial ($x$) gradient, for example, when there is dependence $v_{e\|}(x)$, usual for a circulating beam. Coupling of betatron oscillations may also be significant for the equilibrium characteristics. The minimum equilibrium temperatures may, of course, also depend on other factors, for example, intra-beam scattering (Chapter V).

A disc-like distribution occurs also for electrostatic acceleration of electrons in case of a single-pass beam; however, the cooling process must then be considered including the effect of the accompanying magnetic field.

## 2. Monochromatic instability

Let us next investigate the dependence of the damping rate and established amplitudes on the mismatch of the average velocities of the electron and ion beams. Suppose the electron Maxwell distribution is "shifted" in the co-moving frame by $\langle \vec{v} \rangle = \vec{\Delta}$: $f(\vec{v}_e) = f_T(\vec{v}_e - \vec{\Delta})$ (an error in the electron average velocity). If $\Delta < v_{eT}$ Eqs. (1.40) remain valid since the arising average friction force $\langle \vec{F} \rangle \simeq +\vec{\Delta}/v_{eT}^3$ does not contribute to $Q_i$ and the shift $\Delta < v_{eT}$ does not change the friction characteristics determining the damping decrements.

A completely different situation takes place if $\Delta > v_{eT}$. Suppose an error $\vec{\Delta}$ is directed along the normal degree of freedom 1. Let us find the average friction power in this degree of freedom for small oscillations:

$$Q_1 \simeq -g\frac{1}{m}v_1\frac{v_1 - \Delta}{[(v_1 - \Delta)^2 + v_{eT}^2]^{3/2}} \simeq 2g\frac{1}{m}\frac{\overline{v_1^2}}{\Delta^3}, \tag{1.54}$$

$$(g = 4\pi z^2 e^4 n'_e L)$$

if $\Delta \gg v_{eT}$ and $\Delta \gg |v_1|$.

Thus, $Q_1$ changes sign and oscillations in this degree of freedom grow. The degrees of freedom transverse to the error $\vec{\Delta}$ remain stable:

$$Q_{\alpha\perp} \simeq g\frac{-1}{m}\frac{\overline{v_\alpha^2}}{\Delta^3}, \quad (\dot{I}_\alpha)_{fl} \simeq \frac{g}{M\Delta}. \tag{1.55}$$

The reason for appearance of instability in the direction $\vec{\Delta}$ (it is significant here, however, that $\vec{\Delta}$ is directed along a normal degree of freedom) is that, for a large shift of the average velocity $\Delta > v_{eT}$, small oscillations move to a region where the friction characteristic is negative and oscillation energy accumulates (see Fig. 3).

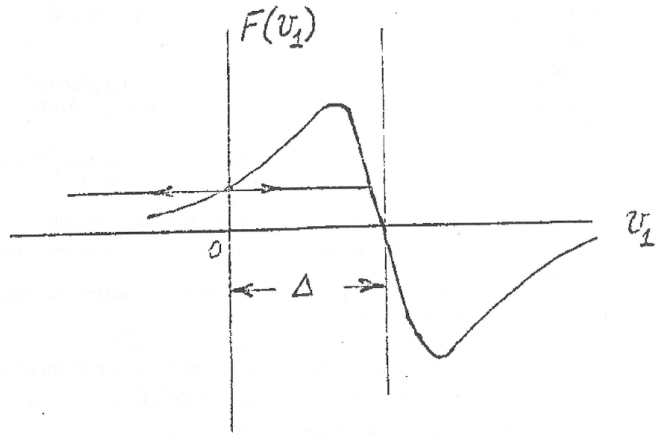

Fig. 3.

Although estimate (1.54) is obtained for the condition $\Delta \gg v_{eT}$, it is clear that, for instability of small oscillations to occur, it is sufficient for the friction characteristic to change sign when the velocity is shifted. For the degrees of freedom transverse to $\vec{\Delta}$, appearance of the error is equivalent to increase in the electron beam temperature with respect to $(\Delta/v_{eT})^2$ while the friction characteristic remains positive.

Under the condition $\Delta \gg v_{eT}$, when considering small ($v < \Delta$) oscillations, non-monochromaticity of the electron beam can be neglected. The induced instability then has a simple interpretation: a pendulum is being acted upon by "wind"; besides, the friction force has a negative characteristic and is small compared to the elastic force. Oscillations then happen about practically unchanged equilibrium but become unstable: energy gain during the half period of motion "along the wind" exceeds the losses during motion "against" it.

Let us now estimate the established average amplitudes. As shown in Fig. 3, buildup continues, at least, until the velocity amplitude $v_0$ approaches $\Delta$:

$$(\Delta - v_0)/v_{eT} \simeq (v_{eT}/\Delta)^2 \ll 1.$$

With further amplitude growth, a phase region $\sin\psi \geq \Delta/a$ comes into play where the friction force has opposite sign and compensates on average the region $\sin\psi \leq \Delta/a$.

In the region $|v_0 - \Delta| \lesssim v_{eT}$, the power $Q_1$ changes in the range $\pm|Q|_{max}$ turning to zero at a certain point $v_{0s} > \Delta$, $v_{0s} - \Delta \approx v_{eT}$. One can estimate the order of magnitude of $|Q|_{max}$:

$$|Q|_{max} \simeq v_{0s}|F|_{max}\frac{\delta\psi}{2\pi} \simeq g\frac{1}{\pi m}\frac{\Delta}{v_{eT}^2}\sqrt{\frac{v_{eT}}{\Delta}} = g\frac{1}{\pi m v_{eT}}\sqrt{\frac{\Delta}{v_{eT}}}, \quad (1.56)$$

where $\delta\psi \simeq \sqrt{v_{eT}/\Delta}$ is the fraction time that the particle spends in the region $|v_{0s} - \Delta| \approx v_{eT}$. This estimate is confirmed by a model calculation given in Appendix III.

In the region $|v_0 - v_{0s}| \lesssim v_{eT}$, oscillation amplitudes thus damp to $v_{0s}$ with a decrement

$$\delta \simeq \frac{|Q|_{max}}{Mv_{eT}\Delta} \simeq \frac{g}{mMv_{eT}^2\sqrt{v_{eT}\Delta}}.$$

Since the rate of diffusion on electrons has an order of magnitude of

$$\left(\frac{dv_0^2}{dt}\right)_{fl} \simeq \frac{g\delta\psi}{2\pi M^2 v_{eT}} \simeq \frac{g}{M^2\sqrt{v_{eT}\Delta}},$$

we get for an equilibrium amplitude spread

$$\langle(v_0 - v_{0s})^2\rangle = \frac{1}{\delta}\left(\frac{dv_0^2}{dt}\right)_{fl} \simeq \frac{m}{M}v_{eT}^2. \quad (1.57)$$

Thus, in the presence of detuning $\Delta \gg v_{eT}$, the equilibrium distribution of oscillators is concentrated near the amplitude $v_{0s} \approx \Delta$ with the same absolute amplitude spread as in thermodynamic equilibrium. At the same time, the relative spread is small:

$$\frac{\langle (v_0 - v_{0s})^2 \rangle}{v_{0s}^2} \simeq \frac{m}{M} \frac{v_{eT}^2}{\Delta^2}.$$

This feature is what distinguishes dissipative "heating" from thermal. The energy is transferred to an oscillator not form thermal but from ordered motion of electrons. For this reason, the amplitude distribution is concentrated in a narrow band near the average value.

For oscillations in a direction transverse to $\vec{\Delta}$

$$Q_\alpha \simeq -\frac{g}{m} \langle \frac{v_\alpha^2}{[(v_1-\Delta)^2 + v_{eT}^2]^{3/2}} \rangle \simeq \frac{g}{m} \frac{\overline{v_\alpha^2}}{\pi v_{eT}^2 v_{0s}}, \tag{1.58}$$

$$\dot{I}_\alpha \simeq \langle \frac{g}{M} \frac{(v_1-\Delta)^2}{[(v_1-\Delta)^2 + v_{eT}^2]^{3/2}} \rangle \simeq \frac{g/M}{\pi v_{0s}} \ln \left( \frac{v_{0s}}{v_{eT}} \right), \tag{1.59}$$

giving

$$\frac{\overline{v_\alpha^2}}{v_{eT}^2} \simeq \frac{m}{M} \ln \left( \frac{v_{0s}}{v_{eT}} \right). \tag{1.60}$$

Practically the same amplitudes are thus established in the transverse degrees of freedom as in thermodynamic equilibrium. Compared to the "start" of the process when $v^2 \simeq v_{eT}^2$, the damping time is reduced with respect to $\Delta^2/v_{eT}^2$.

Note that, from Eqs. (1.54) and (1.55), it follows that the sum of the oscillation decrements is equal to zero if they are defined as $\tau_i^{-1} = (Q_i/I_i)$:

$$\tau_1^{-1} + \tau_2^{-1} + \tau_3^{-1} \sim (-2\Delta^{-3} + \Delta^{-3} + \Delta^{-3}) = 0. \tag{1.61}$$

This approximate result is a particular case of a more general statement obtained in Section 1.4.

Let us now consider the case when $\Delta$ has components of the same order of magnitude in two or all three normal degrees of freedom. For small oscillations

$$Q_\alpha \simeq -\frac{g}{m} \langle \frac{v_\alpha(v_\alpha - \Delta_\alpha)}{[(\vec{v}-\vec{\Delta})^2 + v_{eT}^2]^{3/2}} \rangle \simeq \frac{g}{m} \frac{\Delta^2 - 3\Delta_\alpha^2}{\Delta^5} \overline{v_\alpha^2}; \tag{1.62}$$

appearance of error in other degrees of freedom can thus compensate an instability in the given degree of freedom, since it is equivalent to increase in temperature of the electron gas as noted above. The force characteristic remains positive if $3\Delta_\alpha^2 < \Delta^2$ despite the fact that $\Delta_\alpha^2 > v_{eT}^2$. At the same time, as can be seen from Eq. (1.62), the sum of the decrements remains equal to zero. In reality, as will be shown later, it is difficult to avoid instability with a large error $\Delta \gg v_{eT}$ (although theoretically possible); on the other hand, if $\Delta$ slightly exceeds $v_{eT}$ (but in such a way that small oscillations lie in a region of negative characteristic if $\vec{\Delta}$ is directed along a normal degree of freedom), then all oscillations will damp under the condition that $\Delta_1 \approx \Delta_2 \approx \Delta_3$. Such anisotropy of damping along the direction of $\vec{\Delta}$ is explained by the existence of preferred directions of normal oscillations (a non-degenerate three-dimensional oscillator).

Rise of instability for a distribution of type $\sim \exp[-(\vec{v} - \vec{\Delta})^2/(2v_{eT}^2)]$ is a specific property of oscillatory motion. If the motion is infinite (no auto-phasing in "synchrotron" motion) instability does not occur, instead, one flow is drawn by the other.

Although we considered here the case when the velocity of electron flow in the co-moving frame is different from zero, this is not necessary for appearance of instability. For example, if we take $f(\vec{v}_e)$ as two shifted Maxwell distributions $\sim \exp[-(\vec{v} \pm \vec{\Delta})^2/(2v_{eT}^2)]$ then $\langle \vec{v} \rangle = 0$, however, the friction characteristic at $v = 0$ will be negative in the shift direction if $\Delta > v_{eT}$ that will lead to instability. In general, for emergence of instability, it is necessary that the energy of ordered motion in the electron flow exceeds the thermal one[*], i.e. the distribution must be qualitatively different from Maxwellian. A substantial feature here is that the established value of the oscillator energy, in its order of magnitude, is $M/m$ times greater than the energy of "ordered electron motion", since it is velocities, and not temperatures, that equalize: $v_{0s}^2 \approx \Delta^2$.

S.T. Belyaev and G.I. Budker [29] noted also the case of a spherical distribution

$$f(\vec{v}_e) \sim \delta(v^2 - v_0^2), \tag{1.63}$$

the friction force and momentum transfer are then equal to zero[†] if $v < v_0$ (the field of a charged sphere) and the ion beam heats up: $\overline{v_{st}^2} \simeq L v_0^2 m/M$. This case clearly demonstrates properties of the Coulomb interaction, although it seems to be practically an exception.

Let us also estimate the damping rate in the case when the error $\Delta$ oscillates in time or along the beam. Suppose $w(\vec{\Delta})$ is the error probability distribution: $\int w(\vec{\Delta}) d^3\Delta = 1$. The mean value of the friction force can be written as

$$\langle \vec{F} \rangle = \int \vec{F}(\vec{v} - \vec{\Delta}) w(\vec{\Delta}) d^3\Delta, \tag{1.64}$$

where $f(\vec{v}_e)$ is the electron velocity distribution with respect to the mean velocity value $\langle \vec{v} \rangle_T = \vec{\Delta}$ close in shape to Maxwellian: $\langle (\vec{v} - \vec{\Delta})^2 \rangle \simeq v_{eT}^2$. We are interested in the case $\Delta^2 \gg v_{eT}^2$. If oscillations happen in three dimensions then, obviously, for all degrees of freedom, this is equivalent to an increase in the electron thermal spread to a value of $\overline{\Delta^2}$ (it is assumed that the distribution $w(\Delta)$ is bell shaped). In case of one-dimensional oscillations directed along a normal degree of freedom, for small oscillations of ions

$$Q_1 \simeq -\frac{g}{m}v_1^2 \int d^3\Delta w(\vec{\Delta}) \frac{\partial}{\partial \Delta_1} \frac{\Delta_1}{(\Delta_1^2 + v_{eT}^2)^{3/2}} \simeq \frac{g}{m} \frac{\overline{v_1^2}}{\langle \Delta^2 \rangle^{3/2}} \ . \tag{1.65}$$

---

[*] It is this criterion that justifies the name of the instability.

[†] More precisely, they are reduced by a factor of $L$ in comparison to the case of a Maxwell distribution.

The friction characteristic thus remains positive although the effective temperature of the electron beam increases as in the case of three-dimensional oscillations. For the other degrees:

$$Q_\alpha \simeq -\frac{g}{m}v_1^2 \int \frac{\overline{v_\alpha^2} w(\vec{\Delta}) d^3\Delta}{(\Delta^2 + v_{eT}^2)^{3/2}} \simeq -\frac{g}{m} \frac{\overline{v_\alpha^2}}{v_{eT}^2 \langle \Delta \rangle} \ . \tag{1.66}$$

This result is easily explained: due to a sharp dependence of the "transverse" friction force on the error $\sim \Delta^{-3}$, the main contribution to the power comes from the region $\Delta \simeq v_{eT}$; the fraction of time "spent" by the error in this region is $v_{eT}/\langle \Delta \rangle$.

Note that, for $\delta$-like oscillations $w(\vec{\Delta}) \sim \delta(\vec{\Delta}_\perp) \delta(\Delta_1^2 - \Delta_0^2)$, the integral in Eq. (1.64) gives the previous result (1.54), i.e. instability, as it should be.

Thus, appearance of a varying error with a distribution $w(\Delta)$ under the condition $\overline{\Delta^2} \gg v_{eT}$ is equivalent to establishment of a stationary electron distribution $w(\vec{v}_e)$ (in case of one- or two-dimensional oscillations, the electron spread transverse to them remains equal to $v_{eT}$). Having this analogy in mind, one can generalize the above-stated qualitative instability (or heating) criterion also to the case of a non-stationary electron velocity distribution:

$$\overline{E^2} > \overline{(\Delta E)^2} = \overline{(E - \overline{E})^2}, \quad (E = \frac{mv_e^2}{2}), \tag{1.67}$$

where $\overline{(\ )}$ denotes averaging over the "instantaneous" distribution and over time. This condition is necessary but not sufficient. The strict necessary and sufficient condition is the formal requirement of the friction characteristic being negative (equal to zero) in the direction of normal oscillation.

## 1.3 Effects of spatial non-uniformity

Let us now consider the effect that spatial non-uniformity of an electron distribution $f(\vec{p}_e, \vec{r})$ has on the damping rate of small amplitudes. We will specify the spatial non-uniformity by gradients of the average velocity, temperature and density in the electron flow assuming that, without the gradients, the ion motion damps to an equilibrium $T_s = T_e$. For this purpose, it is sufficient to represent

$$f(\vec{v}_e, \vec{r}) = f_T(\vec{v}_e - \vec{\Delta}(\vec{r})) n'_e(\vec{r}), \tag{1.68}$$

where $f_T$ is a Maxwell-type distribution with a temperature $T_e = T_e(\vec{r})$ and the error $\vec{\Delta}(\vec{r})$ in the order of magnitude does not exceed the thermal velocity:

$$|\vec{\Delta}| < v_{eT}.  \tag{1.69}$$

Under this condition, the friction force for small proton velocities has the form (see Eq. (1.6))

$$\vec{F}(\vec{v},\vec{r}) \simeq \frac{g}{m}\frac{\vec{\Delta} - \vec{v}}{T_e^{3/2}} n_e'.  \tag{1.70}$$

Assuming uniform focusing, using Eq. (1.39), we get

$$\begin{aligned} Q_x &\sim -\overline{v_x^2}\left[\overline{(n_e' T_e^{-3/2})}_s + \frac{1}{\omega_0 v_x^2}\overline{\frac{\partial}{\partial x}(\Delta_\| n_e' T_e^{-3/2})}\right], \\ Q_z &\sim -\overline{v_z^2}\ \overline{n_e' T_e^{-3/2}}. \end{aligned}  \tag{1.71}$$

In the longitudinal direction, it is sufficient to get a force $F_\|$ averaged over betatron oscillations:

$$F_\| \sim -v_\|\left[\overline{(n_e' T_e^{-3/2})} - \frac{1}{\omega_0 v_x^2}\overline{\frac{\partial}{\partial x}(\Delta_\| n_e' T_e^{-3/2})}\right] + \Delta_\|^0 \overline{n_e T_e^{-3/2}},  \tag{1.72}$$

where $\Delta_\|^0 = \Delta_\||_{x=0}$ (in a mode without RF field, one should assume $\Delta_\|^0 = 0$, since the equilibrium velocity is determined from the condition $F_\| = 0$).

Axial oscillations thus always damp if condition (1.69) is satisfied while the radial and longitudinal friction powers contain terms proportional to the gradient of $F_\|(\vec{r})$ in the radial direction on the equilibrium orbit. These terms have to do with coupling of the radial and longitudinal motions, also called a closed-path effect, and give equal in magnitude but opposite in sign changes in the cooling decrements. By comparing Eqs. (1.71) and (1.72), we can obtain the following stability condition

$$\left|\frac{\partial}{\partial x}\left(\frac{n_e' r_\Delta}{T_e^{3/2}}\right)\right| < \left(\frac{n_e'}{T_e^{3/2}}\right), \quad \left(r_\Delta \equiv \frac{\Delta_\|}{v_x^2 \omega_0}\right).  \tag{1.73}$$

Although the influence of spatial non-uniformity disappears if $\Delta_\| = 0$, the condition $|\Delta_\|| > v_{eT}$ is not at all necessary for the emergence of "gradient" instability as in the case of "monochromatic" instability considered above. Suppose, for example, $(\partial/\partial x)(n_e' T_e^{-3/2}) = 0$. Then, from Eq. (1.73), it follows that instability is possible under the condition

$$\left|\frac{\partial}{\partial x}\Delta_\|\right| > v_x^2 \omega_s = \frac{dv_\|(x)}{dx},  \tag{1.74}$$

where $v_\|(x)$ is the azimuthal velocity as a function of the radial deviation of a proton trajectory. If $|v_\|| < v_{eT}$ then Eqs. (1.74) and (1.69) can be compatible since Eq. (1.74) essentially means that $|\Delta_\|| > |v_\||$. In case of $\Delta_\| = const$, the instability condition is

$$|\Delta_\|| > \left|\overline{\left(\frac{n_e'}{T_e^{3/2}}\right)}\bigg/\frac{\partial}{\partial x}\overline{\left(\frac{n_e'}{T_e^{3/2}}\right)}\right|\frac{dv_\|}{dx} \equiv b_x \frac{dv_\|}{dx},  \tag{1.75}$$

where $b_x$ is the size of instability if the relative change $\delta \ln(n'_e/T_e^{3/2}) \sim 1$. Combining the last two conditions, we can formulate a general qualitative criterion of gradient instability: in the size of radial non-uniformity, the error's magnitude $|\Delta_\parallel|$ averaged over radius must exceed the change in $v_\parallel(x)$. If there is no RF field instability can appear only in the presence of a gradient of the average velocity; in an auto-phasing mode, gradients of the density and temperature also contribute to the decrements if there is a velocity error on the equilibrium orbit. In practice, the velocity gradient is the most important.

Let us estimate the maximum amplitudes attainable at gradient instability. To do this without assuming smallness of $|\vec{v} - \vec{\Delta}|/v_{eT}$, let us take the force $\vec{F}$ in the form

$$\vec{F} \sim \frac{\vec{v} - \vec{\Delta}}{[(\vec{v} - \vec{\Delta})^2 + v_{eT}^2]^{3/2}} n'_e, \quad \vec{\Delta} = \{0, \Delta_\parallel, 0\}.$$

If radial betatron oscillations are unstable then one can consider $v_\parallel^2 \ll v_{eT}^2$, since the longitudinal motion will be damping (suppose $\Delta_\parallel^0 < v_{eT}$). Then

$$Q_x \sim -\langle \frac{v_x^2 + \omega_0 x_b \Delta_\parallel}{(\Delta_\parallel^2 + v_x^2 + v_{eT}^2)^{3/2}} \rangle, \quad \Delta_\parallel = \Delta_\parallel^0 + x_b \frac{\partial \Delta_\parallel}{\partial x}.$$

To be specific, let us suppose $\partial v_{eT}/\partial x = 0$, $\partial \Delta_\parallel/\partial x = \Delta_\parallel/r_0$ where $r_0$ is the radial beam size. Considering conditions (1.74) and (1.75), one can conclude that the established amplitude $a_x > r_0$ and is, in general case, determined by the equality $\overline{v_x^2} \simeq \langle \Delta_\parallel^2 \rangle$.

At the same time, it is not necessary that $\Delta_\parallel^2 > v_{eT}^2$. In case of unstable synchrotron motion,

$$F_\parallel \sim -\frac{v_\parallel - \Delta_\parallel}{[(v_\parallel - \Delta_\parallel)^2 + v_{eT}^2]^{3/2}} n, \quad \Delta_\parallel = \Delta_\parallel^0 + \frac{v_\parallel}{v_x^2 \omega_0} \frac{\partial \Delta_\parallel}{\partial x}.$$

If the motion is infinite the "anti-damping" stops when the radial deviation exceeds the beam size: $|v_\parallel| \simeq r_0 v_x^2 \omega_0$. In an auto-phasing mode, oscillations grow limitlessly. This is obvious if $\Delta_\parallel^2 = 0$ and condition (1.74) is satisfied. If $\partial \Delta_\parallel/\partial x = 0$ and condition (1.75) is satisfied then $Q_c > 0$ for all amplitudes, since the energy gain occurs only at "small" velocities $|v_\parallel| < |\Delta_\parallel|$ when the particle trajectory passes through the beam.

To demonstrate the effect of radial non-uniformity, we limited ourselves here to the case of isotropic velocity distribution. The influence of this factor is even more significant in situations when the longitudinal friction force is large compared to the transverse one (a disk-like distribution $f(\vec{v}_e)$). An example of this kind is considered in Section 3.5 in application to the problem of cooling by magnetized, electro-statically accelerated electrons.

# 1.4 Kinetics of small amplitudes.
# Sum of decrements

Let us consider in more detail kinetics in the region of small velocities $v < v_{eT}$ assuming that spatial non-uniformity can be characterized with a good precision by the gradient $f(\vec{v}_e, \vec{r})$. Since the scale of change in the friction force $\vec{F}(\vec{v}, \vec{r})$ is of the order of $v_{eT}$, it can then be expanded into a series:

$$\vec{F}(\vec{v}, \vec{r}) = \vec{F}_0 + (v \frac{\partial}{\partial \vec{v}}) \vec{F} + (\vec{r}_\perp \frac{\partial}{\partial \vec{r}_\perp}) \vec{F} + \ldots$$

Let us make an analogous assumption concerning the quadratic fluctuations:

$$\langle \Delta p_\alpha \Delta p_\beta \rangle = \langle \Delta p_\alpha \Delta p_\beta \rangle_0 + \ldots$$

Keeping the lowest-order terms of these expansions that give a non-vanishing contribution when averaging $Q_i$ and $D_{ik}$ over the phases (see Eqs. (1.33), (1.37), (1.39)), after the averaging, we get:

$$Q_i = -\lambda_i I_i, \tag{1.76}$$

$$D_{ik} = 2\mu_i I_i \equiv D_i, \quad i = k; \quad D_{ik} = 0, \quad i \neq k, \tag{1.77}$$

where

$$\lambda_x = \frac{1}{\gamma M} \langle -\frac{\partial F_x}{\partial x} + \frac{\psi}{\omega_s} \frac{\partial F_\parallel}{\partial x} \rangle_0, \tag{1.78}$$

$$\lambda_\parallel = \frac{1}{\gamma M} \langle -\frac{\partial F_\parallel}{\partial v_\parallel} - \frac{\psi_\omega}{\omega_s} \frac{\partial F_\parallel}{\partial x} \rangle_0, \tag{1.79}$$

$$\lambda_z = -\frac{1}{\gamma M} \langle \frac{\partial F_z}{\partial v_z} \rangle_0; \tag{1.80}$$

$$\mu_x = \frac{1}{2R} \beta_x \langle d_{xx} \rangle_0 + \frac{\gamma^2 \psi^2}{2 \beta_x^2} \langle d_\parallel \rangle_0,$$
$$\mu_\parallel = \frac{1}{2} \langle d_\parallel \rangle_0, \quad \mu_z = \langle \frac{\beta_z}{2R} d_{zz} \rangle_0. \tag{1.81}$$

In a mode without RF field, one should set $D_\parallel = 2\mu_\parallel$, since $I_\parallel = p_\parallel$.

The quantities $\lambda_i$ are the damping decrements of the effective phase-space volumes of the normal degrees of freedom.

The decrements of the radial and longitudinal motions contain terms proportional to the derivative of $F_\parallel$ with respect to the radial coordinate but the sum $\lambda_x + \lambda_\parallel$ does not depend on coupling of the radial and longitudinal motions. An analogous result is known in the accelerator theory for the decrements of radiation damping [21]. The complete sum representing an increment of beam phase-space density growth is determined by the divergence of the friction force with respect to velocity:

$$\sum_{i=1}^{3} \lambda_i = -\frac{1}{\gamma M} \langle \mathrm{div}_{\vec{v}} \vec{F} \rangle = -\frac{1}{\gamma M} \langle \frac{\partial \vec{F}(\vec{v})}{\partial \vec{v}} \rangle_{\vec{v}=0} \tag{1.82}$$

($\vec{v} = 0$ is the equilibrium orbit).

Appendix 2 shows that this relationship is universal when the decrements are defined as

$$\lambda_i \equiv -\frac{\partial}{\partial I_i} \langle \dot{I}_i \rangle$$

and depends neither on the nature of friction nor on coupling of the particle degrees of freedom in an external field.

Let us expand the divergence of the force $\vec{F}(\vec{v})$ for our case:

$$-\frac{\partial \vec{F}}{\partial \vec{v}} = \frac{4\pi z^2 e^4}{m} \frac{\partial}{\partial \vec{v}} \int L(|\vec{v} - \vec{v}_e|) \frac{\vec{v} - \vec{v}_e}{|\vec{v} - \vec{v}_e|^3} f(\vec{v}_e) d^3 v_e$$

$$= \frac{4\pi z^2 e^4}{m} \int \left[ \frac{\partial L(u)}{\partial u} \frac{1}{u^2} + 4\pi L(u) \partial(\vec{u}) \right] f(\vec{v}_e) d^3 v_e$$

$(\vec{u} = \vec{v} - \vec{v}_e)$.

The Coulomb logarithm (see Eq. (1.9a))

$$L(u) = \ln\left(\frac{mu^3 \tau_{eff}}{ze^2}\right)$$

can be considered equal to zero at $u = 0$. Finally, we get:

$$\sum \lambda = \langle \frac{4\pi z^2 e^4}{\gamma m M} 3 \int \frac{f d^3 v_e}{v_e^3} \rangle, \tag{1.83}$$

where the velocity $\vec{v}_e$ is counted at every point of space from the equilibrium ion orbit. The integration over velocity has a logarithmic divergence at $v_e = 0$, which should be cut off at the value

$$(v_e)_{min} = \left(\frac{ze^2}{m\tau_{eff}}\right)^{1/3}. \tag{1.84}$$

The result of Eq. (1.83) depends substantially on the relationship between the velocity spread (including anisotropy of $f(\vec{v}_e)$) and the difference of the average beam velocities $\langle \vec{v}_e \rangle$. In case when $\langle \vec{v}_e \rangle$ falls within the $f(\vec{v}_e)$ distribution width, the integral in Eq. (1.83) becomes equal to

$$\int f d^3 v_e / v_e^3 \approx 4\pi f(0) L(\Delta_{e\,min})$$

and the sum of the decrements becomes

$$\sum \lambda = \frac{16\pi^2 z^2 e^4}{\gamma m M} L(\Delta_{e\,min}) \langle f(0) \rangle. \tag{1.85}$$

For a spatially uniform Maxwellian electron distribution:

$$\lambda_x = \lambda_\parallel = \lambda_z = \frac{16\pi^2 z^2 e^4 L}{3\gamma m M} \langle f(0) \rangle \simeq \frac{8\sqrt{2\pi}}{3} \eta \frac{n_e z^2 e^4 L}{\gamma^2 m M v_{eT}^3} \tag{1.86}$$

that corresponds to Eq. (1.40). In the opposite limiting case, when $|\langle \vec{v}_e \rangle| \gg (\Delta_e)_{max}$, contribution of the "resonant" velocities $v_e \ll |\langle \vec{v}_e \rangle|$ is proportional to the tail of the distribution and then

$$\sum \lambda \approx \frac{12\pi z^2 e^4 n_e \eta}{\gamma m M |\langle \vec{v}_e \rangle|^3}. \tag{1.87}$$

In an intermediate situation, the sum of the decrements depends on partial temperatures of the electron beam as well as on the velocity mismatch.

In particular, from Eq. (1.87), it follows that all $\lambda_i$ can remain positive even if the error $\Delta \gg v_{eT}$. Then, however, their magnitude becomes quite small since it does not contain the Coulomb logarithm. This conclusion agrees with the results of the approximate study given in Section 1.2.

Note also that, under the condition of spatial uniformity, the decrement magnitudes cannot exceed the value of the sum in Eq. (1.85). Meanwhile, in case of strong spatial non-uniformity, as it follows from the results of Section 1.3, the value of $|\lambda_x - \lambda_\parallel|$ can become significantly greater than $\sum \lambda$ (but the maximum of the friction power does not depend on the gradients, of course).

In case of coupling of the $x$ and $y$ motions, redistribution of the decrements of betatron oscillations takes place in accordance with coupling parameters. However, for any variations of particle focusing and electron phase-space distribution, the following statement remains valid: the sum of the decrements is positive and it is independent of coupling of the degrees of freedom in a storage ring and is independent of gradients of the distribution function $\partial f(\vec{v}_e, \vec{r})/\partial \vec{v}_e$, $\partial f(\vec{v}_e, \vec{r})/\partial \vec{r}$. Within the bounds of theorem (1.83), it is practically possible to arrange an arbitrary redistribution of the decrements to optimize the cooling process and equilibrium distribution parameters using gradients of the longitudinal friction and coupling of the degrees of freedom.

Note also that effects of coupling and friction power redistribution may also be important in the region of non-linear friction at large amplitudes.

*The general solution of the kinetic equation* (1.38) in the linear case:

$$\frac{\partial}{\partial t}f - \sum_{i=1}^{3}\frac{\partial}{\partial I_i}(\lambda_i I_i + \mu_i I_i \frac{\partial}{\partial I_i})f = 0, \tag{1.88}$$

if all $\lambda_i > 0$, can be expressed through the fundamental one, or a Green's function

$$G(I \mid I',t) = \prod_i g_i(x_i \mid x_i',t), \tag{1.89}$$

where [30, 31]

$$g_i(x \mid x',t) = \chi_t^i \exp[\chi_t^i(x'e^{-\lambda_i t} - x)] I_0(\frac{\sqrt{xx'}}{\text{sh}(\lambda_i t/2)}),$$

$$\chi_t^i = (1 - e^{-\lambda_i t})^{-1},$$

($I_0$ is the Bessel function of an imaginary argument):

$$f(I,t) = \int d^3I' f(I',0) G(I \mid I',t).$$

The equilibrium distribution and evolution of average amplitudes can also be obtained directly from Eq. (1.88):

$$f_{st} = \prod_i \exp(-I_i/I_{ist}), \quad I_{ist} = \mu_i/\lambda_i,$$

$$\frac{d}{dt}\langle I_i \rangle = -\lambda_i \langle I_i \rangle + \mu_i,$$

as it should be.

In a coasting beam mode (without RF field), the fundamental solution of the equation for $f(p_\parallel, t)$

$$\frac{\partial}{\partial t}f - \frac{\partial}{\partial p_\parallel}(\lambda_\parallel p_\parallel + \mu_\parallel \frac{\partial}{\partial p_\parallel})f = 0$$

is

$$g(x \mid x',t) = \frac{1}{2\sqrt{\pi}}\left\{\sqrt{\text{cth}\frac{\lambda t}{2}} \exp\left[-\frac{(x - x'e^{-\lambda t})^2}{1 - e^{-2\lambda t}}\right]\right. \\ \left. + \sqrt{\text{th}\frac{\lambda t}{2}} \exp\left[-\frac{(x + x'e^{-\lambda t})}{1 - e^{-2\lambda t}}\right]\right\} \tag{1.90}$$

$$\left(x = \frac{p_\| \lambda}{2\mu}\right).$$

The equilibrium solution is $\exp(-x^2)$ while evolution of $\langle p_\|^2 \rangle$ is determined by the equation

$$\langle \frac{dp_\|^2}{dt} \rangle = -2\lambda_\| \langle p_\|^2 \rangle + 2\mu_\| .$$

## 1.5 Relaxation of a large spread. Sweeping

The term "large amplitudes" implies a general case when the kinetic coefficients cannot be linearized in terms of variables $I_i$. This may mainly be related to non-linear behavior of the friction force at velocities $v > v_{eT}$ or to strong spatial non-uniformity of the electron beam (for example, oscillation amplitudes exceed the transverse beam size).

Let us first investigate behavior of the kinetic process in case of special uniformity in the presence of fluctuation background (the diffusion rate due to it is assumed constant: $\langle \dot{I}_i \rangle_{fl} = \mu_i = const$). In the region $v > v_{eT}$, the coefficients $Q_i$ fall of as $v^{-1}$ (or faster, as $\sim v_i^2/v^3$ if $v_i^2 \ll v^2$); for sufficiently large amplitudes

$$|Q_i| < \mu_i ,$$

and the particles are not captured into a damping mode. Let us estimate the region of captured amplitudes in case of $|Q|_{max} \gg \mu$. For one-dimensional oscillations (see Eqs. (1.42)-(1.44)):

$$|Q_1| \simeq \langle \frac{4\pi L e^4 n_e}{m v_0 |\sin \psi|} \rangle \simeq \frac{2}{\pi} \frac{v_{eT}}{v_0} |Q|_{max} \ln \frac{v_0}{v_{eT}},$$

where $v_0$ is the velocity amplitude,

$$|Q|_{max} \simeq \frac{4\pi L e^4 n_e \eta}{m v_{eT}}.$$

Thus,

$$(v_0^{cr}/v_{eT}) \simeq \frac{2}{\pi} \frac{|Q|_{max}}{\mu} \ln \frac{|Q|_{max}}{\mu} .$$

For two- and three-dimensional oscillations, when averaging over the phases, the integrand does not have singularities, therefore

$$(v_0^{cr}/v_{eT}) \simeq |Q|_{max}/\mu .$$

The general picture of amplitude motion approximately described by the equations

$$\dot{I}_i = Q_i + \mu$$

is quite complicated but can be analyzed qualitatively. An idea about the nature of the process is given by Fig. 4, which shows trajectories of two-dimensional motion

$$\frac{dI_1}{\dot{I}_2} = \frac{dI_2}{\dot{I}_2}$$

at constant (or zero) $I_3$.

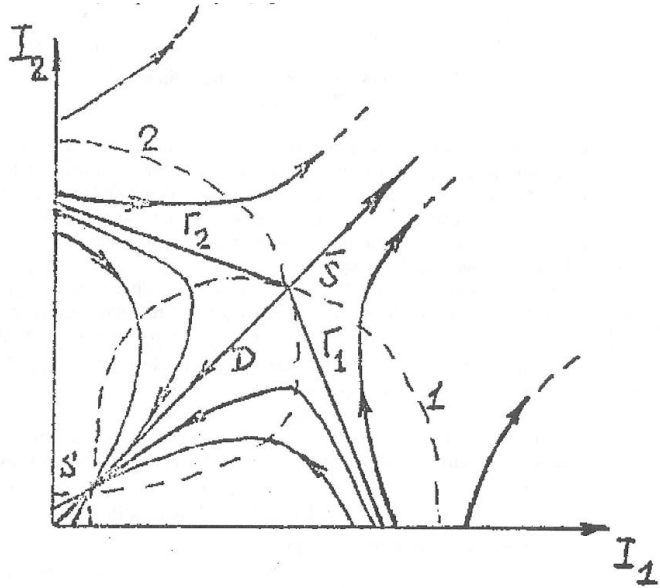

Fig. 4.

The dashed curves 1 and 2 correspond to the equations $\dot{I}_1 = 0$ and $\dot{I}_2 = 0$. Simultaneous damping of the amplitudes occurs only in the region $D$ enclosed by these curves. The curves $\Gamma_1$ and $\Gamma_2$ border the region of captured amplitudes. As seen from the figure, at a large excitation of one degree of freedom in the capture region, the other degree first "heats up" and then the trajectory comes to the region $D$ where both amplitudes damp. The points $S$ and $\bar{S}$ correspond to the stable and unstable equilibrium positions. When "adding" the third degree of freedom, the figure can be considered a projection of a three-dimensional picture onto a plane. The region $D$ becomes a "cocoon" while the general behavior of motion does not change.

Strictly speaking, a stationary distribution does not exist since the region of captured amplitudes is limited. However, one can talk about a quasi-stationary distribution and a particle life time in the capture region (or in a region of allowed amplitudes $I < I_{allow}$) if $|Q|_{max} \gg \mu$. The "equilibrium" distribution can be found from the equation

$$\sum_i \frac{\partial j_i}{\partial I_i} = \sum_i \frac{\partial}{\partial I_i}\left\{Q_i f - \mu_i I_i \frac{\partial f}{\partial I_i}\right\} = 0.$$

A solution can be found in a general form if $\mu_1 = \mu_2 = \mu_3 = \mu$ using properties of the friction force $\vec{F} = -\partial U/\partial \vec{v}$ [1]. Since $v_i = \sqrt{2I_i} \sin \psi_i$, then, under the assumptions made here,

$$Q_i = -v_i \frac{\partial U}{\partial v_i} = -2I_i \frac{\partial}{\partial I_i} \overline{U}.$$

Assuming $j_i = 0$, we get the equations

$$2\frac{\partial \overline{U}}{\partial I_i} f + \mu \frac{\partial f}{\partial I_i} = 0,$$

which have a common solution

$$f = c \exp\left(-\frac{2}{\mu}\overline{U}\right). \tag{1.91}$$

In accordance with what was said above, this solution cannot be normalized since $U \to const$ at infinity. Its use makes sense if, in the interval $0 \leq I \leq I_{allow}$, the majority of particles is concentrated in the region $I \ll I_{allow}$. The exponent can be written as

$$-\frac{2}{\mu} U = 2 \frac{|Q|_{max}}{\mu} \langle \frac{v_{eT}}{u} \rangle,$$

where $u = |\vec{v} - \vec{v}_e|$ while $\langle ... \rangle$ means averaging over the electron distribution. If the distribution is close to Maxwellian there is a solution in the region $v < v_{eT}$

$$f \sim \exp[-\frac{2\lambda}{\mu}(I_1 + I_2 + I_3)],$$

where $2\lambda$ coincides with the expression in Eq. (1.86). In case when $|Q|_{max} \gg \mu$, a "normalized" solution has the form

$$f = \left(\frac{2\lambda}{\mu}\right)^3 \exp\left(2\frac{|Q|_{max}}{\mu} \langle \frac{v_{eT}}{u} - \frac{v_{eT}}{v_e} \rangle\right). \tag{1.92}$$

The solution of Eq. (1.91) can be applied in practice for estimating the "tail" of the distribution and in case when $\mu_i$ greatly differ in magnitude by simply setting $v_i = 0$ for the degrees of freedom with small $\mu_i$.

Note that, from Eq. (1.91), it follows, according to the estimate in Section 1.2, that, with an error $\Delta > v_{eT}$, the distribution near $I = I_{st}$ has a Gaussian form:

$$f \sim \exp\left[-\frac{1}{2I_{st}}\left(\frac{\partial Q}{\partial I}\right)_{st}(I - I_s)^2\right].$$

It is easy to estimate that $\langle (I - I_{st})^2 \rangle \ll I_{st}^2$ if $v_{eT} \ll \Delta \ll v_0^{cr}$, i.e. the distribution in case of "monochromatic" instability is concentrated near $I_{st}$.

Finally, let us also discuss the dependence of the damping rate on the transverse sizes of the electron beam. Suppose the beam is positioned symmetrically with respect to the equilibrium ion orbit. In case of excitation of two-dimensional betatron oscillations, reduction of the transverse sizes $r_{0x}$, $r_{0z}$ is always beneficial, since the power is "accumulated" here at velocities $|v_i| \simeq v_{i0}$ independently of the sizes while the product of the density $n_e$ and the phase fraction, when the particle is in the beam, at least, does not decrease. Therefore, at a fixed current, the integral damping time can only decrease with size reduction.

For one-dimensional oscillations $v_0 \gg v_{eT}$, when the size in the direction of oscillations is reduced from a value $r_0 = v_0/\omega_b$ (oscillation amplitude) to some $r_0 < v_0/\omega_b = a_0$, the power decreases with respect to $\ln(v_0/v_{eT})$, since, at small velocities $v < v_{eT}$, which give the main contribution for $r_0 \gtrsim a_0$, the particle is outside the beam ($v = -v_0 \sin\psi$, $x = a \cos\psi$).

The situation is different when synchrotron motion is excited. In a mode without auto-phasing, there is an obvious effect of "bypass" of the beam when the synchrotron radius deviation exceeds $r_{0x}$. The same effect leads to a sharp reduction of the power $Q_\parallel$ in the oscillation mode as well, since $x_c \sim v_\parallel$. It is easy to estimate that, for amplitudes $v_{\parallel 0} \lesssim v_{eT}$, the power decreases with respect to $\zeta \simeq x_c^2/b_r^2$ when $r_{0x}^2$ is reduced from the value $r_{0x}^2 = x_c^2$. Otherwise, if $1 < (\overline{v_\parallel^2}/v_{eT}^2) < x_c^2/r_{0x}^2$, then $\zeta = (\overline{v_\parallel^2}/v_{eT}^2) \cdot (r_{0x}^2/x_c^2) > 1$. The radial size of the electron beam should thus be kept at the level of $r_{0x}^2 \simeq x_c^2$.

## Sweeping

In stationary conditions, a characteristic property of electron cooling is a cubic dependence of the cooling time on velocity spreads of the beams:

$$\tau \sim [(\Delta \vec{v}_e)^2 + v^2]^{3/2} \, ;$$

in the region $v^2 > \overline{(\Delta v_e)^2}$, efficiency of the method quickly drops with increase of the initial velocity spread in the cooled beam. However, capability to control the beam motion allows one, within certain conditions and limits, to increase the friction power and accelerate the cooling process.

Let us consider a simple example. Suppose the longitudinal velocity spread $\Delta v_\parallel$ in a coasting ion beam is large compared to the transverse one. In this case, for an overwhelming majority of particles, the longitudinal friction is inversely proportional to a square of the relative velocity $v_\parallel - \langle v_{e\parallel} \rangle$ reaching a maximum near $v_\parallel - \langle v_{e\parallel} \rangle \simeq \Delta v_e$ (see Fig. 5):

$$\left|\frac{dv_\parallel}{dt}\right|_{max} \simeq \eta \frac{4\pi z^2 e^4 n_e}{\gamma^2 mM(v_e^2 + v_\perp^2)}. \tag{1.93}$$

Suppose now that velocity of the electron beam changes in time with a derivative $dv_{e\parallel}/dt$ close to but somewhat lower than the value in Eq. (1.93) passing the whole width $\Delta v_\parallel$ of the ion distribution. The ion will then be sequentially captured into a relatively narrow region of maximum friction and will remain there drawn by the electron "broom". Thus, accelerated cooling of the longitudinal spread will take place to a value $v_e^2 + v_\perp^2$ with a gain in time of $\simeq (\Delta v_\parallel)^2/(v_e^2 + v_\perp^2)$ times; further flow of the cooling process is obvious.

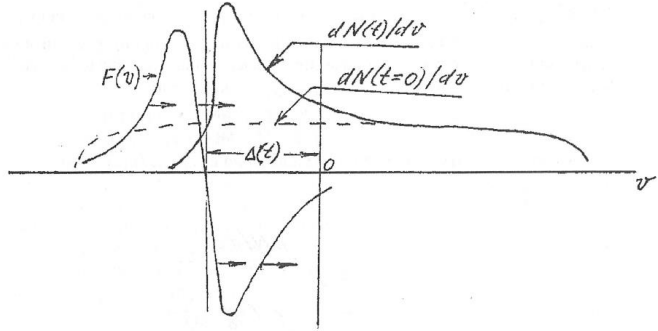

Fig. 5.

The approach of sweeping the longitudinal spread makes it useful to reduce transverse ion velocities in the cooling section through spatial expansion of the beam by defocusing lenses (increase of the beta-functions $\beta_x$, $\beta_z$) if there is a reserve of full electron current at a given density (the size of the cooled beam, when expanded, must not exceed the size of the electron one).

Sweeping can also be used for damping of highly excited oscillatory degrees of freedom although with less efficiency since constant "resonant" tuning of velocities is not possible here. Suppose that one-dimensional transverse oscillations are excited with maximum amplitude $v_{0max}$ and let us introduce an angle in this direction between the electron beam and ion closed orbit corresponding to a velocity $\Delta \simeq v_{0max}$. Then, as shown in Appendix 3 (see also Section 1.2 about monochromatic instability), oscillation amplitudes damp to a value $\Delta$; besides, behavior of the friction power near $v_0 = \Delta$ is similar to the dependence of the friction force on velocity (see Fig. 6); the difference is that the maximum of the friction power is reduced in comparison with Eq. (1.93) by a factor of $\Delta\psi = (\Delta/\sqrt{v_{eT}^2 + u_{tr}^2})^{1/2}$ ($u_{tr}$ is the rms difference of ion and electron velocities transverse to this direction), equal to the fraction of oscillation phases, when the velocity falls within an effective velocity interval $|v - \Delta| \lesssim v_{eT}$. Reducing $\Delta$ from the initial value $v_{0max}$ at a rate of

$$\dot{\Delta} \simeq \eta\sqrt{\frac{u_{tr}}{\Delta}}\frac{4\pi z^2 e^4 n_e L}{\gamma^2 mMu_{tr}^2}$$

can speed up the damping by approximately a factor of $(v_{0max}/u_{tr})^{3/2}$.

A simple transfer of the described approach to the case of two-dimensional excitation does not give an effect since the time here grows not only due to additional reduction of the fraction of effective phases but also due to a necessity to repeat sweeping multiple times for particles with different amplitude relations. Gain can be obtained only under the condition that the initial transverse size of the ion beam exceeds that of the electron beam, in an optimal situation, by a factor of about $(v_{0max}/\sqrt{v_{eT}^2 + (\Delta v_\parallel)^2})$. Correspondingly, to an order of magnitude, the maximum gain will be equal to a square of this factor.

Here we estimated only relatively obvious approaches to accelerated cooling of a large spread but one should also not exclude the possibilities of more sophisticated and effective techniques.

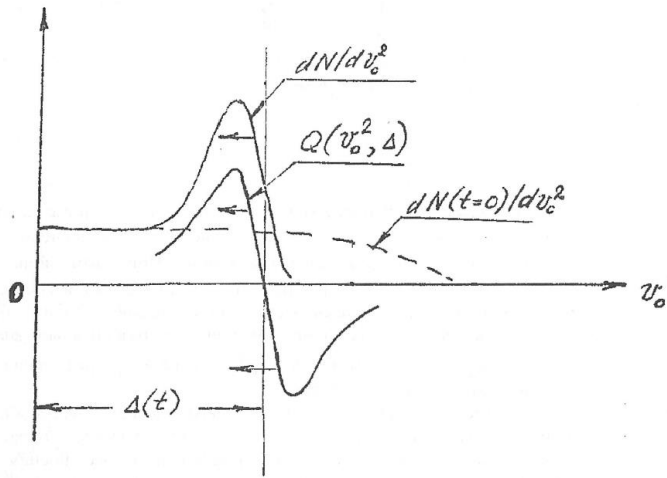

Fig. 6.

# II. INTERACTION OF HEAVY PARTICLES WITH MAGNETIZED ELECTRON BEAM

When studying general properties of electron cooling in the previous chapter, we restricted ourselves to cases of free (straight) electron motion, i.e. we assumed that there is no external field in the cooling section or it is too small to have notable effect on the motion of colliding particles. We will now consider interaction of heavy particles with electrons in an important case when the electron beam is accompanied by longitudinal magnetic field and rederive the collision integral.

## 2.1 Magnetized electron beam

At non-relativistic and moderately relativistic energies ($\gamma - 1 = W/(Mc^2) \lesssim 1$), the most practical way of obtaining a cooling beam is direct electrostatic acceleration of electrons coming out of the cathode of an electron gun [13]. In this case, to compensate repulsion of electrons by the space charge field, as well as an angular beam divergence, one focuses the beam using longitudinal magnetic field accompanying the beam from the cathode to the exit from the cooling section. In the arc sections of the trajectory (merge onto a straight ion orbit), one additionally introduces transverse magnetic field so that electron trajectories remain matched to the accompanying field.

### Larmor rotation. Transverse temperature

In magnetic field, the electron velocity component transverse to the field $\vec{v}_\perp$ rotates with the Larmor frequency

$$\vec{\Omega} = -\frac{e\vec{H}}{mc}. \tag{2.1}$$

In the co-moving frame, electrons then move in circles with radii

$$r_L = \frac{v_\perp}{\Omega}, \tag{2.2}$$

in the lab frame, they move in spirals with radii $r_L$ and steps

$$l_L = \frac{2\pi\gamma\beta c}{\Omega} = \frac{2\pi p_e(s)c}{eH}. \tag{2.3}$$

($s$ is the length along the beam orbit). For example, with $H = 1$ kG, $\beta = 0.3$ ($W_e = 35$ keV), $v_\perp/(\beta c) = 2 \cdot 10^{-3}$, the radius $r_L$ equals $10^{-3}$ cm while the step $l_L \approx 3$ cm.

If the adiabaticity conditions are satisfied for changes in the direction of the $H$ field and in the magnitude and direction of the full electric field acting on electrons, the magnitude of the transverse velocity is preserved: $v_\perp = const$ and the transverse electron temperature in the cooling section remains equal to the cathode temperature. In a more general case, magnetic field can adiabatically change in magnitude along the beam orbit and then, according to invariance of the parameter $v_\perp^2/\Omega$, the transverse temperature and beam density $n_e'$ (in addition to decrease due to acceleration) change proportionally to $H$. For the adiabaticity, it is necessary that characteristic lengths of field change are large compared to $l_L$.

## Longitudinal cooling in case of acceleration in electric potential

Let us now see what happens to the longitudinal spread of electron velocities. Since acceleration is in electron potential, then electron energy change is a single-valued function of position:

$$W_e - W_{0e} = -eU(\vec{r}), \tag{2.4}$$

where $U(\vec{r})$ is the potential (with respect to the cathode) of electro-static field including space-charge field as well (this relation does not, of course, account for fluctuations of potential energy related to discreteness of the charge distribution; more on this in Section 2.6). If one considers a spatial region that is small but contains a large number of electrons, then the spread of electron kinetic energies, as follows from Eq. (2.4), will be equal to the spread at the cathode determined by the temperature: $\Delta W_e = \Delta W_{0e} \simeq T_c \simeq W_{0e}$. Let us write the particle momentum as a sum

$$\vec{p}_e = \vec{p}(\vec{r}) + \delta\vec{p},$$

where $c\sqrt{p^2(\vec{r}) + m^2c^2} = |eU(\vec{r})| + mc^2 + W_0 = mc^2 + W_e$ and $\delta\vec{p}$ is the deviation related to the initial velocity at the cathode. Then, correspondingly,

$$\delta W = W - eU(\vec{r}) = \beta c \delta p_\parallel + \frac{p_\perp^2}{2\gamma m} + \frac{(\delta p_\parallel)^2}{2\gamma^3 m} = W_{0e},$$

which gives

$$\delta p_\parallel \approx (W_{0e} - \frac{p_\perp^2}{2\gamma m})/(\beta c). \tag{2.5}$$

Assuming that the distribution at the cathode is Maxwellian (cut in "half" for the longitudinal direction) and that the transverse momentum spread is preserved, we get for the longitudinal temperature after acceleration[*]:

$$T_\parallel = \frac{\langle(\delta p_\parallel)^2\rangle - \langle\delta p_\parallel\rangle^2}{\gamma^2 m} = \left(\frac{3}{2} + \frac{1}{\gamma^2} - \frac{2}{\gamma}\right)\frac{T_\perp^2}{(\gamma + 1)W_e}. \tag{2.6}$$

---

[*] Strictly speaking, the direction of average "hydrodynamic" velocity $c\langle\vec{\beta}(\vec{r})\rangle$ may not coincide with the direction of magnetic field; however, this difference is negligibly small in the considered aspect.

Thus, the longitudinal temperature (it is, of course, better to talk about the longitudinal velocity spread) turns out to be very small compared to the cathode temperature or, which is practically the same, compared to the transverse beam temperature. This happens due to preservation of the particle density in the phase space $(\vec{p}, \vec{r})$ with large spatial stretching of the beam in the longitudinal direction as a result of acceleration (and with approximate preservation of the transverse momentum spread):

$$\frac{\delta p_\parallel}{(\delta p_\parallel)_C} \simeq \left(\frac{n_{Ce}}{n_e}\right)^{-1} \simeq \left(\frac{\beta c}{\sqrt{T_C/m}}\right)^{-1}.$$

With the cathode temperature of $T_C \simeq 0.2$ eV and the electron kinetic energy of $eU = 35$ keV, Eq. (2.6) gives $T_{e\parallel} \simeq 10^{-7}$ eV. This temperature is so small that it has no real significance for either the electron dynamics or the cooling process of the proton beam. The actual limit on the minimum spread of longitudinal velocities is set by collisional interaction of electrons and, for the problem of cooling, by the interaction of ions with electrons itself.

## Velocity gradients

For the kinetics of electron cooling, what is important is not only the thermal electron velocity spread but the spatial one as well. Change in the average electron velocity $\langle \vec{v}_e \rangle$ as a function of coordinate $\vec{r}$ may occur due to several reasons.

1. A kick (non-adiabatic) impact of transverse electric fields related to imperfections of electron gun optics [32] will cause coherent (in a beam cross section) Larmor twist of electrons along the cooling section: $\vec{v}_L(s)$. Note that, although the coherent velocity can significantly exceed thermal ones, the corresponding Larmor radius is still very small, therefore spatial "vibration" of the beam can be neglected. Larmor velocities are thus determined in general case by the cathode temperature and imperfection of the beam formation system.

2. Adiabatically slow variations of the magnetic field direction in the cooling section lead to oscillations of the electron velocity direction (averaged over Larmor rotation) in phase with the oscillations of the field lines.

3. Transverse electric fields varying over the beam cross section and, in particular, space charge field cause transverse drift of Larmor circles with a velocity

$$\vec{v}_d(\vec{r}) = c\vec{E}(\vec{r}) \times \vec{H}/H^2, \quad (2.7)$$

depending on the coordinate.

4. Another factors related to electric field in the cooling section are the gradients of electrostatic potential $e\nabla U = -e\vec{E}(\vec{r})$ and, consequently, of the electron average longitudinal velocity as well:

$$\nabla v_{e\|} = \frac{e\vec{E}(\vec{r})c}{\sqrt{W_e(W_e + 2mc^2)}} \qquad (2.8)$$

(here $v_{e\|}$ is the velocity in the co-moving frame). "Residual" longitudinal electric field causes a change in $v_{e\|}$ along the beam effectively increasing the relative longitudinal velocities of ions and electrons. Transverse fields lead to dependence of the electron longitudinal velocity on transverse coordinates and, consequently, to coupling of the damping decrements.

5. Gradient of the magnitudes of Larmor velocities excited by the electron gun also leads to a gradient of the longitudinal velocity due to conservation of energy (in the lab frame) in an electrostatic "kick":

$$\nabla_\perp v_{e\|} = -\nabla_\perp v_\perp^2/(\gamma^2 \beta c) . \qquad (2.9)$$

Effect of all of the aforementioned factors is considered in Chapter III; meanwhile we will assume the electron flow to be uniform.

Note that a magnetized electron beam is possible in the option of cooling by a circulating beam at moderately relativistic energies.

## 2.2 Collisions with Larmor circles

Before we start the actual calculation of friction and diffusion of heavy particles in an electron flow including the Larmor rotation of electrons, let us first qualitatively consider the picture of collisions in a strong magnetic field, which can clarify the mechanism of sharp reduction of the effective electron beam temperature when $T_{e\|} \ll T_{e\perp}$. We will also see that smallness of only the longitudinal temperature alone ($T_\perp \simeq 2000°$) leads to situations that are not typical for kinetics of a "hot" plasma, whose description requires going beyond the usual routines.

We know that, in a Coulomb interaction, the momentum and energy exchange of colliding particles diverges logarithmically in the region of large impact parameters and must be cut off at some macroscopic parameter $\rho_{max}$, above which the interaction is effectively reduced. This makes it clear that, in collisions of heavy particles (whose trajectories can be considered straight) with electrons in magnetic field under the conditions when

$$r_L \ll \rho_{max}$$

($r_L$ is the electron Larmor radius), a substantial contribution to the collision integral can come from the region of impact distances $\rho$ satisfying the condition

$$r_L < \rho < \rho_{max} .$$

Duration of such collisions and intensity of the exchange do not depend on the electron Larmor velocity $v_{e\perp}$ and are determined only by the proton velocity with respect the Larmor "circle"

$$\vec{u}_A = \vec{v} - \vec{v}_{e\|} = \vec{v}_\perp + \vec{u}_\| .  \qquad (2.10)$$

Collisions also occur differently depending on the relation between the collision duration

$$\tau = \rho/u_A$$

and the Larmor period $2\pi/\Omega$. In case of $u_A/\rho \gg \Omega$, the collision is "instantaneous" with respect to the Larmor cycle and the collision result is the same as without magnetic field. In the opposite case, when $u_A/\rho \ll \Omega$, collisions occur adiabatically slow capturing a few or many cycles, so that the Larmor degree of freedom is effectively excluded from the kinematic and interaction dynamics and does not take part in exchange of momentum and energy – the exchange is completely due to the electron longitudinal degree of freedom. A particular consequence of the adiabaticity is also that the longitudinal momentum transfer turns to zero when the ion moves strictly along the magnetic field.

Thus, in the "logarithmic" approximation, all collisions in magnetic field are divided into two types:

1. **Fast collisions** with effective interaction time small compared to the electron Larmor period – a region of impact distances $\rho < u_A/\Omega$; contribution of such oscillations is not changed by magnetic field.

2. **Adiabatic collisions** with Larmor circles at distances $\rho > \max\{r_L, u_A/\Omega\}$.

A collision integral in strong magnetic field in this representation was first developed in a paper by S.T. Belyaev [38].

Since velocity $\vec{u}_A$ in adiabatic collisions plays a role of the relative velocity, their contribution to the friction decrement and diffusion rate will be analogous to the contribution of the usual (fast) collisions but with replacement of the relative velocity $\vec{u} = \vec{v} - \vec{v}_e$ with $\vec{u}_A = \vec{v} - \vec{v}_{e\|}$ and of the Coulomb logarithm with the logarithm of adiabatic collisions $L^A(u_A)$:

$$\lambda^A \simeq \frac{4\pi z^2 e^4 n_e L^A}{\gamma^2 m M u_A^3} \eta , \quad \frac{d}{dt}\langle(\Delta\vec{p})^2\rangle^A \simeq \frac{4\pi z^2 e^4 n_e L^A}{\gamma^2 u_A} \eta . \qquad (2.11)$$

For typical conditions, the value of $L^A$ is small compared to the logarithm of close (fast) collisions $L^0$.

Thus, when cooling by an electron beam with a velocity distribution $\Delta_{e\|} \gtrsim \Delta_{e\perp}$ as well as in an initial stage of cooling in an electron flow with a low longitudinal temperature, magnetic field cannot have a strong effect on the cooling process. The situation changes when $T_{e\|} \ll T_{e\perp}$ and $v \ll v_{e\perp}$. While the contribution of fast collisions remains the same, the duration of collisions with impact parameters $\rho > r_L$ becomes increased by a factor of $v_{e\perp}/u_A$ and, despite the relative smallness of the logarithm $L^A$, the contribution of adiabatic collisions becomes dominant. The cooling decrement then increases by a factor of

$$(v_{e\perp}/u_A)^3 (L^A/L^0) ,$$

which is a very strong effect. As it follows from Eq. (2.11), cooling continues until the ion temperature reaches a value determined by a certain effective spread of longitudinal (with respect to the magnetic field) electron velocities:

$$T_s \simeq T_{e\parallel} = m \overline{(\Delta v_{e\parallel})^2_{eff}} .$$

The effect of magnetization is sufficiently well described by the picture of adiabatic collisions while the logarithm $L^A$ is large. At the cooling stage when ion velocity become significantly lower than transverse electron velocities but are still not too small, the value of $L^A$ is a few units. The upper limit on the distances of adiabatic interaction $\rho^A_{max}$ equals either the kinematic parameter $u_A l/(\gamma\beta c)$ (the distance of relative movement during a pass of the cooling section) or the effective screening radius:

$$\rho^A_{max} = \min\{u_A l/(\gamma\beta c), r_{scr}\} . \tag{2.12}$$

Concerning the screening parameter $r_{scr}$, let us note the following. In thermodynamically equilibrium plasma, Coulomb interaction is screened at distances of the order of the Debye one:

$$r_{scr} \simeq r_D = \frac{v_{eT}}{\omega_e} = \sqrt{\frac{T_e}{4\pi n'_e e^2}} . \tag{2.13}$$

Otherwise, if the state of the interacting components (ions and electrons) is far from equilibrium, from general intuitive considerations, it is then clear that the screening radius should in the general case be determined from comparison of the kinematic interaction time $\tau_\rho = \rho/u$ ($u$ is the rms relative velocity of ions and electrons) and the characteristic time of collective electron response, i.e. the Langmuir period:

$$r_{scr}/u \simeq 1/\omega_e . \tag{2.14}$$

For real plasma, a significant difference of Eq. (2.14) from Eq. (2.13) would mean that the ion temperature is $M/m$ times greater than the electron one – a seemingly unlikely situation. However, in our case, the initial thermal velocities of heavy particles can be large, even compared to the transverse spread of electron velocities and definitely exceed manifold the longitudinal spread. Next, magnetization excludes transverse electron motion from the interaction dynamics at distances exceeding the Larmor radii of electrons, since the role of the effective temperature of electron medium is assumed by the longitudinal temperature while the role of relative velocity is assumed by the ion velocity with respect to a Larmor circle $\vec{u}_A = \vec{v} - \vec{v}_{e\parallel}$. The radius of non-equilibrium screening will then be large compared to the Debye radius of magnetized electrons:

$$r_{scr} = u_A/\omega_e \gg r_D = \Delta_{e\|}/\omega_e\,,\tag{2.15}$$

while $u_A \gg \Delta_{e\|}$. In practice, due to the extreme smallness of $\Delta_{e\|}$, the parameter $r_D$ can reach minimum values of the order of $r_L$ or $(n'_e)^{-1/3}$, so that correction of the screening radius proves to be important, not being reducible to corrections of $\simeq 1/L$. Thus, we should adopt

$$\rho^A_{max} = u_A \cdot \min\{l/(\gamma\beta c), 1/\omega_e\}\,.\tag{2.16}$$

Note that the choice of $\tau_{min}$ between $l/(\gamma\beta c)$ and $\omega_e^{-1}$ is also consistent with the initial idea of Eq. (2.14) reflecting the obvious fact that no processes related to collective interaction can occur in times shorter than the Langmuir period.

As a minimum impact parameter of adiabatic collisions, one should adopt the average Langmuir radius of electrons or, for sufficiently small relative velocities $u_A$, the distance $ze^2/(mu_A^2)$ when change in the longitudinal velocity of a Larmor electron becomes of the order of $u_A$:

$$\rho^A_{min} = \max\{r_L, ze^2/(mu_A^2)\}\,.\tag{2.17}$$

With decrease in ion velocities with respect to the electron beam in the process of cooling, the "logarithmic approximation" loses validity when $\rho^A_{max} \lesssim \rho^A_{min}$. This, however, does not mean that the effect of magnetization disappears or even declines. Changes are only in the character of collisions and in the dependence of the cooling process characteristics (friction and diffusion) on velocities $u_A$. Thus, if $\rho^A_{max} < r_L$ but still $\rho^A_{max} > ze^2/(mu_A^2)$, collisions then remain weak (the perturbation theory is applicable) but the main contribution of the interaction with Larmor circles goes to the region $\rho < r_L$. This is a region of <u>multiple cyclic</u> collisions with magnetized electrons: each individual collision is characterized by the relative velocity $u = v_{e\perp}$ and impact parameter $\rho \ll r_L$ but the collisions occur in a correlated fashion, one after another, with a period $2\pi/\Omega$, repeating $\simeq \Omega\rho/u_A$ times during the flight of an ion past a Larmor circle at a distance $\rho$. Obviously, in this region, the intensity of exchange and the cooling rate continue to grow with decrease in velocities $u_A$, although not as quickly because the result of an individual collision no longer depends on the average velocity $u_A$.

Finally, if, in the final stage of the cooling process, the parameter $\rho^A_{max}$, beyond which the Coulomb interaction is effectively screened, reaches the minimum when collisions become strong (for example, $\rho^A_{min} = ze^2/(mu_A^2)$ for sufficiently small Larmor radius), then the system starts to resemble a dense gas – the main contribution comes from interactions with large momentum transfer to Larmor circles ($\Delta v_{e\|} \simeq u_A$) but, at the same time, due to the long range of Coulomb forces, several (or "many") particles can simultaneously take part in an interaction (when $\omega_e l/(\gamma\beta c) \gtrsim 1$). The interaction in this case is saturated but ions continue slowing down in the "gas" of Larmor circles until their thermal energy decreases to a certain minimum determined by field fluctuations of the electron beam.

The described phenomena take place while, in terms of such rough parameters as the density and absolute (transverse) temperature, the electron beam represents a sparse plasma, i.e. $T_{e\perp} \gg ze^2 n_e'^{1/3}$. Thus, smallness of the longitudinal electron temperature alone under the magnetization conditions leads to the electron beam as a cooling medium and electron cooling acquiring interesting and largely unexpected properties. The primary of them is the fact that it becomes possible to cool the beam of heavy particles to temperatures that are many times lower than the transverse temperature of the electron beam (or the temperature of the electron gun). The minimum equilibrium temperature is determined by such extensive parameters as the electron beam density and accompanying magnetic field.

Further in this chapter, we will try to develop the aforementioned aspects more thoroughly.

## 2.3 Friction and diffusion in approximation of weak collisions. Non-equilibrium screening

Let us now construct the collision integral in a situation when the kinetic energies of relative motion of Larmor circles and heavy particles are large compared to the average energy of particle Coulomb interaction; one can then use the perturbation theory and electron interaction can be included through a self-consistent potential. The whole consideration in this chapter will be somewhat idealized: the electron flow and the magnetic field accompanying it are assumed uniform and the field of the electron "space charge" is assumed compensated. We will account for the fact that the time of ion interaction with the electron flow is finite by explicitly turning the interaction on and off at the moments $t = 0$ and $t = l/(\gamma\beta c)$, respectively. Applicability limits of such a model and some general theoretical aspects are discussed after the results are obtained.

In the co-moving frame (with non-relativistic motion of ions and electrons), the change in ion momentum in time $t$ from the moment the ion enters the electron flow is

$$\Delta\vec{p} = ze \int_0^t dt' \left(\frac{\partial\varphi}{\partial\vec{r}}\right)_{\vec{r}=\vec{r}(t')}, \tag{2.18}$$

where $\vec{r}(t) = \vec{r}_0 + \vec{v}t$ is the ion trajectory, $\varphi(\vec{r},t)$ is the sum of Coulomb potentials of all surrounding charges:

$$\varphi(\vec{r},t) = \sum_a \frac{e_a}{|\vec{r} - \vec{r}_a(t)|}. \tag{2.19}$$

where the electron trajectories $\vec{r}_a(t)$ are in general case determined including electron interaction with the ion and between themselves. Perturbation of heavy particle motion (including compensating particles, which are moving fast in the co-moving frame) can be neglected. On the basis of these general expressions, one can determine the friction force and scattering tensor as statistical averages over the initial conditions for the surrounding charges $e_a$:

$$\vec{F} = \frac{d}{dt}\langle\Delta\vec{p}\rangle = ze\frac{\partial}{\partial\vec{r}(t)}\langle\varphi(\vec{r}(t),t)\rangle, \qquad (2.20)$$

$$d_{\alpha\beta} = \frac{d}{dt}\langle\Delta p_\alpha\Delta p_\beta\rangle = z^2e^2\frac{d}{dt}\int_0^t dt_1\int_0^t dt_2\frac{\partial^2}{\partial r_{1\alpha}\partial r_{2\beta}}\langle\tilde{\varphi}(\vec{r}_1,t_1)\tilde{\varphi}(\vec{r}_2,t_2)\rangle, \qquad (2.21)$$

where $\vec{r}_{1,2} = \vec{r}(t_{1,2})$, $\tilde{\varphi} = \varphi - \langle\varphi\rangle$.

The averaging, as usual, means integration over the phase space of particles with a certain given probability distribution $D(\Gamma,t)$:

$$\langle\ldots\rangle = \int \ldots D(\Gamma,t)d\Gamma, \qquad d\Gamma = \prod_a d\Gamma_a.$$

Due to the conservation of particles, the integration can be done at any moment in time but, generally speaking, one then has to account for evolution of the distribution $D(\Gamma,t)$ as a result of the particle interaction itself. We will use a paradigm, in which physical quantities evolving according to exact equations are averaged over initial microscopic conditions with a probability distribution $D(\Gamma,0)$. In the first order of the perturbation theory, the potential $\varphi(\vec{r},t)$ in Eq. (2.19) can be represented as:

$$\varphi(\vec{r},t) = \varphi^0(\vec{r},t) - \frac{\partial}{\partial\vec{r}}\sum_a\frac{e\delta\vec{r}_a(t)}{|\vec{r}-\vec{r}_a(t)|}, \qquad (2.22)$$

where $\varphi^0(\vec{r},t)$ is the potential as a function of time when electrons are moving along the trajectories in an external field while the additional term accounts for perturbation of the trajectories by the interaction:

$$\delta\vec{r}_a = \int_0^t \delta\vec{v}_a(t')dt';$$

in the absence of magnetic field

$$\delta\vec{r}_a = -\frac{e}{m}\int_0^t dt'\int_0^{t'}\nabla\varphi(\vec{r}_a(t''),t'')dt'' = -\frac{e}{m}\int_0^t \tau d\tau\nabla\varphi(\vec{r}_a,t-\tau)_{\vec{r}_a=\vec{r}_a(t-\tau)},$$

while in magnetic field ($\vec{\Omega} = -e\vec{H}/(mc)$)

$$\delta\vec{r}_{a\parallel} = -\frac{e}{m}\int_0^t \tau d\tau\nabla_\parallel\varphi(\vec{r}_a(t-\tau),t-\tau),$$

$$\delta\vec{r}_{a\perp} = -\frac{e}{m}\int_0^t d\tau\left[\frac{\sin(\Omega\tau)}{\Omega}\nabla_\perp\Phi_{t-\tau} + \frac{1-\cos(\Omega\tau)}{\Omega^2}\vec{\Omega}\times\nabla\Phi_{t-\tau}\right], \qquad (2.23)$$

where the potential $\Phi$ includes the ion potential $\varphi_i^0$ as well:

$$\Phi = \varphi - \frac{ze}{|\vec{r} - \vec{r}(t)|} \equiv \varphi + \varphi_i^0 . \tag{2.24}$$

When integrating over time, we already ignore distortion of the trajectory $\vec{r}_a(t)$ by the interaction.

The next step is to replace the summation over electrons in Eq. (2.22) by an integration over an unperturbed stationary distribution while neglecting possible small correlations in the electron distribution at the entrance into the cooling section. From a formal point of view, such a replacement means that fluctuations of the kernel in Eq. (2.22) are neglected. Such coarsening describes well the influence of electrons separated from the point $\vec{r}$ by distances exceeding $n_e^{-1/3}$ – the average distance between electrons, while, for smaller distances, the effect of electron interaction is relatively small and the error introduced by the coarsening is not significant. Making then a spatial Fourier transformation in the obtained equation, we arrive at a time integral equation:

$$\varphi_{\vec{k}}(t) + \omega_e^2 \int_0^t \tau d\tau \left( \frac{k_\|^2}{k^2} + \frac{k_\perp^2}{k^2} \frac{\sin(\Omega\tau)}{\Omega\tau} \right) \langle \exp(-i\vec{k}\Delta\vec{r}_\tau) \rangle \Phi_{\vec{k}}(t - \tau) = \varphi_{\vec{k}}^0(t) , \tag{2.25}$$

where $\omega_e = \sqrt{4\pi n_e' e^2/m}$ is the Langmuir frequency, $k_\|$ and $\vec{k}_\perp$ are the $\vec{k}$ "wave" vector components longitudinal and transverse to the magnetic field, $\Delta\vec{r}_\tau = \vec{r}_t - \vec{r}_{t-\tau}$ is the change in electron coordinate in time $\tau$ in a uniform magnetic field:

$$\Delta\vec{r}_\tau = \frac{\vec{v}_\perp(t)}{\Omega} \sin(\Omega\tau) + \frac{\vec{v}_\perp(t) \times \vec{\Omega}}{\Omega^2} (1 - \cos(\Omega\tau)) + \vec{v}_{e\|} \tau . \tag{2.26}$$

When electron transverse velocity distribution is isotropic, the average of the correlation exponential in Eq. (2.25) equals

$$\langle \exp(-i\vec{k}\Delta\vec{r}_\tau) \rangle = \langle J_0 \left( 2k_\perp r_L \sin \frac{\Omega\tau}{2} \right) \exp(-ik_\| v_{e\|} \tau) \rangle,$$

$$r_L = v_{e\perp}/\Omega , \tag{2.27}$$

where $J_0(2k_\perp r_L \sin(\Omega\tau/2))$ is the Bessel function of the zeroth order, $r_L$ are the Larmor radii of electrons. Let us denote:

$$\varkappa_{\vec{k}}(\tau) = \left( \frac{k_\|^2}{k^2} + \frac{k_\perp^2}{k^2} \frac{\sin(\Omega\tau)}{\Omega\tau} \right) \langle J_0 \left( 2k_\perp r_L \sin \frac{\Omega\tau}{2} \right) \exp(-ik_\| v_{e\|} \tau) \rangle \tag{2.28}$$

and let us write Eq. (2.25) as

$$\varphi_{\vec{k}}(t) + \omega_e^2 \int_0^t \tau d\tau\, \varkappa_{\vec{k}}(\tau) \varphi_{\vec{k}}(t - \tau) = \varphi_{\vec{k}}^0(t) - \omega_e^2 \int_0^t \tau d\tau\, \varkappa_{\vec{k}}(\tau) \varphi_{i\vec{k}}^0(t - \tau) . \tag{2.29}$$

When $\Omega\tau \ll 1$ (and, in particular, in the limit of $\Omega \to 0$), the factor $\varkappa_{\vec{k}}(\tau)$ turns into an expression corresponding to freely moving electrons: $\varkappa_{\vec{k}}(\tau) \to \langle J_0(k_\perp v_{e\perp}\tau)\exp(-ik_\parallel v_{e\parallel}\tau)\rangle = \langle \exp(-i\vec{k}\vec{v}_e\tau)\rangle$, i.e., as it should be, at times $t \ll \Omega^{-1}$, influence of the magnetic field is of no significance. In a specific analysis, we will always be assuming situations when $\Omega t \gg 1$ (but not necessarily $\Omega\tau \gg 1$), since, within the length of the cooling section, electrons must undergo many Larmor cycles.

The right-hand side of Eq. (2.29) is an electron field potential obtained when neglecting electron interaction. The second term on the left-hand side including electron interaction in the approximation of a self-consistent field, accounts for the polarizability of the electron medium. Note that the absolute value of the factor $\varkappa_{\vec{k}}(\tau)$ does not exceed unit: $|\varkappa_{\vec{k}}(\tau)| \leq 1$. Therefore, it can be directly seen from the equation that, when $\omega_e t \ll 1$, the second term on the left-hand side is small compared to the first one. Thus, in times small compared to the Langmuir period, polarization of the electron medium is small and has no significant effect on the interaction of electrons with an ion; and there is no dynamic screening for any distances. This does not mean, of course, that the friction force and scattering tensor have a logarithmic divergence at large distances ($\vec{k} \to 0$): the integrals will be constrained by the finite extent of the interaction time (from the moment of ion arrival) explicitly present in Eqs. (2.29) and (2.21). In stationary conditions $\langle \varphi_{\vec{k}}^0(t)\rangle = 0$, then, from the last equation, we get equations determining the potentials $\langle \varphi_{\vec{k}}(t)\rangle$ and $\tilde{\varphi}_{\vec{k}}(t)$:

$$\langle \varphi_{\vec{k}}(t)\rangle + \omega_e^2 \int_0^t \tau d\tau\, \varkappa_{\vec{k}}(\tau)\langle \varphi_{\vec{k}}\rangle_{t-\tau} = \omega_e^2 \int_0^t \tau d\tau\, \varkappa_{\vec{k}}(\tau)\varphi_{i\vec{k}}^0(t-\tau)\,, \qquad (2.30)$$

$$\tilde{\varphi}_{\vec{k}}(t) + \omega_e^2 \int_0^t \tau d\tau\, \varkappa_{\vec{k}}(\tau)\tilde{\varphi}_{\vec{k}}(t-\tau) = \varphi_{\vec{k}}^0(t)\,. \qquad (2.31)$$

The "forcing" potentials on the right-hand sides of the equations are equal, respectively,

$$\varphi_{i\vec{k}}^0(t) = \frac{e^{-i\vec{k}\vec{r}(t)}}{2\pi^2 k^2}(-ze)\,, \qquad (2.32)$$

$$\varphi_{\vec{k}}^0(t) = e\sum_a \frac{e^{-i\vec{k}\vec{r}_a(t)}}{2\pi^2 k^2}\,. \qquad (2.33)$$

As seen from the equation, the fluctuation field $\tilde{\varphi}_{\vec{k}}$ is a superposition of fields $\tilde{\varphi}_{a\vec{k}}$ corresponding to individual charges. Due to the fact that the ion velocity with respect to the compensating charges is high, contribution of the latter to the diffusion tensor will be negligibly small. In the absence of coherent fluctuations (induced by "external sources") and with the relative velocities of ions and electrons (Larmor circles) being not too small, one can also neglect the correlation of electron initial positions and trajectories. As a result, the friction force and scattering tensor will be determined by the integrals:

$$\vec{F} = ze \int i\vec{k} \langle \varphi_{\vec{k}}(t) \rangle e^{i\vec{k}\vec{r}(t)} d^3k, \qquad (2.34)$$

$$d_{\alpha\beta} = 2(2\pi)^3 z^2 e^2 n'_e \int k_\alpha k_\beta d^3k \int_0^t \langle \varphi_{a\vec{k}}(t) \varphi_{a\vec{k}}(t-\tau) \rangle e^{i\vec{k}(\vec{r}_t - \vec{r}_{t-\tau})} d\tau, \qquad (2.35)$$

where $\langle \varphi_{\vec{k}}(t) \rangle$ is the statistically average potential induced by the ion, $\varphi_{a\vec{k}}$ is the average field of an individual electron including the polarizability of the electron beam:

$$\langle \varphi_{\vec{k}}(t) \rangle + \omega_e^2 \int_0^t \tau d\tau \, \varkappa_{\vec{k}}(\tau) \langle \varphi_{\vec{k}} \rangle_{t-\tau} = -\frac{ze}{2\pi^2 k^2} \omega_e^2 \int_0^t \tau d\tau \, \varkappa_{\vec{k}}(\tau) e^{-i\vec{k}\vec{r}_{t-\tau}}, \qquad (2.36)$$

$$\varphi_{a\vec{k}}(t) + \omega_e^2 \int_0^t \tau d\tau \, \varkappa_{\vec{k}}(\tau) \varphi_{a\vec{k}}(t-\tau) = \frac{e}{2\pi^2 k^2} e^{-i\vec{k}\vec{r}_a(t)}. \qquad (2.37)$$

### Interaction without screening

Before finding solutions of these equations in general case, let us do a certain preliminary analysis for the situations $\omega_e t \ll 1$. Then, as noted above, one can neglect the integral terms in the left-hand sides of the equations ($|\varkappa_{\vec{k}}(\tau)| \leq 1$), so that the potentials $\langle \varphi_{\vec{k}}(t) \rangle$ and $\varphi_{a\vec{k}}(t)$ will be simply equal to the right-hand sides. We then get (see Eq. (2.27)):

$$\vec{F}(t) = z^2 e^2 \omega_e^2 \frac{\partial}{\partial \vec{v}} \int \frac{d^3k}{2\pi^2 k^2} \int_0^t d\tau \left( \frac{k_\parallel^2}{k^2} + \frac{k_\perp^2}{k^2} \frac{\sin(\Omega\tau)}{\Omega\tau} \right) \langle J_0(2k_\perp r_L \sin\frac{\Omega\tau}{2}) e^{i\vec{k}\vec{u}_A \tau} \rangle, \qquad (2.38)$$

$$d_{\alpha\beta} = \frac{2}{\pi} z^2 e^4 n'_e \int \frac{k_\alpha k_\beta}{k^4} d^3k \int_0^t d\tau \langle J_0(2k_\perp r_L \sin\frac{\Omega\tau}{2}) e^{i\vec{k}\vec{u}_A \tau} \rangle. \qquad (2.39)$$

The obtained expressions contain contribution from the interactions at all distances starting with the smallest ones $\rho \simeq ze^2/(mu^2)$. It is not difficult to establish a correspondence of the structure of the integral over $k$ discussed in Section 2.2 to the picture of collisions with electrons in a magnetic field. The whole integral can be split into two regions.

1. $ku_A > \Omega$, a region of collisions with impact parameters, for which the cyclicity of electron motion has no significance: during the pass of an ion through an interaction region of a size $\simeq \rho$ near an electron Larmor trajectory, the latter has time to enter this region not more than once: $\rho/u_A < \Omega^{-1}$. The result of the interaction (collision) does not then depend on the magnetic field and is determined by the full relative velocity $\vec{u} = \vec{v} - \vec{v}_e$. Indeed, under the condition $ku_A > \Omega$, the integrals over $\tau$ converge at times $\simeq 1/(ku_A)$; then $\Omega\tau < 1$ and one can replace $\sin(\Omega\tau) \to \Omega\tau$, $\sin(\Omega\tau/2) \to \Omega\tau/2$. It is also convenient to return from the Bessel function to the average of the exponential $\exp(i\vec{k}_\perp \vec{v}_\perp \tau)$. The contribution of the fast collision region thus equals:

$$\vec{F}^0 = \frac{2}{\pi} \frac{z^2 e^4 n'_e}{m} \frac{\partial}{\partial \vec{v}} \int \frac{d^3 k}{k^2} \langle \frac{\sin(\vec{k}\vec{u}t)}{\vec{k}\vec{u}} \rangle \approx \frac{4\pi z^2 e^4 n'_e}{m} \frac{\partial}{\partial \vec{v}} \langle \frac{L^0}{u} \rangle, \qquad (2.40)$$

$$d_{\alpha\beta} = 4\pi z^2 e^4 n'_e \langle L^0 \frac{\delta_{\alpha\beta} - u_\alpha u_\beta}{u^3} \rangle \, ; \quad L^0 = \ln\left(\frac{mu^2 u_A}{e^2 \Omega}\right). \qquad (2.41)$$

In obtaining the final expression, we used the assumption $\Omega t \gg 1$ allowing the substitution $\sin(\vec{k}\vec{u}t/(\vec{k}\vec{u})) \to \pi\delta(\vec{k}\vec{u})$ in the whole integration region. Note that the contribution of fast singular collisions is described by the obtained formula for an arbitrary relation between $u_A$ and $v_{e\perp}$.

2. $ku_A < \Omega$

In the region corresponding to distances $\rho > u_A/\Omega$, the duration of an interaction exceeds the electron Larmor period. Contrary to the previous situation, the factor $\sin(\Omega\tau)$ oscillates faster than the exponential. Therefore, the integration over $\tau$ in the considered region of $k$ can be done in two stages: first by averaging over the quickly-oscillating dependence on $\Omega\tau$ and then by integrating $\exp(i\vec{k}\vec{u}_A\tau)$. We finally arrive at the expressions:

$$\vec{F}^L = \frac{2z^2 e^4 n'_e}{\pi m} \frac{\partial}{\partial \vec{v}} \int \frac{d^3 k}{k^2} \frac{k_\parallel^2}{k^2} \langle J_0^2(k_\perp r_L) \frac{\sin(\vec{k}\vec{u}_A t)}{\vec{k}\vec{u}_A} \rangle, \qquad (2.42)$$

$$d^L_{\alpha\beta} = \frac{2z^2 e^4 n'_e}{\pi} \int \frac{d^3 k}{k^2} \frac{k_\alpha k_\beta}{k^2} \langle J_0^2(k_\perp r_L) \frac{\sin(\vec{k}\vec{u}_A t)}{\vec{k}\vec{u}_A} \rangle, \qquad (2.43)$$

$$1/\Omega \ll t \ll 1/\omega_e.$$

Note that, due to the factor $J_0^2(k_\perp r_L)$, the integral over $k$ does not contain any singularities; however, in a general case, the integration is cut off at the upper end by the condition $ku_A < \Omega$. This cut is not significant in the situation $u_A \ll r_L \Omega = v_{e\perp}$, since the integral actually converges before the integration limit is reached. For the opposite relation of the velocities $u_A$ and $v_{e\perp}$, the convergence boundary ($k_\perp \simeq r_L^{-1}$) lies beyond the integration limit (in the integration region $J_0(k_\perp r_L) \approx 1$) but the integral only logarithmically depends on the "exact" value of the integration limit. Thus, Eqs. (2.42) and (2.43) are a general approximation well describing the contribution of collisions with magnetized electrons under the conditions $\Omega t \gg 1$, $\omega_e t \ll 1$.

The region $kr_L < 1$ where $J_0(k_\perp r_L)$, corresponds to adiabatic collisions of an ion with Larmor circles. The region $kr_L > 1$ accounts for the contribution of multiple cyclic collisions with electrons moving along Larmor trajectories; this contribution is inversely proportional to the Larmor velocity: $\langle J_0^2(k_\perp r_L) \rangle \approx 1/(\pi k_\perp r_L)$.

The factor $k_\parallel^2/k^2$ in the friction force reflects the fact that, in the region of distances $\rho > u_A/\Omega$, the significance of the electron mobility is mainly in the longitudinal direction. This is obvious for the adiabatic interaction at distances $\rho > r_L$. For distances $\rho < r_L$ (and, consequently, $u_A < v_{e\perp}$), the electron momentum transfers $\Delta p_e$ in the longitudinal and Larmor degrees of freedom as a result of each singular collision have the same orders of magnitude; however, shifts of the electron velocity and coordinate (determining the friction effect) during the complete interaction time are $\simeq (\Delta p_e/m)\Omega(\rho/u_A)$ and $(\Delta p_e/m)\Omega(\rho/u_A)^2$ in the longitudinal direction while, in the transverse one, they are $\simeq \Delta p_e/m$ and $(\Delta p_e/m) \cdot (\rho/u_A)$ (drift motion). The average changes in the kinetic energies are $\simeq (\Delta p_e \Omega \rho/u_A)^2$ and $(\Delta p_e)^2 \Omega \rho/(m u_A)$, respectively. The ratio of the effects equals exactly $\rho\Omega/u_A$.

We postpone the derivation of the final explicit expressions for $\vec{F}$ and $d_{\alpha\beta}$ and their discussion to Sections 2.4 and 2.5.

## Non-equilibrium screening

When $\omega_e t \gtrsim 1$, the interaction of electrons can no longer be neglected; Eqs. (2.36) and (2.37) need to be solved exactly. Before that, however, it is convenient to somewhat unify the definition of the force $\vec{F}$ by introducing, analogously to $\varphi_{a\vec{k}}$, a full effective ion field

$$\langle \phi_{\vec{k}}(t) \rangle = -\frac{ze}{2\pi^2 k^2} \exp(-i\vec{k}\vec{r}(t)) + \langle \varphi_{\vec{k}}(t) \rangle$$

satisfying the equation

$$\langle \phi_{\vec{k}}(t) \rangle + \omega_e^2 \int_0^t \tau d\tau \varkappa_{\vec{k}}(\tau) \langle \phi_{\vec{k}}(t-\tau) \rangle = -\frac{ze}{2\pi^2 k^2} \exp(-i\vec{k}\vec{r}(t)) . \qquad (2.44)$$

Formally defining the integral $\int (\vec{k}/k^2) d^3k$ to equal zero, one can write the force $\vec{F}$ in the form:

$$\vec{F} = ze \int i\vec{k} \langle \phi_{\vec{k}}(t) \rangle \exp(i\vec{k}\vec{r}(t)) \, d^3k . \qquad (2.45)$$

Equations (2.37) and (2.44) (as well as Eq. (2.36)) belong to the category solvable by the Laplace-Mellin transformation method:

$$\langle \phi_{\vec{k}}(t) \rangle = -\frac{ze}{2\pi^2 k^2} \int_C d\omega \exp \frac{-i\omega t}{2\pi \varepsilon_{\vec{k}}(\omega)} \left[ \exp(-i\vec{k}\vec{r}(t)) \right]_\omega , \qquad (2.46)$$

$$\varphi_{a\vec{k}}(t) = \frac{e}{2\pi^2 k^2} \int_C \frac{d\omega \exp(-i\omega t)}{2\pi \varepsilon_{\vec{k}}(\omega)} \left[ \exp(-i\vec{k}\vec{r}_a(t)) \right]_\omega , \qquad (2.47)$$

where the Fourier transforms of the exponentials

$$\left[ \exp(-i\vec{k}\vec{r}(t)) \right]_\omega = \int_0^\infty d\tau \exp(i(\omega - \vec{k}\vec{v})\tau) = \frac{i}{\omega - \vec{k}\vec{v}}, \qquad (2.48)$$

$$\left[\exp(-i\vec{k}\vec{r}_a(t))\right]_\omega = \int_0^\infty d\tau \exp(i(\omega - k_\| v_{e\|})\tau - i\vec{k}_\perp \vec{r}_\perp(\tau))$$
$$= \sum_{l=-\infty}^\infty \frac{i J_l(k_\perp r_L) \exp(i\psi_L)}{\omega - k_\| v_{e\|} - l\Omega} \quad (2.49)$$

and the **electric permittivity** of the electron flow

$$\varepsilon_{\vec{k}}(\omega) = 1 + \omega_e^2 \int_0^\infty \tau d\tau \varkappa_{\vec{k}}(\tau) \exp(i\omega\tau)$$

$$= 1 + \omega_e^2 \int_0^\infty \tau d\tau \times \quad (2.50)$$

$$\times \left(\frac{k_\|^2}{k^2} + \frac{k_\perp^2}{k^2}\frac{\sin(\Omega\tau)}{\Omega\tau}\right) \langle J_0\left(2k_\perp r_L \frac{\sin(\Omega\tau)}{2}\right) \exp(i(\omega - k_\| v_{e\|})\tau)\rangle$$

are determined by integrals with Im $\omega > 0$, and accordingly the integration over $\omega$ in Eqs. (2.46) and (2.47) is done along a path lying in the upper half-plane (above the zeros of $\varepsilon_{\vec{k}}(\omega)$) as shown in Fig. 7:

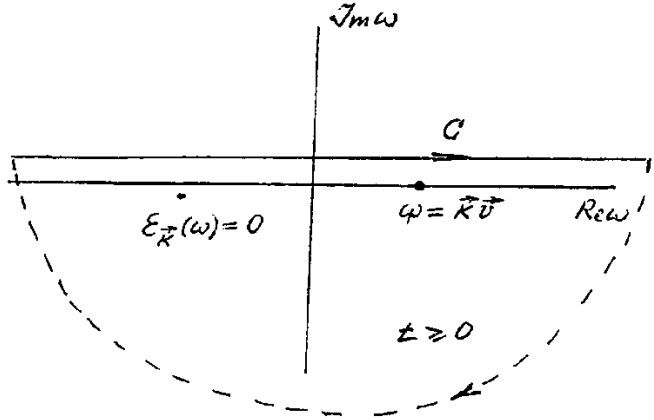

Fig. 7.

This choice of the path satisfies the initial conditions following from Eqs. (2.36), (2.37), and (2.44) themselves.

We will do further analysis only for the friction force in order to make it less cumbersome. The main conclusions will be equally relevant to the scattering tensor as well, whose explicit expressions will be given as needed.

Assuming for certainty that $\varepsilon_{\vec{k}}(\omega)$ has no special point other than simple zeros, let us perform the integration over $\omega$ in Eq. (2.46):

$$\langle \phi_{\vec{k}}(t) \rangle = -\frac{ze}{2\pi^2 k^2} \left\{ \frac{\exp(-i\vec{k}\vec{v}t)}{\varepsilon_{\vec{k}}(\vec{k}\vec{v})} + \sum_s \left[ \frac{\exp(-i\omega t)}{(\omega - \vec{k}\vec{v})\partial \varepsilon_{\vec{k}}(\omega)/\partial \omega} \right]_{\omega=\omega_s} \right\}, \quad (2.51)$$

where the summation is done over the eigen oscillations of the electron plasma $\varepsilon_{\vec{k}}(\omega_s) = 0$. Substituting Eq. (2.51) in the definition of Eq. (2.45) and considering the symmetry properties of the function $\varepsilon_{\vec{k}}(\vec{k}\vec{v})$ when changing the sign of $\vec{k}$, we get:

$$\vec{F}(t) = -\frac{z^2 e^2}{2\pi^2} \int d^3k \frac{\vec{k}}{k^2} \left\{ \frac{\text{Im}\, \varepsilon_{\vec{k}}(\vec{k}\vec{v})}{|\varepsilon_{\vec{k}}(\vec{k}\vec{v})|^2} + i \sum_s \left[ \frac{\exp(-i(\omega - \vec{k}\vec{v})t)}{(\omega - \vec{k}\vec{v})\partial \varepsilon_{\vec{k}}(\omega)/\partial \omega} \right]_{\omega=\omega_s} \right\}. \quad (2.52)$$

The first, time-independent term in Eq. (2.52) in general represents an expression for the friction force on a particle moving in a plasma; this expression is widespread in the plasma theory and was first obtained by R. Balesku [17, 18]. It is related to the "forced" solution of Eq. (2.36) proportional to $\sim \exp(-i\vec{k}\vec{v}t)$ (the pole $\omega = \vec{k}\vec{v}$ in Eq. (2.46)). This solution can be obtained directly from the original Eq. (2.36) by substituting into it the dependence $\langle \varphi_{\vec{k}}(t) \rangle$ as $\sim \exp[-i(\vec{k}\vec{v} - i0)t]$ (the imaginary term $-i0$ is introduced in addition to the frequency in accordance with the principle of damping or causality). Such a recipe is usually used when determining the response of a medium to the motion of a "test" particle in the plasma theory or, in a more general case, when constructing a two-particle correlator in the method of the BBKGI equation chain [33]. The justification is usually the consideration that, after a time exceeding the characteristic time of interaction in a collision, the memory of the initial "microscopic" conditions in the system is lost and therefore the average statistical characteristics related to the interaction become functionals of the single-particle distributions and do not depend explicitly on time. In case of long-distance Coulomb forces, the notion of a characteristic interaction time becomes indeterminate due to the familiar logarithmic divergence of the collision integral at large distances. The divergence is eliminated when accounting for the collective interaction of electrons that, in the thermodynamic limit, leads to Debye screening of the Coulomb potential. However, screening is the result of dynamic processes and cannot be established instantly after the start of interaction (after "preparation" of the system). Therefore, for finite times, besides the equilibrium forced part, the response must also contain a non-equilibrium component in terms of free plasma oscillations (whose contribution is what satisfies the initial condition $\vec{F}(t = 0) = 0$). Contribution of this component is described by the second terms in Eqs. (2.51) and (2.52).

Note that the division itself of the response into the stationary and non-stationary parts has physical meaning only for times exceeding the Langmuir period $\omega_e^{-1}$, since no processes related to collective interaction can take place in a plasma at shorter times. When $\omega_e t \ll 1$, use of the general formula, Eq. (2.52) is not adequate and it is more reasonable to use Eq. (2.38).

Let us consider in more detail the relative role of the two terms in the force in Eq. (2.52). For simplicity, suppose first that the electron flow is not magnetized. Then

$$\varepsilon_{\vec{k}}(\omega) = 1 + \omega_e^2 \int_0^\infty \tau d\tau \langle e^{i(\omega - \vec{k}\vec{v}_e)\tau} \rangle = 1 - \frac{\vec{k}}{k^2} \langle \frac{\partial}{\partial \vec{v}_e} \frac{\omega_e^2}{\omega - \vec{k}\vec{v}_e} \rangle, \quad (2.53)$$

$$\varepsilon_{\vec{k}}(\vec{k}\vec{v}) = 1 + \omega_e^2 \frac{\vec{k}}{k^2} \int \frac{d^3 v_e}{\vec{k}\vec{u} + i0} \frac{\partial f}{\partial \vec{v}_e},$$

$$\text{Im } \varepsilon_{\vec{k}}(\vec{k}\vec{v}) = -\pi \omega_e^2 \frac{\vec{k}}{k^2} \int \delta(\vec{k}\vec{u}) \frac{\partial f}{\partial \vec{v}_e} d^3 v_e.$$

Accounting for the $\delta$ function in Im $\varepsilon_{\vec{k}}(\vec{k}\vec{v})$, one can write the equilibrium part of the friction force as

$$\vec{F}_{st} = -\frac{z^2 e^2 \omega_e^2}{2\pi} \int d^3 k \frac{\vec{k}}{k^4} \int d^3 v_e \left( \vec{k} \frac{\partial f}{\partial \vec{v}_e} \right) \frac{\delta(\vec{k}\vec{v} - \vec{k}\vec{v}_e)}{|\varepsilon_{\vec{k}}(\vec{k}\vec{v}_e)|^2}, \quad (2.54)$$

where, for characteristic electron velocities, the electric permittivity

$$\varepsilon_{\vec{k}}(\vec{k}\vec{v}_e) = 1 + \omega_e^2 \frac{\vec{k}}{k^2} \int \frac{d^3 v_e'}{\vec{k}(\vec{v}_e - \vec{v}_e')} \frac{\partial f(\vec{v}_e')}{\partial \vec{v}_e'}$$

has an order of magnitude of

$$|\varepsilon_{\vec{k}}(\vec{k}\vec{v}_e)| \simeq 1 + \frac{\omega_e^2}{(kv_{eT})^2}.$$

This shows that, for any ion velocities, in particular, for $v \gg v_{eT}$, the radius of equilibrium screening equals the Debye one $r_D = v_{eT}/\omega_e$. Including the factor $\varepsilon^{-2}$ in Eq. (2.54), the Coulomb logarithm equals

$$L_D \approx \int_0^{k_{max}} \frac{dk}{k \left(1 + (\omega_e^2/k^2 v_{eT}^2)\right)^2} \approx \ln(k_{max} \cdot r_D) = \ln\left(\frac{mu^2 v_{eT}}{z\omega_e e^2}\right),$$

where $k_{max}^{-1} = ze^2/(mu^2)$.

Let us now estimate the contribution to the force in Eq. (2.52) of the collective oscillations $\omega(\vec{k})$ where $\varepsilon_{\vec{k}}(\omega(\vec{k})) = 0$, excited by an ion. Short-wave excitations $kr_D \gtrsim 1$ damp in times of the order of $\omega_e^{-1}$, therefore, in the situation $\omega_e t \gg 1$ that we are interested in, this region can be neglects. On the contrary, the damping decrements of long-wave excitations $k \ll r_D^{-1}$ are exponentially small according to parameter $(kr_D)^{-2}$ [19]:

$$\text{Im } \omega(k) \approx -\frac{\pi}{2} \omega_e^3 \frac{\partial}{\partial \omega_e} \langle \delta(\omega_e - \vec{k}\vec{v}_e) \rangle, \quad \omega(k) \approx \omega_e + i \text{ Im } \omega(k).$$

Thus, contribution of the non-stationary component belongs to the region of large distances exceeding the Debye radius. Substantial harmonics in the second term of the friction force, Eq. (2.52), are determined by the condition

$$\text{Im } \omega(k)t \lesssim 1.$$

For not too large $\omega_e t$, effective $k$ have an upper limit, in its order of magnitude, close to $r_D^{-1}$. In the region $kr_D \ll 1$,

$$\left.\frac{\partial \varepsilon_{\vec{k}}(\omega)}{\partial \omega}\right|_{\omega=\omega(k)} \simeq \pm \frac{2}{\omega_e},$$

so that the non-equilibrium component can approximately be written as

$$\vec{F}_W \simeq -\frac{z^2 e^2 \omega_e}{2\pi^2} \int d^3k \, \frac{\vec{k}}{k^2} \frac{\sin[(\omega_e - \vec{k}\vec{v})t]}{\omega_e - \vec{k}\vec{v}} e^{\text{Im } \omega(k)t}. \tag{2.55}$$

At velocities $v \lesssim v_{eT}$, in the order of magnitude,

$$\vec{F}_W \simeq -\frac{z^2 e^2 \omega_e^2}{v_{eT}^3} \vec{v};$$

from comparison to Eq. (2.54), it can be seen that the non-stationary part can be neglected ($L \gg 1$).

For high velocities $v \gg v_{eT}$, the argument of sine in Eq. (2.55) may be large also in the region $k \ll r_D^{-1}$ with a small damping index, one can then make a substitution

$$\frac{\sin[(\omega_e - \vec{k}\vec{v})t]}{\omega_e - \vec{k}\vec{v}} \to \pi\delta(\omega_e - \vec{k}\vec{v}) \approx \pi[\delta(\vec{k}\vec{v}) - \omega_e \delta'(\vec{k}\vec{v})];$$

the expansion of the $\delta$ function is valid up to $\vec{k}\vec{v} \simeq \omega_e$. Finally, for the force in Eq. (2.55), we get the formula:

$$\vec{F}_W = z^2 e^2 \omega_e^2 \frac{\partial}{\partial \vec{v}} \frac{1}{v} \int_{\omega_e/v}^{k_{max}} \frac{dk}{k} \simeq z^2 e^2 \omega_e^2 \frac{\partial}{\partial \vec{v}} \frac{1}{v} \ln \frac{v}{v_{eT}}. \tag{2.56}$$

Here we neglect the difference of $k_{max}$ from $r_D^{-1}$ due to the weak logarithmic dependence on the "exact" values of the integration limits. Formally, the upper limit $k_{max}$ goes to zero when $t \to \infty$, therefore, it may seem that, for sufficiently large $\omega_e t$, the "non-stationary" part can be neglected after all. Here one should note the following. Firstly, the Landau damping decrements for waves $\omega_e/v \lesssim k \lesssim k_D$ (giving a resonant contribution in Eq. (2.55)) are proportional to the tales of the distribution and, therefore, there may be no substantial damping even for quite large values of $\omega_e t$. Secondly, there is a conceptual limitation of Landau damping for long-wave oscillations due to finite size of the wave amplitudes [34]. As known, this damping is related to absorbtion of the wave energy by electrons having the same velocity as the phase velocity of the wave: $\vec{k}\vec{v}_e = \omega(k) \simeq \omega_e$. However, the notion of electron moving with a constant velocity is valid, strictly speaking, only in the limit of infinitesimally small wave amplitude. In reality, influence of the wave field changes the particle velocity and the phase of the wave action $\omega_e t - \vec{k}\vec{r}(t)$ shifts along with it; as a result, the energy absorbtion is replaced by deceleration, the resonance is passed in the opposite direction and so forth – there appears an auto-phasing mode well-known in the nonlinear mechanics. In the field of a wave with a constant amplitude $E_k$, the particle velocity and phase oscillate with a frequency

$$\delta\omega \simeq \sqrt{\dot{p}_{max} \frac{\partial(kv)}{\partial p}} \simeq \sqrt{eE_k \frac{k}{m}}.$$

This frequency should be compared to the decrement of Landau damping $|\text{Im }\omega(k)|$. When $\delta\omega < |\text{Im }\omega(k)|$, the coherent motion damps before the wave phase has time to shift. In the opposite case, the velocities and phases of particle motion are modulated by the coherent field, so that, on average, there is no absorbtion of energy of the collective motion and damping vanishes along with it.

In our case, the characteristic field strength can be estimated from the friction force expression, Eq. (2.56) itself:

$$E \simeq \frac{ze\omega_e^2}{v^2} \ln\frac{v}{v_{eT}},$$

since the same excited collective field acts both on an ion and on individual electrons. Thus, we arrive at a damping criterion:

$$-\sqrt{\frac{ze^2 k}{mv^2} \ln\frac{v}{v_{eT}}} < -\frac{\pi}{2} \omega_e^2 \frac{\partial}{\partial \omega_e} \langle \delta(\omega_e - \vec{k}\vec{v}_e) \rangle. \tag{2.57}$$

The damping region happens to be relatively narrow, since, with decrease in $|\vec{k}|$ (increase in distances), $\text{Im }\omega(k)$ goes to zero at an exponential rate. We are interested in the region $k_D \ll k \lesssim \omega_e/v$, since the maximum effective distances are limited by the value of $\simeq v/\omega_e$, beyond which the interaction is screened. The limitations due to nonlinearity are significant if the condition in Eq. (2.57) is violated for $k = \omega_e/v$. For example, for the Maxwell distribution of the critical velocity $v$, Eq. (2.57) gives an equation:

$$\left(\frac{v}{v_{eT}}\right)^9 e^{-\left(\frac{v}{v_{eT}}\right)^2} = \frac{8z}{\pi n r_D^3}.$$

When a particle is moving with a velocity $v < v_{cr}$, the plasma oscillations it excites experience Landau damping in the whole range of distances $\omega_e/v < k < \omega_e/v_{cr}$. Otherwise, if $v > v_{cr}$, then this range subdivides into a region with damping $\omega_e/v_{cr} < k < r_D^{-1}$ and one without damping $\omega_e/v < k < \omega_e/v_{cr}$ (for this region one should set $\text{Im}\,\omega(k) = 0$ in Eq. (2.55)).

In a hot plasma (or in a non-magnetized electron flow), the parameter $nr_D^3$ is large, then

$$\frac{v_{cr}}{v_{eT}} \simeq \sqrt{\ln A} + \frac{9}{4}\frac{\ln \ln A}{\sqrt{\ln A}} + \dots, \qquad A = \frac{\pi n r_D^3}{8z}.$$

With a density $n = 10^8$ and a temperature $T_e = 0.2$ eV, we get $v_{cr} \simeq 5 v_{eT}$. With a decrease in temperature, the damping region narrows and, in the limiting situation $nr_D^3 \simeq 1$ ($\overline{mv_e^2} \simeq e^2 n^{-1/3}$), the collective response in the region $v_e/\omega_e < r < v/\omega_e$ turns out to have no damping at all. In this case, the relative contribution of the region becomes maximal and equal to a half of the contribution of pair-wise collisions $r < r_D$:

$$L_p = \ln\left(\frac{r_D}{r_{min}}\right) = \ln\frac{mv^2}{T_e} = 2\ln\frac{v}{v_{eT}} = 2L_W, \quad (z=1),$$

we used here the limiting relations $e^2/r_D = T_e = e^2 n^{-1/3}$. We come to a conclusion that, in case of a non-magnetized plasma, the implemented correction of the collision integral reduces mainly to a correction of the Coulomb logarithm; moreover, its magnitude increases by not more than a factor of one and a half:

$$\ln\left(\frac{r_D}{r_{min}}\right) \to \ln\left(\frac{u}{\omega_e r_{min}}\right) = \ln\left(\frac{mu^3}{e^2 \omega_e}\right).$$

Note also that, despite the time dependence of the general expression for the friction force, Eq. (2.52), contribution of the non-equilibrium part of the collective response at velocities $v \gg v_{eT}$ in practice turns out to be constant in time. The effective steadiness is due to the smallness (or absence) of damping and the resonant nature of the interaction: $\vec{k}\vec{v} = \omega_e$.

Let us now turn to our case of a magnetized electron flow with a low longitudinal temperature. The relation between the two components of the friction force is determined by the properties of the electric permittivity and by the ion velocity. Let us first consider how the contribution of the stationary part determined by the permittivity at $\omega = \vec{k}\vec{v}$, is modified in a magnetic field:

$$\varepsilon_{\vec{k}}(\vec{k}\vec{v}) = 1 + \omega_e^2 \int_0^\infty \tau d\tau \left(\frac{k_\parallel^2}{k^2} + \frac{k_\perp^2}{k^2}\frac{\sin(\Omega\tau)}{\Omega\tau}\right) \langle J_0(2k_\perp r_L \sin\frac{\Omega\tau}{2}) e^{i\vec{k}\vec{u}_A\tau}\rangle. \qquad (2.58)$$

As in the case of $\omega_e t \ll 1$, let us split the integral over $k$ into regions $k u_A > \Omega$ and $k u_A < \Omega$ (we assume $\omega_e \ll \Omega$, since otherwise influence of the magnetic field has no significance at all).

In the region $ku_A > \Omega$, characteristic times $\tau$ in $\varepsilon_{\vec{k}}(\vec{k}\vec{v})$ do not exceed $\Omega^{-1}$; then the difference of the electric permittivity from unit is small: $\varepsilon = 1 - \langle \omega_e^2/(\vec{k}\vec{u} + i0)^2 \rangle$. Therefore, contribution of this region is the same as with $\omega_e t \ll 1$.

In the region $ku_A < \Omega$

$$\varepsilon_{\vec{k}}(\vec{k}\vec{v}) \approx 1 - \omega_e^2 \frac{k_\parallel^2}{k^2} \langle \frac{J_0^2(k_\perp r_L)}{(\vec{k}\vec{u}_A + i0)^2} \rangle. \tag{2.59}$$

By analogy with Eq. (2.54), the equilibrium force component from the interaction with magnetized electrons can be written as:

$$\vec{F}_{st}^L = -\frac{z^2 e^2 \omega_e^2}{2\pi} \int \frac{\vec{k} k_\parallel}{k^4} d^3k \int d^3v_e \frac{\partial f}{\partial v_{e\parallel}} \frac{\delta(\vec{k}\vec{u}_A) J_0^2(k_\perp r_L)}{\left|\varepsilon_{\vec{k}}(k_\parallel v_{e\parallel})\right|^2}, \tag{2.60}$$

where

$$\left|\varepsilon_{\vec{k}}(k_\parallel v_{e\parallel})\right| = \left|1 - \frac{\omega_e^2}{k^2} \int f(\vec{v}_e') d^3v_e' \frac{J_0^2(k_\perp r_L)}{\left(v_{e\parallel} - v_{e\parallel}' + i0\right)^2}\right|. \tag{2.61}$$

The integration over $k$ in Eq. (2.60) with velocities $u_A < v_{e\perp}$ can be done in the limits $0 \leq k \leq \infty$. Without accounting for the electric susceptibility of the electron flow, the integral diverges logarithmically at the lower limit. The parameter of effective cut off is determined from the equation $\left|\varepsilon_{\vec{k}}(k_\parallel v_{e\parallel})\right| - 1 \simeq 1$ for characteristic velocities $v_{e\parallel} \lesssim \Delta_{e\parallel}$:

$$\frac{\omega_e^2 \langle J_0^2(k_\perp r_L) \rangle}{k^2 \Delta_{e\parallel}^2} \simeq 1, \tag{2.62}$$

from which it follows

$$k_{st} \simeq \frac{\omega_e}{\Delta_{e\parallel}} \cdot \min\left\{1, \left(\frac{\Delta_{e\parallel}}{\pi \omega_e r_L}\right)^{1/3}\right\}.$$

As can be seen, the stationary screening in the magnetization region is determined by the longitudinal temperature while Larmor rotation of electrons can only effectively reduce the interaction force ($\omega_e^2 \to \omega_e^2 \langle J_0^2(k_\perp r_L) \rangle$). With a decrease in the longitudinal temperature, the screening radius shrinks and can even become smaller than the Larmor one.

At the same time, the minimum impact parameters either are equal (in the order of magnitude) to Larmor radii or become comparable to $k_{min}^{-1}$. Then the corresponding Coulomb logarithm can become of the order of unit or the region of integral's logarithmic behavior vanishes at all if $\Delta_{e\parallel} \lesssim \omega_e r_L$. Therefore, there is a substantial increase in the role of interaction with a collective response of magnetized electrons, which now becomes dominant for a wide range of conditions and ion velocities. To estimate its contribution, one should consider free oscillations of the electron flow determined by the equation $\varepsilon_{\vec{k}}(\omega) = 0$, in the region of weak damping. For oscillations with a frequency $\omega \ll \Omega$ in the region $k \ll \Omega/\Delta_{e\|}$, the electric permittivity equals

$$\varepsilon_{\vec{k}}(\omega) \approx 1 - \omega_e^2 \frac{k_\|^2}{k^2} \langle \frac{J_0^2(k_\perp r_L)}{(\omega - k_\| v_{e\|} + i0)^2} \rangle . \tag{2.63}$$

Due to magnetization of the transverse motion, absorbtion of the wave energy can be related only to the spread of electron longitudinal velocities. Ignoring the spread

$$\omega(\vec{k}) = \pm \omega_M(\vec{k}) \equiv \pm \omega_e \left|\frac{k_\|}{k}\right| \langle J_0^2(k_\perp r_L) \rangle^{1/2} , \tag{2.64}$$

the oscillation frequency equals the Langmuir one in the order of magnitude or slowly drops with an increase of $k$ in the region $k > r_L^{-1}$. In the long-wave limit, in accordance with the equation $\varepsilon_{\vec{k}}(\omega) = 0$, Im $\omega(k)$ equals

$$\text{Im } \omega = -\frac{\pi}{2} \frac{\omega_M^3(\vec{k})}{k_\| |k_\||} \frac{\partial f(v_{e\|})}{\partial v_{e\|}}\bigg|_{v_{e\|} = \pm \omega_M(k)/k_\|} , \tag{2.58}$$

where $f(v_{e\|})$ is the electron longitudinal velocity distribution; besides, $\langle v_{e\|} \rangle = 0$ in accordance with the definition of $\omega(\vec{k})$. Due to the small size of the longitudinal spread, it follows from the resonance condition $k_\| v_{e\|} = \omega(k)$ that the damping will be very weak (or absent completely) already for distances $r \gg k_{st}^{-1}$ (see Eq. (2.62)), which can themselves be small compared to the former Debye radius $\simeq \Delta_{e\perp}/\omega_e$. The criterion in Eq. (2.57) in our case takes the form

$$-\left(\frac{ze^2 k}{mv^2} \ln \frac{v}{\Delta_{e\|}}\right)^{1/2} < -\frac{\pi}{2} \omega_M^2 \frac{\partial}{\partial \omega_M} \langle \delta(\omega_M - k_\| v_{e\|}) \rangle . \tag{2.66}$$

The non-equilibrium component of the friction force can be written in a form analogous to Eq. (2.55):

$$\vec{F}_W \simeq -\frac{z^2 e^2}{2\pi^2} \int d^3 k \frac{\vec{k}}{k^2} \omega_M \frac{\sin[(\omega_M - \vec{k}\vec{v})t]}{\omega_M - \vec{k}\vec{v}} e^{-\lambda(\vec{k})t} ; \tag{2.67}$$

moreover, one should again keep in mind the above discussed limitations of damping by nonlinear effects, whose role grows due to the small longitudinal temperature of electrons. The decrement $\lambda(k)$ should be assumed equal to zero if the following condition is violated

$$\left|\frac{1}{z} F_W \frac{k_\|}{m}\right|^{1/2} < \frac{\pi}{2} \left|\frac{\omega_M^3(\vec{k})}{k_\|^2} \frac{\partial f(v_{e\|})}{\partial v_{e\|}}\right|_{v_{e\|} = \pm \omega_M(k)/k_\|} . \tag{2.68}$$

If one sets aside the subtleties related to damping, integration over $k$ in Eq. (2.67) extends to $k_{max} \simeq k_{st}$, Eq. (2.62), instead of $k_{max} \simeq \omega_e/\Delta_{e\perp}$ as in case of non-magnetized electrons. At the same time, the range of distances contributing to $\vec{F}_{st}$ sharply narrows due to reduction in the radius of stationary equilibrium screening at $\Delta_e \ll \Delta_{e\perp}$ ($\rho_{max} = k_{st}^{-1}$ while $\rho_{min}$ cannot be less than $u_A/\Omega$). Thus, the relation between the friction force $F_{st}$ obtained using the canonical recipe and the "non-stationary" part including the interaction with the excited collective electron motion in the considered conditions can change in favor of the latter.

In real conditions, the longitudinal temperature of an electron flow (accelerated electrostatically) turns out to be so small that the condition in Eq. (2.68) is not satisfied even for the shortest waves $k \simeq n_e'^{1/3}$ specifically because the Debye radius reaches its minimum value of $\simeq n_e^{-1/3}$ (the spread of longitudinal velocities at the entrance is small compared to the fluctuations caused by the Coulomb interaction of neighboring particles). The time of electron beam thermalization, i.e. of energy transfer from the Larmor degrees of freedom, greatly exceeds the time of flight through the cooling section, so that the longitudinal temperature has time only to get up to the level of Coulomb energy fluctuations $\simeq e^2 n_e^{1/3}$ (see Section 2.6). In fact, in these circumstances, it is not appropriate to include the longitudinal spread in the consideration as an independent parameter and, within the frame of the perturbation theory, it should be neglected at all: $\Delta_{e\|} = 0$. There is no non-collisional damping (relaxation) of the collective excitation caused by an ion moving with a velocity $v > (e^2 n_e^{1/3}/m)^{1/2}$. Transfer processes are also not significant since their rate cannot exceed the thermal velocity spread (longitudinal in our case). The electric permittivity for interaction with Larmor circles ($ku_A < \Omega$) can be written as

$$\varepsilon_{\vec{k}}(\omega) = 1 - \frac{\omega_e^2}{\omega^2}\frac{k_\|^2}{k^2}\langle J_0^2(k_\perp r_L)\rangle \equiv 1 - \frac{\omega_M^2(\vec{k})}{\omega^2}, \tag{2.69}$$

and then, after integration over $\omega$, Eqs. (2.45), (2.46) and (2.35), (2.47) give the following expressions for the friction force and scattering tensor due to interaction with the circles:

$$\vec{F}^L = -\frac{z^2 e^2}{2\pi^2}\int_{k<\Omega/v} d^3k \frac{\vec{k}}{k^2}\omega_M(\vec{k})\frac{\sin[(\omega_M - \vec{k}\vec{v})t]}{\omega_M - \vec{k}\vec{v}}, \tag{2.70}$$

$$d_{\alpha\beta}^L = \frac{2}{\pi}z^2 e^4 n_e' \int \frac{k_\alpha k_\beta}{k^4} d^3k \langle J_0^2(k_\perp r_L)\rangle \frac{\sin[(\omega_M - \vec{k}\vec{v})t]}{\omega_M - \vec{k}\vec{v}}, \tag{2.71}$$

$$\omega_M = \omega_e \langle J_0^2(k_\perp r_L)\rangle^{1/2}\frac{|k_\||}{k} = \left[\frac{4\pi n_e' e^2}{m}\frac{k_\|^2}{k^2}\langle J_0^2(k_\perp r_L)\rangle\right]^{1/2}. \tag{2.72}$$

At velocities $v$ that are not small compared to the electron Larmor ones $v_{e\perp}$, the complete expressions should include a contribution of the fast collisions determined by Eqs. (2.40) and (2.41). One can also obtain unified formulae encompassing contributions of all distances down to $\rho_{min} = e^2/(mv^2)$ and explicitly accounting for magnetization under the single condition of $\omega_e \ll \Omega$. To do this, let us expand the Bessel function in the expression for the electric permittivity, Eq. (2.50), while neglecting the electron longitudinal velocity spread:

$$J_0(2k_\perp r_L \sin\frac{\Omega\tau}{2}) = \sum_{l=-\infty}^{\infty} J_l^2(k_\perp r_L) e^{-il\Omega\tau}.$$

Completing integration over $\tau$, we get:

$$\varepsilon_{\vec{k}}(\omega) = 1 - \sum_l \left[\frac{\omega_e^2 k_\parallel^2/k^2}{(\omega - l\Omega)^2} + \frac{\omega_e^2 k_\perp^2/k^2}{(\omega - l\Omega)^2 - \Omega^2}\right]\langle J_l^2(k_\perp r_L)\rangle. \tag{2.73}$$

Integration over $\omega$ in Eqs. (2.46) and (2.47) reduces to calculation of the residues at the points $\varepsilon_{\vec{k}}(\omega) = 0$. When $\omega_e \ll \Omega$, zeros of $\varepsilon_{\vec{k}}(\omega)$ are located at the points

$$\omega \approx l\Omega \pm \omega_{Ml} \equiv l\Omega \pm \omega_e \frac{|k_\parallel|}{k} \langle J_l^2(k_\perp r_L)\rangle^{1/2}. \tag{2.74}$$

As a result, we arrive at the following general expression:

$$\vec{F} = -\frac{z^2 e^2}{2\pi^2} \int d^3k \frac{\vec{k}}{k^2} \sum_l \omega_{Ml} \frac{\sin[(\omega_{Ml} - \vec{k}\vec{v})t]}{\omega_{Ml} - \vec{k}\vec{v}}. \tag{2.75}$$

The scattering tensor can be written analogously as well.

Finally, let us add that the ultimate characteristic of the friction effect can be an integral of the force $F$ along the cooling section, or the change in momentum $\langle\Delta\vec{p}\rangle$:

$$\langle\Delta\vec{p}\rangle = \int_0^{t_0} \vec{F}(t)dt, \quad t_0 = \frac{l}{\gamma\beta c}, \tag{2.76}$$

for example,

$$\langle\Delta\vec{p}\rangle = -\frac{z^2 e^2}{\pi^2} \int d^3k \frac{\vec{k}}{k^2} \omega_M \frac{\sin^2[(\omega_{Ml} - \vec{k}\vec{v})t_0/2]}{(\omega_M l - \vec{k}\vec{v})^2}. \tag{2.77}$$

The same applies to scattering as well.

Sections 2.4 and 2.5 are devoted to a study of the behavior of the obtained expressions for $\vec{F}(\vec{v})$ and $d_{\alpha\beta}(\vec{v})$. Interaction with magnetized electrons at low velocities $v < (\Delta_{e\parallel})_{eff}$, when the perturbation theory is not applicable, is considered in Section 2.6.

## 2.4 Integral of adiabatic collisions

Let us consider in more detail properties of the friction and diffusion in the case, when the determining contribution to the integral of collisions with magnetized electrons comes from the distances $\rho > r_L$. For the corresponding $k$ in the formulae for $\vec{F}^L$ and $d_{\alpha\beta}^L$, one can set $J_0^2(k_\perp r_L) = 1$. Let us separate the cases of $\omega_e t \ll 1$ and $\omega_e t \gg 1$, which can later be joined.

1. $\omega_e t \ll 1$.

In this case, for the friction force, one can use Eq. (2.42):

$$\vec{F}^L = \vec{F}^A = \frac{2z^2 e^4 n'_e}{\pi m} \frac{\partial}{\partial \vec{v}} \langle \int_0^{k_{max}} \frac{k_\parallel^2 d^3k}{k^4} \frac{\sin(\vec{k}\vec{u}_A t)}{\vec{k}\vec{u}_A} \rangle,$$

where $k_{max} = \min\{\Omega/u_A, r_L^{-1}\}$. Integration over $\vec{k}$ can be conveniently done in spherical coordinates with the $z$ axis along $\vec{u}_A$:

$$\int_0^{k_{max}} \frac{k_\parallel^2 d^3k}{k^4} \frac{\sin(\vec{k}\vec{u}_A t)}{\vec{k}\vec{u}_A} = 4\pi \int_0^{k_{max}} \frac{dk}{k} \int_0^1 dx \left( \frac{u_\parallel^2}{u_A^2} x^2 + \frac{1-x^2}{2} \frac{v_\perp^2}{u_A^2} \right) \frac{\sin(k u_A t x)}{u_A x}.$$

Assuming the argument $u_A t k_{max}$ to be large, one can first integrate over $k$ in the limits $(0, \infty)$. The first term of the resulting integral over $x$ has no irregularities, while, in the second integral $\int dx/x$, one should set the lower limit equal to $1/(u_A t k_{max})$, i.e. to the value, below which the replacement $k_{max} \to \infty$ in the integral over $k$ is no longer valid. Finally, we get:

$$\vec{F}^A \approx \frac{2\pi z^2 e^4 n'_e}{m} \frac{\partial}{\partial \vec{v}} \langle \left[ \frac{v_\perp^2}{u_A^3} L^A + \frac{1}{u_A} \right] \rangle, \qquad (2.78)$$

where

$$L^A = \ln \frac{u_A t}{\rho_{min}^A}, \quad \rho_{min}^A = \max\{r_L, \frac{u_A}{\Omega}\} = \frac{1}{\Omega} \max\{v_{e\perp}, u_A\}. \qquad (2.79)$$

To calculate the scattering tensor in the considered case (Eq. (2.43))

$$d_{\alpha\beta}^A = \frac{2}{\pi} z^2 e^4 n'_e \int \frac{k_\alpha k_\beta}{k^4} d^3k \frac{\sin(\vec{k}\vec{u}_A t)}{\vec{k}\vec{u}_A},$$

we present its structure as

$$d_{\alpha\beta}^A = S(u_A^2 \delta_{\alpha\beta} - u_{A\alpha} u_{A\beta}) + \Pi(u_A^2 \delta_{\alpha\beta} - 3 u_{A\alpha} u_{A\beta}).$$

After calculating the simple integrals determining $S$ and $\Pi$, we get:

$$d_{\alpha\beta}^A = 4\pi z^2 e^4 n'_e \langle \frac{u_A^2 \delta_{\alpha\beta} - u_{A\alpha} u_{A\beta}}{u_A^3} L^A - \frac{1}{2} \frac{u_A^2 \delta_{\alpha\beta} - 3 u_{A\alpha} u_{A\beta}}{u_A^3} \rangle. \qquad (2.80)$$

Let us now switch to the case of a high density or a long-length cooling section, when screening manifests itself:

2. $\omega_e t \gg 1$.

Let us first estimate contribution of the stationary component of the friction force. For the region $k r_L < 1$, Eq. (2.60) takes the form

$$\vec{F}_{st}^A = \frac{2z^2 e^4 n}{m} \int \frac{\vec{k} k_\parallel}{k^4} d^3k \int d^3v_e \frac{\delta(\vec{k}\vec{u}_A)}{\left|\varepsilon_2(\vec{k}, k_\parallel v_{e\parallel})\right|^2} \frac{\partial f}{\partial v_{e\parallel}}, \quad (2.81)$$

$$\varepsilon_2 = 1 + \frac{\omega_e^2}{k^2} \int \frac{\partial f(\vec{v}_e')}{\partial v_{e\parallel}'} \frac{d^3 v_e'}{v_{e\parallel} - v_{e\parallel}' + i0}. \quad (2.82)$$

The electric permittivity in the denominator cuts off the logarithmic divergence in the region of small $k < k_D \simeq \omega_e/\Delta_{e\parallel}$. In case of

$$\frac{\Delta_{e\parallel}}{\omega_e} \gg r_L, \frac{u_A}{\Omega},$$

the integration over $\vec{k}$ and over electron velocities (with transfer of the derivative to the $\delta$-function) leads to an answer:

$$\vec{F}_{st}^A = \frac{2\pi z^2 e^4 n}{m} \frac{\partial}{\partial \vec{v}} \langle \frac{v_\perp^2}{u_A^3} L_{st}^A \rangle, \quad L_{st}^A = \ln \frac{\Delta_{e\parallel}}{\omega_e \rho_{min}^{st}}, \quad \rho_{min}^{st} = \frac{1}{\Omega} \max\{v_{e\perp}, u_A\}. \quad (2.83)$$

The condition required for validity of this expression is

$$r_D = \frac{\Delta_{e\parallel}}{\omega_e} \gg \rho_{min}^{st}.$$

The scattering tensor $(d_{\alpha\beta}^A)_{st}$ differs from Eq. (2.80) by the replacement $L^A \to L_{st}^A$ and the absence of the second term with a trace equal to zero

$$\left(d_{\alpha\beta}^A\right)_{st} = 4\pi z^2 e^4 n \langle \frac{u_A^2 \delta_{\alpha\beta} - u_{A\alpha} u_{A\beta}}{u_A^3} L_{st}^A \rangle. \quad (2.84)$$

The wave component of the force $\vec{F}_W^A$ significant at velocities $v \gg \Delta_{e\parallel}$ is determined by the integral in Eq. (2.67) in the region $k < r_L^{-1}$:

$$\vec{F}_W^A = -\frac{z^2 e^2 \omega_e}{4\pi} \frac{\partial}{\partial \vec{v}} \int d^3 k \frac{k_\parallel}{k^3} \left[\theta\left(\vec{k}\vec{v} - \omega_e \frac{k_\parallel}{k}\right) - \theta\left(\vec{k}\vec{v} + \omega_e \frac{k_\parallel}{k}\right)\right].$$

Remembering that the integration limit in $k$ for the wave component in any case does not exceed $k_D = \omega_e/\Delta_{e\parallel}$, after integration, we get:

$$\vec{F}_W^A \approx \frac{2\pi z^2 e^4 n}{m} \frac{\partial}{\partial \vec{v}} \frac{1}{v} \left[\frac{v_\perp^2}{v^2} \ln\left(\frac{2v^2}{v_\perp \omega_e \rho_{min}^W}\right) + 1\right], \quad (2.85)$$

where $\rho_{min}^W = \max\{\Delta_{e\parallel}/\omega_0, r_L, v/\Omega\}$ with the necessary condition $v/\omega_0 \gg \rho_{min}^W$. Note that, with $\Delta_{e\parallel}/\omega_e < r_L$, i.e., when the Debye radius becomes smaller than the Larmor one, adiabatic collisions do not contribute to the stationary component of the friction force and their contribution is completely contained in the wave component $\vec{F}_W^A$.

A formula for the scattering tensor $(d^A_{\alpha\beta})_W$ will have a form analogous to Eq. (2.80) only with a modified Coulomb logarithm.

All explicit expressions for the friction force and, correspondingly, the scattering tensor obtained in this section can be combined into one with a logarithmic accuracy sufficient for practical purposes. One should then account for an additional lower-limit constraint on the impact parameters related to a possible violation of the applicability of the perturbation theory: $\rho > e^2/(mu_A^2)$. We write the general formulae as:

$$\vec{F}^A = \frac{2\pi z^2 e^4 n'_e}{m} \frac{\partial}{\partial \vec{v}} \langle \frac{1}{u_A} \left( \frac{v_\perp^2}{u_A^2} L^A + 1 \right) \rangle, \tag{2.86}$$

$$d^A_{\alpha\beta} = 4\pi z^2 e^4 n'_e \langle L^A \frac{u_A^2 \delta_{\alpha\beta} - u_{A\alpha} u_{A\beta}}{u_A^3} \rangle, \tag{2.87}$$

$$L^A = \ln(\rho^A_{max}/\rho^A_{min}), \quad \vec{u}_A = \vec{v} - \vec{v}_{e\|}, \tag{2.87}$$

$$\rho^A_{max} = \min\{u_A t, u_A/\omega_e\}, \tag{2.89}$$

$$\rho^A_{min} = \max\{r_L, \frac{u_A}{\Omega}, \frac{ze^2}{mu_A^2}\}, \tag{2.90}$$

moreover, the following condition must be satisfied

$$\rho^A_{max} \gg \rho^A_{min}. \tag{2.91}$$

In practice, the conditions $\Omega t \gg 1$ and $\Omega \gg \omega_e$ are necessarily met with a large margin. Therefore, the condition in Eq. (2.91) can be written as a limit on the velocity of ion motion with respect to Larmor circles:

$$u_A \gg (u_A)_{min}, \tag{2.92}$$

where

$$(u_A)_{min} = \max\left\{\frac{r_L}{\tau_{eff}}, \left(\frac{ze^2}{m\tau_{eff}}\right)^{1/3}\right\}, \quad \tau_{eff} = \min\left\{\omega_e^{-1}, \frac{l}{\gamma\beta c}\right\}.$$

In experimental conditions, in particular, at the NAP-M installation, there are regions of parameters and proton velocities where these conditions are well satisfied. The interaction of heavy particles with magnetized electrons will be considered more generally in Sections 2.5 and 2.6; meanwhile, let us discuss the behavior of the force as a function of the ion velocity $\vec{v}$ with respect to the average electron velocity.

Let us first consider the situation when

$$v \gg \Delta_{e\|}. \tag{2.93}$$

Then

$$\vec{F}_\perp^A = -\frac{2\pi z^2 e^4 n'_e L^A(v)}{m} \frac{(v_\perp^2 - 2v_\parallel^2)}{v^2} \frac{\vec{v}_\perp}{v^3}, \quad (2.94)$$

$$F_\parallel^A = -\frac{2\pi z^2 e^4 n'_e}{m}\left(3\frac{v_\perp^2}{v^2}L^A + 1\right)\frac{v_\parallel}{v^3}, \quad (2.95)$$

$$\frac{d}{dt}\langle \Delta p_\alpha \Delta p_\beta \rangle^A = 4\pi n'_e z^2 e^4 L^A(v) \frac{v^2 \delta_{\alpha\beta} - v_\alpha v_\beta}{v^3}. \quad (2.96)$$

As can be seen, the logarithmic term of the longitudinal friction occurring due to collisions with Larmor circles in case of Eq. (2.93) has the feature that it disappears when $v_\perp \ll v_\parallel$. This fact has an obvious reason: with adiabatic motion of an ion and a Larmor circle along a magnetic field line, the integral momentum transfer in the longitudinal direction equals zero. This "defect" of the adiabatic collisions is partly compensated by the contribution of far-range collisions $\rho \gtrsim \rho_{max}$, for which the transferred momentum does not vanish at $v_\perp = 0$ (the term without the Coulomb logarithm in Eq. (2.95)).

Properties of the transverse friction are particularly unusual: when $v_\perp < \sqrt{2}|v_\parallel|$, $\vec{F}_\perp^A$ is directed along (rather than against) $\vec{v}_\perp$, i.e., there appears an anti-friction. The change of the friction sign at small $v_\perp \ll |v_\parallel|$ (compared to friction on free electrons) can be understood from the following considerations: when an ion approaches the "field line" of a Larmor circle, the electron integral longitudinal velocity decreases, while, when the ion moves away, it increases; the resulting difference in the times of effective interaction leads to acceleration of the ion (for same-sign charges, analogous considerations with obvious modifications lead, of course, to the same result).

Let us now evaluate the friction and diffusion at velocities

$$v < \Delta_{e\parallel}.$$

To be specific, let us choose the distribution $f(v_{e\parallel})$ of the form

$$f = \left\{(2\pi)^{3/2}\Delta_{e\perp}\Delta_{e\parallel}\exp\left(\frac{v_{e\perp}^2}{2\Delta_{e\perp}^2} + \frac{v_{e\parallel}^2}{2\Delta_{e\parallel}^2}\right)\right\}^{-1}. \quad (2.97)$$

Then, from Eqs. (2.86) and (2.87), we get

$$\vec{F}_\perp^A \approx -2\sqrt{2\pi}\frac{n'_e z^2 e^4}{m\Delta_{e\parallel}^3}\vec{v}_\perp \ln\left(\frac{\Delta_{e\parallel}}{v_\perp}\right)L^A(\Delta_{e\parallel}), \quad (2.98)$$

$$F_\parallel^A \approx -2\sqrt{2\pi}\frac{n'_e z^2 e^4}{m\Delta_{e\parallel}^3}v_\parallel L^A(v_\perp), \quad (2.99)$$

$$\frac{d}{dt}\langle(\Delta p_\perp)^2\rangle \approx 8\sqrt{2\pi}\frac{n'_e z^2 e^4}{\Delta_{e\|}}\ln\left(\frac{\Delta_{e\|}}{v_\perp}\right)L^A(\Delta_{e\|}),  \tag{2.100}$$

$$\frac{d}{dt}\langle(\Delta p_\|)^2\rangle \approx 4\sqrt{2\pi}\frac{n'_e z^2 e^4}{\Delta_{e\|}}L^A(v_\perp).  \tag{2.101}$$

With a precision of up to numerical and logarithmic multipliers, these expressions are similar to the usual friction and diffusion in a non-magnetized electron flow with isotropic distribution of electron velocities at a temperature of $T_e = m\Delta_{e\|}^2$.

## 2.5 Integral of cyclic collisions

Let us continue investigation of the collision integral switching to situations when collisions still remain weak, i.e., the perturbation theory is applicable, but the effectiveness of the Coulomb interaction at the distances of the order of or greater than the electron Larmor radius becomes reduced. This happens when the parameter $\rho_{max}^A = \min\{u_A t, u_A/\omega_e\}$ becomes smaller than $r_L$ but, at the same time, the velocities $u_A = |\vec{v} - v_{e\|}|$ are not so small that the condition $\rho_{max}^A > ze^2/(mu_A^2)$ is violated. From here it follows that such a situation is possible if (see Eqs. (2.89) – (2.92))

$$r_L > \left(\frac{ze^2 \tau_{eff}^2}{m}\right)^{1/3}, \quad \tau_{eff} = \min\{\omega_e^{-1}, t\}.  \tag{2.102}$$

Note that, in this case, the limits of applicability of the perturbation theory with further reduction in velocities $u_A$ should be determined accounting for the changed character of the collisional interaction at distances $\rho < r_L$, which is what we will be considering.

Thus, suppose

$$r_L \gg \min\{u_A t, u_A/\omega_e\}.  \tag{2.103}$$

Let us again start with a simpler situation when $\omega_e t \ll 1$ (but, of course, $\Omega t \gg 1$).

The friction force is then determined by the integral in Eq. (2.42) where the argument of the Bessel function is large in comparison to the argument of the sine; therefore, $J_0^2(k_\perp r_L)$ can be replaced by its average asymptote: $J_0^2(k_\perp r_L) \to 1/(\pi k_\perp r_L)$.

The integral then simplifies to

$$I = \int \frac{d^3k}{k^4}\frac{k_\|^2}{\pi k_\perp r_L}\frac{\sin(\vec{k}\vec{u}_A t)}{\vec{k}\vec{u}_A} \equiv \int_0^t d\tau \int \frac{d^3k}{k^4}\frac{k_\|^2}{\pi k_\perp r_L}\cos(\vec{k}\vec{u}_A \tau).$$

We first integrate over $k_\|$ and then over the directions $\vec{k}_\perp$:

$$I = \frac{\pi}{r_L} \int_0^t d\tau \int \frac{dk_\perp}{k_\perp} (1 - k_\perp |u_\|| \tau) e^{-k_\perp |u_\|| \tau} J_0(k_\perp v_\perp \tau)$$

$$\equiv \frac{\pi}{r_L} \int_0^t d\tau \int \frac{dx}{x} (1 - x) e^{-x} J_0(v_\perp x / |u_\||) .$$

The divergence at $k \to 0$ is fictitious, since it disappears when taking the derivative $\partial/\partial \vec{v}$ entering the expression for the force. It can be eliminated by subtracting unit from $J_0(v_\perp x/|u_\||)$:

$$I = \frac{\pi t}{r_L} \int_0^\infty \frac{dx}{x} [J_0(\frac{v_\perp}{|u_\||} x) - 1](1 - x) e^{-x}$$

$$= -\frac{\pi t}{r_L} [\frac{1}{\sqrt{1 + \frac{v_\perp^2}{u_\|^2}}} + \ln(1 + \sqrt{1 + \frac{v_\perp^2}{u_\|^2}}) - \ln 2 - 1] ;$$

the integral was calculated using differentiation with respect to a parameter and the integral

$$\int_0^\infty dx J_0(\alpha x) e^{-x} = \frac{1}{\sqrt{1 + \alpha^2}} .$$

Finally, the force $\vec{F}$ can be written as:

$$\vec{F} = -\frac{2z^2 e^4 n_e'}{m} \Omega t \frac{\partial}{\partial \vec{v}} \langle \frac{1}{v_{e\perp}} (\frac{|u_\||}{u_A} + \ln(|u_\|| + u_A)) \rangle . \tag{2.104}$$

In the region $u_A t < r_L$, the growth in friction with reduction in velocity slows down being proportional to only $u_A^{-1}$ instead of the "usual" law of $\sim u_A^{-2}$. It is interesting that the friction force is then proportional to the magnetic field and the path travelled by the particle from the entrance into the cooling section and is inversely proportional to the electron Larmor velocities.

The condition $u_A t < r_L$ means that the velocity $u_A$ is at least $\Omega t$ times lower than the electron Larmor velocity $v_{e\perp}$. Hence, it is easy to understand the proportionality of the friction force to the number of Larmor cycles in time $t$. When a proton with an impact parameter $\rho$ is moving near a Larmor circle, the electron passes through the interaction area $\simeq \Omega \rho / u_A$ times; moreover, the momentum kicks received by the electron in each pass add up coherently:

$$\Delta p_\| \simeq \frac{ze^2}{\rho v_{e\perp}} \cdot \Omega \frac{\rho}{u_A} = \frac{ze^2 \Omega}{v_{e\perp} u_A} ,$$

so that the shift of the circle coordinate along the magnetic field equals

$$\Delta S \simeq \frac{ze^2 \Omega}{m v_{e\perp} u_A} \frac{\rho}{u_A}$$

and the average force of proton interaction with the circles is

$$F \simeq \left(\frac{ze^2}{\rho^3}\Delta S\right) \cdot n\rho^3\big|_{\rho \simeq \rho_{eff}=u_A t} \simeq \frac{z^2 e^4 n \Omega t}{m v_{e\perp} u_A}$$

in accordance with Eq. (2.104).

With the scattering tensor, we constrain ourselves to estimation of its trace $d_{\alpha\alpha}$:

$$\frac{d}{dt}\langle(\Delta\vec{p})^2\rangle = \frac{2z^2 e^4 n}{\pi}\int \frac{d^3k}{k^2}\langle\frac{1}{\pi k_\perp r_L}\frac{\sin(\vec{k}\vec{u}_A t)}{\vec{k}\vec{u}_A}\rangle \ ;$$

the logarithmic divergence at zero should be cut off at $k \simeq 1/r_L$. After calculations similar to those that were just completed, we get:

$$\frac{d}{dt}\langle(\Delta\vec{p})^2\rangle = 4z^2 e^4 n'_e \Omega t \langle\frac{1}{v_{e\perp}}\ln\left(\frac{r_L}{u_A t}\right)\rangle. \tag{2.105}$$

The obtained formulae are applicable with velocities $u_A > \sqrt{ze^2/(mr_L)}$ (see Section 2.6).

Let us move to the case

2. $t > t_{scr}$ where $t_{scr}$ is the screening time for distances $\rho < \rho_L$, which should be determined accounting for the Larmor rotation of electrons.

Since the maximum impact parameter of adiabatic collisions $\rho_{max}^A$ equals $u_A/\omega_e$, their contribution is suppressed if $u_A/\omega_e \ll r_L$; interaction at distances $\rho < r_L$, with the upper limit determined by dynamic screening, then becomes dominant.

In case of $u_A/\omega_e \ll r_L$, the screening parameter does not remain equal to $u_A/\omega_e$ but increases somewhat due to effective weakening of the interaction at distances $\rho < r_L$ as a result of the fast Larmor motion of electrons. The screening distance corresponding to the stationary component of the interaction is $r_{scr} = k^{-1}$ determined from the equation (see Eq. (2.62))

$$k\Delta_{e\|} \simeq \omega(k),$$

while the screening distance corresponding to the wave component is determined by (Eq. (2.67))

$$kv \simeq \omega(k), \quad (v \gg \Delta_{e\|}),$$

where $\omega(k) \simeq \omega_e \langle J_0^2(kr_L)\rangle^{1/2}$ is the frequency of electron collective oscillations accounting for magnetization. For estimates, these two equations can be combined:

$$k_{scr} u_A \simeq \omega_e \langle J_0^2(k_{scr} r_L)\rangle^{1/2} \simeq \omega(k_{scr}) \tag{2.106}$$

with $u_A$ implying the average value of the ion velocity with respect to Larmor circles. In general case, to have a meaningful discussion of screening, the condition $\omega(k_{scr})t \gg 1$ must be satisfied, which is a stronger requirement than simply $\omega_e t \gg 1$, since the frequency $\omega(k_{scr})$ cannot exceed $\omega_e$. In the long-wave limit, $J_0^2(kr_L)$, Eq. (2.106) then gives the screening parameter value of $k_{scr} \simeq \omega_e/u_A$ used in Section 2.4. This dependence is valid for the region $u_A < \omega_e r_{scr}$, while, for lower velocities, $k_{scr} r_L \gg 1$; hence, one has to use the asymptote: $\langle J_0^2(k r_L) \rangle \to 1/(\pi k r_L)$. We then get:

$$k_{src}^{-1} = \left(\frac{\omega_e^2}{\pi r_L u_A^2}\right)^{-1/3} = \frac{u_A}{\omega_e}\left(\frac{\pi \omega_e r_L}{u_A}\right)^{1/3}, \tag{2.107}$$

$$\omega(k_{scr}) = \omega_e \left(\frac{u_A}{\pi r_L \omega_e}\right)^{1/3}.$$

Thus, when $u_A \ll \omega_e r_L$, the characteristic time of dynamic screening is

$$t_{scr} = \omega_e^{-1}\left(\frac{\pi r_L \omega_e}{u_A}\right)^{1/3} \tag{2.108}$$

and one should distinguish between the cases $t < t_{scr}$ and $t > t_{scr}$.

When $t > t_{scr}$, the friction force is determined by Eqs. (2.60) and (2.67), where we replace the Bessel function with its asymptote:

$$\vec{F} = \vec{F}_{st} + \vec{F}_W ;$$

$$\vec{F}_{st} = \frac{2z^2 e^4 n}{m}\int \frac{\vec{k} k_\parallel d^3 k}{k^4 k_\perp}\int d^3 v_e \frac{\partial f}{\partial v_{e\parallel}}\delta(\vec{k}\vec{u}_A)\frac{\Omega}{v_{e\perp}}\frac{1}{|\varepsilon_M|^2}, \tag{2.109}$$

where

$$\varepsilon_M = 1 + \frac{\omega_e^2 \Omega}{\pi k^2 k_\perp}\int \frac{d^3 v_e'}{v_{e\perp}'}\frac{\partial f(\vec{v}_e')}{\partial v_{e\parallel}'}\frac{1}{v_{e\parallel} - v_{e\parallel}' + i0};$$

$$\vec{F}_W = -2\frac{z^2 e^2}{4\pi}\int d^3 k \frac{\vec{k}}{k^2}\omega(\vec{k})\delta[\omega(\vec{k}) - \vec{k}\vec{v}], \tag{2.110}$$

where

$$\omega(k) = \omega_e \frac{|k_\parallel|}{k}\langle J_0^2(k_\perp r_L)\rangle^{1/2} \approx \omega_e \frac{|k_\parallel|}{k}\frac{1}{\sqrt{\pi k r_L}}.$$

Recall that the force $\vec{F}_W$ can be (and should be) neglected at velocities $v \lesssim \Delta_{e\parallel}$, while, at $v \gg \Delta_{e\parallel}$, the full force is the sum of the two components.

Due to the combined effect of magnetization and screening, the formulae for the friction force contain no divergences and the main contribution comes exactly from the distances $\rho \simeq k_{scr}^{-1} \simeq (u_A/\omega_e)(\pi \omega_e r_L/\Delta_{e\parallel})^{1/3}$ and $(u_A/\omega_e)(\pi \omega_e r_L/v)^{1/3}$, respectively. The integral over $|\vec{k}|$ in $\vec{F}_{st}$ can be calculated in terms of elementary functions; however, the resulting answer is too cumbersome. For an estimate, one can do without explicit integration, using dimensionality considerations.

Note that the integrals over electron (longitudinal) velocities are concentrated in the region $v_{e\|} - v'_{e\|} \lesssim \Delta_{e\|}$, then

$$\varepsilon_M = 1 + \frac{\omega_e^2}{\pi r_L \Delta_{e\|}^2 k^3}.$$

Changing to a dimensionless variable $X = k(\pi r_L \Delta_{e\|}^2/\omega_e^2)^{1/3}$, we gen an estimate (not paying attention to the non-isotropy of $\vec{F}_{st}$ as a function of $\vec{v}$):

$$\vec{F}_{st} \simeq -\frac{z^2 e^4 n}{m} \cdot \left(\frac{\Delta_{e\|}}{\omega_e r_L}\right)^{2/3} \langle\frac{\vec{u}_A}{u_A^3}\rangle, \tag{2.111}$$

(compare to Eq. (2.104)).

One can obtain more accurate estimates of the force $\vec{F}_W$. Introducing angle $\theta$ between $\vec{k}$ and $\vec{v}$ and angle $\chi$ between $\vec{k}$ and $\vec{\Omega}$, we remove integration over $|\vec{k}|$ in Eq. (2.110):

$$\vec{F}_W = -\frac{z^2 e^2}{3\pi}\left(\frac{\omega_e^2}{\pi r_L v^2}\right)^{2/3} \int \frac{\vec{k}}{k} d\Omega \frac{|\cos\chi|}{\cos\theta}\left|\frac{\cos\chi}{\cos\theta \sin^2\chi}\right|^{1/3}, \tag{2.112}$$

where $d\Omega = \sin\theta\, d\theta d\varphi$ and $\cos\chi = (v_\|/v)\cos\theta - (v_\perp/v)\sin\theta\cos\varphi$. For the force along the velocity, we get a formula:

$$(\vec{F}_W)_v = -\frac{4z^2 e^2}{3}\left(\frac{\omega_e^2}{\pi r_L v^2}\right)^{2/3} \frac{\vec{v}}{v} \int \frac{d\Omega}{4\pi}|\cos\chi|\left|\frac{\cos\chi}{\cos\theta \sin^2\chi}\right|^{1/3}. \tag{2.113}$$

The angular integral as a function of the velocity direction has no special points and has an order of magnitude of one.

The component of the force in Eq. (2.112) that is transverse to the velocity, lies in the plane $(\vec{v}, \vec{\Omega})$:

$$F_{tr} = \frac{4z^2 e^2}{3}\left(\frac{\omega_e^2}{\pi r_L v^2}\right)^{2/3} \int \frac{d\Omega}{4\pi}\sin\theta\cos\varphi \frac{|\cos\chi|}{\cos\theta}\left|\frac{\cos\chi}{\cos\theta \sin^2\chi}\right|^{1/3}.$$

Comparing Eq. (2.112) to Eq. (2.111), we see that, at velocities $v \gg \Delta_{e\|}$, the friction force exceeds the force $\vec{F}_{st}$ corresponding to the usual method of obtaining the collision integral, by a factor of $\simeq (v/\Delta_{e\|})^{2/3}$.

For estimates and qualitative analysis, one can use an interpolation formula

$$\vec{F} \simeq -\frac{4\pi z^2 e^4 n'_e}{m} \langle\frac{\vec{u}_A}{u_A^3}\left(\frac{u_A}{\pi\omega_e r_L}\right)^{2/3}\rangle, \tag{2.114}$$

when

$$u_A < \omega_e r_L, \quad \omega_e \left(\frac{u_A}{\pi\omega_e r_L}\right)^{1/3} t \gg 1.$$

What is noteworthy is the fractional-power dependence of the friction force on the electron beam density, Larmor velocities, and magnetic field:

$$F \sim \left(\frac{n_e' H}{v_{e\perp}}\right)^{2/3}. \tag{2.115}$$

Using Eq. (2.71), one can estimate the momentum scattering rate under the same condition as in Eq. (2.114):

$$\frac{d}{dt}\langle(\Delta\vec{p})^2\rangle \simeq 4\pi z^2 e^4 n_e' \langle \frac{1}{u_A}\left(\frac{u_A}{\pi\omega_e r_L}\right)^{2/3}\rangle. \tag{2.116}$$

As shown in Section 2.6, the applicability of Eqs. (2.114) and (2.116) is limited to the region of $mu_A^2 > e^2(n_e'/r_L)^{1/4}$, which gives that, assuming the condition $u_A < \omega_e r_L$, the dependence in Eq. (2.114) can be realized only in case of $nr_L^3 > 1$, i.e., when Larmor radii exceed the average distance between electrons.

## 2.6 Extreme relaxation regimes. Maximum decrements and minimum temperatures

In the earlier narrative, we limited ourselves to the domain of the perturbation theory and also neglected a possible change in the velocity distribution of electrons due to their collisions with each other. Realistically, the velocity spread of Larmor circles can be so small that it becomes necessary to study the collisional kinetics without these limitations. Regrettably, one is not able to do a rigorous quantitative consideration of the extreme situations in an analytic form due to obvious difficulties. Nevertheless, starting with the perturbation theory formulae and involving additional physics considerations, one can obtain reliable estimates of the main characteristics of the relaxation process. The results allow one to get an idea of the limiting capabilities of the electron cooling method and the requirements, which an electron cooling system must satisfy, so that these capabilities could be realized.

Logically, it is natural to first consider the collisional relaxation of the electron beam itself and then move on to the estimation of friction and diffusion for heavy particles.

### Relaxation of electron flow

The single question of practical importance here is that of evolution of the electron longitudinal velocity spread in view of its relative smallness as a result of electrostatic acceleration. Its increase or, in general, increase in the entropy of the longitudinal motion can be related to energy transfer from the transverse degrees of freedom during scattering of electrons on each other as well as, at a sufficiently small velocity spread, to interaction without energy exchange with the Larmor degree of freedom, which can lead only to thermalization of the circles' motion.

Collisional relaxation, in general case, starts already at the stage of particle acceleration from the cathode. Assuming for now that the role of the acceleration section comes down to preparation of the "initial" bam state, let us estimate the effect of electron interaction in different cases.

In a non-magnetized electron flow, growth of the longitudinal temperature due to collisions would be equal to (see Chapter I):

$$\Delta T_{e\|} = \frac{2\pi^{3/2} L e^3 j_e S}{(\gamma + 1) W_e} \sqrt{\frac{m}{T_k}}, \qquad (2.117)$$

where $j_e$ is the current density of the beam, $S$ is the path travelled after acceleration, and $W_e = (\gamma - 1)mc^2$ is the electron energy. For example, for $j_e = 300$ mA/cm², $W_e = 35$ keV, $T_k = 0.2$ eV, and $S = 2$ m, we get $T_{e\|} \simeq 3 \cdot 10^{-4}$ eV (while damping due to deformation of the phase volume during acceleration gives an initial temperature of $T_{e\|}^{(a)} \simeq 10^{-7}$ eV).

Influence of magnetization will be small at a sufficiently weak magnetic field and not-too-small longitudinal temperatures. Let us consider an opposite extreme situation when the Larmor radii are small compared to the characteristic distance between neighboring electrons:

$$r_L \ll n_e'^{-1/3} \qquad (2.118)$$

and the kinetic energies of relative (longitudinal) motion of the circles are of the order of or smaller than the energy of the Coulomb interaction of neighbors:

$$mu_\|^2 \lesssim e^2 n_e'^{1/3}. \qquad (2.119)$$

In this case, energy transfer from the transverse (Larmor) degree of freedom is hindered by the clearly pronounced finite extent of the electron transverse motion. Moreover, due to mutual repulsion, correlations are established in the spatial distribution of circles that prohibit electrons from approaching each other at distances small enough that a collision with energy transfer from the Larmor motion could take place. There is an extreme situation when the longitudinal temperature is small compared to "fluctuations" of the Coulomb interaction: $m\Delta_{e\|}^2 \ll e^2 n_e'^{1/3}$ and the spatial distribution of circles is ordered into a crystal lattice. Practically, a more probable initial (post-acceleration) state is that with chaotic positioning of circles stationary with respect to each other, which, in a time of the order of

$$\tau_{rel} \simeq \sqrt{\frac{n_e'^{-1/3}}{2(e^2 n_e'^{2/3}/m)}} \simeq \omega_e^{-1} = \tau_{scr}, \qquad (2.120)$$

i.e., in a Langmuir period (or several periods), relaxes to a distribution of the Gibbs type with a longitudinal velocity spread corresponding to the approximate equality

$$m\Delta_{e\parallel}^2 \simeq e^2 n_e'^{1/3} . \tag{2.121}$$

At a density of $10^8$ cm$^{-3}$, this is about $5 \cdot 10^{-5}$ eV. The radius of Debye screening in such states reaches a minimum value approaching the average distance between particles of $\simeq n_e'^{-1/3}$, i.e., interaction with "far" neighbors is screened. Note that the steadiness of Larmor circles impedes the establishment of true Gibbs correlations (including the crystal lattice one when $m\Delta_{e\parallel}^2 \ll e^2 n^{1/3}$); it is easy to estimate that fluctuations of the circles' drift velocity relate to the characteristic velocity of $\simeq \sqrt{e^2 n^{1/3}/m}$, as this velocity itself relates to the Larmor one of $\sqrt{T_k/m}$.

Arising as a result of acceleration, the considered meta-stable states can exist during a time exponentially large (in terms of the parameter $1/(n_e' r_L^3)$) compared to the Langmuir period and, almost certainly, along the whole length of the cooling section.

Let us next consider another characteristic situation, one of large Larmor radii

$$n_e' r_L^3 \gg 1 , \tag{2.122}$$

but again, with an infinitesimally small (at first, zero) initial longitudinal velocity spread. In this case, increase in the spread occurs as a result of multiple cyclic collisions of electrons moving in Larmor circles with the distance between their centers shorter than $r_L$ and with a small phase difference relative to the "crossing point" of the circles. If an individual collision is characterized by an impact parameter $\rho \ll r_L$, then, in time $t$, the circles exchange momentum

$$\Delta p_\parallel \simeq \frac{e^2}{\rho v_{e\perp}} \cdot \frac{\Omega t}{2\pi} = \frac{e^2}{\pi \rho r_L} t ; \tag{2.123}$$

the number of electrons interacting with a given one in an interval $d\rho$ equals

$$dN = 2\pi\rho d\rho \cdot 2\pi r_L \cdot n$$

and thus

$$\langle (\Delta p_\parallel)^2 \rangle \simeq \frac{4 n_e' e^4}{r_L} t^2 \ln \frac{\rho_{max}}{\rho_{min}} . \tag{2.124}$$

The time $t$ in this relation, as in the case of $n r_L^3 \ll 1$, is limited by the screening time, which, as was shown in Sections 2.3 and 2.4, with a spread $\Delta_{e\parallel} < \omega_e r_L$, equals

$$\tau_{scr} \simeq \omega_e^{-1} \left( \frac{\pi \omega_e r_L}{\Delta_{e\parallel}} \right)^{1/3} ;$$

substituting time $t$ from Eq. (2.124) instead of $\tau_{scr}$ here, we get a quasi-equilibrium temperature

$$m\Delta_{e\|}^2 \simeq e^2 \left(\frac{n}{r_L}\right)^{1/4} = \frac{e^2}{r_L}(nr_L^3)^{1/4}, \quad nr_L^3 \gg 1, \tag{2.125}$$

(then $\rho_{max} \simeq \rho_{min} \simeq t_{scr}\Delta_{e\|}$, so that $\ln(\rho_{max}/\rho_{min}) \simeq 1$), which is established in a characteristic time

$$\tau_{rel} \simeq \omega_e^{-1}(nr_L^3)^{1/8}, \tag{2.126}$$

and, correspondingly, the screening distance is

$$r_{scr} = \Delta_{e\|}\tau_{scr} \simeq \left(\frac{r_L}{n'_e}\right)^{1/4} < r_L. \tag{2.127}$$

Thus, interaction without energy exchange with the Larmor motion leads only to a fast thermalization of the longitudinal motion with a temperature very small compared to the transverse one.

Let us now consider processes with energy exchange accounting for magnetization. First of all, at $u_{e\|} \ll v_T$, circle collisions with impact parameters exceeding their diameters do not result in an exchange, since they are adiabatic with respect to the Larmor rotation. Next, collision with "crossing" of circles with relative Larmor phases corresponding to impact parameters

$$\rho \gg u_{e\|}/\Omega$$

are also not effective, since the collisions occur multiple times (with a period $2\pi/\Omega$) in a characteristic time $\tau_{tot} \simeq \rho/u_A$ with symmetric (opposite-sign) contributions to the longitudinal momentum change before and after the crossing of the circle planes; as a result, the integral effect vanishes. Only collisions with $\rho < u_{e\|}/\Omega$, which can happen only once, lead to a transfer into the longitudinal motion, independent of magnetization. As we can see, the situation is quite analogous to the ion-electron collisions.

For a single collision to take place, the particles must overcome a potential barrier of the longitudinal interaction (see Eq. (2.123))

$$U(\rho) = \int_{-\infty}^{\rho} \frac{dp_\|}{dt} d\rho \simeq \frac{e^2}{\pi r_L} \ln \frac{\rho_{max}}{\rho}, \tag{2.128}$$

where $\rho_{max} = \min(r_L, (r_L/n)^{1/4})$ and one should set $\rho = u_{e\|}/\Omega$; thus, we arrive at a condition, which can allow the start of energy transfer from the Larmor degree of freedom:

$$m\Delta_{e\|}^2 \gtrsim \frac{e^2}{2\pi r_L} \ln \frac{\varkappa r_L T_k}{e^2}, \quad \varkappa = \min\{1, \frac{1}{\sqrt{nr_L^3}}\}. \tag{2.129}$$

Relative to the considered above relaxation regimes with a zero initial spread, this condition, as can be seen, is satisfied only with sufficiently large values of the parameter $n'_e r_L^3 \gg 1$:

$$(n'_e r_L^3)^{1/4} \gtrsim \frac{1}{2} \ln \frac{r_L T_k}{e^2}. \tag{2.130}$$

When the condition in Eq. (2.129) is satisfied, the longitudinal spread growth rate in times $t \gg t_{scr}$ (in shorter times, singular collisions are definitely of no significance) can be obtained using Eq. (2.35) where $\vec{r}(t)$ should be replaced by the electron trajectory in the magnetic field. Furthermore, since we are considering relaxation of the relative motion of particles constituting the medium, screening at $t \gg \tau_{scr}$ will already be equilibrium, i.e., one can omit the wave terms. We then get:

$$\frac{d}{dt}\frac{\langle(\Delta p_\parallel)^2\rangle}{m} = \frac{4n'_e e^4}{m}\int d^3k \frac{k_\parallel^2}{k^4}\sum \left\langle\left\langle\frac{J_l^2(k_\perp r_L)J_{l'}^2(k_\perp r_L)\delta[k_\parallel u_\parallel + (l-l')\cdot\Omega]}{\left|\varepsilon_{\vec{k}}(\omega)\right|^2_{\omega=k_\parallel v_{e\parallel}+l\Omega}}\right\rangle\right\rangle, \tag{2.131}$$

where $J_l$ is the Bessel function, $l, l' = 0, \pm 1, \pm 2, \ldots$, the brackets $\langle\langle\ldots\rangle\rangle$ denote averaging over the variable of the two interacting electrons, and the electric permittivity is determined by the expression in Eq. (2.50).

All $l' = l$ terms in Eq. (2.131) become identically zero $k_\parallel^2 \delta(k_\parallel u_\parallel) = 0$). Generally speaking, their sum describes a mutual diffusion of the circles during interaction averaged over the Larmor rotation. As can be seen, contribution of the $l' \neq l$ terms differs from zero only in the region of $k \gtrsim \Omega/\Delta_{e\parallel}$. Remember that we are interested in the situation $\Delta_{e\parallel} \ll v_T$, therefore, the border distance is small compared to the average Larmor radius. As it should be, the impact parameters $\rho > \Delta_{e\parallel}/\Omega$ do not contribute to the energy exchange. Note that the considered zeroing happens specifically due to screening (moreover, $r_D = \Delta_{e\parallel}/\omega_e \gg \Delta_{e\parallel}/\Omega$); without the electric permittivity in the denominator of Eq. (2.131), the integral over $\vec{k}$ for the $l' = l$ terms is finite and is inversely proportional to $|u_\parallel|$ that corresponds to the contribution of the "no-pass" distances $\rho \sim |u_\parallel|$, which, in reality, are cut off by screening.

Let us now consider contribution of the region $k > \Omega/\Delta_{e\parallel}$ corresponding to the impact parameters, at which an effective collision (with a relative velocity $\vec{v}_\perp - \vec{v}'_\perp$) occurs not more than once. The respective distances are small compared to the Debye and Larmor radii, so that one can set $\varepsilon = 1$. It is convenient to do the summation over $l, l'$ using the representation

$$\pi\delta(x) = \int_0^\infty \cos(x\tau)\,d\tau.$$

For estimates, we will assume the electron distribution to be factorized in transverse and longitudinal velocities and the transverse distribution to be Maxwellian. After calculations using all of the indicated conditions, we get

$$m^2\frac{d}{dt}\Delta_{e\parallel}^2 = \frac{2\pi\sqrt{\pi}n'_e e^4}{v_T}\ln\frac{\Delta_{e\parallel}}{\Omega\rho_{min}}. \tag{2.132}$$

As one should expect, the result is close to the diffusion coefficient in a non-magnetized plasma differing only by the Coulomb logarithm. The upper impact parameter equals $\Delta_{e\parallel}/\Omega$ analogously to the case of singular proton collisions with Larmor electrons; the limit on the parameter $\rho_{min}$ is the value of $e^2/(m\Delta_{e\parallel}^2)$ but, for finite, not-too-large values of $t/\tau_{scr}$, improvement in precision is required. It is based on the consideration that, after time $t$, probable collisions are those with impact parameters satisfying the condition

$$w(\rho) = \pi\rho^2 n'_e v_T t \gtrsim 1,$$

while collisions in the region of $w(\rho) \ll 1$ do not yet occur and should not be included. The subtlety is that Eq. (2.132) with $\rho_{min} = e^2/(m\Delta_\parallel^2)$ correctly describes the change in such a rough parameter as $\langle(\Delta v_{e\parallel})^2\rangle$ but not that in the temperature in general case, since this formula is, in fact, an average over a large number of electrons, among which there is a small fraction that experienced scattering with large momentum transfers, giving the main contribution to the logarithm. However, for interaction with ions (with $v \lesssim \Delta_{e\parallel}$), what is significant is the main mass of electrons that experienced scattering with a relatively small transfer. Thus, one should set

$$\rho_{min} = \frac{1}{\sqrt{\pi n'_e v_T \tau}} \gtrsim \frac{e^2}{m\Delta_{e\parallel}^2} \tag{2.133}$$

and then Eq. (2.132) overall is meaningful when

$$w(\tau) = \frac{\Delta_{e\parallel}^2 \pi n'_e v_T \tau}{\Omega^2} \gg 1. \tag{2.134}$$

Substituting here the law of longitudinal temperature evolution, Eq. (2.132), we get that the energy inflow from the transverse motion should be included if

$$w \equiv 2\pi^2 \sqrt{\pi} \left(\frac{mcjS}{(\gamma+1)W_e H}\right)^2 \gg 1, \tag{2.135}$$

where $S$ is the path travelled by particles. The change in temperature as a function of $S$ then equals

$$\Delta T_{e\parallel}(S) = \frac{\pi\sqrt{\pi} mcj_e S}{(\gamma+1)W_e H} \frac{e^2}{r_L} \ln w(S) = \frac{e^2}{r_L} \sqrt{\frac{\omega(S)}{2}} \sqrt{\pi} \ln w(S). \tag{2.136}$$

It is instructional that the Coulomb logarithm reduced simply to a probability logarithm. The length of complete relaxation is

$$S_{tot} \simeq \frac{(\gamma+1)W_e T_\perp^{3/2}}{2\pi\sqrt{\pi m} e^3 j_e \ln(T_\perp r_L/e^2)}. \tag{2.137}$$

Based on the above, it is now easy to establish a dependence of the state arising a result of acceleration on the basic parameters: $j_e$, $T_k$, $W$, $H$, and the length of the acceleration path $S_{acc}$. In discussions, the density and acceleration time in the co-moving frame can be more convenient parameters than $j_e$ and $S_{acc}$:

$$n'_e = \frac{j_e}{e}\sqrt{\frac{m}{(\gamma+1)W_e}},$$

$$\tau_{acc} = \frac{S_{acc}}{c}\frac{mc^2}{W_e}\ln(\gamma+\sqrt{\gamma^2-1}).$$
(2.138)

The density can be considered to have an order of magnitude equal to its value starting approximately from the middle of the path $S_{acc}$. The parameter $n'_e r_L^3$ does not depend on the electron interaction and can also be expressed through $j_e$, $W_e$, $H$, and $T_k$.

In practice, the acceleration time is always sufficiently small that one can neglect the processes of energy transfer from the transverse degrees of freedom, so that the final state is determined by the competition of freezing due to spatial stretching and diffusion due to longitudinal repulsion of circles. One can distinguish the following characteristic cases.

1. $$T_{\parallel}^a \simeq \frac{T_k^2}{(\gamma+1)W} \gg \min\{e^2 n'_e{}^{1/3}, e^2(n'_e/r_L)^{1/4}\}.$$

In this case, the interaction of circles is a small perturbation in comparison to their relative longitudinal motion, and not only in the acceleration section but in the whole subsequent electron path; the initial longitudinal temperature can be considered equal to

$$T_{\parallel}^a \simeq \frac{T_k^2}{(\gamma+1)W}.$$

2. $$\frac{T_k^2}{(\gamma+1)W} \ll \min\{e^2 n'_e{}^{1/3}, e^2(n'_e/r_L)^{1/4}\}.$$

Practically, this is a more probable situation. The evolution here depends on the relation of the acceleration time and the time of longitudinal relaxation $\tau_{scr} = \omega_e^{-1}\cdot\max\{1,(n'_e r_L^3)^{1/8}\}$. When $\tau_{acc} \ll \tau_{scr}$, the interaction is again not significant and the initial spread corresponds to the temperature $T_{e\parallel}^a$; since spatial correlations at the cathode are small, then the distribution of circles at the end of acceleration will also be completely chaotic. With the opposite relation of the times, in case of $n'_e r_L^3 \ll 1$, positioning of circles develops correlations. The particles undergo oscillations in the longitudinal direction with a frequency of $\simeq \omega_e$, while the temperature changes from the value $T_{e\parallel}^a(S) = e^2 n^{1/3}(S)$ following the law of the adiabatic invariance $\Delta_{e\|}^2/\omega_e = const$ and drops below the level of $e^2 n'^{1/3}$. In case of $n'_e r_L^3 \gg 1$, a temperature of $T_{e\|} = e^2(n'_e/r_L)^{1/4}$ is maintained during acceleration without notable spatial correlations.

## Extreme cooling regimes

In the cooling process of a heavy particle beam, ion velocities with respect to the electron flow can reach values small (due to the large mass difference) compared to the minimum spread of electron longitudinal velocities determined by fluctuations of the Coulomb interaction energy. The relaxation process can then be envisioned in the following way. At the initial moment, at the entrance into the cooling section, a heavy particle is surrounded by a frozen cloud of electron Larmor circles without notable correlations in their mutual positions and, of course, in their positions with respect to the heavy particles. The circles closest to the ion exchange with it an energy of the order of the initial Coulomb interaction energy (in general, averaged over the Larmor rotation), while contribution of remote particles is limited by a finite time of the pass or by polarization of the cloud under the influence of the ion itself. Due to long-distance Coulomb interaction, the main contribution will come from those particles that happen to be at distances of the order of the effective screening radius. The slow moving ion will experience deceleration (on average) linear in its velocity with respect to the electron beam while the diffusion rate is already independent of its motion; in this respect, the situation is analogous to the Brownian motion.

Despite the extreme smallness of relative velocities, in practice, one can exclude from the consideration the stage of particle approach when an ion enters the electron beam, since the entrance takes a relatively short time, after which the particles having ended up near each other interact during a significantly longer time. This question is even less substantial in case of a sufficiently long interaction section, when the distribution of circles has enough time to mix during the pass.

We will also neglect the processes of energy transfer from the Larmor degree of freedom to the longitudinal one, since, in practice, as we saw, increase in the longitudinal temperature during a pass is too small.

We will base the estimates of friction decrements and diffusion coefficient of heavy particles on the perturbation theory formulae but cutting off the contribution of "small" distances at the boundary where the energy (momentum) exchange during the maximum time of correlated interaction reaches the initial Coulomb energy. Let us start with Eqs. (2.42) and (2.43) where time $t$ should be limited by the time of pass or by the time of non-equilibrium screening $t_{scr}$, while the integration over $|\vec{k}|$ should be limited by the value of $k_{max} \simeq 1/\rho_{eff}$ where $\rho_{eff}$ is the distance from a Larmor circle, at which the shift $\Delta\rho$ during the interaction time reaches the initial distance $\rho$. Correspondingly, the role of an effective relative velocity should be taken by the velocity $|\vec{v} - \Delta\vec{v}_{e\parallel}|$ where $\Delta\vec{v}_{e\parallel} \simeq \rho_{eff}/t_{eff}$ is the circle velocity arising due to the interaction in a time $t_{eff}$ ($\vec{v} \ll \Delta v_{e\parallel}$). The argument of sine then is of the order of unit, so that, for estimates, one can use the expansion

$$\frac{\sin x}{x} \simeq 1 - \frac{x^2}{6} + \ldots$$

The orders of magnitudes of the friction force and momentum diffusion rate are

$$\vec{F} \simeq -\frac{16z^2 e^4 n'_e \vec{v}}{3m} \int_0^{1/\rho_{eff}} dk \langle J_0^2(k_\perp r_L)\rangle k^2 t_{eff}^3, \tag{2.140}$$

$$d \equiv \frac{d}{dt}(\Delta\vec{p}_c)^2 \simeq 8z^2 n'_e e^4 \int_0^{1/\rho_{eff}} dk \langle J_0^2(k_\perp r_L)\rangle t_{eff}. \tag{2.141}$$

Let us introduce a characteristic of friction $\mathcal{K}$

$$\vec{F} \simeq -\mathcal{K}\vec{v};$$

then the friction decrement (in the lab frame) and the equilibrium ion temperature can be written as

$$\lambda_{max} = \frac{\mathcal{K}}{\gamma M}\eta, \quad T_{st} = \frac{d}{6\mathcal{K}}. \tag{2.142}$$

According to Eqs. (2.140) and (2.141), we have:

$$\mathcal{K} \approx \frac{z^2 e^4 n'_e}{m(\Delta v_{e\parallel})^3}\begin{cases} 1, & \rho_{eff} \gg r_L \\ \rho_{eff}/r_L, & \rho_{eff} \ll r_L \end{cases} \tag{2.143}$$

$$d \simeq \frac{z^2 e^4 n'_e}{\Delta v_{e\parallel}}\begin{cases} 1, & \rho_{eff} \gg r_L \\ \frac{\rho_{eff}}{r_L}\ln\frac{r_L}{\rho_{eff}}, & \rho_{eff} \ll r_L \end{cases} \tag{2.144}$$

where $\Delta v_{e\parallel} = \rho_{eff}/t_{eff}$. Then

$$T_{st} \simeq \frac{1}{6}\begin{cases} m(\Delta v_{e\parallel})^2, & \rho_{eff} \gg r_L \\ m(\Delta v_{e\parallel})^2 \ln\frac{r_L}{\rho_{eff}}, & \rho_{eff} \ll r_L \end{cases} \tag{2.145}$$

After determining the characteristic value of $\Delta v_{e\parallel}$, these formulae allow one to estimate the maximum decrements (achievable at velocities $v < \Delta v_{e\parallel}$) and minimum equilibrium ion temperatures. These quantities no longer depend on the electron longitudinal velocity spread as a free parameter but are rather determined by such parameters as the density, electron Larmor radius, ion charge and the time of pass through the cooling section.

Let us start with the cases when electron interaction can definitely be neglected:

I. $\omega_e t_0 < 1$ ($t_0 = l/(\gamma\beta c)$).

With a sufficiently strong magnetic field (or low density $n'_e$), the effective interaction will be taking place at distances exceeding the Larmor radii; then $\Delta\rho \simeq ze^2 t_0^2/(m\rho^2) \simeq \rho$, which gives $\rho_{eff} \simeq (ze^2 t_0^2/m)^{1/3}$, and the condition of $r_L$ being small is thus

1) $r_L < (ze^2 t_0^2/m)^{1/3}$;

then $(\Delta v_{e\parallel})^2 = (\rho_{eff}/t_0)^2 = (ze^2/(mt_0))^{2/3}$ and, using Eqs. (2.142) – (2.145), we find:

$$\lambda \simeq \frac{ze^2 n_e l}{\gamma^3 M \beta c} \eta, \quad T_{st} \simeq \left(\frac{z^2 e^4 m}{t_0^2}\right)^{1/3}. \tag{2.146}$$

In a "weak" magnetic field

2) $r_L > (ze^2 t_0^2/m)^{1/3}$

the Larmor radius exceeds the parameter $\rho_{eff}$ and then

$$\Delta\rho \simeq \frac{1}{m}\frac{ze^2}{\rho v_{e\perp}}\Omega t^2 \simeq \rho, \quad \rho_{eff} \simeq t\left(\frac{ze^2}{mr_L}\right)^{1/2},$$

$$\lambda = \frac{ze^2 n_e l}{\gamma^3 M \beta c}\eta, \quad T_{st} \simeq \frac{ze^2}{r_L}. \tag{2.147}$$

Let us switch to the case of an extended cooling section.

II. $\omega_e t_0 > 1$.

1) With a small $r_L$, the interaction time equals the Langmuir period (screening time), then

$$\Delta\rho \simeq \frac{ze^2}{m\rho^2}\frac{1}{\omega_e^2} \simeq \rho, \quad \rho_{eff} \simeq \left(\frac{z}{n}\right)^{1/3},$$

$$\lambda = \frac{\eta}{\gamma}\frac{m}{M}z\omega_e, \quad T_{st} \simeq e^2(z^2 n'_e)^{1/3}, \tag{2.148}$$

under the condition that $r_L < (z/n'_e)^{1/3}$ or $n'_e r_L^3 < z$.

2) $r_L > (z/n'_e)^{1/3}$.

Since, in this situation, the interaction time itself can depend on the parameter $\rho_{eff} > r_L$ and the latter is determined by the interaction, then, generally speaking, one also has to account for the electron interaction that thermalizes the electron longitudinal motion to a temperature of $\simeq e^2(n'_e/r_L)^{1/4}$ in a time $t \simeq \omega_e^{-1}(n'_e r_L^3)^{1/8}$.

Suppose first that the e-e interaction is small. Then, with a sufficiently large $\omega_e t$, one has

$$t_{eff}^{-1} \simeq \frac{\omega_e}{\sqrt{\pi r_L/\rho_{eff}}} \simeq \omega_e \sqrt{\frac{\rho_{eff}}{r_L}}, \quad \Delta\rho \simeq \frac{1}{m}\frac{ze^2}{\rho v_{e\perp}}\Omega t_{eff}^2 \simeq \rho,$$

which gives

$$\rho_{eff} \simeq \left(\frac{z}{n}\right)^{1/3}, \quad t_{eff}^2 \simeq \omega_e^{-2} r_L \left(\frac{n}{z}\right)^{1/3},$$

if $t > \omega_e^{-1}(n_e' r_L^3/z)^{1/6}$. Otherwise, if $t < \omega_e^{-1}(n_e' r_L^3/z)^{1/6}$, then

$$t_{eff} = t, \quad \rho_{eff} = t\left(\frac{ze^2}{mr_L}\right)^{1/2}.$$

Thus, interaction with ions gives the circle an energy of

$$\frac{1}{2}m(\Delta v_{e\parallel})^2 \simeq \frac{ze^2}{r_L},$$

from which it follows that

$$T_{st} \simeq \frac{ze^2}{r_L}, \quad \lambda = \eta \frac{zn_e' e^2}{\gamma M} \cdot \max\left\{t_0, \frac{1}{\omega_e}\left(\frac{n_e' r_L^3}{z}\right)^{1/6}\right\}. \tag{2.149}$$

At the same time, interaction with neighbors gives an energy increase of (see Eq. (2.124))

$$\frac{1}{2}m(\Delta v_{e\parallel})^2 \simeq \frac{n_e' e^4}{mr_L} t^2 \lesssim e^2\left(\frac{n}{r_L}\right)^{1/4}, \quad \text{since } t^2 \lesssim \frac{1}{\omega_c^2}(nr_L^3)^{1/4}.$$

Therefore, when

a) $n_e' r_L^3 < z^4$,

interaction of circles is not significant.

Otherwise, in case of

b) $n_e' r_L^3 > z^4$,

when $\omega_e^2 t_0^2 < z$, the decrement and equilibrium temperature are still equal to the expressions in Eq. (2.149) and, when $\omega_e^2 t_0^2 > z$, the parameters $t_{eff}$ and $\rho_{eff}$ are determined by interaction of circles:

$$t_{eff} = \min\{t_0, \omega_e(n_e' r_L^3)^{1/8}\},$$

$$\rho_{eff} = \min\left\{\sqrt{\frac{n_e' e^4}{m^2 r_L} t_0^2}, \left(\frac{r_L}{n}\right)^{1/4}\right\}, \quad \omega_e^2 t_0^2 > z,$$

then

$$\lambda = \left\{\eta \frac{z^2}{t_0} \frac{m}{\gamma M}, \frac{\eta}{\gamma} \frac{z^2}{t_{eff}} \cdot \frac{m}{M}\right\} = \eta \frac{m}{\gamma M} \frac{z^2}{t_{eff}},$$

$$T_{st} = \frac{n_e' e^4}{mr_L} t_{eff}^2 \approx \left\{(\omega_e t_0)^2 \frac{e^2}{r_L}, e^2\left(\frac{n_e'}{r_L}\right)^{1/4}\right\}. \tag{2.150}$$

The constructed hierarchy of relationships between the main parameters and the values of $\lambda_{max}$ and $T_{min}$ completes the description of extreme regimes under the general conditions of $\Omega t_0 \gg 1$ and $(\Delta_{e\parallel})_{t=0} \lesssim (\rho/t)_{eff}$. Recall that, at velocities $v > \Delta v_{e\parallel}$, one can apply the perturbation theory formulae from the previous sections.

Our consideration is, to a large degree, phenomenological and does not claim to provide more than an order-of-magnitude estimate. Besides, we did not take into account such processes as capture of Larmor circles by ions with subsequent accompanying and break off at the exit from the cooling section. However, one may expect that such processes cannot be the defining ones (and it is even more likely that their contribution is small) due to a long range of the Coulomb forces.

Next, since we started from the domain of the perturbation theory, in our consideration, there was no distinction between the cases of attraction and repulsion (antiprotons-electrons), although such a distinction should appear under the considered conditions. However, again, since, due to the long range of the interaction, the main contribution to the deceleration processes of interest should come from the range of distances on the border between the region of small transfers $((\Delta p)^2 \ll mze^2/r)$ and the region of "strong" interaction $((\Delta p)^2_{max} \to mze^2/r)$, a change in the sign of the interaction cannot lead to large differences in the cooling rates and equilibrium temperatures. Nevertheless, apparently, due to some overall "distancing" of electrons under the effect of repulsion, cooling of antiprotons at the final stage ($v < \Delta v_{e\parallel} = (\rho/t)_{eff}$) will go on slower than of protons.

Finally, note that, at sufficiently high densities of the heavy particle beam, one will need to account for perturbation of the motion of a circle by several (or many) ions at once. In case of protons or antiprotons, this interaction becomes significant at densities of $n_i \gg \rho_{eff}^{-3}$ while, for ions with large $z$, this happens even before that and, apparently, limits the achievable $\lambda_{max}$ and $T_{min}$. This factor can be included into the consideration in the same semi-phenomenological way as the one used above but we will not dwell on it. Practically, more substantial limitations on $T_{min}$ (or the beam size) can be those related to the ion space charge effect on the beam orbit in a storage ring.

Let us give a numeric example characteristic to operating facilities NAP-M and EPOKHA. In magnetic field of $H = 1$ kG, at a cathode temperature of $T_k = 2000°$ (0.2 eV), the Larmor radius is $\simeq 10^{-3}$ cm. At electron current of $\simeq 300$ mA, beam radius of $\simeq 0.5$ cm and electron energy of 35 keV, the density $n_e$ and frequency $\omega_e$ are $n_e \simeq 10^8$ cm$^{-3}$ and $\omega_e \simeq 5 \cdot 10^8$ s$^{-1}$ while the time of pass with a length of $l = 1$ m equals $t_0 \simeq 10^{-8}$ s. We then get $nr_L^3 \simeq 0.1$ and $\omega_e t_0 \simeq 5$, i.e., the situation of Eq. (2.148) takes place. Then, $T_{st} \simeq e^2 n^{1/3} \simeq 7 \cdot 10^{-5}$ eV while the relative momentum spread of protons is

$$\left(\frac{\Delta p}{p}\right)_{min} \simeq \sqrt{\frac{T_{st}}{W}} = \sqrt{\frac{7 \cdot 10^{-5}}{65 \cdot 10^6}} \simeq 10^{-6}. \tag{2.151}$$

The maximum "instantaneous" damping decrement is

$$\lambda_{inst} \simeq \frac{m}{M}\omega_e \simeq 2 \cdot 10^5 \text{ s}^{-1}, \tag{2.152}$$

while the average one at $\eta = l/(2\pi R) \simeq 0.02$ is

$$\lambda_{max} = 4 \cdot 10^3 \text{ s}^{-1}. \tag{2.153}$$

This decrement can be realized in the range of

$$\frac{v}{\beta c} \lesssim \sqrt{\frac{e^2 n^{1/3}}{mc^2 \beta^2}} \simeq 3 \cdot 10^{-5}$$

playing the role of the electron effective longitudinal velocity spread (while the spread at the entrance resulting from electrostatic acceleration, as estimated in Section 2.1, is $(\Delta v_{e\parallel}/(\beta c))_0 \simeq 10^{-6})$.

## 2.7 Process of cooling in uniform stationary flow

Let us now describe in general terms the process of cooling of an ion beam by a magnetized electron flow assuming its spatial uniformity and stationarity and neglecting aspects related to the cyclic nature of motion in a storage ring. Suppose, at the initial moment, the ion velocity spread $\Delta_i$ exceeds the electron spread $\Delta_{e\perp}$. At the initial stage, all collisions can be divided with a logarithmic accuracy into fast and adiabatic. While $\Delta_i > \Delta_{e\perp}$, their contributions scale as the respective logarithms, so that the damping decrement has an order of magnitude of

$$\lambda \simeq \frac{4\pi}{3}\eta \frac{n'_e z^2 e^4 L}{\gamma m M \Delta_i^3}, \quad L = L^0 + L^A, \quad \text{when} \quad \Delta_i > \Delta_{e\perp}.$$

Then, after $\Delta_i$ becomes smaller than the electron transverse spread $\Delta_{e\perp}$, friction due to fast collision starts to decrease while the contribution of adiabatic ones continues to quickly grow:

$$\lambda \simeq \frac{4\pi n'_e z^2 e^4}{\gamma m M}\left(\frac{L^0}{\Delta_{e\perp}^3} + \frac{L^A(\Delta_i)}{\Delta_i^3}\right), \quad \text{when} \quad \Delta_i < \Delta_{e\perp}, \quad L^A(\Delta_i) > 1.$$

Fairly quickly there comes a moment when the first term can be neglected. In the logarithm of adiabatic collisions $L^A = \ln(\rho_{max}^A/\rho_{min}^A)$, the parameter $\rho_{max}^A = \min\{u_A t_0, u_A/\omega_e\}$ decreases during the cooling process while $\rho_{min}^A$ remains constant or increases: $\rho_{min}^A = \max\{r_L, ze^2/(mu_A^2)\}$ (see Section 2.4).

After the value of $L^A$ decreases to unit, growth of the decrement slows down becoming proportional to $\sim 1/v^2$ or $1/v^{7/3}$ (Section 2.5, $r_L > \max\{(ze^2 t_0^2/m)^{1/3}, (z/n)^{1/3}\}$) or ceases if $r_L$ does not satisfy this condition. Cooling of the heavy particle beam then continues as described in Sections 2.5 and 2.6 until the equilibrium temperature $T_{st}$ is established unless cooling is limited by interaction of particles (space charge, intra-beam scattering).

# III. EFFECTS OF NON-UNIFORMITY AND NON-STATIONARITY WHEN COOLING BY MAGNETIZED BEAM

The positive features of cooling in a magnetized electron flow with a low longitudinal temperature can be limited or deformed by a number of factors creating spatial or temporal variations of electron velocities. On the other hand, simulation of these factors in an experiment can be used to determine the physical conditions, in which the cooling process takes place, and to optimize parameters of the cooling system.

## 3.1 Collision integral in non-uniform flow

The factors of spatial non-uniformity were already described when discussing properties of the electron beam in Section 2.1. To include them in the consideration, let us first modify the collision integral obtained in Section 2.3. We lift the conditions of complete spatial uniformity of the magnetic field and electron flow and of the absence of average electric field not related to heavy particles. Let us represent the electric field potential of electrons as

$$\varphi = \langle \varphi^0 \rangle + \Delta\varphi \,,$$

where $\langle \varphi^0 \rangle$ is a statistical average of the potential for a state no perturbed by interaction with heavy particles, or self-consistent potential, and $\Delta\varphi$ is a deviation accounting for all "fluctuations". In magnetic and average electric fields, an electron would move along some trajectory $\vec{r}^0(t)$ with a velocity $\vec{V}^0(t)$, which can be presented as

$$\vec{V}^0 = \vec{V}(\vec{r}) + \vec{v} \,,$$

where $\vec{V}(\vec{r})$ is the "hydrodynamic" velocity while $\vec{v}$ is a deviation corresponding to the thermal spread. The trajectory $\vec{V}(\vec{r})$ includes the following components of motion:

1) coherent Larmor twist of the beam by transverse electric fields that are non-adiabatic with respect to the Larmor rotation of electrons (for example, due to defects of electron gun optics). The spatial shift of the beam is then negligible but the distribution of rotation phases becomes strongly non-uniform;

2) motion together with the field lines of the guiding magnetic field whose change of direction along the cooling section is assumed adiabatically slow in comparison to the Larmor rotation;

3) drift of the Larmor spiral across the magnetic field (practically, in the direction transverse to the beams) under the effect of transverse electric field;

4) change in the longitudinal velocity along the path in the presence of a "residual" longitudinal electric field.

The average longitudinal velocity $V_\parallel(\vec{r})$ can change along with the particle kinetic energy over the beam cross section due to a transverse gradient of the electric potential related to space charge of the beam (the energy gradient is accumulated in the acceleration section where the density varies along the length). Besides, the transverse gradient of a kick Larmor excitation also leads to a gradient of the longitudinal velocity due to conservation of energy in an electrostatic kick.

Within the framework of the perturbation theory, one can write an equation for the potential $\Delta\varphi$ as

$$\Delta\varphi(\vec{r},t) = \Delta\varphi^0 - e\frac{\partial}{\partial\vec{r}}\sum_a \frac{\delta\vec{r}_a(t)}{|\vec{r}-\vec{r}_a(t)|}, \tag{3.1}$$

where $\Delta\varphi^0$ represents the statistical fluctuations with respect to $\langle\varphi^0\rangle$ unperturbed by the interaction while $\delta\vec{r}_a(t)$ is the deviation of the electron trajectory from the motion trajectory $\vec{r}^0(t)$ in the external field including the self-consistent potential as well.

Equation (3.1) does not conceptually differ from Eq. (2.22) and this means that systems with space charge, stabilized by an external (in our case, magnetic) field, possess the same properties as quasi-neutral systems such as a real plasma: Langmuir oscillations, electric polarizability, Debye (or dynamic) screening of the interaction of colliding particles. Certain complications for a "pure" theory may be related to spatial (or temporal) non-uniformity, which makes finding the solution $\Delta\varphi(\vec{r},t)$ more difficult because the polarizability is no longer a function of only the coordinate difference. However, in practical cases, the sizes of non-uniformities are large compared to the size of an effective particle interaction region. The interaction with a heavy particle involves a relatively small group of electrons (due to the smallness of relative velocities) experiencing the same "external" conditions that allows one to neglect the gradient of the "hydrodynamic" velocity within the size of the interaction region. The picture can be clearly seen from the point of view of the coordinate system locally connected to the average electron velocity $\langle\vec{v}_e\rangle(\vec{r})$: the electron medium is weakly non-uniform while a heavy particle moves relative to the medium at a low (varying) velocity producing a perturbation in a small vicinity. The relative ion velocity at a point $\vec{r}$ can be written as

$$\langle\vec{u}\rangle = \vec{v} - \langle\vec{v}_e\rangle(\vec{r}),$$

while the change of the relative coordinate with time can be written in the form

$$\vec{R}(t) = \vec{r}_0 + \int_0^t [\vec{v} - \langle\vec{v}_e\rangle(\vec{r}(t'))]dt'. \tag{3.2}$$

As in the spatially-uniform case, the Fourier transformation can be applied to Eq. (3.1) with a "quasi-classical" accuracy. Thus, Eqs. (2.34) and (2.35) for the friction force and scattering tensor along with Eqs. (2.36) and (2.37) for the harmonics $\langle\varphi_{\vec{k}}(t)\rangle$ and $\varphi_{a\vec{k}}(t)$ directly transfer to the case of a beam with electron velocity gradients by replacement of the heavy particle trajectory $\vec{r}(t) = \vec{r}_0 + \vec{v}t$ with the relative trajectory $\vec{R}(t)$ where

$$\dot{\vec{R}}(t) = \vec{v} - \langle \vec{v}_e \rangle(\vec{r}).$$

The density of the electron beam along the ion trajectory can practically be considered constant. Then, to obtain an answer, one can use Eqs. (2.45) – (2.47) with the only generalization

$$(e^{-i\vec{k}\vec{r}(t)})_\omega \to (e^{-i\vec{k}\vec{R}(t)})_\omega = \int_0^\infty e^{-i\vec{k}\vec{R}(t)+i\omega t} dt.$$

With spatial non-uniformity, it is adequate to directly determine the integral quantities

$$\langle \Delta \vec{p} \rangle = \int_0^{t_0} F(t) dt, \quad \langle \Delta p_\alpha \Delta p_\beta \rangle = \int_0^{t_0} d_{\alpha\beta}(t) dt.$$

For example,

$$\begin{aligned}
\Delta \vec{p} &= -\frac{z^2 e^2}{2\pi^2} \int \frac{i\vec{k} d^3 k}{k^2} \int_0^{t_0} dt \int_C \frac{d\omega}{2\pi \varepsilon_{\vec{k}}(\omega)} \int_0^\infty e^{i\omega(t'-t)+i\vec{k}(\vec{R}_t - \vec{R}_{t'})} dt' \\
&= -\frac{z^2 e^2}{2\pi^2} \int \frac{\vec{k} d^3 k}{k^2} \int_0^{t_0} dt \int_0^t dt' e^{i\vec{k}(\vec{R}_t - \vec{R}_{t'})} \sum_s \left[ \frac{e^{i\omega_s(t'-t)}}{\partial \varepsilon_{\vec{k}}(\omega)/\partial \omega} \right]_{\omega = \omega_s}.
\end{aligned} \quad (3.3)$$

When $\omega_e \ll \Omega$ and neglecting the thermal spread of electron longitudinal velocities, one can explicitly write out the sum $\Sigma$ (see Eq. (2.75)). Using the symmetry properties of $\varepsilon_{\vec{k}}(\omega)$, we transform $\Delta \vec{p}$ to the form:

$$\Delta \vec{p} \approx -\frac{z^2 e^4}{4\pi^2} \int d^3 k \frac{\vec{k}}{k^2} \sum_l \omega_{Ml} \left| \int_0^{t_0} dt e^{i\vec{k}\vec{R}(t) - i(l\Omega + \omega_{Ml})t} \right|^2,$$

$$\omega_{Ml}^2 = \omega_e^2 \frac{k_\parallel^2}{k^2} \langle J_l^2(k_\perp r_L) \rangle. \quad (3.4)$$

The Larmor radii in the argument of the Bessel function here are determined by electron thermal velocities. The coherent Larmor rotation can also be included by extracting the component $\vec{R}_L(t)$ from $\vec{R}(t)$:

$$\vec{k}\vec{R}_L(t) = k_\perp R_L \cos \widehat{\vec{k}\vec{R}_L}(t).$$

In the region where magnetization is significant ($v < v_{e\perp}$), one can average over the Larmor rotation in Eq. (3.4):

$$\Delta \vec{p} \approx -z^2 e^4 \int d^3 k \frac{\vec{k}}{4\pi^2 k^2} \sum_l \omega_{Ml} J_l^2(k_\perp R_L) \left| \int_0^{t_0} dt e^{i\vec{k}\vec{R}_A(t) - i\omega_{Ml} t} \right|^2, \quad (3.5)$$

where $\vec{R}_A(t) = \int \vec{u}_A(\vec{r}(t))dt$ and $\vec{u}_A(\vec{r})$ is the ion velocity with respect to the electron medium averaged over the Larmor oscillations.

Similarly,

$$\langle \Delta p_\alpha \Delta p_\beta \rangle \approx z^2 e^4 n'_e \int d^3k \frac{k_\alpha k_\beta}{k^4} \\ \times \sum_l J_l^2(k_\perp R_L) \langle J_l^2(k_\perp r_L) \rangle \left| \int_0^{t_0} dt e^{i\vec{k}\vec{R}_A(t) - i\omega_M t} \right|^2, \quad k u_A < \Omega. \quad (3.6)$$

At velocities $u_A$ that are not small relative the full Larmor velocity $v_{e\perp} = \sqrt{R_L^2 \Omega^2 + 2T_{e\perp}/m}$, one should add to these expressions the contributions of fast collisions from Eqs. (2.40) and (2.41). Note that it is only in this case that the limitation of the integration region $ku_A < \Omega$ is significant (and even then only with a logarithmic dependence on $k_{max}$), since, when $u_A \ll v_{e\perp}$, the integrals over $|\vec{k}|$ converge much earlier.

Equations (3.5) and (3.6) are applicable while velocities $u_A$ exceed the minimum effective spread $(\Delta_{e\parallel})_{eff}$ determined by the interaction itself (Section 2.6). Note that the Larmor twist can somewhat reduce the value of $(\Delta_{e\parallel})_{eff}$ due to a decrease of the averaged interaction with an increase of $R_L$. For example, for $|z| = 1$, a consideration similar to that done in Section 2.6 gives

$$T_{min} = m(\Delta_{e\parallel})^2_{eff} \simeq \frac{e^2}{R_L}, \quad \text{when } r_L \lesssim n'^{-1/3}_e \ll R_L.$$

Let us estimate the effect of a coherent Larmor twist when $R_L \gg r_L$ assuming that the region of effective interaction lies beyond the "thermal" Larmor radii. Them the main contribution in the sum in Eq. (3.5) comes from the term with $l = 0$, $\omega_{M0} \approx \omega_e |k_\parallel|/k$. For the constant velocity $\vec{u}_A = \vec{v}$, we get:

$$\Delta \vec{p} = -z^2 e^2 \int d^3k \frac{\vec{k}|k_\parallel|}{\pi^2 k^3} \omega_e \frac{\sin^2[(\vec{k}\vec{v} - \omega_e|k_\parallel|/k)t_0/2]}{(\vec{k}\vec{v} - \omega_e|k_\parallel|/k)^2} J_0^2(k_\perp R_L).$$

Let us consider the case of $\omega_e t_0 \gg 1$. Then the integral converges in the region where the argument of sine is large, therefore, one can make a substitution

$$\frac{\sin^2(\alpha x)}{x^2} \to \alpha \pi \delta(x), \quad \alpha = \frac{t_0}{2}.$$

Removing then integration over $|\vec{k}|$, we arrive at a two-dimensional integral over the angles (for the component of $\Delta \vec{p}$ along the velocity $\vec{v}$):

$$(\Delta \vec{p})_{\vec{v}} = -\frac{z^2 e^2 \omega_e^2 t_0}{v^2} \int \frac{dO \cos^2 \chi}{4\pi |\cos \theta|} J_0^2 \left( \frac{\omega_e R_L}{2v} \frac{\sin(2\chi)}{\cos \theta} \right), \quad (3.7)$$

where $dO = \sin\theta\, d\theta d\varphi$, $\cos\chi = (v_\parallel/v)\cos\theta + (v_\perp/v)\sin\theta\cos\varphi$, and $v_\parallel$ and $v_\perp$ are the velocity components relative the magnetic field direction. In the region $v \gg \omega_e R_L$, the friction force has the behavior described in Section 2.4 while, in the opposite case,

$$(\Delta\vec{p})_{\vec{v}} \approx -\frac{z^2 e^2 \omega_e^2 t_0}{v^2} \int \frac{dO}{4\pi} \left|\frac{\cos\chi}{\sin\chi}\right| \cdot \frac{v}{\pi\omega_e R_L} = -\frac{z^2 e^2 \omega_e t_0}{\pi v R_L}. \tag{3.8}$$

The inverse damping time in the latter case equals

$$\lambda = \tau^{-1} \approx \frac{2z^2 e^2 \omega_e \Omega}{\gamma \pi v_i^2 v_V M} \eta, \quad v_L = \Omega R_L, \quad v_i = v(t=0), \tag{3.9}$$

when $R_L \gg r_L$, $v \lesssim R_L \omega_e$, and $\omega_e t_0 \gg 1$.

Note the difference of the time dependence $\tau(v, R_L)$ from the corresponding formula in Eq. (2.113) and agreement of this dependence with the case of Eq. (2.104). The explanation is that, when $R_L \gg r_L$, the thermal Larmor motion is not involved in screening (in electron interaction), unlike in the case of $v < \omega_e r_L$ and $r_L \gg R_L$.

## 3.2 Modulation of electron energy

One of the factors increasing the effective temperature of the electron flow can be temporal and spatial (longitudinal) variation of electric potential in the cooling section causing modulation of the electron longitudinal velocities. In its pure form, this factor will appear when the conditions over the electron beam cross section are uniform or, in general case, when there is no coupling of the friction decrements of the longitudinal and transverse motions of heavy particles. In a non-magnetized flow, this modulation could have an effect on the transverse decrements only starting with the amplitudes $\Delta v_{e\parallel}$ exceeding the transverse thermal spread $\Delta_{e\perp}$.

Let us first consider the influence of temporal potential modulation (periodic or noise one) at frequencies small compared to the inverse time of pass through the interaction section but large compared to the friction decrements. Since the decrements themselves will depend on the modulation amplitude, the second condition can be explicitly formulated after estimating the decrements. The estimates can use the formulae for the friction force in Eq. (2.70), in which one can neglect the thermal spread of electron velocities and consider the friction force as a function of the difference $v_\parallel - v_{e\parallel}(t)$ with averaging over time. It may also be convenient to introduce a probability distribution $w(v_{e\parallel})$ of different values of $v_{e\parallel}$, then

$$\langle\vec{F}\rangle = \int \vec{F}(\vec{v}_\perp, v_\parallel - v_{e\parallel}) w(v_{e\parallel}) dv_{e\parallel}.$$

Obviously, if the distribution width $w$, i.e., the rms spread $\Delta v_{e\parallel}$, does not exceed the thermal spread $\Delta_{e\parallel}$ or the effective minimum spread determined by the dynamics of the interaction itself of heavy particles with Larmor circles (Section 2.6), then oscillations of the potential have no significance at all. The latter value $(-(\Delta U)_{min}/U \simeq 10^{-4} - 10^{-5})$ is what determines the level of potential's stability, which it makes sense to strive for to create optimal cooling conditions and obtain minimum temperatures.

When $\Delta v_{e\|} > (\Delta_{e\|})_{eff}$ but $\Delta v_{e\|} \ll \Delta_{e\perp}$, for velocities $v < \Delta v_{e\|}$, the contribution of adiabatic collisions changes proportionally to $\sim 1/(\Delta v_{e\|})^3$. Observation of such a dependence of the decrements when artificially modulating the potential (with the noted limitations) may indicate that the damping takes place in conditions when the main contribution comes from adiabatic collisions if, of course, the decrements themselves greatly exceed the values corresponding to cooling in a non-magnetized beam (recall that, in the latter case, the transverse decrements and the maximum size of the longitudinal friction force would not depend on $\Delta v_{e\|}$ when $\Delta v_{e\|} < \Delta_{e\perp}$). It should be noted that, for the longitudinal degree of freedom, effect of the oscillations substantially depends on the shape of the distribution. For instance, a distribution of the type $\sim \delta(\Delta v_{e\|}^2 - \Delta_0^2)$ ($\Delta U(t)$ is a step-like function with a small amplitude dispersion) creates a negative characteristic of the longitudinal friction, i.e., a "monochromatic" instability leading to the same equilibrium distribution of the longitudinal velocities for heavy particles. The same effect is produced by harmonic oscillations of the potential $\Delta U = (\Delta U_0)_{max} \cdot \cos(\omega_U t)$, since such a dependence corresponds to the distribution

$$w(\Delta v_{e\|}) = \frac{1}{\pi \sqrt{\Delta_0^2 - \Delta v_{e\|}^2}}$$

having a minimum at the center. Indeed, a calculation for $f(v_{e\|}) = w(\Delta v_{e\|})$ in this case (with velocities $v_\| < \Delta_0$) gives

$$F_\|^A = \frac{4 n_e' z^2 e^4 L^A}{m} \frac{v_\|}{\Delta_0^3},$$

i.e., one gets anti-friction.

Note also a curious feature of a plateau distribution $w \approx \Theta(\Delta v_{e\|}^2 - \Delta_0^2)$: the "main" logarithmic term in the longitudinal friction force here is relatively small $\simeq v_\perp^2/\Delta_0^2$ and the decrement is determined by the second term including the contribution of the non-pass interaction ($\rho \simeq \rho_{max}$).

Oscillations with a normal bell-shaped distribution $w(\Delta v_{e\|})$ only increase the effective temperature to a value of $\simeq m \Delta v_{e\|}^2$ not changing the sign of the friction characteristic.

The influence of spatial potential modulation along the path in the cooling section can be more complicated, since a heavy particle can experience multiple correlation collisions with the same Larmor circle alternatively accelerating and decelerating its motion (an effect similar to cyclic collisions with an electron moving in a circle). Effects of this kind can be easily accounted for using Eqs. (3.5) and (3.6) and substituting the dependence $\langle v_{e\|}\rangle(\vec{r})\sim \nabla_\| U(\vec{r})$ into the velocity $u_{A\|}$.

## 3.3 Deviations of magnetic field lines

Deviation of the accompanying magnetic field direction from the heavy ion closed (equilibrium) orbit is conveniently described by an angle $\vec{\alpha} = (\alpha_x, \alpha_z)$; the relative transverse velocity of a particle and a circle can represented as

$$\vec{u}_{A\perp} = \gamma\beta c(\vec{\theta} - \vec{\alpha}(s)), \quad \vec{\theta} = (\theta_x, \theta_z). \tag{3.10}$$

where $s$ in the length along the closed orbit and $\vec{\theta}$ is the angular deviation of the particle velocity from this orbit oscillating from turn to turn at the betatron frequencies.

Obviously, changes in the damping process can only occur in conditions when the value of $\gamma\beta c\alpha$ exceeds the effective spread of electron longitudinal velocities $(\Delta_{e\|})_{eff}$. Let us first consider a relatively simple case when $\vec{\alpha}$ is constant along the cooling section: $d\vec{\alpha}/ds = 0$ but can change in time. The effect can then be described by Eqs. (2.70) and (2.71) averaged over the distribution $w(\vec{\alpha})d^2\alpha$:

$$\langle \vec{F}^L \rangle = \int \vec{F}^L(\vec{u}_A) w(\vec{\alpha}) d^2\alpha, \quad \langle d^L_{\alpha\beta} \rangle = \int d^L_{\alpha\beta}(\vec{u}_A) w(\vec{\alpha}) d^2\alpha.$$

Let us estimate the effect for small amplitudes of the vertical and radial oscillations ($\overline{\theta^2} < \overline{\alpha^2}$) assuming the electron flow to be radially uniform and the particle longitudinal velocities to be damping towards the inside of the effective spread of longitudinal velocities.

When $\theta < \alpha$, the friction force $\vec{F}^A$ equals

$$\vec{F}^A_\perp \approx -\frac{2\pi n'_e z^2 e^4 L^A}{m(\gamma\beta c)^2} \cdot \frac{\vec{\theta} - \vec{\alpha}}{|\vec{\theta} - \vec{\alpha}|^3} \approx \frac{2\pi n'_e z^2 e^4 L^A}{m(\gamma\beta c)^2 \alpha^3} [\vec{\alpha} - \vec{\theta} + 3\vec{n}(\vec{n}\vec{\theta})],$$

$$F^A_\| \approx -\frac{6\pi n'_e z^2 e^4 L^A}{m(\gamma\beta c)^2 \alpha^2} \cdot \frac{v_\|}{\gamma\beta c\alpha},$$

where $\vec{n} = \vec{\alpha}/\alpha$. Note that the friction characteristics proportional to $\partial \vec{F}^A/\partial \theta_{x,z}$ do not change when changing the sign of $\vec{\alpha}$.

Suppose that $\vec{\alpha}$ is directed along a normal degree of freedom, for example, $\alpha_x = 0$. Then

$$\dot{\varepsilon}_z \sim 2\theta_z^2, \quad \dot{\varepsilon}_x \sim -\theta_x^2,$$

i.e., oscillations along $\vec{\alpha}$ grow, while those transverse to $\vec{\alpha}$ damp; in addition, the sum of the decrements is negative. One can show (see Appendix 2) that, if the electron flow is uniform in the $x$ and $z$ directions near the equilibrium ion orbit, the sum of the decrements of transverse oscillations with arbitrary coupling of the $x$ and $z$ motions does not depend on the orientation of $\vec{\alpha}(s)$ and equals

$$\lambda_1 + \lambda_2 = -\frac{1}{2M\gamma}\frac{\partial \vec{F}_\perp^A}{\partial \vec{v}_\perp} = -\eta\frac{\pi z^2 e^4 L^A}{mM\gamma^2}(\gamma\beta c\alpha_0)^{-3} n_e,$$

where averaging is done along the ion closed orbit. Thus, transverse oscillations in this case happen to be unstable. Meanwhile, the average vector value $\langle \vec{\alpha}(t) \rangle$ does not play a substantial role. As illustrations, one can give simple examples of behavior of $\vec{\alpha}$ with $|\vec{\alpha}| = const$:

1) $\vec{\alpha} = const$;
2) $\vec{\alpha}$ has an instantaneous sign change;
3) $\vec{\alpha}(t)$ rotates at a constant rate about the closed orbit.

We earlier considered an effect (see Section 1.2) of the so-called monochromatic instability – growth of ion oscillations occurring, when the difference of the average velocities exceeds the spread. The cause of the instability is a change in sign of the friction characteristic (reduction in the friction force with velocity) for velocities $|\langle \vec{u} \rangle| > \Delta_{e\perp}$. Obviously, this effect can also appear in the kinetics of collisions "frozen" electrons and, moreover, not only in the region of $\alpha > \theta_e$ but also at substantially smaller "detunings" $\Delta_{e\perp} > v_\perp > \Delta_{e\parallel}$.

However, when $v_\parallel \ll \gamma\beta c\alpha$, the total sum of the decrements is positive, since the magnitude of $\lambda_\parallel$ exceeds the sum of the transverse decrements by a factor of three:

$$(\lambda_x + \lambda_z + \lambda_\parallel)^A = -\frac{1}{2M\gamma}\langle\frac{\partial \vec{F}_\perp^A}{\partial \vec{v}}\rangle = \frac{2\pi z^2 e^4 L^A}{mM\gamma^2}\langle (\gamma\beta c\alpha)^{-3} n_e \rangle.$$

According to the general theorem about the complete sum of the decrements, this sum is independent of coupling of the ion degrees of freedom and, in general case, is determined by the divergence of the friction force as a function of the particle velocity (Appendix 2). This property can be used to suppress the considered instability of betatron oscillations by redistributing the decrements between the longitudinal and transverse particle motions through the introduction of $z - x$ coupling and a transverse gradient of the longitudinal friction (for example, a gradient of the electron longitudinal velocity $dv_{e\parallel}/dx$).

With the decrements being positive, the quantity $\gamma\beta c\alpha$ plays the role of an effective electron velocity spread and the ion beam is cooled to the temperature of $\simeq m(\gamma\beta c\alpha)^2$; then

$$\theta_{st} \simeq \sqrt{\frac{m}{M}}\alpha.$$

If the decrements are negative, the angular oscillations of ions grow to amplitudes

$$\theta_0 \simeq \alpha,$$

i.e. there occurs equalization not of the effective temperatures but of the ion and electron velocities with respect to the closed orbit. This conclusion can be made on the basis of the monochromatic instability study done in Chapter I.

Let us now consider the situation when the size of $\alpha(t)$ oscillates in time passing small values of $\alpha \lesssim \Delta_{e\parallel}/(\gamma\beta c)$. We are interested in a situation when

$$\theta^2 + v_\parallel^2/(\gamma\beta c)^2 < \langle \alpha^2 \rangle;$$

in the opposite case, $\vec{F}^A$ and $\alpha^A_{\alpha\beta}$ do not depend on $\vec{\alpha}$. It is easy to estimate that, for a two-dimensional distribution $w(\vec{\alpha})$ of Maxwell type with a width of

$$\alpha_0 = \sqrt{\langle \alpha^2 \rangle/2} \gg \Delta_{e\parallel}/(\gamma\beta c)$$

equal in both transverse directions, the transverse and longitudinal friction forces, respectively, equal ($\theta < \alpha_0$):

$$\langle \vec{F}^A_\perp \rangle \simeq -\pi \sqrt{\frac{\pi}{2}} \frac{n'_e z^2 e^4 L^A(v_0)}{m v_0^2} \frac{\vec{\theta}}{\alpha_0}, \tag{3.11}$$

$$\langle F^A_\parallel \rangle \simeq -\frac{4\pi n'_e z^2 e^4}{m v_0^2} \begin{cases} \dfrac{v_\parallel}{|v_\parallel|} L^A(v_\parallel) - \dfrac{3}{2}\sqrt{\dfrac{\pi}{2}} \dfrac{v_\parallel}{v_0} L^A(v_0), & v_0 > v_\parallel > \Delta_{e\parallel}, \\ 2 \dfrac{v_\parallel}{\Delta_{e\parallel}} \sqrt{\dfrac{2}{\pi}} L^A(\Delta_{e\parallel}), & v_\parallel < \Delta_{e\parallel}, \end{cases} \tag{3.12}$$

where $v_0 \equiv \gamma\beta c \alpha_0$.

With an accuracy of up to numerical factors, these expressions are analogous to the friction force due to fast collisions in Eqs. (1.49) and (1.50) where the role of the $\Delta_{e\perp}$ spread is played by the parameter $v_0$.

Thus, for a "normal" distribution $w(\vec{\alpha})$, as it should be expected, oscillations of $\vec{\alpha}$ do not lead to an instability creating only an effective temperature of the transverse motion of Larmor circles, which enters the expressions for the decrements. Then, if the following condition is satisfied:

$$\alpha_0 \ll \theta_e \equiv \Delta_{e\perp}/(\gamma\beta c),$$

contribution of the adiabatic collisions to friction and diffusion in the region

$$\theta^2 + (v_\parallel/(\gamma\beta c))^2 < \theta_e$$

remains dominant.

Let us also give expressions for the friction force in case of one-dimensional oscillations of $\vec{\alpha}$. The friction force in the direction of oscillations differs from Eq. (3.11) only by a logarithmic multiplier:

$$\langle F^A_l \rangle \approx -2\sqrt{2\pi} \frac{n'_e z^2 e^4 L^A(v_0)}{m v_0^2} \frac{\theta_l}{\alpha_0} \ln\left(v_0 / \sqrt{v_{tr}^2 + \Delta_{e\parallel}^2}\right); \tag{3.13}$$

in the directions transverse to $\vec{\alpha}$ (interpolation formula):

$$\langle \vec{F}_{tr}^A \rangle \approx -2\sqrt{2\pi} \frac{n'_e z^2 e^4 L^A(\sqrt{v_{tr}^2 + \Delta_{e\parallel}^2})}{m(v_{tr}^2 + \Delta_{e\parallel}^2)} \cdot \frac{\vec{v}_{tr}}{v_0}, \quad \theta < \alpha_0, \quad v_{tr} < v_0, \qquad (3.14)$$

where $v_{tr} = (\gamma\beta c\theta_{tr}, v_\parallel)$.

From comparison to the previous situation, when there were only large values of $|\vec{\alpha}(s)| \gg \Delta_{e\parallel}/(\gamma\beta c)$, one can derive a qualitative criterion: the decrements of transverse ion oscillations become negative only in the situations, when the sizes of angular deviations of magnetic field from the closed orbit direction are localized near a certain value of $|\vec{\alpha}| > \Delta_{e\parallel}/(\gamma\beta c)$ with a relatively small spread:

$$\langle |\vec{\alpha}|^2 \rangle - \langle |\vec{\alpha}| \rangle^2 < \langle |\vec{\alpha}| \rangle^2 \,.$$

Equations (3.13) and (3.14) assume a bell-shaped one-dimensional distribution $w_1(\alpha)$ ("noise" oscillations). Note in this respect that sinusoidal temporal oscillations of the magnetic field direction corresponding to a distribution $w_1(\alpha) = 1/(\pi\sqrt{\alpha_0^2 - \alpha^2})$ lead to a negative friction characteristic in the direction of oscillations:

$$F_{(1)}^A \approx 4 \frac{n'_e z^2 e^4 L^A(v_0)}{m v_0^2} \frac{\theta_1}{\alpha_0} \ln(v_0/\sqrt{v_{tr}^2 + \Delta_{e\parallel}^2}) \,; \qquad (3.15)$$

the force component transverse to this direction is analogous to Eq. (3.14).

From comparison of Eq. (3.13) or (3.15) with Eq. (3.14), one can see that, for one-dimensional oscillations of $\alpha$, the friction decrements in the directions of $\alpha$ and transverse to it have a ratio of $\simeq \theta^2/\alpha_0^2$. If, however, betatron oscillations are coupled the difference of the damping decrements of normal oscillations decreases all the way to a complete equalization of the decrements at a strong (resonant) coupling. In this case, the decrements will be equal to a half of the value corresponding to the force in Eq. (3.14) and, thus, will be inversely proportional to the first power of the oscillation amplitude $\alpha_0$ (including also a harmonic law of $\alpha(t)$, easier realizable experimentally) and to the second power of the particle oscillation amplitudes $\Delta_{e\parallel}^2/(\gamma\beta c)^2 < \theta^2 < \alpha_0^2$.

We considered in sufficient detail the influence of the deviation of magnetic field direction from the closed orbit for the range of angles when the main contribution to friction comes from adiabatic collisions:

$$(\gamma\beta c)^2(\alpha^2 + \theta_b^2) > \omega_e^2(r_L^2 + R_L^2) \,.$$

In case of the opposite relation, in accordance with the change of the dependence $\vec{F}^L(v) \sim 1/v^2$ to the law $\sim 1/(v v_{eL})$ (Eq. (3.7)), the dependence of the friction decrements on $\theta$ also changes:

$$|\lambda| \sim \frac{1}{\alpha^2} \cdot \frac{1}{v_{e\perp}}; \quad \theta < \alpha < \frac{\omega_e \sqrt{r_L^2 + R_L^2}}{\gamma \beta c}. \tag{3.16}$$

The presented above model calculations for $\langle \vec{F}^A \rangle$ can also be done for the region of cyclic collisions $v < \omega_e \sqrt{r_L^2 + R_L^2}$ as well as for the general formula, Eq. (3.5):

$$\langle \Delta \vec{p} \rangle = -z^2 e^4 \int d^3 k \frac{\vec{k}}{4\pi^2 k^2} \sum_l \omega_{Ml} J_l^2(k_\perp R_L) \int w(\vec{\alpha}) d^2 \alpha$$
$$\times \left| \int_0^{t_0} dt \exp\{i \vec{k}_\perp \gamma \beta c (\vec{\theta} - \vec{\alpha}) t + i k_\parallel v_\parallel t - i \omega_{Ml} t \} \right|^2. \tag{3.17}$$

**Effect of spatial oscillations of the field lines.** To make an estimate, as in the case of variation of the longitudinal electron velocity, one has to use Eq. (3.5), which accounts for the possibility of repeating collisions during oscillations of the circle velocity:

$$\Delta \vec{p} = (ze)^2 \int \omega(\vec{k}) \frac{\vec{k}}{k^2} d^3 k J_0^2$$
$$\times \int_0^{t_0} dt \int_0^t d\tau \sin(\omega(\vec{k})\tau) \exp(i\vec{k}[\vec{v}\tau - \int_{t-\tau}^t \vec{v}_M(t') dt']), \tag{3.18}$$

where $\vec{v}_M(t) = \gamma \beta c \vec{\alpha}(s(t))$ while the integral $\int \vec{v}_M(t) dt$ describes motion of the Larmor circle along a field line in the direction transverse to the closed orbit. Let us consider some characteristic situations.

1. The dependence $\vec{\alpha}(t)$ is adiabatic with respect to the electron Langmuir oscillations. Since the maximum time of particle interaction with individual electrons in any case does not exceed the order of magnitude of $\omega_e^{-1}$ ($\tau_{eff} \lesssim \omega_e^{-1}$), the integral in the exponential can be simply replaced with $\vec{v}_M(t) \cdot \tau$ and we actually arrive at the previous case, when the friction force is determined as a function of the relative velocity $\vec{v} - \vec{v}_M(t)$ and the effect in general is the result of averaging over time (in this case, over the length of the cooling section). If $\omega_e t_0 \lesssim 1$ the adiabaticity condition simply reduces to $\vec{\alpha}(s)$ being approximately constant.

2. $\vec{\alpha}(t)$ is a rapidly oscillating periodic function, for example:

$$\vec{\alpha}(t) = \vec{\alpha}_0 \cos(\omega_\alpha t + \varphi)$$

(one-dimensional oscillations)

or

$$\vec{\alpha}(t) = \alpha_0 (\vec{e}_1 \cos(\omega_\alpha t + \varphi) + \vec{e}_2 \sin(\omega_\alpha t + \varphi))$$

(uniform spiraling of the field lines)

besides, $\omega_e \ll \omega_\alpha \ll \Omega$ and, of course, $\overline{\theta^2} < \alpha_0^2$.

As in the analysis of the "primary" cycle effects, i.e. of the Larmor rotation, the integration over $|\vec{k}|$ can be split into the regions of $k > \omega_\alpha/v$ and $k < \omega_\alpha/v$. In the first region, the cyclic behavior is not significant and its contribution can again be described by a formula of the same type as Eqs. (3.11) and (3.12). In the second region, one can average the exponential over the oscillations of $\alpha$:

$$\Delta \vec{p} = (ze)^2 \int_{k<\frac{\omega_\alpha}{v}} \frac{\vec{k}}{k^2} d^3k \omega(\vec{k}) J_0^2(\zeta) \frac{\sin^2([\vec{k}\vec{v} - \omega(\vec{k})]t_0/2)}{[\vec{k}\vec{v} - \omega(\vec{k})]^2} J_0^2(k_\perp R_L), \quad (3.19)$$

where $\zeta = \vec{k}\vec{v}_0/\omega_\alpha$ or $\zeta = k_\perp v_0/\omega_\alpha$ and $v_0 = \gamma\beta c\alpha_0$. The amplitude of the field line beating $\alpha_0 = v_0/\omega_\alpha$ can be considered large in comparison to the Larmor radius, then $\omega(\vec{k}) = \omega_e |k_\parallel|/k$.

We will not provide a more expanded answer for different combinations of parameters. We only note that, in case of the spiraling field lines, Eq. (3.17) gives the dependence ($\omega_e t_0 \gg 1$):

$$\begin{aligned} \lambda &\sim \frac{\omega_e^2}{v^3} \ln(\frac{v}{\omega_e} \cdot \frac{1}{a_0}), \quad \text{when } \frac{v}{\omega_e} > a_0, \\ \lambda &\sim \frac{\omega_e}{v^2 a_0}, \quad \text{when } \frac{v}{\omega_e} < a_0. \end{aligned} \quad (3.19a)$$

As one can see, the cyclic behavior of the interaction leads to a significant change in the dependence of decrements on velocities $v$ and $v_M$ compared to the case of $\vec{v}_M = const$.

## 3.4 Drift of Larmor circles

In a transverse electric field (of the space charge), the Larmor circles drift with a velocity

$$\vec{v}_d(\vec{r}) = c \frac{\vec{E}(\vec{r}) \times \vec{H}}{H^2}$$

moving on average in a spiral about the beam axis with a frequency

$$\omega_d = \frac{v_d}{|\vec{r} - \vec{r}_0|} = c \frac{2\pi n e}{H} = \frac{1}{2} \frac{\omega_e^2}{\Omega}$$

($\vec{r}_0$ is the coordinate of the axis of a cylindrical beam).

Since the condition $\Omega^2 \gg \omega_e^2$ must be satisfied for the electron beam stability, it follows that the ratio $\frac{\omega_d}{\omega_e}$ will always be small. With a sufficiently long cooling section, electrons could go through a few drift cycles and it may seem that, in this case, one would have to account for the cyclic nature of this motion when considering interaction with heavy particles. However, the relationship $\omega_d/\omega_e \ll 1$ eliminates such an effect, since it means that the time $\simeq \omega_e^{-1}$ of correlated interaction is small compared to the drift period. On the other hand, this allows one to account for the drift effect locally by simply introducing the velocity $\vec{v}_d(\vec{r})$ as a function of the transverse coordinate of a heavy particle into the relative velocity $\vec{v} - \langle\vec{v}_e\rangle(\vec{r})$ in Eqs. (3.5) and (3.6) or by making the substitution $\vec{v} \to \vec{v} - \vec{v}_d(\vec{r}_\perp)$ in all formulae with explicit expressions of $\vec{F}(\vec{v})$ and $d_{\alpha\beta}(\vec{v})$ with subsequent averaging of the friction strength and diffusion over the phases of heavy particle oscillations.

For a cylindrical beam, the drift velocity can be written as

$$\vec{v}_d = \frac{1}{2} \frac{\omega_e^2}{\Omega^2} \vec{\Omega} \times (\vec{r}_\perp - \vec{r}_0).$$

The impact of the drift is minimal if the beam "axes" coincide. The relative velocity then equals:

$$v_x - v_{xd} = v_x - \omega_d z; \quad v_z - v_{zd} = v_z + \omega_d x.$$

In the cooling process, the relationship between the amplitude values of the transverse velocities of heavy particles and the drift velocities will be maintained (in case of approximate isotropy of the process) and, therefore, the condition when the drift can be completely neglected, reduces a relationship between $\omega_d$ and the betatron oscillation frequency $\nu\omega_0$:

$$\omega_d < \nu\gamma\omega_0$$

(in a general case, $\nu \to 1/\beta_x, 1/\beta_z$).

If the opposite relationship takes place then the drift motion will be slowing the cooling process down while (and if) the drift velocity within the size of the cooled beam exceeds the effective spread of electron velocities and will not be significant starting with amplitudes (in the region of linear friction):

$$\omega_d a < |\Delta_{e\|}|_{eff},$$

where $(\Delta_{e\|})_{eff}$ is determined by the factors considered above. If the equilibrium closed orbit does not coincide with the electron beam axis $(\beta(r_0) \neq \beta_e)$ then the drift motion will cause a shift of the relative beam velocity in the direction transverse to $\Delta\vec{r}$:

$$\Delta\vec{v}_e = \vec{\omega}_d \times \Delta\vec{r}.$$

Such an effect will be present, in particular, in case of a ribbon beam aligned with the plane of the closed orbit with an expansion of the electron longitudinal velocity in radius (in accordance with the dependence $r(\beta)$ of heavy particles). Then ($\Delta z/\Delta x$ is the size ration of the ribbon beam):

$$v_{xd} = -\frac{\omega_e^2}{\Omega}(z - z_0), \quad v_{zd} = \frac{\omega_e^2}{\Omega} \cdot \frac{2\Delta z}{\pi\Delta x}(x - x_0),$$

so that a shift $\Delta x$ of the closed orbit results in a vertical electron velocity of the order of $(\omega_e^2/\Omega) \cdot (2/\pi) \cdot (\Delta z/\Delta x) \cdot \Delta x$. Besides, presence of an average (for a given closed orbit) electron velocity $v_{zd}(x_s)$ can cause a monochromatic instability of the vertical oscillations when the equilibrium amplitude of the betatron oscillations equals $|v_{zd}(x_s)|$. It may also, of course, appear in case of shift of an axially-symmetric beam.

## 3.5 Radial gradient of longitudinal velocity

A gradient of the electron velocity in the plane of the closed orbits leads, as in the case of a non-magnetized beam, to redistribution of the strength or decrements of friction between the longitudinal and transverse degrees of freedom. When the friction forces are known, an estimate of these effects does not present a significant difficulty.

Redistribution of the decrements can be especially important and useful in cases when the friction characteristics (or "eigen" decrements) differ largely in different directions, since equalization of the decrements would speed up damping of the degrees of freedom with small friction. Since redistribution really takes place due to a gradient of the longitudinal friction, the speed-up effect can appear when the longitudinal friction force dominates over the transverse ones. In collisions with non-magnetized electrons (or fast collisions) such a situation can occur if the electron velocity distribution is squeezed in the longitudinal direction. The same situation takes place in adiabatic singular collisions with the circles (the Coulomb behavior of the effective interaction is then preserved) if deviations of magnetic field lines create an effective spread of the circles' transverse velocities large compared to the longitudinal one.

The main factors creating a radial gradient of the longitudinal friction force can be the space charge field [35, 36] and the gradient of the electron coherent Larmor rotation velocity $v_L(r)$. Let us consider in some detail effects of the former. In that case,

$$\frac{dv_{e\parallel}}{dx} = \frac{1}{m\beta c}eE = \frac{2\pi n e^2 (x - x_0)}{m\beta c} = \frac{\omega_e^2 (x - x_0)}{2\beta c} \; ;$$

electrons at the beam center ($x = x_0$) move somewhat slower than on the periphery. The gradient at the center is zero, therefore there is no redistribution when the equilibrium closed orbit is aligned with the electron beam axis. The position of the equilibrium orbit is determined by the equation

$$v_\parallel(x_0) + \frac{dv_\parallel}{dx}(x - x_0) = \frac{\omega_e^2}{4\beta c}(x - x_0)^2 + v_{e\parallel}(x_0) \, ,$$

which can be graphically represented as an intersection of a line with a parabola.

Two equilibrium orbits are theoretically possible: on one slope of the parabola when

$$0 < -\Delta v_\parallel < \frac{\beta c}{\omega_e^2}\left(\frac{dv_\parallel}{dx}\right)^2, \quad (\Delta v_\parallel = v_\parallel(r_0) - v_{e\parallel}(r_0))$$

and on the opposite ones when $\Delta v_\parallel > 0$.

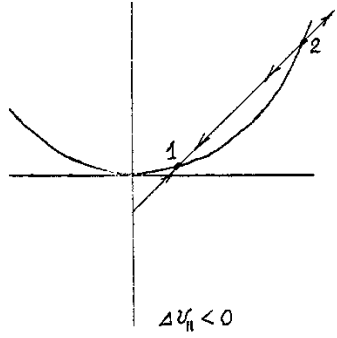 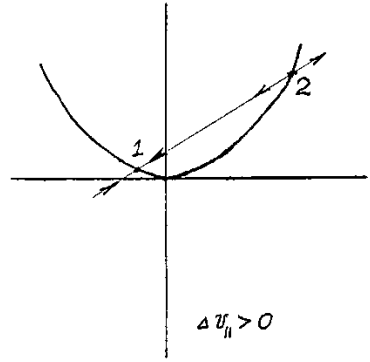

Fig. 8.                             Fig. 9.

Only orbit 1 can be stable, in whose vicinity

$$\frac{dv_{e\|}}{dx} < \frac{dv_\|(x)}{dx}$$

(algebraically), so that the heavy particles are "drawn" by the electrons to a point $r_1$ rather than being pushed away as near point 2. Finally, a stable orbit must go through the electron beam (of radius $a$), which requires fulfillment of the following conditions:

$$|\Delta v_\||  < a\left(\frac{dv_\|}{dx} - \frac{a\omega_e^2}{4\beta c}\right), \quad \text{when } \Delta v_\| < 0, \quad a < \frac{2\beta c}{\omega_e^2}\frac{dv_\|}{dx}$$

or

$$\Delta v_\| < a\left(\frac{dv_\|}{dx} + \frac{a\omega_e^2}{4\beta c}\right), \quad \text{when } \Delta v_\| > 0.$$

A similar picture (but with a changed sign of $dv_{e\|}/dx$) can arise due to a gradient of the electron coherent Larmor velocity caused by a non-adiabatic impact of transverse forces related to imperfections of the electron gun optics. It is natural that such forces are large near the edge of the beam; then the sign of the gradient

$$\frac{dv_{e\|}}{dx} = \frac{1}{2\gamma^2 \beta c}\frac{dv_\perp^2}{dx}$$

is opposite to the sign of $(x - x_0)$.

To illustrate the effect of a gradient, let us consider a case when oscillations of the magnetic field direction create an effectively pancake-shaped electron velocity distribution (Eqs. (3.11) and (3.12)) with a transverse width of $v_0$. For the velocity region $v < v_0$, the longitudinal friction force and the friction power of radial oscillations respectively equal (see Eq. (1.39)):

$$F_\parallel^A = -F_0 \frac{v_\parallel - v_{e\parallel}(x)}{|v_\parallel - v_{e\parallel}(x)|},$$

$$\left\langle \frac{da_x^2}{dt} \right\rangle = -\frac{\eta}{4}\sqrt{\frac{\pi}{2}} F_0 \cdot \frac{a_x^2}{\gamma M v_0} - \eta \frac{\psi}{p} \langle x_b F_\parallel^A \rangle,$$

where

$$F_0 = \frac{4\pi n_e' z^2 e^4 L^A(v_0)}{m v_0^2}, \quad x = \psi \frac{v_\parallel}{\beta c} + x_b.$$

Assuming that the longitudinal velocities damp to the equilibrium one and averaging explicitly over the betatron oscillations, we get:

$$\frac{da_x^2}{dt} = -\lambda_0 a_x^2 - \frac{2}{\pi}\frac{\psi}{p} F_0 a_x \cdot \text{sign}\left(\frac{dv_{e\parallel}}{dx}\right); \qquad (3.20)$$

the answer is applicable within the limits of

$$\Delta_{e\parallel} < \left|\frac{dv_{e\parallel}}{dx}\right| a_x < v_0.$$

For a positive effect, the gradient $dv_{e\parallel}/dx$ must be greater than zero that corresponds to Fig. 8, i.e. the electron velocity at the beam center must exceed the closed orbit's one. Note that, at the above conditions, the change in the damping rate of the radial amplitudes does not explicitly contain the parameter $dv_{e\parallel}/dx$ [36], which is related to a discontinuous nature of the longitudinal friction force in case of a pancake-like distribution. Let us assume $\psi = const = 1/\nu^2$, then Eq. (3.20) gives that the amplitudes

$$v_{x0} < \frac{8}{\pi}\sqrt{\frac{2}{\pi}\frac{1}{\nu}} v_0 \equiv v_0'$$

damp at a constant absolute rate exceeding the "natural" one by the ratio of $\simeq v_0'/v_{x0}$.

In the presence of resonant coupling of $x$ and $z$ oscillations, the latter get also involved in the accelerated damping. To determine the established amplitudes, one needs to take into account that the jump in the longitudinal friction when $v \to 0$ occurs within some width $\Delta_{e\parallel}$ determined in general by a combination of factors considered earlier. Within this width, $F_\parallel \simeq -F_0 v_\parallel/\Delta_{e\parallel}$, redistribution of the decrements is proportional to $dv_{e\parallel}/dx$ and, in the optimal case, the decrements are equal having an order of magnitude value of $\simeq \eta[F_0/(M\Delta_{e\parallel})] \cdot (1/3)$. The equilibrium temperature of the heavy particles is then equal to $T_s \simeq m\sqrt{v_0 \Delta_{e\parallel}}$ (see Eq. (1.39)).

Thus, a small longitudinal spread in the presence of a gradient $dv_{e\parallel}/dx > 0$ can lead to an enhanced cooling of transverse oscillations in those cases when the effect of magnetization is limited (or suppressed completely) by the factors creating large relative transverse velocities of heavy particles and electron Larmor circles.

## 3.6 Discussion of experimental dependencies

The cooling effect has been confidently demonstrated already in the first cycle of experiments with a proton beam in the NAP-M storage ring [3, 5, 43, 44]. The obtained results were in a qualitative agreement with the theoretical perception of the cooling process as relaxation of a proton beam in a co-moving gas of free electrons, only including accelerator specifics (Chapter I).

Table 1. Typical experimental parameters and results of cooling protons in the first cycle of studies (1974-75).

| | |
|---|---|
| Proton energy | 35-80 MeV |
| Electron energy | 19-43.6 keV |
| Electron beam diameter | 10 mm |
| Electron current $I_e$ | 0.1-0.25 A |
| Proton current $I_p$ | 20-100 μA |
| Average vacuum | 5·10$^{-10}$ Torr |
| At an energy of 65 MeV | |
| Established size of the proton beam | 0.8 mm |
| Cooling time (at an electron current of 0.1 A) | 5 s |
| Lifetime in the cooled mode | 5,000 s |
| Lifetime without cooling | 900 s |

Positive effects (damping of oscillations and energy spread, existence of an established size, increase in the lifetime) were observed when the average beam velocities were brought together with a precision better than $1 \cdot 10^{-3}$. When separating the velocities by a relative distance of $\simeq 2 \cdot 10^{-3}$, the listed effects disappeared. This can be interpreted as a manifestation of the monochromatic instability (for the transverse degrees of freedom (Section 1.2)) or this could happen due to a shift of the proton equilibrium orbit as a function of energy and deterioration of the cooling conditions. The increase in the lifetime is related to suppression of particle multiple scattering on the residual gas by friction in the electron beam. The orders of magnitudes of the cooling time and established proton beam size are consistent with the values, which can be obtained using the "plasma relaxation" formulae (Section 1.1) for the given parameters of the electron beam ($T_{e\perp} \simeq 0.2$ eV).

Table 2. Typical experimental parameters and results of cooling protons in the second cycle of experiments (1976).

| | |
|---|---|
| Proton energy | 65 MeV |
| Electron energy | 35 keV |
| Electron gun's cathode diameter | 20 mm |
| Electron current $I_e$ | 0.1-0.8 A |
| Proton current $I_p$ | 20-100 μA |
| Average vacuum | 5·10$^{-10}$ Torr |
| Established size (diameter) of the proton beam in the middle of the straight | 0.47 mm |
| Cooling time ($I_e = 0.8$ A) | 83 ms |
| Proton lifetime in the cooled mode | under 8 h |
| Effective electron temperature | 0.25 eV |

The next stage of studies was done starting in 1976 at an upgraded facility with improved parameters and monitoring system. Towards the very beginning of experiments, there has been a revision of views on the cooling process properties accounting for magnetization of the electron cooling flow with a small longitudinal velocity spread[*] [37, 38].

Therefore, the effect of fast cooling [7] independently discovered in first experiments of the new cycle could be explained by the influence of the collinear magnetic field on the particle collisions.

To verify the theoretical views and determine the conditions, in which the cooling process takes place, there was a series of experiments [8, 9, 32], which measured characteristics of the process in greater detail and studied its sensitivity to changes in a number of parameters. The observation methodology involved excitation (absolute or relatively to the electron beam) of individual degrees of freedom of a previously cooled proton beam and subsequent measurement of the rate of damping to an equilibrium.

---

[*] V.V. Parkhomchuk even earlier noted (1975) the presence of such an important factor as compression of the electron velocity in the longitudinal direction. Reference [37] also played a stimulating role. It considered coherent interaction of a proton beam with a magnetized electron flow (see Chapter IV). One should also note that Ref. [2] gave a kinetic equation that accounted for electron magnetization using an integral of collisions in a strong magnetic field first obtained by S.T. Belyaev [38]. However, the specific analysis of the cooling process in Ref. [2] was done neglecting magnetization, since the initial study of electron cooling focused mainly on a situation, when the electron velocity spread is approximately equal in all directions and, in such conditions, magnetic field does not substantially change the relaxation process.

At the present time, there are data on cooling protons at the energies of 65, 35, and 1.5 MeV at the electron beam currents of 0.5 A to 2 mA. The following dependencies were studied.

1. Dependence of the transverse oscillation damping decrement (inverse time) $\lambda$ and of the longitudinal friction force $F_\parallel$ on the electron Larmor velocities. The Larmor velocities (or coherent spiraling of electrons) were excited (before the electrons entered the cooling section) by an electrostatic field by sending the beam through a small-size capacitor. When cooling by a non-magnetized beam, $\lambda$ and $F_\parallel$ should be inversely proportional to $v_{eL}^3$. For a magnetized beam, according to the results of Sections 2.4 and 3.1, the dependence can be only logarithmic or inversely proportional to the first power of $v_{eL}$.

2. Dependence of $\lambda$ and $F_\parallel$ on modulation of the Larmor circle velocities (i.e. electron velocities averaged over the Larmor rotation). The modulation was created by adiabatically waving the magnetic field lines by introducing transverse fields in the cooling section. The dependence on the average difference of the magnetic field and proton closed orbit directions was also studied. According to the theory (Section 3.3), these dependencies should be substantially sharper than those on the Larmor velocity, which is inversely proportional to the cube or square of the modulation amplitude (or the rms deviation angle). In contrast to a non-magnetized beam, a sharp dependence should take place also in the region $v_M < v_L$ and not only in the region $v_M > v_L$.

3. Dependence on electron energy modulation. This dependence should have a similar behavior. In fact, its observation in the first experiments with the upgraded system [7] was the first experimental evidence of the deviation of the cooling process from the "classical" views.

4. Dependence of the cooling time on the longitudinal and transverse proton velocities (the excitation amplitudes) themselves. In a case without magnetization, the damping decrements in the region $v < v_{eL}$ should be constant ($\lambda \sim v_{eL}^{-3}$) while the longitudinal friction force should saturate at $v \lesssim v_{eL}$.

The main results of the experimental studies are shown in Figs. 10-18. Figures 10 and 11 illustrate the difference in the dependence of the cooling rate on the electron Larmor velocities and on the difference, in general, of the average (over the Larmor rotation) proton and electron velocities. These data quite clearly express the fact that cooling takes place in conditions when spatial variations of the average (over the thermal spread and Larmor motion) electron velocities are small compared to the thermal (Larmor) velocities. Otherwise, $\lambda$ and $F_\parallel$ would remain practically constant, at least, in the interval $\overline{\Delta v}/(\beta c) \lesssim (2-4) \cdot 10^{-3}$ corresponding to the electron transverse temperature. In general, the sharp dependence on the average velocity difference and the weak one on the Larmor velocity confirm the determining (or enhanced) role of the far collisions ($\rho > r_L$) when $u_A < v_{eL}$. The curves in Figs. 10 and 11 were taken at an electron current of 0.3 A, an electron beam diameter of 10 mm and a proton energy of 62.5 MeV.

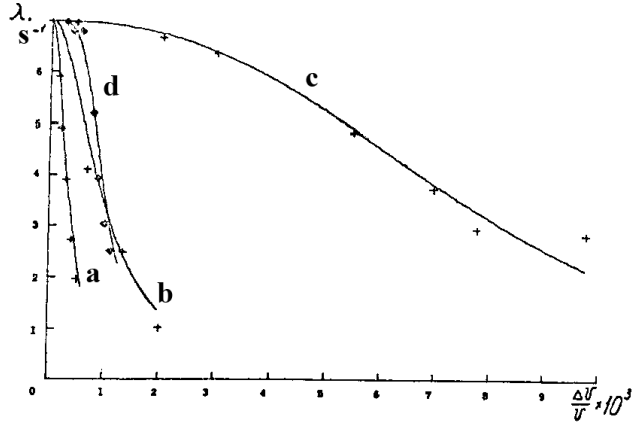

Fig. 10. Dependence of the damping decrement of transverse oscillations on:
a – the amplitude of electron energy modulation,
b – the amplitude of modulation of the transverse velocities of Larmor circles,
c – the Larmor rotation velocity,
d – the proton transverse velocities.

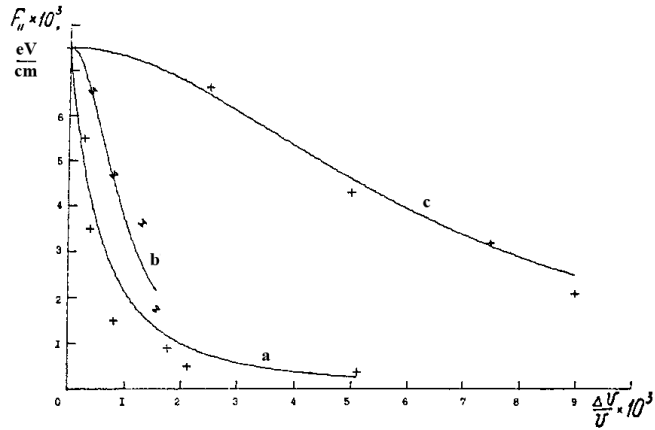

Fig. 11. Dependence of the longitudinal component of the friction force on:
a – the difference of the proton and electron longitudinal velocities,
b – the amplitude of modulation of the transverse velocities of Larmor circles,
c – the Larmor rotation velocity.

Figure 12 shows the dependence of the transverse oscillation decrement $\lambda$ on the amplitude of modulation of electron energy (or longitudinal velocity) and Fig. 13 shows the dependence of $\lambda$ on the excited Larmor radius on a logarithmic scale. One can see that the decrement is inversely proportional to the square of the difference of the average velocities $\lambda \sim 1/(\Delta v)^2$ (for velocities $10^{-4} \lesssim \Delta v \lesssim 10^{-3}$) and only to the first power of $v_L$ ($v_{eL} > 10^{-3}$).

Figure 14 shows the dependence of the density-normalized decrement on the transverse proton velocity with respect to the electron beam. This dependence with a good accuracy matches the curve $\sim 1/(\Delta v_\perp)^2$ for different values of the proton density (current) and energy and for different methods of "exciting" the relative transverse (averaged over the Larmor rotation) velocities.

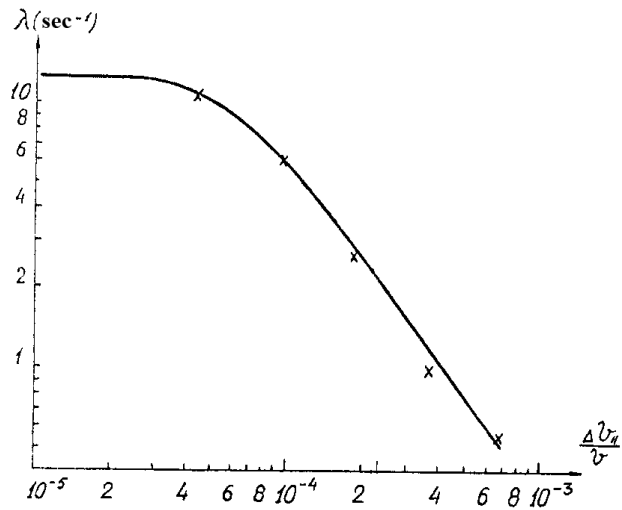

Fig. 12. Dependence of the damping decrement of transverse proton oscillations on the amplitude of longitudinal electron velocity modulation (at an electron current of 300 mA and a proton energy of 65 MeV).

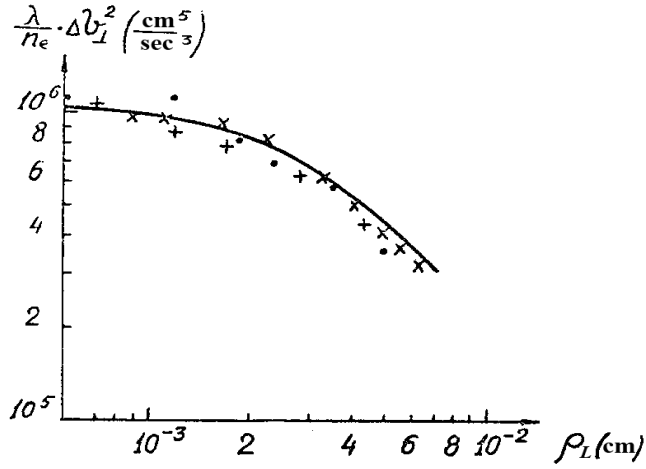

Fig. 13. Effect of the electron Larmor rotation on the damping decrement of transverse oscillations (the definition of the symbols is the same as in Fig. 14).

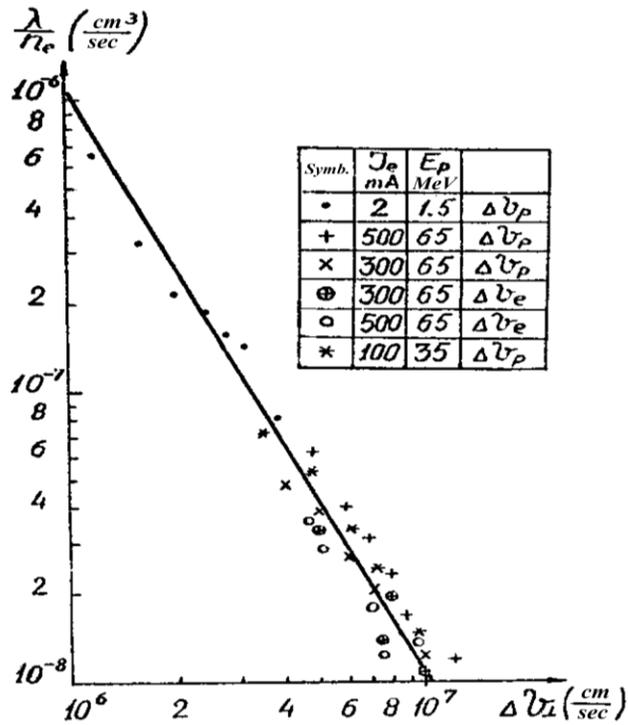

Fig. 14. Dependence of the damping decrement of transverse oscillations on the relative velocity of protons and electrons for different proton energies. The relative transverse velocity was created by exciting proton betatron oscillations ($\Delta v_p$) or tilting the electron beam with respect to the proton trajectory ($\Delta v_e$).

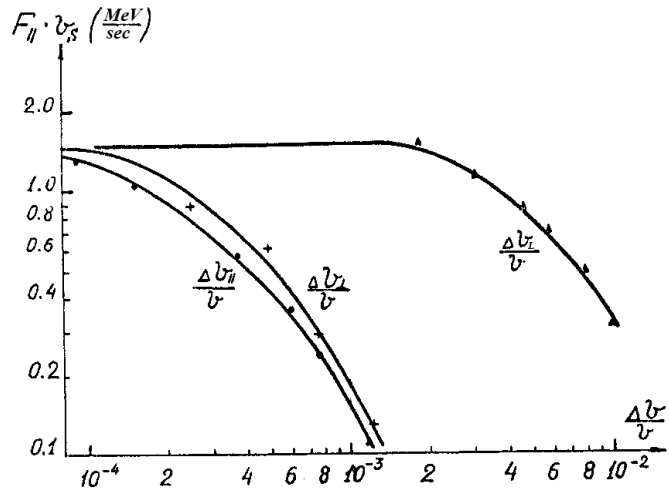

Fig. 15.  Dependence of the longitudinal friction force on the detuning of the longitudinal velocities $\Delta v_\parallel$, transverse velocities $\Delta v_\perp$, and Larmor rotation velocity $\Delta v_L$ (at an electron current of 300 mA and a proton energy of 65 MeV).

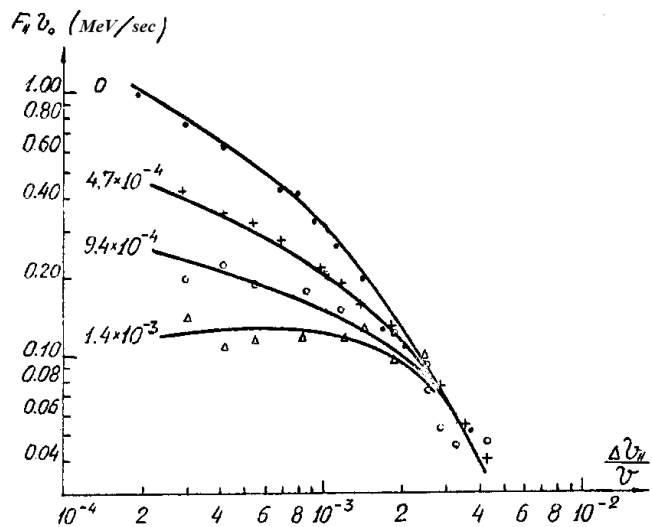

Fig. 16.  Dependence of the longitudinal friction force on the detuning of the proton velocity from the average electron velocity ($\Delta v_\parallel$) at different relative transverse velocities ($\Delta v_\perp / v_0$). The electron current was 300 mA, the proton energy was 65 MeV.

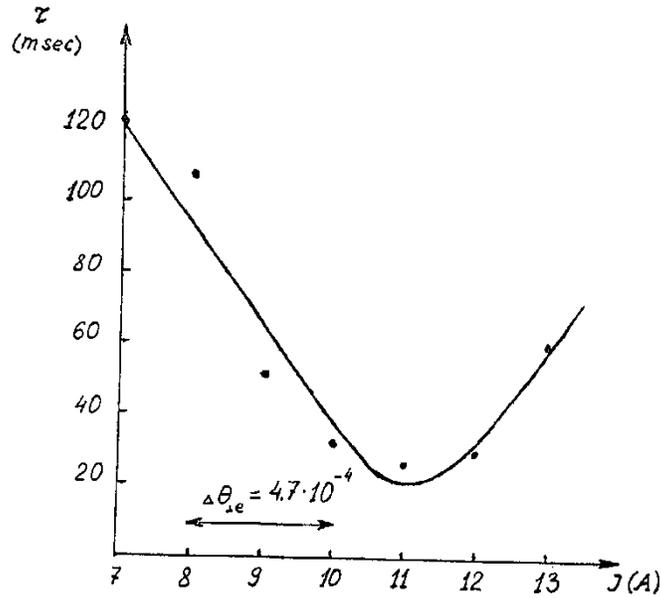

Fig. 17.  Dependence of the damping time of the proton energy spread on the current in corrector coils changing the direction of the accompanying magnetic field (the arrow shows the scale of change in the field inclination angle towards the orbit with the current).

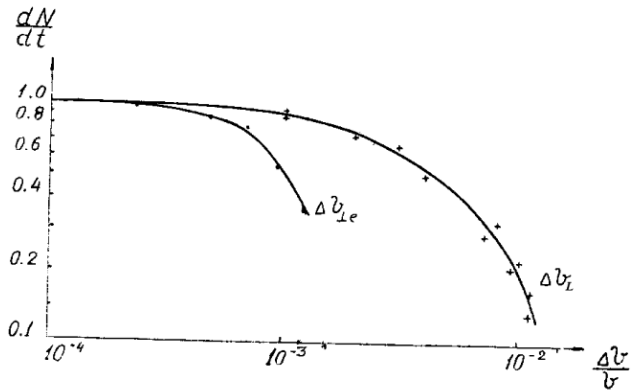

Fig. 18.  Normalized dependence of the neutrals' exit rate on the Larmor velocities and the average transverse electron velocity.

One should note the equality of the radial and vertical oscillation decrements observed in all cases, which was apparently caused by a resonant coupling of the betatron oscillations ($|v_x - v_z| \approx 0.1$).

Figure 15 shows the dependence of the longitudinal friction on $v_L$ and the "average" velocities $\Delta v_\parallel$ and $\Delta v_\perp$ while Fig. 16 shows the dependence on $v_\parallel$ for different angles between the proton orbit and magnetic field ($\Delta v_\perp = \beta c \alpha$).

The minimum cooling time (the time of damping of proton small oscillations) achieved in the experiments is about 40 ms (Fig. 17) while the theoretical limit (Section 2.6) (for a density of $\simeq 10^8$ cm$^{-3}$) is approximately 1 ms. Apparently, the practical limit is currently set by non-uniformity of the magnetic field ($|\Delta \vec{H}/H| \simeq 4 \cdot 10^{-4}$).

The most important feature of the experimental results is the fact that the friction force $\vec{F}(\vec{v})$ as a function of velocity quickly grows with decrease in $v$ in the region $v < v_L$ (and not only with $v > v_L$), which quite certainly indicates a strong positive effect of the magnetic field.

For now, however, one can mainly talk only about qualitative agreement of the experimental results with the theoretical description. Preliminary estimates show good quantitative agreement between the theoretical formulae and experimental curves for the longitudinal friction force $F_\parallel(v_\parallel)$ in the velocity region of $\Delta v_\parallel/(\beta c) > |\delta \vec{H}/H| \simeq 4 \cdot 10^{-4}$; at lower velocities, if using formulae not including the limiting effect of various "imperfections" of the system, one significantly overestimates the experimental values. The agreement is less satisfactory between the theoretical formulae under the "clean" conditions and the experimental data for the transverse cooling decrements. Nevertheless, even in this case, the differences at maximal velocities $\Delta v_\perp$ (but remaining small compared to $v_L \simeq 3 \cdot 10^7$ cm/s) in each experimental cycle are relatively small; for small velocities, theoretical values exceed the experimental ones by a factor of a few.

All experimental values of the transverse decrements normalized to the density match with a satisfactory accuracy the dependence $\lambda/n_e \approx const/v^2$ [9] while, for the clean conditions, the theory gives $\simeq 1/v^3$ when $v > \omega_e R_L$ and the dependence of $\sim 1/v^2$ occurs only in the region $v \ll \omega_e R_L$, Eq. (3.9). The theory then gives $\lambda \sim \sqrt{n_e}$ (without intentional excitation of the Larmor motion, this region is not typical for the conditions of the performed experiments; in practice, deviations of the field lines are important here). The dependence of $\lambda \sim 1/v^2$ may be related to a redistribution of the friction power due to radially-longitudinal coupling at a disk-like effective distribution of the Larmor circle velocities [36] (see Section 3.5). The source of the gradient in $v_{e\parallel}(x)$ may be the space charge of electrons. However, experiments with a compensated beam [9] gave the same values of the $\lambda_\perp$ decrements. A third reason for the $\sim 1/v_\perp^2$ dependence may be oscillations of the magnetic field lines, Eqs. (3.19a) and (3.14).

It should be noted that study of the longitudinal damping may be more effective for extracting information about the dependence of the friction force on the proton velocity and other parameters, since the process in this degree of freedom is complicated less by the influence of the motion cyclicity and coupling effects.

One feature of the experimental conditions discovered during the measurements themselves was the relatively narrow region of fast damping in the radial position ($\Delta r \simeq 3$ mm) of the equilibrium orbit in the electron beam. A reason for this, besides a strong gradient in $v_{e\parallel}(r)$ due to the space charge, may also be non-uniformity (in the direction) of the accompanying magnetic field when moving away from a certain field symmetry axis.

Finally, there is yet no explanation for one other feature: the observed dependence of the hydrogen atom yield (recombination) on the change in the average direction of magnetic field, which is sharper than the dependence on the electron Larmor velocity (Fig. 18). Since the recombination process is related to microscopic distance scales ($10^{-8} - 10^{-7}$ cm), it appears unlikely that the magnetic field or the average motion of a proton at a low velocity with respect to the Larmor circles can influence the probability of the process [28]. Possibly, the reason for reduction in $dN/dt$ is simply a spatial shift of the orbit (when tilting it) into a region of larger Larmor radii in the presence of a velocity gradient $v_L(r)$ excited by imperfection of the electron gun optics. However, this question still remains unresolved.

As can be seen from everything discussed here and earlier, a direct comparsion of the theoretical and experimental dependencies is complicated by the fact that, at low relative velocities, the cooling process is is sentisive to quite a large number of parameters, some of which are hard to control. For quantiative verification of the theoretical prections and determination, on the other hand, of the cooling conditions, one must next complete a significant amount of analytic and numerical calculations for various model situations. Additional experimental studies will also be needed. Of course, one also should not exclude the possibility of progress or changes in certain aspects of the theoretical consideration. There is a particular interest in improving the qualities of a cooling system to such a degree (Section 2.6) that the considered unique properties of electron cooling can be realized to the full extent.

### 3.7 On possibilites of optimizing cooling of large spread

Let us briefly discuss the possibility of using the special properties of cooling in a magnetized electron flow in the sweeping technique (Section 1.5). Let us assume that the electron flow is "cooled" to an ultimately low longitudinal temperature ($(T_{e\parallel})_{eff} \simeq e^2 n^{1/3}$) and is strongly magnetized, i.e. the Larmor radii do not exceed the average distance between electrons: $r_L < n^{-1/3}$. $r_L < n^{-1/3}$. For example, at a density of $10^9$ cm$^{-3}$ and a transverse temperature of 0.2 eV, this requires a field of the order of

$$H_\parallel > \frac{mc}{e}\sqrt{\frac{2T_\perp}{m}} n_e'^{1/3} \simeq 500 \text{ G} \qquad (3.20)$$

and, at a density of $10^{12}$ cm$^{-3}$, the required field is $H_\| > 5$ kG. From the point of view of suppressing the space-charge "defocusing", the required field is ($\Omega \gg \omega_e$):

$$H_\| \gg \frac{mc}{e}\sqrt{\frac{4\pi n'_e e^2}{m}} \ ; \tag{3.21}$$

the first and second criteria relate as $(T_\perp/(2\pi e^2 n^{1/3}))^{1/2}$; in practice, always $e^2 n^{1/3} \ll T_\perp$, therefore fulfilment of the first condition is sufficient.

A proton moving in such an electron flow at a velocity of $v_{opt} \simeq (e^2 n_e^{1/3}/m)^{1/2}$ will experience maximum friction with a force (see Eq. (2.148)) of

$$F_{max} \simeq \lambda_{max} \cdot M v_{opt} \simeq \eta \omega_e m \sqrt{\frac{e^2 n_e'^{1/3}}{m}} \simeq e^2 n^{2/3} \eta \ . \tag{3.22}$$

Suppose that the proton transverse velocity spread has been preveiously reduced (for example, by an adiabatic beam expansion in the interaction region) to a value of $v_{opt}$ while the longitudinal velocity spread is large: $\Delta v_\| \gg v_{opt}$. Then applying the sweeping technique in the longitudinal direction, i.e. changing the electron beam velocity with time at a rate of

$$\frac{dv_{e\|}}{dt} \simeq \frac{F_{max}}{M} \simeq \frac{e^2 n^{2/3}}{M}\eta \tag{3.23}$$

and passing through the whole width $\Delta v_\|$, one can cool the longitudinal degree to a spread of $\Delta v_\| \simeq v_{opt}$ in time

$$\tau \simeq \frac{M\Delta\theta}{\eta e^2 n^{2/3}} \ ; \tag{3.24}$$

next, if there are no impeding factors, the proton beam cools to a temperature of $(T_{e\|})_{eff}$, or to a spread of

$$\Delta v \simeq \sqrt{e^2 n_e'^{1/3}/M} \ ,$$

already in a small time $\tau_{min} \simeq (M/m) \cdot [1/(\eta\omega_e)]$.

Note that the velocity $v_{opt}$ is practically not too small. For example, with $n'_e = 10^9$ cm$^{-3}$ and $\beta = 0.35$ (65 MeV proton energy), we get

$$v_{opt}/(\beta c) \simeq 5 \cdot 10^{-5} \ .$$

If the initial proton angular spread $\theta$ exceeds this value, then the time $\tau$ increases by a factor of $(\theta\beta c/v_{opt})^2$.

Overall, with use of sweeping being generally beneficial, magnetization can increase its efficiency manyfold.

# IV. COLLECTIVE SATIBILITY OF COOLED BEAM

With increase in intensity of the cooling beam, heavy particle interaction, collective and collisional, may start playing a noticeable role. An interest to the former, besides the usual limitations on space charge, is related to the fact that the passing electron flow introduces dissipation in the particle collective oscillations and thereby leads to their increase or damping. An extensive study of these effects is not our goal. Thus, we will limit ourselves to estimates illustrating the significance of collective interaction based on the approach developed in Refs. [39, 40].

## 4.1 Space charge effect

The space charge of a heavy particle beam weakens beam focusing and leads, generally speaking, to increase in the beam size at a given temperature. If one neglects non-uniformity of focusing along the orbit, then the minimum size of an unbunched beam in an equilibrium state with zero temperature can be found by setting the focusing and defocusing forces equal. For a cylindrical beam with equal betatron tunes $v_x = v_z = v$, we get the maximum density and the dependence of the size on the current or the number of particles:

$$n_{max} = \frac{\gamma^3 \beta^2 v^2 M}{z^2 r_e 2\pi R^2 m}, \qquad (4.1)$$

$$r_0^2 = \frac{N z^2 m r_e R}{\pi \gamma^3 \beta^2 v^2 M}. \qquad (4.2)$$

A complete compensation of the focusing forces inside the beam can take place when the particle thermal energy, i.e. the beam temperature, is small compared to the potential energy of the electrostatic interaction:

$$T \ll \pi z^2 e^2 \frac{n_{max}}{\gamma} r_0^2 = \frac{z^2 e^2 N}{2\pi R \gamma} \equiv T_0. \qquad (4.3)$$

For the NAP-M storage ring ($2\pi R \simeq 50$ m, $\gamma \approx 1$) with $N = 10^7$, the temperature $T_0$ is $\simeq 3 \cdot 10^{-4}$ eV while the minimum temperature achievable with electron cooling at $n_e = 10^8$ cm$^{-3}$ is $T_{min} \simeq 10^{-4}$ eV, i.e. it has a similar value. Thus, even with idealized focusing, space charge imposes practical limitations on the achievable size.

In real cases, the rigidity of the focusing field varies along the orbit and, as a result of that, the transverse density distribution $n(x, z)$ is azimuthally modulated leading to the corresponding modulation of the space charge field. Thus, the transverse degrees of freedom experience the effect of non-conservative forces $\vec{f}(x, z, \theta)$, which may come in resonance with particle oscillations in the main field. The general resonant condition is

$$m_x \nu_x + m_z \nu_z = n, \tag{4.4}$$

where $m_x$, $m_z$, and $n$ are positive and negative integers and the tunes $\nu_x$ and $\nu_z$ include the shift due to the space charge field. A characteristic feature of this field is its substantial nonlinearity as a function of $x$ and $z$ creating a significant number of possible resonances and giving a tune shift $\Delta\nu(a^2)$ depending on the oscillation amplitudes. With $\Delta\nu(0) \ll 1$, it is relatively easy to detune from linear resonances $|m_x| + |m_z| \leq 2$; however, the practical limitations on $\Delta\nu_{max}$ are related to nonlinear resonances $|m_x| + |m_z| > 2$. The collective interaction (in a stationary state) leads to formation of parasitic separatrices in the transverse motion with oscillations of amplitudes near the equilibrium values (or the curves $F(a_x, a_z) = 0$) determined by Eq. (4.4). The beating size increases with growth of the beam density and, after the separatrices of neighboring resonances overlap, the motion loses stability and becomes stochastic. These phenomena were investigated quite thoroughly for the case of colliding beams [22, 23, 25]. Some weakening of the influence of nonlinear resonances compared to the case of colliding bunches (with the same $\Delta\nu(0)$) may be related to the fact that the strengths of higher-order resonances (large $n$) are relatively small, since the interaction is not pulsed. Therefore, although $\Delta\nu_{max}$ is significantly less than a unit, it can substantially exceed the limit achievable in colliding beams ($10^{-2} - 10^{-3}$).

A similar influence, although with less pronounced effects of nonlinear resonances, can come from the field of the electron beam. It mainly results in a constant (independent of amplitudes) $\nu_x$, $\nu_z$ tune shift, since, in the case of straight electron acceleration, the electron beam size usually exceeds the size of the cooled beam. For the NAP-M conditions

$$(\Delta\nu)_{ep} \simeq \frac{\pi n e^2 R^2}{\gamma M \nu \omega_0^2} \eta \approx 0.015. \tag{4.5}$$

The tune shift $\Delta\nu_{pp}$ with $N_p = 10^8$ and a beam size of $0.5 \times 0.5$ mm² (the minimum observed size) is:

$$\Delta\nu_{pp} \simeq \frac{\pi n_p e^2 R^2}{\gamma M \nu \omega_0^2} \simeq 0.06 \quad (n_p \simeq 10^7 \text{ cm}^{-3}), \tag{4.6}$$

but, with $N_p = 10^7$, $\Delta\nu_{pp} \simeq 6 \cdot 10^{-3}$.

The tune shift $\Delta\nu(a^2)$ can play a positive role as well, since it creates a spread of tunes in the beam stabilizing potential collective instabilities.

## 4.2 Dispersion equations for small coherent oscillations

A stationary state of a cooled beam with the temperatures determined by equilibrium of the friction and diffusion processes is realized if it happens to be stable towards small collective excitations. We will consider that additional influence on the collective modes of particle motion, which is introduced by the cooling beam itself.

We will describe the interaction of collective oscillations with the electron flow by an equation (written in the co-moving frame)

$$\frac{\partial \tilde{F}}{\partial t} + \omega_\alpha \frac{\partial \tilde{F}}{\partial \psi_\alpha} - ze\left(\frac{\partial \tilde{\varphi}^e}{\partial \psi_\alpha} + \frac{\partial \tilde{\varphi}}{\partial \psi_\alpha}\right)\frac{\partial F_{st}}{\partial I_\alpha} = 0, \qquad (4.7)$$

where $\tilde{F}$ is the deviation of the distribution function from $F_{st}(I)$ in the $(I,\psi)$ phase space, $I$ and $\psi$ are the action-phase variables, and $\tilde{\varphi}^e$ and $\tilde{\varphi}$ are the "electrostatic" potentials related to excitations of the electron and proton beams. The $I, \psi$ variables are determined, generally speaking, including the Coulomb interaction in the stationary state. We assume that interaction with external surroundings is small or makes an additive contribution to the resulting decrements of collective excitations.

As usually done in the linear theory of stability, we apply the Laplace transformation in time to Eq. (4.7) and expand it in a series in the phases:

$$(\omega - \vec{m}\vec{\omega})F_m - ze\left(\vec{m}\frac{\partial F_{st}}{\partial \vec{I}}\right)(\varphi_m^e + \varphi_m)_\omega = 0, \qquad (4.8)$$

where $\vec{m} = (m_x, m_z, m_\parallel)$ are the indices of phase harmonics (of the betatron and azimuthal motion). While the collective tune shifts are relatively small, one can assume that separate harmonics oscillate independently ($\Delta\omega \ll \omega_0, \omega_x, \omega_z$) and then, despite the fact that the interaction is not a function of simply the phase difference, we can assume that the harmonics $\varphi_m^e$ and $\varphi_m$ are proportional to $F_m$ [39]. The definition of $\varphi_\omega$ is obviously:

$$\varphi_\omega = \int \frac{F_\omega(\Gamma')d\Gamma'}{|\vec{r} - \vec{r}'|}.$$

The potential $\varphi_\omega^e$ can be determined from the equation describing excitation of the electron flow (we limit ourselves to a non-relativistic case, generalization is trivial):

$$n_q = \int f_q(\vec{v}_e)d^3v_e, \quad \varphi_q^e = \frac{4\pi n_q e}{q^2},$$
$$-i\omega f_{\vec{q}} + iq_\parallel v_{e\parallel}f_{\vec{q}} - i\frac{\omega_e^2 q_\parallel}{q^2}\frac{\partial f_{st}}{\partial v_{e\parallel}}n_q^e = \frac{ei}{m}q_\parallel \varphi_{\vec{q}}\frac{\partial f_{st}}{\partial v_{e\parallel}}, \qquad (4.9)$$

where index $q$ denotes harmonics $\sim e^{i\vec{q}\vec{r}}$ determined in the quasi-local approximation for the coordinate system connected to the hydro-dynamic electron velocity averaged over the Larmor rotation. Thus, we switched the transverse electron mobility off limiting ourselves to the region of frequencies (in the co-moving frame) small compared to the Larmor one. Equation (4.9) gives the connection:

$$\varphi_{\vec{q}}^{e} = [\varepsilon_{\vec{q}}^{-1}(\omega) - 1]\varphi_{\vec{q}}, \quad \varepsilon_{\vec{q}}(\omega) = 1 - \omega_{e}^{2}\frac{q_{\|}^{2}}{q^{2}}\langle\frac{1}{(\omega - q_{\|}v_{e\|})^{2}}\rangle. \tag{4.10}$$

For a continuous beam, the normal excitation is close to $\sim e^{in\theta}$; using this behavior, after a series of transformations, Eq. (4.8) gives an integral dispersion equation for density oscillations [40]:

$$\rho_{\vec{k}_{\perp}'} = \frac{N(ze)^{2}}{R}\int \rho_{\vec{k}_{\perp}'}\frac{d^{2}k_{\perp}}{2\pi^{2}k_{n}^{2}}\left[1 + \int[\varepsilon_{\vec{q}}^{-1}(\omega) - 1]b_{q_{\|}n}^{2}dq_{\|}\right] \\ \times \langle\left(\vec{m}\frac{\partial}{\partial I}\right)(e^{i\vec{k}_{\perp}\vec{r}_{\perp}})_{m_{\perp}}(e^{i\vec{k}'\vec{r}_{\perp}})_{m_{\perp}}^{*}\rangle, \quad (\vec{m} = \vec{m}_{\perp}, n), \tag{4.11}$$

where $\rho_{\vec{k}_{\perp}}$ is the spatial Fourier component of the heavy particle beam density, $k_{n} = \sqrt{k_{\perp}^{2} + (n/R)^{2}}$, $\vec{q} = (\vec{k}_{\perp}, q_{\|})$, the factor $b$ accounts for the finite length of the interaction:

$$b_{q_{\|}n} = \frac{1}{2\pi}\left|\int_{0}^{\theta_{0}}d\theta e^{i(n - q_{\|}R)\theta}\right| = \frac{1}{\pi}\left|\frac{\sin[(n - q_{\|}R)\theta_{0}/2]}{n - q_{\|}R}\right|, \quad (\theta_{0} = 2\pi\eta),$$

and the betatron harmonics of the $e^{i\vec{k}_{\perp}\vec{r}_{\perp}}$ Fourier components are:

$$(e^{i\vec{k}_{\perp}\vec{r}_{\perp}})_{m_{\perp}} = J_{m_{x}}(k_{x}a_{x})J_{m_{z}}(k_{z}a_{z})e^{ik_{x}\psi\frac{\Delta p}{p}},$$

where $a_{x}$ and $a_{z}$ are the amplitudes of particle (incoherent) betatron oscillations. In defining factor $b$, we neglected the possibility of variation in the electron hydrodynamic velocity along the length of the interaction section. In a more general case,

$$b = \frac{1}{2\pi}\left|\int_{0}^{\theta_{0}}d\theta e^{in\theta - i\vec{k}\vec{r}(\theta)}\right|,$$

where $d\vec{r}(\theta)/dt = \vec{v}_{\Gamma}(\theta)$ is the average electron velocity as a function of the azimuthal angle. In practice, such effects may play a role only for very high harmonics $n$ with a sufficiently small transverse size of the cooled beam. Hence, in the following, we neglect the spread of electron longitudinal velocities.

In case of a bunched ion beam, or in a mode with RF field, excitations, generally speaking, do not decompose into azimuthal harmonics and are, instead, characterized by a harmonic number $m_{s}$ $m_{s}$ of the synchrotron oscillation phase. Interaction with the electron beam is represented by a sum over all of the azimuthal harmonics $n$, since, in the representation $\{\psi_{x}, \psi_{z}, \psi_{s}\}$, it becomes an explicit (periodic) function of time. Besides, when the tunes of betatron oscillations depend on the energy (total momentum), one needs to account for synchrotron modulation of the $\psi_{x}$ and $\psi_{z}$ phases. Following the approach of Ref. [39], one can obtain the following equation for the density harmonics $n_{\vec{k}}$:

$$\rho_{\vec{k}'} = N(ze)^2 \int \rho_{\vec{k}} \frac{d^3k}{2\pi^2 k^2} \sum_{\vec{m}} \langle (\vec{m} \frac{\partial}{\partial \vec{I}}) \frac{(e^{i\vec{k}\vec{r}})_{\vec{m}} (e^{i\vec{k}'\vec{r}})^*_{\vec{m}}}{\omega - \vec{m}\vec{\omega}} \rangle$$
$$\times \left\{ 1 + \frac{1}{2} \int dq_\parallel \left[ [\varepsilon_{\vec{q}}^{-1}(\omega + \zeta\omega_0 + k_\parallel v) - 1] b^2 q_\parallel - k_\parallel \right. \right. \tag{4.12}$$
$$\left. \left. + [\varepsilon_{\vec{q}}^*(\omega + \zeta\omega_0 - k_\parallel v) - 1] b^2 q_\parallel + k_\parallel \right] \right\},$$

Where $\vec{m} = \{m_x, m_z, m_s\}$, $\vec{k} = (\vec{k}_\perp, n/R)$, $\vec{q} = (\vec{k}_\perp, q_\parallel)$, and

$$\zeta = \vec{m}_\perp \frac{d\vec{\omega}(p)}{d\omega_0(p)}.$$

One can assume that the numbers $n$ and $n'$ have continuous spectra due to the spectrum of interaction with the electron flow being continuous. The harmonics $(e^{i\vec{k}\vec{r}})_{\vec{m}}$ equal

$$(e^{i\vec{k}\vec{r}})_{\vec{m}} = J_{m_x}(k_x a_x) J_{m_z}(k_z a_z) J_{m_s}(\varphi \sqrt{n^2 + (k_x \psi \frac{v_s}{\alpha})^2}) e^{-im_s \chi},$$
$$\alpha = \frac{p}{\omega_0} \frac{d\omega_0}{dp}, \quad \chi = \arctan(k_x \psi \frac{v_s}{\alpha}/n),$$

where $\varphi$ is the angular amplitude of particle phase oscillations.

Interaction of heavy particles through the electron beam described by the integral factors in the square brackets cannot exceed the direct interaction and, in practice, is usually relatively small. Since excitation of the electron beam has continuous spectrum, then the frequency $\omega$ in $\varepsilon_{\vec{k}}(\omega)$ can be considered equal to the own frequency of ion beam oscillations: $\omega \to \vec{m}\vec{\omega}$, and, when integrating over $q_\parallel$, one can neglect the real part of the integral. One can also assume the velocities $v_{e\parallel}$ and $v_\parallel$. For a continuous beam, we then get:

$$\left[ 1 + \int [\varepsilon_{\vec{q}}^{-1}(\vec{m}\vec{\omega}) - 1] b^2 q_\parallel n dq_\parallel \right] \approx 1 + i\mu(n)$$
$$\equiv 1 + i\frac{\omega_e^2}{\pi v^2} \frac{\theta_0^2}{2k_n^2} \left[ \frac{k_\perp^2}{k_n^2}(n + \vec{m}_\perp \vec{v}_\perp) - \frac{\theta_0^2}{12}(\vec{m}_\perp \vec{v}_\perp)(n + \vec{m}_\perp \vec{v}_\perp)^2 \right]. \tag{4.13}$$

For a bunched beam (in a mode with RF field), the factor in the curly brackets in Eq. (4.13) comes from Eq. (4.12) by replacing $\vec{m}_\perp \vec{v}_\perp \to \vec{m}\vec{v} = \vec{m}_\perp \vec{v}_\perp + m_s v_s$, $n \to n + \zeta$ and subsequently adding the two terms with the opposite signs of $n$ ($n = \pm|n|$), which corresponds to averaging of the interaction over the synchrotron motion.

We next estimate the role of coherent interaction assuming a lack of the beam's own dynamical instabilities and neglecting the tune spread. The stability criterion may be a comparison of the increments of possible instabilities with the damping decrements due to incoherent friction and tune spread.

## 4.3 Coherent stability of coasting beam

I. <u>Transverse excitations</u> are characterized by the indices $\vec{m}_\perp \neq 0, n$. For these excitations, the interaction is necessarily a function of the particle oscillation amplitudes, so that contribution of the denominator in the derivative $\langle \partial/\partial I \dots \rangle$ in Eq. (4.11) can be neglected according to the assumed condition of the tune spread being small (equal to zero). Besides, one can neglect the longitudinal forces. The equation then takes the form ($\Delta\omega = \omega - n\omega_0 - \vec{m}_\perp \vec{\omega}_\perp$):

$$\Delta\omega \rho_{\vec{k}'_\perp} = \frac{Nz^2e^2}{R} \int (1 + i\mu) \langle (\vec{m}_\perp \frac{\partial}{\partial I_\perp})(e^{i\vec{k}_\perp \vec{r}_\perp})_{m_\perp} (e^{i\vec{k}'_\perp \vec{r}_\perp})^*_{m_\perp} \rho_{\vec{k}_\perp} \frac{d^2 k_\perp}{2\pi^2 k_n^2}. \tag{4.14}$$

This equation can be brought to the type of equations with a Hermitian symmetric kernel [39, 45] having same-sign (positive) eigen values. If the trace on the right-hand side converges, the sum of the eigen values is a good estimate of the maximum of them. Thus,

$$\begin{aligned}
\operatorname{Im} \Delta\omega &\simeq \frac{Nz^2 e^2}{2\pi^2 R} \int \mu \langle (\vec{m}_\perp \frac{\partial}{\partial \vec{I}_\perp}) |(e^{i\vec{k}_\perp \vec{r}_\perp})_{m_\perp}|^2 \rangle \frac{d^2 k_\perp}{k_n^2} \\
&= \frac{Nz^2 e^2}{4\pi^2} \cdot \frac{\omega_e^2 \theta_0^2}{R v^2} \int \frac{d^2 k_\perp \langle (m_\perp \frac{\partial}{\partial I_\perp}) |(e^{i\vec{k}_\perp \vec{r}_\perp})_{m_\perp}|^2 \rangle}{[k_\perp^2 + (n/R)^2]^2} \\
&\quad \times \left[\frac{\gamma^2 k_\perp^2}{k_n^2}(n + \vec{m}_\perp \vec{v}) - \frac{(n + \vec{m}_\perp \vec{v})^2}{12} \vec{m}_\perp \vec{v} \theta_0^2 \right].
\end{aligned} \tag{4.15}$$

As can be seen, the harmonics $|n + \vec{m}_\perp \vec{v}| < 12/(|\vec{m}_\perp \vec{v}|\theta_0^2)$ happen to be unstable but all shorter-wave modes damp. For example, in the NAP-M storage ring, $\theta_0 \simeq 1/8$, the critical wave length is then

$$\lambda_{cr} = \frac{2\pi R}{n_{cr}} = \frac{\pi R v \theta_0^2}{6} \simeq 7 \text{ cm}, \quad (R = 8 \text{ m}, \ n_{cr} \simeq 700).$$

Despite the very high number $n_{cr}$, the excitation length still remains large compared to the transverse size of the proton beam $\sigma$ and even of the electron one ($r_0 \simeq 1$ cm), so that $k_\perp \gg n/R$. For the main oscillation type $|\vec{m}_\perp| = 1$, Eq. (4.15) gives

$$(\operatorname{Im} \Delta\omega)_{max} = \frac{3Nz^2 r_e^2 n_e \ln(r_0/\sigma)}{2\pi \beta^3 v^2} c \frac{m}{M}. \tag{4.16}$$

With $N = 10^8$, $n_e = 2 \cdot 10^8$ cm$^{-3}$, $\beta = 0.3$, and $v \simeq 1$, we get $(\operatorname{Im} \Delta\omega)_{max} \simeq 1$ sec$^{-1}$. If one compares this value with the friction decrement, one can see that the instability could be dangerous only in the initial acceleration stage (when $\theta_p \simeq 3 \cdot 10^{-3}$). However, here the instability is definitely suppressed by the spread: indeed, Eq. (4.16) corresponds to a very high azimuthal harmonic, whose frequency spread is $n_{cr} \Delta\omega_0/2$, let alone the betatron tune spread.

We arrive at a conclusion that, in transverse oscillations, the instability practically does not manifest itself.

<u>Longitudinal excitations</u> ($\vec{m}_\perp = 0$). At a sufficiently small longitudinal spread, one may account only for the frequency dispersion $\omega_0(p)$ and Eq. (4.11) then takes the form:

$$(\Delta\omega)^2 \rho_{\vec{k}'_\perp} = z^2 e^2 N \int \langle (e^{i\vec{k}'_\perp \vec{r}_\perp})_0 (e^{i\vec{k}_\perp \vec{r}_\perp})_0^* \rangle (1 + i\mu) \frac{n^2}{R^2} \frac{d\omega_0}{dp} \rho_{\vec{k}} \frac{d^2 k}{2\pi^2 k_n^2}.$$

Estimating the maximum eigen value of this equation similarly to the previous one, we get:

$$(\Delta\omega)^2 \simeq z^2 e^2 N \int \frac{d^2 k_\perp}{2\pi^2 k_n^2} \langle |(e^{i\vec{k}_\perp \vec{r}_\perp})_0|^2 \rangle \frac{n^2}{R^2} \frac{d\omega_0}{dp} (1 + i\mu). \qquad (4.17)$$

Let us consider only the case of $d\omega/dp > 0$; in the opposite case, there emerges an intrinsic negative mass instability. As can be seen, in the presence of dissipation ($\mu \neq 0$), in contrast to the case of transverse oscillations, an instability appears for any sign of $\mu$, so that one can consider any (large) values of $n$. The maximum increment is realized for $n \simeq R/\sigma$ and has an order of magnitude of

$$(\text{Im } \Delta\omega)_{max} \simeq \sqrt{\frac{\alpha N}{2} \frac{r_e}{\beta^2 R} \frac{m}{M} \frac{\omega_e^2}{\omega_0} \frac{\pi}{3} \eta^2}. \qquad (4.18)$$

For typical conditions in NAP-M, $(\text{Im } \Delta\omega)_{max} \simeq 100$ sec$^{-1}$. This is a sufficiently strong effect, which will appear if the dynamic frequency shift exceeds the spread. An estimate of Re $\Delta\omega$ using Eq. (4.17) gives $\simeq 10^5$ sec$^{-1}$ (with $n \simeq R/\sigma$), while, with a momentum spread of $\Delta p/p \simeq 10^{-5} - 10^{-6}$ (expected and observed values), the spread of this harmonic is $10^5 - 10^4$ sec$^{-1}$, i.e. one may expect appearance of a longitudinal instability of an unbunched beam. There are experimental indications of this effect.

Note that, in contrast to transverse instabilities, longitudinal ones cannot be suppressed by friction because the spatial position of equilibrium makes no difference for the longitudinal degree of freedom of a coasting beam. In Ref. [41], it is also noted that the effect of the spread is reduced in the presence of strong friction, such that the friction decrement exceeds the spread size. Nevertheless, in practice, for the considered short-wave instabilities, the spread of the harmonics $n \simeq R/a$ greatly exceeds the friction decrement, therefore the usual criteria remain valid. Overall, apparently, the theory of coherent longitudinal stability of a beam with cooling is presently not sufficiently developed in specific aspects, so that one could make final conclusions.

## 4.4 Stability of bunched beam

I. An equation for <u>transverse excitations</u> is obtained from Eq. (4.12) similarly to Eq. (4.14):

$$\Delta\omega\rho_{\vec{k}'} = Nz^2 e^2 \int \frac{d^3k\rho_{\vec{k}}}{2\pi^2 k^2}\langle(\vec{m}\frac{\partial}{\partial \vec{I}})(e^{i\vec{k}\vec{r}})_m(e^{i\vec{k}'\vec{r}})^*_m\rangle[1+i\mu^+], \qquad (4.19)$$

where

$$\mu^+ \approx \frac{\omega_e^2}{\pi v^2}\frac{\theta_0^2}{2k^2}\left[\frac{k_\perp^2}{k^2}\vec{m}_\perp(\frac{d\vec{\omega}_b}{d\omega_0}+\vec{v}) - \frac{n^2\theta_0^2}{12}(\vec{m}_\perp \vec{v})\right]. \qquad (4.20)$$

Due to averaging over the particle synchrotron oscillations, Eq. (4.19) is lacking the main term, which was earlier leading to instability (compare to Eqs. (4.13) and (4.15)). For dipole oscillations, the residual effect will, in practice, be negligibly small (significantly smaller than the result of Eq. (4.16)). It may appear that, due to the summation (integration) over $n$ in Eq. (4.19), there may occur addition of the increments from individual azimuthal modes instead of compensation of the opposite-sign terms ($\pm n$) in Eq. (4.20). However, such a summation of the increments (calculation of a trace) in this case is not valid and does not actually take place. In reality, as shown in Ref. [39], for a bunched beam, in a given normal oscillation, there may occur addition of no more than $\Delta n \sim R/l_b$ ($l_b$ is the beam length) azimuthal harmonics. This means that "short-wave" oscillations are close to plane waves averaged over the particle (incoherent) phase oscillations $f_m(\varphi) \sim J_{m_s}(n\varphi)$ with an uncertainty of $\Delta n \sim 1/\bar{\varphi} = R/l_b$, i.e. of the order of the system's inverse size (in units of $R$). The addition of harmonics actually reduces to increase in the density due to bunching that practically gives no significant effect. Besides, interaction in the region of $n\bar{\varphi} \gg 1$ is weakened again due to averaging over the phase oscillations ($\sim 1/n$). Thus, in general, bunching has the tendency to increase the transverse stability, although, as one could see above, there is no real danger even for a coasting beam.

For <u>longitudinal excitations</u> of a bunched beam ($\vec{m}_\perp = 0$, $m_s \neq 0$), the imaginary part of the kernel in Eq. (4.19) vanishes completely ($\mu = 0$). This, of course, is not quite an exact result, since we actually neglected the contribution of the synchrotron harmonics when substituting the frequencies $\vec{m}\vec{\omega} = \vec{m}_\perp\vec{\omega}_\perp + m_s\omega_s$ into the factor $\mu$ in Eq. (4.13) but it means that the decrements will be very small while averaging over the synchrotron oscillations is valid (while the excitation splits into phase harmonics). The criterion is the comparison between the shift in the short-wave excitation frequency and the frequency of synchrotron oscillations, which is actually equivalent to the requirement of providing bunching when space charge is taken into account.

# V. INTRABEAM SCATTERING

In a sufficiently dense and cooled heavy particle beam, the processes of mutual scattering may become significant. The collisional interaction may be especially important at the final stage of cooling due to fast reduction of the internal relaxation time with decrease in temperature at a given current ($\tau_{in} \sim T^{5/2}$). Thus, at the same orbit-<u>averaged</u> densities of the electron and ion beams (the density ratio in the cooling section is then $n_p/n_e = l/\Pi$) and thermal equilibrium ($T = T_{e(eff)}$), the internal relaxation time turns out to be $\sqrt{M/m}$ time <u>shorter</u> than the cooling time. This happens because the proton velocity spread is just $\sqrt{M/m}$ times lower than the effective electron one. The intrabeam scattering may be especially intense if cooling in a magnetized electron flow with $T_{e\parallel} \ll T_{e\perp}$ when very small proton beam sizes are reached.

## 5.1 Collisional beam kinetics without cooling

Inner scattering in an intense beam is significant not only from the point of view of its effect on the cooling process but has an independent importance for the dynamics of a circulating beam. The Touschek effect is well known in the kinetics of ultra-relativistic beams in storage rings: particles get kicked out of the beam as a result of mutual single scattering with longitudinal momentum exchange. Due to relativism, this leads to a strong energy exchange in the lab frame (whose limit reaches the full energy). At moderately relativistic and non-relativistic energies, particles do not leave the beam but multiple scattering may lead to increase in the energy spread and to excitation of betatron oscillations due to energy kicks, i.e., heating of the beam in general. Then there is a question of whether this process stops at a certain level or heating continues indefinitely (but, of course, with a slowing rate due to reduction of the density). The existence of this type of criterion can be seen when considering a collision of two particles with zero relative longitudinal velocity and initial radial velocities equal in size scattering at 90°, so that the transverse motion completely converts into the longitudinal one. However, a jump-like change in energy again leads to excitation of radial oscillations. Let us find the resulting change in the sum of the squares of the transverse oscillation amplitudes of the two particles. In the limit of an azimuthally uniform beam path,

$$a_x^2 \sim p_x^2 + \gamma^2 \left( M\omega_x x - \frac{p_\parallel}{\nu_x} \right)^2 ; \qquad (5.1)$$

using the fact that, in the scattering, $|\Delta p_\parallel| = |p_r|$ and $\Delta x = 0$, we get:

$$\Delta(a_{x1}^2 + a_{x2}^2) \sim -2p_x^2 + 2\frac{\gamma^2}{\nu_x^2}p_x^2 = -2\gamma^2 p_x^2 \left( \frac{1}{\gamma^2} - \frac{1}{\nu_x^2} \right) = -2\gamma^2 p_x^2 \frac{p}{\omega_0} \frac{d\omega_0}{dp'}.$$

As one can see, when $\gamma > \nu_x$ (or $\alpha \equiv (p/\omega_0)(d\omega_0/dp) < 0$), scattering leads to increase of the oscillation amplitudes while, in the opposite case, the energy of the transverse motion is reduced. Note that this criterion matches the criterion of longitudinal stability of a bunched beam (the "negative mass'' effect), which is not, of course, a random coincidence, since, in both cases, the same dynamic mechanism is in action – dependence of the closed orbit on the energy.

The intrabeam scattering was considered by a number of authors, most comprehensively, as we know, in Ref. [46]; a series of estimates of model nature is included in Ref. [47]. We base our consideration on a kinetic equation with the Landau collision integral:

$$\frac{\partial F}{\partial t} + \{\mathcal{H}; F\} = \frac{2\pi z^4 e^4 L}{\gamma} \frac{\partial}{\partial p_\alpha} \int d^3 p' \frac{u^2 \delta_{\alpha\beta} - u_\alpha u_\beta}{u^3} \left( \frac{\partial}{\partial p_\beta} - \frac{\partial}{\partial p'_\beta} \right) F F', \tag{5.2}$$

where the momenta $\vec{p}$ and the velocities $\vec{u} = (\vec{p} - \vec{p}')/M$ are related to the co-moving frame. The most interesting question appears to be that about an equilibrium distribution and the existence itself of such a distribution. We know that a Maxwell distribution makes the collision integral identically zero (substituting $F \sim \exp(-\beta p^2)$ or $\exp(-\beta \mathcal{H})$, we get the convolution $(u^2 \delta_{\alpha\beta} - u_\alpha u_\beta) u_\beta \equiv 0$). Note that, in magnetic field, a Maxwell distribution makes the Poisson bracket zero as well for any dependence of the field on the coordinates, since the Hamiltonian coincides with the kinetic energy. However, we cannot be satisfied with a distribution of this type, since the only distributions that make practical sense are those localized in the transverse coordinates near a certain closed equilibrium orbit $\vec{r}_s(\theta)$ and accordingly having a small momentum spread with respect to an equilibrium (average) value $\vec{p}_s(\theta)$. In case of an azimuthally-symmetric magnetic field, along with the Hamiltonian, there exists an exact integral of motion – the generalized momentum

$$\mathcal{P}_\theta = r(p_\theta + eA_\theta), \tag{5.3}$$

whose derivative with respect to the usual momentum does not depend on the velocities:

$$\frac{\partial \mathcal{P}}{\partial p_\theta} = r;$$

therefore, one can write a distribution of a more general form than Maxwellian, also making the right-hand side of Eq. (5.2) identically zero:

$$F \sim e^{-\varkappa(\mathcal{H} - \omega_s \mathcal{P}_\theta)}. \tag{5.4}$$

Obviously, the combination $\mathcal{H} - \omega_s \mathcal{P}_\theta$ represents a Hamiltonian with respect to a rotating frame connected to a certain closed orbit $\vec{r}_s(\theta), \vec{p}_s(\theta)$ corresponding to the frequency $\dot{\theta} = \omega_s = Const$. For the distribution in Eq. (5.4) to be able to describe the beam state in a storage ring, it must possess the property of normalizability, i.e. the Hamiltonian $\mathcal{H} - \omega_s \mathcal{P}_\theta$ $= \mathcal{H}(p) - \omega_s r(p_\theta + eA_\theta(r,z))$ must be a single-sign function (a positive quadratic form near the equilibrium orbit).

For small deviations from the "equilibrium" orbit, the Hamiltonian $\mathcal{H} - \omega_s \mathcal{P}_\theta$ transforms to the form (see Appendix I)

$$\mathcal{H}_c = \mathcal{H} - \omega_s \mathcal{P}_\theta \approx \frac{1}{\gamma}\left[\varepsilon_x + \varepsilon_z + \frac{1}{2}Mv_\parallel^2\left(1 - \frac{\gamma^2}{v_x^2}\right)\right], \tag{5.5}$$

where

$$\begin{aligned}\varepsilon_x &= \frac{1}{2}Mv_x^2 + \frac{1}{2}Mv_x^2\omega_0^2\left(x - \frac{1}{v_x^2}v_\parallel/\omega_0\right)^2, \\ \varepsilon_z &= \frac{1}{2}Mv_z^2 + \frac{1}{2}Mv_z^2\omega_0^2 z^2.\end{aligned} \tag{5.6}$$

$\varepsilon_x$ and $\varepsilon_z$ are the energies of "betatron oscillators". As one can see, in case of uniform focusing ($v_x^2 + v_z^2 = 1$), the distribution in Eq. (V.5.4) is not normalizable and this means that the intrabeam scattering leads to unlimited growth of the internal beam energy.

A real situation of non-uniform (strong) focusing is more complicated and, unfortunately, even a stationary solution cannot be found in an analytic form. A simplified qualitative consideration can be made assuming an approximation of constant $\beta_x$, $\beta_z$ and $\psi$ functions. Such an approximation is similar to uniform focusing but with independent betatron tunes $v_x$ and $v_z$ (then $\psi = v_x^{-2}$). The distribution $\sim \exp(-\varkappa \mathcal{H}_c)$ then makes the collision integral zero and is normalizable if $\gamma < v_x$. The value of $\gamma = v_x$ approximately corresponds the storage ring's transition energy, when the dispersion factor changes sign

$$\alpha = \frac{p}{\omega_0}\frac{d\omega_0}{dp} = \frac{1}{\gamma^2} - \overline{k\psi} \approx \frac{1}{\gamma^2} - \frac{1}{v_x^2}.$$

Thus, it seems plausible that the criterion for the existence of an equilibrium state stable in regard to intrabeam scattering is the comparison of the particle energy in the storage ring with its transition value [46]:

$$\gamma < \gamma_{cr} \simeq \gamma_{tr}, \tag{5.7}$$

in the opposite case, scattering leads to heating of the beam.

In the stability region ($\gamma < \gamma_{cr}$), as one can see from the form of the Hamiltonian $\mathcal{H}_c$, the equilibrium velocity distribution in the co-moving frame is approximately isotropic as long as the energy is not too close to the critical one; in the latter case, the longitudinal spread is relatively large. Concerning the momentum spread in the laboratory frame, due to the relativistic transformation, the longitudinal spread is increased by a factor of $\gamma$ while the transverse one is conserved. The temperature (or the parameter $\varkappa$) of the equilibrium state in the smooth approximation can be determined as a function of the initial state using conservation of the average value of the Hamiltonian $\mathcal{H}_c$:

$$\langle \mathcal{H}_c \rangle = \int \mathcal{H}_c F d\Gamma = Const.$$

Let us multiply Eq. (5.2) by $\mathcal{H}$ and integrate it over the phase space, then, on the right-hand side, we get a double integral over the velocities

$$\frac{d}{dt}\langle \mathcal{H}_c \rangle \sim -\int d^3v d^3v' \frac{\partial \mathcal{H}_c}{\partial p_\alpha} \frac{u^2 \delta_{\alpha\beta} - u_\alpha u_\beta}{u^3}\left(\frac{\partial}{\partial p_\beta} - \frac{\partial}{\partial p'_\beta}\right) FF' d^3r \; ;$$

symmetrizing the integrand in $\vec{p}$ and $\vec{p}'$ and noting that $\partial \mathcal{H}_c/\partial \vec{p}_c - \partial \mathcal{H}'_c/\partial \vec{p}'_c = \vec{u}$, we again arrive at the convolution $u_\alpha T_{\alpha\beta} \equiv 0$. Thus, the temperature can be determined in the following way (we define $\langle \Delta p_\parallel \rangle = 0$):

$$\langle \mathcal{H}_c \rangle = \int \mathcal{H}_c F(t=0) d\Gamma = \langle \varepsilon_x \rangle_0 + \langle \varepsilon_z \rangle_0 + \frac{\langle p_\parallel^2 \rangle_0}{2M}\left(1 - \frac{\gamma^2}{v_x^2}\right) ;$$

integration over the normalized equilibrium distribution gives

$$\langle \mathcal{H}_c \rangle = \frac{5}{2\varkappa} \equiv \frac{5}{2} T_{eq}$$

and thus

$$T_{eq} = \frac{2}{5}\langle \frac{p_\perp^2}{M} + (1 - \frac{\gamma^2}{v^2})\frac{p_\parallel^2}{2M}\rangle_{t=0} \quad (\gamma < \gamma_{cr} \simeq v_x) \tag{5.8}$$

(in RF mode, $2/5 \to 1/3$).

Let us now make some estimates of the relaxation process (heating when $\gamma > \gamma_{cr}$). Multiplying Eq. (5.2) by $\varepsilon_x$, $\varepsilon_z$ and $\varepsilon_\parallel = p_\parallel^2/(2M)$ and integrating it, we get:

$$\langle \dot\varepsilon_z \rangle = \frac{2\pi (ze)^4 L}{\gamma M} \int d^3p\, d^3p'\, d^3r FF' \frac{u^2 - 3u_z^2}{u^3}, \tag{5.9}$$

$$\langle \dot\varepsilon_\parallel \rangle = \frac{2\pi (ze)^4 L}{\gamma M} \int d^3p\, d^3p'\, d^3r FF' \frac{u^2 - 3u_\parallel^2}{u^3}, \tag{5.10}$$

$$\langle \dot\varepsilon_x \rangle = \frac{2\pi (ze)^4 L}{\gamma M} \int d^3p\, d^3p'\, d^3r FF' \left[\frac{u^2 - 3u_x^2}{u^3} + \frac{\gamma^2}{v^2}\frac{u^2 - 3u_\parallel^2}{u^3}\right]. \tag{5.11}$$

These expressions, in particular, give

$$\frac{d}{dt}\langle \mathcal{H}_c \rangle = 0,$$

while

$$\langle\dot{\varepsilon}\rangle \equiv \langle\dot{\varepsilon}_z\rangle + \langle\dot{\varepsilon}_x\rangle + \langle\dot{\varepsilon}_\parallel\rangle = \frac{2\pi(ze)^4 L}{\gamma M^2} \int d^3p\, d^3p'\, d^3r\, FF'\, \frac{\gamma^2}{v^2}\frac{u^2 - 3u_\parallel^2}{u^3}.$$

Besides, for isotropic distributions of the relative velocities $\vec{u}$, Eqs. (5.9) – (5.11) give:

$$\langle\dot{\varepsilon}_z\rangle = \langle\dot{\varepsilon}_x\rangle = \langle\dot{\varepsilon}_\parallel\rangle = 0\,;$$

the distribution $\sim \exp(-\varkappa\mathcal{H}_c)$ with $\gamma < \gamma_{cr}$ belongs to those, since the sum of the Hamiltonians of two particles reduces to the form:

$$\mathcal{H}_c(\vec{v},\vec{r}_\perp) + \mathcal{H}_c(\vec{v}',\vec{r}_\perp) = \frac{M}{4}u^2 + 2\mathcal{H}_c(\frac{\vec{v}+\vec{v}'}{2},\vec{r}_\perp)\,.$$

It is instructive also to compare the values of Eqs. (5.9) – (5.11) and

$$\langle\dot{\varepsilon}_\perp\rangle \equiv \langle\dot{\varepsilon}_x\rangle + \langle\dot{\varepsilon}_z\rangle = \frac{2\pi(ze)^4 L}{\gamma M} \int d^3p\, d^3p'\, d^3r\, FF'\, \left(\frac{\gamma^2}{v^2} - 1\right)\frac{u^2 - 3u_\parallel^2}{u^3}\,;$$

as one can see, when $\gamma < \gamma_{cr}$, simultaneous increase of the longitudinal and transverse temperatures is not possible while, when $\gamma > \gamma_{cr}$, the rates of their change have the same sign.

Let us estimate the change in time of the energy spread and beam size at energies significantly exceeding the transition one. As one can see from Eqs. (5.9) – (5.11), the most rapidly changing beam temperature is the radial one, therefore, in these formulae, one can assume $u_\parallel, u_z \ll u_x$, then

$$\langle\dot{\varepsilon}_z\rangle \approx \langle\dot{\varepsilon}_\parallel\rangle \approx g \int d^3p\, d^3p'\, d^3r\, \frac{FF'}{|u_x|} \simeq \frac{v_x^2}{\gamma^2}\langle\dot{\varepsilon}_x\rangle, \qquad (5.12)$$

where

$$g = \frac{2\pi(ze)^4 L}{\gamma M}.$$

At the order-of-magnitude level, one can assume

$$\int d^3p\, d^3p'\, d^3r\, \frac{FF'}{|u_x|} \simeq \frac{n}{\gamma}\sqrt{\frac{M}{T_x}}\ln\sqrt{\frac{T_x}{T_z}}, \qquad (5.13)$$

where $n$ is the beam density in the laboratory frame. Writing the density $n$ in terms of the beam sizes $\sigma_x$ and $\sigma_z$

$$n = \frac{N}{2\pi R \sigma_x \sigma_z}$$

and considering the relations

$$\sigma_x^2 = \frac{T_x}{M\gamma^2\omega_0^2 v^2} + \frac{T_\parallel}{M\omega_0^2 v^4}, \qquad \sigma_z = \frac{T_z}{M\gamma^2\omega_0^2 v^2}, \qquad (5.14)$$

we arrive at three equations:

$$\frac{d}{dt}\sigma_x^2 = \frac{2g}{M\omega_0^2 v^4} \cdot \frac{N}{\gamma 2\pi R}\sqrt{\frac{M}{T_x}}\frac{1}{\sigma_x \sigma_z} = \frac{2\gamma^2}{v^2}\frac{d}{dt}\sigma_z^2 = \frac{2}{M\omega_0^2 v^2 \gamma^2}\frac{dT_x}{dt}. \qquad (5.15)$$

Asymptotically in time, these equations give

$$T_x \approx \frac{1}{2}M\omega_0^2 v^2 \gamma^2 \sigma_x^2, \quad \sigma_z^2 \approx \frac{v^2}{2\gamma^2}\sigma_x^2, \qquad (5.16)$$

and, correspondingly,

$$\frac{d}{dt}\sigma_x^2 \approx \frac{4(ze)^4 LN}{\gamma^2 M^2 v^6 R \omega_0^3 \sigma_x^3}, \qquad (5.17)$$

which give

$$\sigma_x \simeq \left[\frac{10 z^4 r_p^2 L N c R^2}{\gamma^2 v^6 \beta^3}t\right]^{1/5}, \qquad (5.18)$$

$$\sigma_z \simeq \frac{v}{\sqrt{2}\gamma}\sigma_x, \qquad (5.19)$$

$$\frac{\Delta p}{p} = \frac{v^2}{\sqrt{2}}\frac{\sigma_x}{R}. \qquad (5.20)$$

The formulae suppose that the changes in the sizes and energy spread exceed the initial values. Note that the contributions of the energy and angular spreads to the radial size are equal. Also, the longitudinal and vertical temperatures are $(\gamma/v)^2$ times smaller than the radial one. As one can see, the beam expansion rate goes down quite quickly with time, which has to do with a fall off of the Rutherford cross section with increase in the relative velocities and, additionally, with decrease of the collision probability due to the density reduction.

To avoid misunderstanding, let us make one comment. When estimating the expansion process $(\gamma > \gamma_{cr})$, we used the assumption that the transverse (radial) relative velocities dominate over the longitudinal ones. It may seem that an opposite assumption is possible and then, instead of beam expansion, we would get self-cooling, which is absurd. In reality, what happens is the following. Equations (5.9) – (5.11) are obtained from the Landau collision integral, which considers particle interaction and collisions as occurring locally, i.e. when particle "macroscopic" coordinates coincide. This circumstance is reflected in the fact that the product of the distribution functions $F$ and $F'$ everywhere enters the general formulae at one spatial point, in other words, the kinetics is determined by a local velocity spread. On the other hand, in storage rings or accelerators, spatial "mixing" of particles with different energies (different longitudinal velocities) is related only to transverse oscillations, in which the particles cross the closed orbits of "others". So, if the transverse beam temperatures are equal to zero, then, from a "thermodynamic" point of view, such a beam is already absolutely cold, since the relative longitudinal velocities $u_\parallel$ at each point are equal to zero. Accordingly, when exciting radial betatron oscillations, the difference of the longitudinal velocities of colliding particles cannot exceed the value determined by the relation

$$\langle|\Delta x_b|\rangle = \frac{\langle|\Delta u_x|\rangle}{\gamma v \omega_0} \simeq \psi \frac{\langle|\Delta p|\rangle}{p} = \frac{\langle|\Delta u_\parallel|\rangle}{\omega_0 v^2},$$

i.e.

$$\langle|\Delta u_\parallel|\rangle \leq \frac{v}{\gamma} \langle|\Delta u_x|\rangle,$$

and thus, with $\gamma > v$, the assumption $u_\parallel > u_\perp$ would be contradictory.

Of course, due to the long-range action of Coulomb forces, strictly speaking, one cannot completely ignore the "non-local" nature of particle interaction. It is possible to generalize the collision integral including particle interaction also at the distances of the order of the (transverse) system size together with the finiteness of particle motion. However, the structure of the collision integral then changes in such a way, that there are no consequences contradicting to those following from the Landau integral.

Let us also try to improve the estimates by taking into account azimuthal non-uniformity of the magnetic field in real storage rings. For the average rates of change of the variables $\varepsilon_z$, $\varepsilon_x$, and $\varepsilon_\parallel$, we then get the expressions ($\varepsilon_z = v_z I_z$ and $\varepsilon_x = v_x I_x$):

$$\begin{aligned}
\langle \dot{\varepsilon}_\parallel \rangle &= g \int d^3p\, d^3p'\, d^3r\, FF' \frac{u^2 - 3u_\parallel^2}{u^3}, \\
\langle \dot{\varepsilon}_z \rangle &= g v_z \int d^3p\, d^3p'\, d^3r\, FF' \beta_z \frac{u^2 - 3u_z^2}{u^3}, \\
\langle \dot{\varepsilon}_x \rangle &= g v_x \int d^3p\, d^3p'\, d^3r\, FF' \Bigg[ \beta_x \frac{u^2 - 3u_x^2}{u^3} \\
&\quad + \gamma^2 \left( \frac{\psi^2}{\beta_x} + \beta_x \chi^2 \right) \frac{u^2 - 3u_\parallel^2}{u^3} + 6\gamma \beta_x \chi \frac{u_x u_\parallel}{u^3} \Bigg],
\end{aligned}$$

(5.21)

where

$$\chi(\theta) = \frac{d\psi}{d\theta} - \frac{\psi}{2\beta_x} \frac{d\beta_x}{d\theta}, \quad \langle \frac{1}{\beta_z} \rangle = v_z, \quad \langle \frac{1}{\beta_x} \rangle = v_x.$$

As in the uniform approximation, it is natural to assume that the radial component of relative velocities should dominate in an expanding beam; one can then assume

$$|u_x|_\theta = \sqrt{\frac{\varepsilon_x}{M\nu_x\beta_x}}, \quad \sigma_z^2(\theta) = \frac{\beta_z \varepsilon_z}{\nu_z \gamma^2 M \omega_0^2},$$

$$\sigma_x^2(\theta) = \left(\frac{\beta_x \varepsilon_x}{\gamma^2 \nu_x} + \psi^2 \varepsilon_\|\right)/(M\omega_0^2), \quad u_x^2(\theta) \simeq \frac{\varepsilon_x}{M\nu_x\beta_x} + \gamma^2 \chi^2 \frac{\varepsilon_\|}{M}.$$

Under these approximations, we get the equations:

$$\dot\varepsilon_x \approx \frac{g\nu_x N}{2\pi R\gamma} \cdot \gamma^2 \left\langle \left(\frac{\psi^2}{\beta_x} + \beta_x \chi^2\right) \frac{1}{\sigma_x(\theta)\sigma_z(\theta)u_x(\theta)} \right\rangle,$$

$$\dot\varepsilon_\| \approx \frac{gN}{2\pi R\gamma} \left\langle \frac{1}{\sigma_x(\theta)\sigma_z(\theta)u_x(\theta)} \right\rangle, \qquad (5.22)$$

$$\dot\varepsilon_z \approx \frac{gN}{2\pi R\gamma} \nu_z \left\langle \frac{\beta_z}{\sigma_x \sigma_z u_x} \right\rangle.$$

As before, asymptotically in time, the relation between $\varepsilon_x$, $\varepsilon_\|$, and $\varepsilon_z$ is constant and we get the same beam expansion law of $\varepsilon \sim t^{1/5}$ as in the smooth approximation (the temperature relations can be found from the same equations as before). To make the solution consistent with the initial assumption of $u_\|, u_z < u_x$, it must satisfy the requirement $v_x^2 > v_\|^2 + v_z^2$, which is the criterion of heating. As expected, this criterion does not significantly differ from that obtained in the smooth approximation if, of course, focusing does not have prominent singularities (usually, $\psi \simeq 1/\nu$, $\beta \simeq 1/\nu$, and $\chi \simeq 1/\nu$). Apparently, studying the asymptotic solution of the kinetic equation (or of the moments' equations) more carefully, one can analytically formulate the exact expansion criterions; however, we will skip this.

The equilibrium distribution existing at energies below the threshold does not have the form of Eq. (5.4) and cannot even be represented in the more general form of the Boltzmann distribution with different temperatures $T_x$, $T_z$, and $T_\|$. Such an attempt leads to three ordinary equations for the three temperatures not having a solution in a general case. From the general point of view, the failure to satisfy Eq. (5.2) by a Boltzmann-type solution in case of azimuthal non-uniformity is related to non-stationarity of external conditions with respect to the beam's inner degrees of freedom. In principle, an equilibrium distribution can be found as a stationary solution of Eq. (V.5.2); one should transform it to the variables $\varepsilon_x$, $\varepsilon_z$, and $p_\|$ and average the collision integral over the phases $\psi_x$ and $\psi_z$ and the azimuthal angle. It should also not significantly differ from the solution in the smooth approximation.

## 5.2 Cooling at energies below transition

At energies below the threshold value $\gamma_{cr}$, inner scattering tends to only thermalize the beam to the state $\sim \exp(-\varkappa \mathcal{H}_c)$ where the parameter $\varkappa$ is determined by an "external" thermostat, for example, a cooling electron beam. At a sufficiently high intensity of the heavy particle beam and as the temperature drops and the density increases, the thermalization rate quickly grows and, from the moment, when the decrement of thermalization (mixing of particles in the beam) becomes greater than the cooling decrements, the distribution

$$F \sim e^{-\mathcal{H}_c/T}$$

is established (we limit ourselves to the uniform focusing approximation); after that, the process reduces to change (reduction) of the temperature $T$ in time to a certain equilibrium value. Let us write a kinetic equation for the cooled beam including internal relaxation:

$$\gamma \frac{\partial \mathcal{F}}{\partial t} + (St)_{intr}\mathcal{F} = -\frac{\partial}{\partial p_\alpha}\left(F_\alpha \mathcal{F} - \frac{1}{2}d_{\alpha\beta}\frac{\partial \mathcal{F}}{\partial p_\beta}\right).$$

Let us multiply this equation by $\mathcal{H}_c$ and integrate it over the momenta and coordinates. Since the average value of the Hamiltonian is not changed by internal scattering, we get

$$\gamma \frac{d}{dt}\langle \mathcal{H}_c \rangle = \int d\Gamma \frac{\partial \mathcal{H}_c}{\partial p_\alpha}\left(F_\alpha \mathcal{F} - \frac{1}{2}d_{\alpha\beta}\frac{\partial \mathcal{F}}{\partial p_\beta}\right). \quad (5.23)$$

If, at a sufficient beam intensity, thermalization already happened the distribution $\mathcal{F}$ on the right-hand side can be assumed to be proportional to $\sim \exp(-\mathcal{H}_c/T)$; note that the average value $\langle \mathcal{H}_c \rangle$ is expressed through the temperature according to Eq. (5.8). Let us further transform the right-hand side. The first term is:

$$\int d\Gamma \frac{\partial \mathcal{H}_c}{\partial p_\alpha} F_\alpha e^{-\frac{\mathcal{H}_c}{T}} = -T \int d\Gamma F_\alpha \frac{\partial}{\partial p_\alpha} e^{-\frac{\mathcal{H}_c}{T}} = T \int d\Gamma \frac{\partial F_\alpha}{\partial p_\alpha} e^{-\frac{\mathcal{H}_c}{T}},$$

the second term is:

$$\int d\Gamma \frac{\partial \mathcal{H}_c}{\partial p_\alpha} d_{\alpha\beta} \frac{\partial \mathcal{F}}{\partial p_\beta} = -\int \mathcal{F}d\Gamma \frac{\partial}{\partial p_\beta}\frac{\partial \mathcal{H}_c}{\partial p_\alpha}d_{\alpha\beta} = -\langle \frac{\partial^2 \mathcal{H}_c}{\partial p_\alpha \partial p_\beta}d_{\alpha\beta} + \frac{\partial \mathcal{H}_c}{\partial p_\alpha}\cdot \frac{\partial d_{\alpha\beta}}{\partial p_\beta}\rangle.$$

The term with the divergence of $d_{\alpha\beta}$ can be omitted, since its contribution reduces to a correction of the friction force by a value of $\simeq (m/M)F$ (replacement of the electron mass with the normalized one). Taking into account that the part of the Hamiltonian quadratic in momenta is simply the kinetic energy in the co-moving frame, we get

$$\int d\Gamma \frac{\partial \mathcal{H}_c}{\partial p_\alpha} d_{\alpha\beta} \frac{\partial \mathcal{F}}{\partial p_\beta} = -\frac{1}{M}\langle d_{\alpha\alpha}\rangle.$$

Finally, Eq. (5.23) gives the following equation determining temperature change with time:

$$\gamma \frac{dT}{dt} = \frac{2}{q}[T \langle \frac{\partial F_\alpha}{\partial p_\alpha}\rangle + \frac{1}{2M}\langle d_{\alpha\alpha}\rangle]. \quad (5.24)$$

Recall that the divergence of the friction force with respect to momentum (with a minus sign) is, by definition, the damping decrement (in the co-moving frame) of the beam's effective phase-space volume (see Appendix 2) equal to a sum of the damping decrements in the normal degrees of freedom while the second term equals the average diffusional growth of the particle energy in the co-moving frame. At the stage of the cooling process, when the heavy particle velocity spread becomes small compared to the electron one (the true one, i.e. the thermal one or the effective one due to a spatial spread) or, in a general case, compared to the rms relative particle velocity of the two beams, the decrement and diffusion do not depend on the temperature of the heavy particles. Making the right-hand side equal to zero in this region, we get the equilibrium temperature:

$$T_s = \frac{1}{2M}\left[\frac{1}{\Lambda}\frac{d}{dt}\langle(\Delta\vec{p})^2\rangle\right]_c^{T=0}, \quad \Lambda = \lambda_1 + \lambda_2 + \lambda_3. \tag{5.25}$$

Thus, with intense inner scattering, the equilibrium temperature of the cooled beam is determined by the ratio of the momentum diffusion rate (on electrons) in the co-moving frame to the sum of the friction decrements.

Let us now explicitly specify the conditions when thermalization can be a faster process than electron cooling. The corresponding relaxation decrements relate as:

$$\lambda_{intr} : \lambda_{cool} \simeq \frac{z^2 L_i n}{\frac{M}{2}(2\frac{T}{M})^{3/2}} : \frac{n_e \eta L_e}{m(\frac{T_e}{m} + \frac{T}{M})^{3/2}}. \tag{5.26}$$

The density of the heavy particle beam should be expressed through the temperature:

$$n = \frac{N}{2\pi R \cdot 2\pi \sigma_x \sigma_z} \simeq \frac{N}{(2\pi)^2 R}\frac{(\gamma+1)W_i}{\beta_x \beta_z T},$$

where $W_i = (\gamma - 1)Mc^2$ is the kinetic energy of the heavy particles in the storage ring and $\beta_{x,z}$ are the beta functions of the storage ring ($\beta_{x,z} \simeq R/\nu_{x,z}$). Since the minimum spread of heavy particle velocities corresponds to the equality of the temperatures $T = T_e$, the relation in Eq. (5.26) then gives that inner scattering (when $\gamma < \gamma_{cr}$) can interfere with the relaxation process when

$$\zeta \equiv N\frac{z^2 L_i(\gamma+1)W_i}{2\pi^2 \beta_x \beta_z} \Big/ \Big(\sqrt{\frac{2m}{M}} L_e n_e L T_e\Big) > 1, \tag{5.27}$$

where $l$ is the length of the cooling section.

Recall that the effective temperature $T_e$ of the electron beam can be much <u>lower</u> than the true electron transverse temperature or the cathode temperature.

If the condition of Eq. (5.27) takes place then the inner relaxation comes into play in the cooling process starting with the temperature

$$T = \zeta^{2/5} T_e, \quad 1 < \zeta < \left(\frac{M}{m}\right)^{5/2} \tag{5.28}$$

and

$$T = \left(\frac{m}{M}\right)^{3/2} \zeta T_e, \quad \zeta > \left(\frac{M}{m}\right)^{5/2}. \tag{5.29}$$

A practically plausible case is that of Eq. (5.28).

The provided estimates pertain to situations when cooling in the electron beam is approximately isotropic. When the effective electron velocity spread is drastically non-isotropic, the decrements in the different degrees of freedom can significantly differ. Then the decrement of internal relaxation (weakly depending on non-isotropicity of the particle velocity distribution) should be compared not only with the maximum but also with the minimum of the three electron cooling decrements, since collisional mixing can speed up cooling of weakly-damping degrees of freedom. The case, when the thermalization decrement exceeds the sum of the electron cooling decrements, was described above. One can estimate in a similar way the situation, when the thermalization decrement is small compared to the maximum of the friction decrements but significantly exceeds the small decrements (one or two). In this case, the degree of freedom with the strong friction quickly damps practically to the equilibrium temperature, while the other degrees of freedom will damp due to the energy exchange with the first one through particle collisions during the time of internal relaxation determined by the initial velocity spread in these degrees of freedom. During the cooling process, the intensity of exchange quickly grows and, in the final stage, the thermalization decrement may exceed the maximum friction decrement, so that an isotropic equilibrium distribution is established according to Eq. (5.25). If the thermalization decrement for such a distribution happens to be less than the sum of the cooling decrements, then the degrees of freedom with weak friction will have higher temperature. Making estimates for possible situations does not present any difficulties.

Collisions of heavy particles may also play a useful role in a situation, when new particles are added in small portions to an already accumulated (damped) beam. Due to collisions with the particles of the stored beam, their cooling may happen to be a faster process than cooling directly in an electron beam.

Situations are also possible when, due to some source of diffusion, a large spread is maintained in one or two degrees of freedom. Inner scattering will then be heating the other degrees of freedom as well. Such a case is considered in Ref. [52]. In the NAP-M machine, the transverse proton temperature obtained in measurements greatly exceeds the longitudinal one (which is possible to interpret as a space charge effect) and, due to inner scattering, the energy spread may increase as well. It should be noted that, with the tune $\nu \simeq 1.2$, the parameter $\alpha$ in the NAP-M conditions is small ($\alpha \simeq 0.1$), which enhances the negative effect of inner scattering.

## 5.3 Relaxation at energies above transition

When $\gamma > \gamma_{cr}$, the cooling process will stop at temperatures, when the friction power becomes equal to the beam diffusion rate due to internal scattering. Let us estimate the established sizes based on the equations of Section 5.1:

$$\frac{d}{dt}\sigma_x^2 = -\lambda\sigma_x^2 + 2\frac{A}{\sigma_x\sigma_z}\sqrt{\frac{M}{T_x}}, \tag{5.30}$$

$$\frac{d}{dt}\sigma_z^2 = -\lambda\sigma_z^2 + \frac{A\sqrt{M/T_x}}{\sigma_x\sigma_z}\frac{v_x^2}{\gamma^2}, \tag{5.31}$$

$$\frac{dT_x}{dt} = -\lambda T_x + \frac{A}{\sigma_x\sigma_z}\sqrt{\frac{M}{T_x}} \cdot \frac{v_x^2 M \omega_0^2 \gamma^2}{2}, \tag{5.32}$$

where

$$A = \frac{(ze)^4 L_i N}{\gamma^2 M^2 \cdot 2\pi\omega_0^2 \nu^4 R}, \quad \lambda = \frac{4\pi z^2 e^4 L_e n \eta}{M\gamma^2 m(\frac{T_e}{M} + v^2)^{3/2}}. \tag{5.33}$$

For convenience in estimating the temperature $T_\parallel$, let us complement this system with an equation (no having an independent content):

$$\frac{dT_\parallel}{dt} = -\lambda T_\parallel + \frac{A}{\sigma_x\sigma_z}\sqrt{\frac{M}{T_x}}\frac{M\omega_0^2 v^4}{2}. \tag{5.34}$$

We limit ourselves to the case of isotropic friction and approximation of azimuthal uniformity. Since the radial velocities will be the largest, one can set $v^2 \simeq v_x^2 = T_x/M$.

Equations (5.30) – (5.34) give that the relations of the equilibrium values of the parameters $\sigma_x^2$, $\sigma_z^2$, and $T_x$ will be the same as the relations of these parameters in the asymptotic heating mode without (electron) cooling. Suppose that an equilibrium regime is realized in the region of $v_x^2 < v_e^2 = T_e/m$. We then get the following equilibrium values:

$$T_x = \left(\zeta\frac{\gamma^3}{\nu_x^3}\right)^{2/5} T_e,$$

$$T_\parallel = T_z = \left(\frac{v_x}{\gamma}\right)^2 T_x = \left(\zeta\frac{v^2}{\gamma^2}\right)^{2/5} T_e, \quad \left(\frac{\gamma}{\nu}\right)^2 < \left(\zeta\frac{\gamma^3}{\nu^3}\right)^{2/5} < \frac{M}{m}, \tag{5.35}$$

where $\zeta$ is the parameter in Eq. (5.27). The conditions of solution applicability, on one (right) hand (side), ensure validity of the assumption $v_x^2 < T_e/m$ and, on the other hand, limit the equilibrium temperatures ($T_\parallel$ and $T_z$) to the minimum possible value $T_e$ (in the equations, we neglected diffusion on electrons, which prevents the temperatures from going below the value $T_e$). However, there is a possible intermediate situation when

$$\zeta < \left(\frac{\gamma}{\nu}\right)^2$$

but still $T_x > T_e$; one should then set

$$T_\parallel = T_z = T_e$$

(inner scattering is not significant for these degrees of freedom). Since the longitudinal temperature is limited here by diffusion on electrons, then, as it follows from Eq. (5.14), the size in $x$ is determined by the longitudinal temperature:

$$\sigma_x^2 = \frac{T_e}{M\omega_0^2 \nu^4}.$$

The radial temperature (although it no longer presents an independent interest) can be found from Eq. (5.32) (the other equations are not suitable for this, since diffusion on electrons plays a significant role in them):

$$T_x = \frac{1}{2}\left(\zeta\frac{\gamma}{\nu}\right)^{2/3} T_e,$$
$$T_z = T_\parallel = T_e;$$

the region of these values is $\nu/\gamma < \zeta < (\gamma/\nu)^2$. Thus, inner scattering is of no significance at all if

$$\zeta < \frac{\nu}{\gamma} \quad (\gamma > \nu(!)).$$

Similarly, one can determine the dependence of the equilibrium temperatures on the beam parameters (not depending on the cooling process) for the region $v_x^2 > T_e/m$, which we will not go into.

The estimates can also be generalized for the cases when the size of the cooled beam exceeds that of the electron beam.

In this regard, let us note one more feature of the self-heating effect. Taking into account that the transverse size of an electron beam is finite, a stable equilibrium state of a heavy particle beam with $v_x^2 > T_e/m$ can be realized only in case when the beam size is smaller than that of the electron beam (at least, in one dimension). Indeed, it is easy to estimate that, in the opposite case, the friction decrements $\lambda$ are inversely proportional to the fifth power of the size and, as it follows from Eqs. (5.30) – (5.32), the beam in such a state will be either constantly expanding or damping inside the electron beam depending on the relation of the beam currents.

## CONCLUSIONS

The main results of this work are the following:

1. We investigated the dependence of the relaxation process and equilibrium state of a heavy particle beam on the electron velocity distribution.

We discovered the effect of "monochromatic" dissipative instability, i.e. oscillation growth due to change in sign of the friction characteristic, and formulated a criterion of optimal cooling regime: the difference of the average velocity of the electron and cooled beams should not exceed the electron velocity spread.

2. We studied the effect of transverse non-uniformity of the electron beam parameters on the cooling process occurring due to coupling of the transverse and longitudinal degrees of freedom of particle motion in storage rings. We obtained critical values of the gradients of average velocity, density, and temperature. Exceeding those leads to change in the sign of the partial friction powers.

3. We obtained decrements of the beam relaxation in the region of small spread ($v < \Delta_e$) for arbitrary coupling of the heavy particle degrees of freedom. We established that the sum of the decrements is positive and is independent of the gradients of electron distribution in the phase space.

4. We studied the nature of relaxation of beams with a large initial velocity spread and proposed techniques for accelerated damping of the spread.

5. We found a strong positive effect of accompanying magnetic field on the relaxation process when cooling by a flow of electrostatically accelerated electrons. Smallness of the electron longitudinal temperature and magnetization of the transverse motion lead to a sharp acceleration of cooling in the region $v < v_{e\perp}$ and to the possibility of cooling a heavy particle beam to temperatures that are many times lower than the temperature of the electron gun cathode.

6. We obtained an integral of collisions of a heavy particle with magnetized electrons for an arbitrary relation between the average Larmor radius and the screening distance of Coulomb interaction. For the first time, a generalization was made including the transitional regime of electron collective response (non-stationary screening). This generalization is significant when one considers motion of a fast particle in a low-temperature plasma ($v > \Delta_{e\parallel}$) and is especially important in this work due to magnetization of the transverse motion and ultimate smallness of the electron longitudinal velocity spread. The true screening radius is determined by the particle velocity and is large compared to the Debye one ($r_{scr} \simeq r_D \cdot v/\Delta_{e\parallel}$), which usually plays the role of the maximum impact parameter in the theory of plasma collisional kinetics.

7. We identified and investigated two regions of characteristic behavior of the friction force as a function of the heavy particle velocity and electron flow parameters: the region of far adiabatic collisions with electron Larmor circles and the region of multipole cyclic collisions with magnetized electrons.

8. We investigated relaxation processes in the conditions of ultimate smallness of the relative velocities of heavy particles and Larmor circles when the perturbation theory is not applicable. We obtained estimates of maximum cooling decrements and minimum equilibrium temperatures achievable with cooling in magnetized electron flow. It was shown that these parameters are determined by the density of the flow, the size of the electron Larmor radii, and the length of the beam interaction section.

9. We considered the cooling process including spatial non-uniformity and non-stationarity of a magnetized electron flow. We studied the limiting effect of such factors as deviation of magnetic field lines, drift of Larmor circles, gradients and instability of electric potential, transverse gradient of electron Larmor velocities. We formulated requirements towards the quality of implementation and control of the cooling system parameters. When the requirements are met, the effect of the indicated factors becomes negligible.

10. We considered limitations coming from space charge of the cooled beam. We investigated beam stability in case of its coherent interaction with the electron flow.

11. We studied collisional kinetics of a heavy particle beam in a storage ring. We found a critical dependence of internal relaxation on the parameter $d\omega/dp$ characterizing the dependence of the circulation frequency on the particle energy in the storage ring. When $d\omega/dp > 0$, mutual scattering of the particles leads to thermalization without significant increase in the thermal energy of the beam while, in the opposite case, it leads to self-heating of the beam.

12. We investigated cooling of an intense beam including the processes of internal scattering. We made estimates of the cooling decrements and equilibrium temperatures depending on the number of particles, the energy and focusing rigidity of a storage ring. The effects of internal scattering must be taken into account when choosing an optimal regime of cooling and storage of heavy particles.

The general result of the completed investigation is the conclusion of high efficiency of electron cooling. The theoretical and experimental studies not only confirmed the initial predictions but also revealed new positive properties of the method significantly enhancing its capabilities.

# APPENDIX 1. ACTION-PHASE VARIABLES IN STORAGE RING

Let us expand the Hamiltonian of a particle moving near an equilibrium orbit $\vec{p} = \vec{p}_s(\vec{r}_s)$ into a power series in $\Delta \vec{p}/p_s$:

$$h(\hat{p}, \hat{q}) = \sqrt{\vec{p}^2 + m^2} + e\varphi = \mathcal{E}_s + \beta_s \Delta p + \frac{p_\perp^2}{2\mathcal{E}_s}\left(1 - \frac{\Delta \mathcal{E}}{\mathcal{E}_s}\right) + \frac{(\Delta p)^2}{2\gamma^2 \mathcal{E}_s} + e\varphi + \ldots$$

Conversion to generalized momenta is done, as well known, using the transformation $\vec{p} = \vec{\nabla} S - e\vec{A}$ where, by definition, $\hat{p}_i = \partial S/\partial \hat{q}_i$. Let us represent the potential $\vec{A}$ as $\vec{A} = \vec{A}_{id} + \tilde{\vec{A}}$ where $\vec{A}_{id}$ is the potential of an "ideal" part of the magnetic field while $\tilde{\vec{A}}$ includes various perturbations including the potential of radio-frequency (RF) field. Then $h(\hat{p}, \hat{q})$ can be written as $h = h_0 + \tilde{h}$ where $h_0$ corresponds to the unperturbed motion described by linear equations while the perturbation $\tilde{h}$ is

$$\tilde{h} = -\beta_s \left[ e\tilde{A}_y + \frac{\hat{\vec{p}}_\perp}{p_s e \vec{A}_\perp} + \frac{p_\perp^2}{2p_s} \frac{\Delta \mathcal{E}}{\mathcal{E}_s} + \ldots \right] + e\varphi .$$

The ideal solution is well known [21]:

$$r = r_b + r_s; \quad r_b = Re(A_r \tilde{f}_r); \quad r_s = R\frac{\Delta p}{p_s}\psi(\theta);$$

$$z = Re(A_z \tilde{f}_z); \quad p_z = \frac{p_s}{R} z'; \quad p_r = \frac{p_s}{R} r'; \quad \Delta p = const; \quad \text{(A.1)}$$

$$\theta = \omega_s t + \tilde{\theta} + \varphi_s; \quad \tilde{\theta} = \psi\frac{r_b'}{R} - \psi'\frac{r_b}{R} .$$

Here $f_z = \tilde{f}_z \exp(-i\psi_z)$ and $f_r = \tilde{f}_r \exp(-i\psi_r)$ are the Floquet functions, $\psi_z$ and $\psi_r$ are the betatron oscillation phases: $\psi_z' = \nu_z$, $\psi_r' = \nu_r$, and $\psi(\theta)$ is the "psi function", which is the forced solution of the equation (we use unitless time $\theta_s = \omega_s t$):

$$\psi'' + (1 - n)\psi = R/R(\theta) .$$

$\varphi_s$ is the synchrotron motion phase:

$$\varphi_s' = -\left(\frac{R}{R(\theta)}\psi - \frac{1}{\gamma^2}\right)\frac{\Delta p}{p_s} .$$

The constants $A_z$ and $A_r$ specify the complex amplitudes of the betatron oscillations.

The expressions in Eq. (A.1) can be considered as a canonical transformation

$$\begin{Bmatrix} \hat{p}_r, & \hat{p}_y, & \hat{p}_z \\ r, & \theta, & z \end{Bmatrix} \rightarrow \begin{Bmatrix} I_r, & I_y, & I_z \\ \psi_r, & \varphi, & \psi_z \end{Bmatrix},$$

where

$$I_{r,z} = \frac{1}{2}\frac{p_s}{R}|C_{r,z}|^2, \quad I_y = \hat{p}_y = R\Delta p$$

with the generating function

$$\mathcal{F} = S_0(I, \hat{q}, t) + (\nu_z I_z + \nu_r I_r)\theta_s + \frac{I_y^2}{2} \int \frac{dt}{\mu(\theta)}.$$

The new Hamiltonian is

$$\mathcal{H} = h + \frac{\partial}{\partial t}\mathcal{F} = (\nu_z I_z + \nu_r I_r)\omega_s + \frac{I_y^2}{2\mu(\theta)} + \tilde{h} \quad \left(h_0 + \frac{\partial S_0}{\partial t} \equiv 0\right).$$

The "mass" coefficient $\mu^{-1}(\theta)$ and the potential $\vec{A}_{RF}$ can be replaced in advance with their values overaged over the trajectory

$$\mu_s^{-1} = \overline{\mu^{-1}(\theta)} = -\frac{1}{\mathcal{E}_s \beta_s^2}\left(R\overline{R^{-1}(\theta)} - \frac{1}{\gamma^2}\right); \quad -\beta_s e\bar{A}_{RF} = \frac{\mu_s}{2}\Omega_s^2 \varphi^2 .$$

# APPENDIX 2. THEOREM ABOUT SUM OF FRICTION DECREMENTS

Suppose $I_\nu$ and $\psi_\nu$ ($\nu = 1, 2, 3$) are the action-phase variables of a particle moving in external electro-magnetic field: $I_\nu = const$ and $\dot\psi_\nu = \omega_\nu(I) = const$. They are related to the generalized momentum $\vec{\mathcal{P}} = \vec{p} + e\vec{A}(\vec{r},t)/c$ and the coordinate $\vec{r}$ by a certain canonical transformation. Under the action of friction $\vec{F}(\vec{p},\vec{r})$, the variables $I_\nu$ are slowly changing:

$$\dot I_\nu = \frac{\partial I_\nu}{\partial \vec{p}}\vec{F} = \frac{\partial I_\nu}{\partial \vec{\mathcal{P}}}\vec{F}.$$

Let us determine the sum of the decrements as

$$\Lambda = -\frac{1}{2}\frac{\partial}{\partial I_\nu}\langle \dot I_\nu\rangle,$$

where the brackets $\langle\ \rangle$ indicate averaging over the phases. Taking the averaging into account, the quantity $\Lambda$ can be written as

$$\Lambda = -\frac{1}{2}\langle\left(\frac{\partial}{\partial I_\nu}\frac{\partial I_\nu}{\partial \vec{\mathcal{P}}} + \frac{\partial}{\partial \psi_\nu}\frac{\partial \psi_\nu}{\partial \vec{\mathcal{P}}}\right)\vec{F}\rangle. \qquad (A.2.1)$$

Using the canonical relations

$$\frac{\partial I_\nu}{\partial \vec{\mathcal{P}}} = \frac{\partial \vec{r}}{\partial \psi_\nu},\quad \frac{\partial \psi_\nu}{\partial \vec{\mathcal{P}}} = -\frac{\partial \vec{r}}{\partial I_\nu},$$

let us transform Eq. (A.2.1) to

$$\Lambda = -\frac{1}{2}\langle\left(\frac{\partial}{\partial I_\nu}\frac{\partial \vec{r}}{\partial \psi_\nu} - \frac{\partial}{\partial \psi_\nu}\frac{\partial \vec{r}}{\partial I_\nu}\right)\vec{F}\rangle = -\frac{1}{2}\langle\frac{\partial \vec{r}}{\partial \psi_\nu}\frac{\partial \vec{F}}{\partial I_\nu} - \frac{\partial \vec{r}}{\partial I_\nu}\frac{\partial \vec{F}}{\partial \psi_\nu}\rangle = -\frac{1}{2}\langle\{\vec{F},\vec{r}\}\rangle.$$

Thus, the sum of the decrements can be expressed through the Poisson brackets of the friction force and the particle's radius vector; writing it explicitly in terms of the variables $\vec{\mathcal{P}}$ and $\vec{r}$, we get

$$\Lambda = -\frac{1}{2}\langle\partial \vec{F}(\vec{p},\vec{r})/\partial \vec{p}\rangle, \qquad (A.2.2)$$

Q.E.D.

In situations, when the longitudinal (with respect to the ion closed orbit) friction force does not depend on the transverse coordinates $x$ and $z$, the longitudinal decrement obviously equals $-\langle\partial F_\parallel/\partial p_\parallel\rangle/2$; Eq. (A.2.2) then gives that, with other conditions being arbitrary, the sum of the betatron oscillation decrements equals

$$\lambda_1 + \lambda_2 = -\frac{1}{2}\langle\partial \vec{F}_\perp/\partial \vec{p}_\perp\rangle,$$

# APPENDIX 3. FRICTION POWER IN CASE OF MONOCHROMATIC INSTABILITY

Let us find the friction power averaged over an oscillation period when there is a "mismatch" of the proton and electron average velocities. Let us take the dependence of the friction force on the velocity as

$$\vec{F}(\vec{v}) = -\frac{g}{m} \frac{\vec{v} - \vec{\Delta}}{\left[(\vec{v} - \vec{\Delta})^2 + v_{eT}^2\right]^{3/2}}.$$

Let us consider a situation, when the mismatch $\vec{\Delta}$ is directed along one of the three normal (or spatially-orthogonal) degrees of freedom. Then, to study the equilibrium state of the proton beam, we can assume that the proton velocity transverse to $\Delta$ is small compared to $v_T$:

$$v_\perp \ll v_{eT}$$

and, for the degree of freedom along $\Delta$,

$$\langle \frac{d}{dt} \frac{1}{2} M a^2 \rangle = -\frac{g}{m} \frac{1}{\pi} \int_{-a}^{a} \frac{v - \Delta}{[(v - \Delta)^2 + v_{eT}^2]^{3/2}} \frac{v dv}{\sqrt{a^2 - v^2}} \equiv -\frac{g}{m\pi} \mathcal{J} \quad \text{(A.3.1)}$$

We are interested in a situation when

$$|a - \Delta| \ll \Delta.$$

Considering this condition, one can replace the integral in Eq. (A.3.1) with an approximate one:

$$\mathcal{J} = \sqrt{\frac{\Delta}{2}} \int_{-\infty}^{a-\Delta} \frac{-x}{(x^2 + v_{eT}^2)^{3/2}} \frac{dx}{\sqrt{a - \Delta - x}}.$$

Let us introcude a notation $\zeta = (a - \Delta)/v_{eT}$ and rewrite $\mathcal{J}$ as

$$\mathcal{J} = -\sqrt{\frac{\Delta}{2v_T^3}} \int_{-\infty}^{\zeta} \frac{x dx}{(x^2 + 1)^{3/2} \sqrt{\zeta - x}}.$$

For the final estimate of the integral, it is convenient to transform it to the form

$$\mathcal{J} = \sqrt{\frac{\Delta}{2v_T^3}} \left[ \int_{\zeta}^{\infty} \frac{x dx}{(x^2 + 1)^{3/2} \sqrt{\zeta - x}} - 2 \int_{0}^{\zeta} \frac{x^2 dx}{(x^2 + 1)^{3/2} \sqrt{\zeta^2 - x^2}(\sqrt{\zeta + x} + \sqrt{\zeta - x})} \right].$$

Obviously, when $\zeta = 0$, $\mathcal{J} \sim 1$, while, when $\zeta \gg 1$,

$$\mathcal{J} \sim \sqrt{\frac{\Delta}{2 v_T^3}} \left( \frac{1}{\sqrt{\zeta}} \int_\zeta^\infty \frac{dx}{x^2} - \frac{1}{\sqrt{\zeta}} \int_1^\zeta \frac{dx}{x\zeta} \right) \sim - \sqrt{\frac{\Delta}{(a-\Delta)^3}} \ln \zeta < 0.$$

At the same time, the maximum values of each of the integrals lie in the region of $\zeta \sim 1$ and have orders of magnitude also equal to 1. This mean that the power $Q$ in this region changes within the limits of $\sim \pm (g/m)\sqrt{\Delta/v_T^3}$ and its derivative at the point $Q(\zeta) = 0$ equals

$$\frac{dQ}{da} \sim -\frac{|Q|_{max}}{v_T} \sim -\frac{g}{m v_T^2} \sqrt{\frac{\Delta}{v_T}}.$$

# APPENDIX 4. DERIVATION OF BELYAEV RELATION FOR KINETIC MOMENTS

To prove Eq. (1.36), let us provide a derivation of the general relation for kinetic moments contained in Ref. [51]. Suppose $C_\nu$ is a set of canonical integrals of motion of the action and phases of two particles in an external field and $V(C, t)$ is the interaction between the particles. The change in $C_\nu$ with time during collision can be written as the Poisson bracket

$$\dot{C}_\nu(t) = \{V; C_\nu\}_t \, .$$

Let us find $\dot{C}_\nu(t)$ as a function of the initial conditions at $t = 0$ to the 2nd order in $V$:

$$\dot{C}_\nu^{(2)} = \Delta C_{\nu'} \frac{\partial}{\partial C_{\nu'}} \{V; C_\nu\} = \{\tilde{V}; C_{\nu'}\} \frac{\partial}{\partial C_{\nu'}} \{V; C_\nu\} = \{\tilde{V}; \{V; C_\nu\}\} \, . \tag{A.4.1}$$

where

$$\tilde{V} = \int_0^t V(C, t') dt' \, , \qquad \Delta C_\nu = \{\tilde{V}, C_\nu\} \, .$$

Splitting Eq. (A.4.1) into the parts symmetric and antisymmetric in $V$ and $\tilde{V}$ and using the Jacobi identity, we get:

$$\dot{C}_\nu^{(2)} = \frac{1}{2} \frac{\partial}{\partial C_{\nu'}} \frac{d}{dt} \{\tilde{V}; C_\nu\} \{\tilde{V}; C_{\nu'}\} + \frac{1}{2} \{\{\tilde{V}; V\}; C_\nu\} \, .$$

We are interested in the rate of change of the action variables $I_\nu$ averaged over the initial phases $\psi_\nu$. In averaging, the second term that has the form of a Poisson bracket $I_\nu$ with a "potential" $\{\tilde{V}; V\}/2$, vanishes and we arrive at the relation:

$$\frac{d}{dt} \overline{\Delta I_\nu^{(2)}} = \frac{1}{2} \frac{\partial}{\partial I_{\nu'}} \frac{d}{dt} \overline{\Delta I_\nu \Delta I_{\nu'}} \, . \tag{A.4.2}$$

# ADDITION 1.
# On cooling by circulating beam

In the region of ultra-relativistic energies ($\gamma \gtrsim 10$), the option of cooling by single-pass electron flow becomes technically unfeasible and it is more sensible to switch to cooling by a beam circulating in an adjacent storage ring [13]. Electrons in the storage ring can be periodically replaced by injecting a new portion of electrons after they are heated up by heavy particles or can be cooled by radiation friction at sufficiently high energies. Currently, USA laboratories and CERN are already considering options for storage rings with a cooling beam for energies of hundreds of MeV for maintaining small sizes of proton and antiproton beams. Cooling by a circulating beam has a number of features differing it from the technique with direct electron acceleration.

1. Obtaining the need (or possible) cooling rate of a heavy particle beam is associated with such a necessarily present effect as internal scattering of electrons in the beam, which will either tend to thermalize the velocity distribution ($\gamma > \nu$) or lead to net heating ($\gamma > \nu$, Chapter V). The relaxation time $\tau_{ee}$ due to internal scattering is related to the (proton) cooling time $\tau$ in the following way:

$$\frac{\tau_{ee}}{\tau} \simeq \frac{m}{M} \eta \frac{1}{1 + \frac{\gamma^2}{\nu^2}}$$

(we assume that the angular spread $\theta \lesssim \theta_e$; in the opposite case, the situation can only get worse), i.e. $\tau_{ee}$ is, at least, 2000 times smaller than $\tau$ and, in practice, is even significantly smaller than that, since it is unlikely that the relative length of the interaction section $\eta$ can be made greater than 0.1. Therefore, even when operating below the transition energy ($\gamma < \nu$), $\tau_{ee}/\tau \lesssim 5 \cdot 10^{-5}$. This means that, with reasonable cooling times (of at least a few hours), internal relaxation of the electron beam can be the fastest of all processes and must be considered when choosing optimal conditions. Internal scattering plays an especially important (negative) role in the region of $\gamma \gg \nu$ expanding the beam, most significantly, in the radial direction. For this reason, it is possible that the only acceptable options will be those in the region of $\gamma < \nu$. In a general case, a complete theory of the electron beam state must include radiative processes (radiation and its quantum fluctuations), internal scattering and collisions with protons heating the beam. Internal scattering in the heavy particle beam must also be taken into account.

2. There may be significant effects of radial-longitudinal coupling especially if the electron (and proton) velocity distribution in the co-moving frame still turns out to be disk-like (for electrons, such a state is natural due to radiative effects). In a storage mode, the radial gradient of the longitudinal velocity (of the energy in the lab frame) leading to redistribution of the friction decrements is a natural (and controlled) factor and can have a detrimental as well useful effect (Sections 1.3 and 3.5).

3. There is an interesting question about the possibility of using the special properties of cooling by magnetized electrons with a small longitudinal spread in case of a circulating beam. The option with a longitudinal magnetic field can be practically realized only in the region of moderately relativistic energies ($\gamma \lesssim 20$), although, even here, one can encounter significant difficulties. The initial conditions must provide a small longitudinal spread, although one may hope to use the radiation effect as well. The magnetic field must be sufficiently strong (tens of kG) and have a significant length so that the electrons could complete a large number of Larmor cycles in the interaction region. An ideal case would be when the longitudinal magnetic field fills the whole orbit of the electron beam in order to suppress its thermalization (Section 2.6). Although implementation of such a system is apparently quite complex technically, there seem to be no conceptual obstacles.

4. The final feature of the cyclic option that we would like to note is that the finite character of particle motion (for heavy particles as well as electrons) manifests itself in the collisional interaction itself. Imagine two particles moving on adjacent (but different) orbits with close (or integer-multiple) circulation frequencies and frequencies of transverse oscillations. Due to the long-range action of Coulomb forces, it makes sense to consider particle collisions (approaches) not only at "microscopic" distances but also at distances comparable to the oscillation amplitudes (or the transverse beam sizes): since we know that the contribution of "large" distances diverges logarithmically. Besides, for such distances, with a sufficient "tune up" of the frequencies, a correlation between consecutive (from turn to turn) approaches will be preserved for a long time, i.e. a resonant interaction will take place. For two particles, a sufficiently long interaction time would simply lead to beating without irreversible changes in the state of motion. However, in the presence of a large number of other particles, the interaction (or correlation) time of two particles in a favorable phase becomes finite; this then results in an irreversible exchange but already amplified by the multiplicity of the correlated collisions. The criterion for transition from the dynamic mode of reversible beating to the stochastic walk and irreversible relaxation is the Chirikov criterion [23, 24, 26] that is well satisfied in practice. The limiting factors here are the widths of the frequency spreads (the probability of getting into a resonance is inversely proportional to the width of the spread) as well as effective dynamic screening that conceptually has the same nature as screening in plasma (see also Addition 2). The possibilities of realistic use of the effects of this kind are still not clear and concrete theoretical work must be done in this direction.

Recall that, in case of magnetization, we also deal with an effect of cyclic motion but only on a significantly less macroscopic scale and with an explicit limitation of the correlation time by the length of the interaction section (at least, in case of a single-pass beam). In contrast, in our case, interaction lasts "forever"; therefore, to validate the approach using the kinetic equations, from a theoretical point of view, one must involve extensive results of ergodic theory. In the area of beam dynamics, this question can today be considered solved in its substantial physical aspects.

# ADDITION 2.
# On capabilities of stochastic cooling

The technique of stochastic cooling proposed by S. van der Meer [48] is based on application of wide-band feed-back. The effect of cooling, or phase-space volume reduction, is conceptually single-particle (which is the only kind it can be) and appears due to the fact that the signal induced by a particle in a monitoring system is amplified and delayed as needed and then acts on the particle in a control element (kicker). Of course, the particle then also experiences the effect of amplified signals from neighboring particles and this mutual influence has an adverse effect limiting the maximum achievable decrement. Conceptually, this kind of incoherent particle interaction through the electron beam is present in electron cooling as well; however, the interaction area here is microscopically small and therefore there are no notable practical limitations. The capability of the cooling effect itself in the feed-back technique with a large number of particles (the average distance between the particles is small compared to the size of the interaction area) as well as its inherent conceptual limitations can be understood from a simple consideration of two particles closely-spaced in frequency with some small initial phase difference. The self-action and mutual effect experienced by the particles are signals of the same amplitudes but different frequencies. The distance between the particles (in the phase space) obviously cannot be reduced under the action of these signals before the time equal to the inverse frequency separation elapses. If this separation exceeds the frequency shift due to the interaction (equal to the self-action decrement) the mutual influence "quickly" averages out and damping on average occurs the same way as if there were no interaction between the particles (non-resonant interaction). However, in the opposite case, the interaction couples the particle motion further decreasing the average frequency difference and, with increase in the signal amplitudes, damping of the relative motion is weakened instead of getting stronger (only the damping rate of the coherent state component increases).

Let us provide a general description of the method's theory using the results of work in Refs. [49, 50] completed in collaboration with S.A. Heifetz.

Interaction between particles in the considered cooling technique occurs though an external amplifying system; thus, the particle system is not conservative. A consistent description of such a system can be based on a kinetic equation accounting for non-Hermeticity of the interaction $V(a,b) \neq V(b,a)$, $a,b = 1,2,...N$. In case, when one can neglect resonances (see below), it has the form [49]:

$$\frac{\partial f}{\partial t} + \Sigma \left(\vec{n}\frac{\partial}{\partial \vec{I}}\right) \frac{\mathrm{Im}\, V_{nm}(I,I)}{|\mathcal{E}_n|^2} f(I) = St, \qquad (1)$$

where $f(I,t)$ is the zero-integer harmonic of the distribution function, $V_{nm}(I_1\,I_2)$ is the phase harmonic of the interaction $V(1,2)$, and $\vec{I}$ is the action (the square of the amplitude) canonically conjugate to the oscillation phase $\vec{\psi}$. The quantity $\mathcal{E}_n$ given by

$$\mathcal{E}_n(q) = 1 + i\pi N \int d\Gamma_2 \delta_+[\vec{n}(\vec{\omega}_2 - \vec{\omega})] V_{nn}(I_2\,I_2)\vec{n}\frac{\partial f}{\partial \vec{I}_2} \tag{2}$$

plays a role analogous to that of the dielectric permittivity in plasma, $q = p - p_s$, and $p_s$ is the equilibrium momentum.

The right-hand side of Eq. (1) corresponds to the usual collisional term, which only gives redistribution of the decrements. The second term on the left-hand side of Eq. (1) related to non-Hermiticity of the interaction and determines the damping decrements. When $|\mathcal{E}_n| \simeq 1$, it describes the single-particle damping effect occurring due to the "self-action" of the particle. Difference of $|\mathcal{E}_n|$ from unit is related to mutual influence of the particles. This difference is significant when the decrement becomes of the order of frequency separation between effectively interacting particles. With a fixed amplification coefficient $\varkappa$, the decrement drops with increase in the number of particles $N$ as $N^{-2}$. At a fixed $N$, there is an optimal amplification coefficient and its corresponding ultimately achievable damping decrement.

**Damping of transverse oscillations**. It is easy to get from Eq. (1) that, with a time-constant momentum distribution, damping of the betatron oscillations happens exponentially in time with a decrement

$$\lambda_\perp = \frac{\varkappa e^2}{p_s \nu_\perp l_\perp^2}, \tag{3}$$

where $\nu_\perp$ is the betatron oscillation frequency and $l_\perp$ is the quantity determining the transverse field gradient in a pickup ($l_\perp$ is of the order of the pickup transverse size).

The limiting value of $\lambda_\perp$ is determined by the condition $|\mathcal{E}_n| \simeq 1$, which gives

$$\lambda_\perp < \lambda_{\perp max} = \frac{\Delta\omega}{\pi N \theta_0^2}, \tag{4}$$

where $\Delta\omega$ is the circulation frequency spread and $\theta_0$ is the parameter determined either by the azimuthal length where there is an effective interaction between the particles (i.e. the pickup length $\theta_0 = l_\parallel / R$) or by the operational bandwidth of the amplifier $n_0$ (so that the signal harmonics with frequencies $n\omega$ where $n > n_0$ are not amplified):

$$\theta_0 = \max\left(\frac{l_\parallel}{R}, n_0^{-1}\right).$$

The condition $\lambda_\perp < \lambda_{\perp max}$ has an order of magnitude consistent with the condition of stability of coherent oscillations.

The difference of Eq. (4) from the estimate by van der Meer (by a factor of $\omega\theta_0/\Delta\omega$) has to do with the fact that it takes into account that the phase correlations are maintained over many turns in contrast to van der Meer's assumption. However, with an optimal choice of the quantity $\theta_e$,

$$\theta_0 \sim \theta_{0min} = \frac{L}{R} \cdot \frac{\Delta\omega}{\omega}, \tag{5}$$

where $L$ is the distance from the pickup to the kicker, the difference is small. The pickup length $l_\parallel$ cannot be made smaller than given by Eq. (5); otherwise, a part of the particles will pass through the system without experiencing its effect.

**The decrement of longitudinal oscillations** is related to the excitation of a field by a non-equilibrium particle at the entrance and exit of the pickup

$$\lambda_p = \frac{\varkappa e^2 \psi_n}{\beta \, l_\perp^2}, \tag{6}$$

where $\psi_n$ is the value of the $\psi$ function at the pickup. Damping of the energy spread has a special feature that, with decrease in the frequency spread $\Delta\omega$, the mutual influence of the particles gets stronger ($\mathcal{E}$ grows). Therefore, damping does not behave exponentially. In particular, if the pickup's signal for the equilibrium particle is not zero, then damping leads to a distribution with a finite but different from zero equilibrium size. In the opposite case, the amplitude damps to zero. However, damping has exponential behavior only at a large initial frequency spread $\Delta\omega(0) \gg \Delta\omega_{cr} = N\lambda_p(\Delta\theta)^2$, $\Delta\theta = l_\perp/R$ and goes to $\Delta\omega_{cr}$. The rms momentum spread then decreases inversely proportionally to time. If the initial spread is small then damping goes inversely proportionally to time from the start. The value of the decrement $\lambda_p$ optimal for damping of the energy spread is

$$\lambda_{opt} = \frac{4\Delta\omega(0)}{\pi N(\Delta\theta)^2}. \tag{8}$$

The dependence of $q = p - p_s$ on time then has the form:

$$\langle q^2(t) \rangle = \langle q^2(0) \rangle \cdot \left[1 + \lambda_{opt} \cdot t\right]^{-1}. \tag{9}$$

The system damping the energy spread can have an effect on damping of the transverse motion. In an optimal case, systems damping the different degrees of freedom should be independent.

**The equilibrium beam size** is determined by several effects. First is the possibility of resonances between the particle frequencies 1 (generating the signal) and 2 (experiencing the action of the kicker) of the form $\bar{n}_1\bar{\omega}(1) = \bar{n}_2\bar{\omega}(2)$ with $\bar{n}_1 \neq \bar{n}_2$. Resonances with $|n_{1\perp}| = 1$ and $n_{2\perp} = 0$ result in a finite beam size:

$$\frac{\langle a_\perp^2 \rangle}{l_\perp^2} = \frac{\lambda_\perp}{\lambda_{\perp max}} \ln\left(\frac{\pi \Delta \omega}{\theta_0 \delta \omega}\right),$$

under the condition that the argument of the logarithm is greater than one. Here $\delta\omega$ is the detuning of the frequency $\omega_\perp$ from the nearest integer resonance. The effect decreases with reduction in $\Delta\omega$. Another way of dealing with this is the choice of a monitoring pickup such that it gives no signal for the equilibrium particle. Then $\langle a^2 \rangle$ is determined by the setting precision of such a pickup.

Another reason is thermodynamic fluctuations of the field in the monitoring pickup $\langle \mathcal{E}^2 \rangle \simeq 4\pi l_\perp^{-3} T$ where $T$ is the temperature of the pickup (or a matching element) in energy units. This leads to a beam size of the order of

$$\langle a_\perp^2 \rangle \simeq \varkappa \left(\frac{T\theta_0}{\nu_r E_s}\right) \cdot R^2 . \tag{10}$$

Finally, if there is significant intrinsic amplifier noise, it must also be considered.

**Optimization of the technique** comes down to reducing the mutual influence of the particles while preserving the self-action effect. To achieve this, it is desirable to reduce the longitudinal size of the pickup to the value of $\theta_{0min}$ (see Eq. (5)). Since $l_\parallel \gtrsim l_\perp$, this may require reduction of the transverse size of the pickup $l_\perp$, which imposes a limitation on the size of the $\psi$ function at the pickup (on the beam size) $\psi_n < L\langle\psi\rangle/R$. At the same time, one must appropriately reduce the $\beta$ function at the pickup.

The effect of field fluctuations can be reduced by increasing the number $n$ of the pickup-kicker pairs on the orbit and correspondingly reducing the amplification coefficient of each pair. Since phase correlations are preserved over many turns, the whole system works as a single pair with an efficient amplification coefficient of $n\varkappa$. Therefore, with $n\varkappa = Const$, the maximum achievable decrement $\lambda_{\perp max}$ does not change but the effect of field fluctuations given by Eq. (10) is reduced by a factor of $n$. It is implied here that the signal from a pickup is transmitted only to its matching kicker; otherwise, mutual influence of the particle will increase.

Note also that, if cooling of the transverse oscillations occurs with $\lambda_{max} \sim N^{-1}$, the accumulation time then does not depend on whether the beam is cooled as a whole or in separate portions if the total number of cooled particles is the same in both approaches.

The problem of damping of the complete phase-space volume can be solved in two ways. In the first approach, one starts by damping the betatron oscillations with $\lambda_{max} \sim \Delta\omega(0)$ and then damps the energy spread according to Eq. (9). In the second approach, cooling occurs in parallel in all degrees of freedom and the amplification coefficient of the system damping the betatron oscillations is adjusted according to change in $\Delta\omega(t)$, so that $\lambda_\perp$ remains optimal. The damping time is generally speaking the same in both approaches.

It is unlikely that the damping times of the considered technique can reach the record values already obtained in the electron cooling technique. However, van der Meer's technique may have advantages for a large initial velocity spread. The use of which technique is more appropriate should depend on the considered task. In a number of cases, it may be optimal to use stochastic cooling at the initial damping stage with a subsequent switch to electron cooling.